\documentclass[11pt,oneside,final]{huthesis}

\usepackage{epsfig,bm,epsf,float}
\usepackage{amssymb,amsmath,amsfonts}
\usepackage{multirow}
\usepackage{subfigure}
\usepackage{placeins}
\usepackage{color}
\usepackage{graphicx}

\newcommand{\mybold}[1]{{\mbox{\bf\boldmath ${#1}$}}}
\newcommand{\mybolds}[1]{\mbox{\tiny \bf\boldmath ${#1}$}}		
\newcommand{\myboldm}[1]{\mbox{\scriptsize \bf\boldmath ${#1}$}}		
\newcommand{\hodge}[1]{{{}^*{#1}}}

\newcommand{\bfr}{{\mybold{r}}}

\newcommand{\bfR}{{\mybold{R}}}

\newcommand{\bfF}{{\mybold{F}}}

\newcommand{\bfv}{{\mybold{v}}}

\newcommand{\bfw}{{\mybold{w}}}

\newcommand{\bfz}{{\mybold{z}}}

\newcommand{\bfx}{{\mybold{x}}}

\newcommand{\bfX}{{\mybold{X}}}
\newcommand{\bfXs}{{\mybolds{X}}}
\newcommand{\bfA}{{\mybold{A}}}

\newcommand{\bfnabla}{{\mybold{\nabla}}}

\newcommand{\bfomega}{{\mybold{\omega}}}

\newcommand{\bfxi}{{\mybold{\xi}}}

\newcommand{\bfalpha}{{\mybold{\alpha}}}

\newcommand{\bfbeta}{{\mybold{\beta}}}

\newcommand{\bfrho}{{\mybold{\rho}}}
\newcommand{\bfrhos}{{\mybolds{\rho}}}
\newcommand{\bff}{{\mybold{f}}}

\newcommand{\bfp}{{\mybold{p}}}

\newcommand{\bfJ}{{\mybold{J}}}

\newcommand{\bfs}{{\mybold{s}}}

\newcommand{\bfS}{{\mybold{S}}}

\newcommand{\bfimath}{{\mybold{\imath}}}

\newcommand{\mybar}[1]{{\overline{{#1}}}}
\newcommand{\bracketnewln}[1]{\right.\\#1&\left.{}}
\newcommand{\sgn}{\operatorname{sgn}}

\newcommand{\leftexp}[2]{{\vphantom{#2}}^{#1}{#2}}	

\newcommand{\myman}{\mathcal{M}^{2N}}
\newcommand{\mymans}{\mathcal{M}_s}			
\newcommand{\mymanL}{\mathcal{M}_L}			
\newcommand{\mymanLa}{\mathcal{M}_{L^{\ast}}}			
\newcommand{\mymanH}{\mathcal{M}_{H}}

\newcommand{\lrp}[1]{\left({#1}\right)}	
\newcommand{\lrbk}[1]{\left\{{#1}\right\}}	
\newcommand{\lrbc}[1]{\left[{#1}\right]}	
\newcommand{\lrang}[1]{\langle{#1}\rangle}	

\newcommand{\intprod}[2]{\bfimath_{#1}{#2}}
\newcommand{\extprod}[2]{{#1}\wedge{#2}}
\newcommand{\Lieder}[2]{\mathcal{L}_{#1}{#2}}

\newcommand{\pten}{{\mybold{\mathbb{J}}}}	
\newcommand{\ptenm}{{\myboldm{\mathbb{J}}}}	
\newcommand{\HamVec}[1]{\pten d{#1}}		
\newcommand{\HamVecm}[1]{\ptenm d{#1}}		

\newcommand{\deriv}[2]{\frac{d#1}{d#2}}
\newcommand{\derivc}[3]{\left. \frac{d#1}{d#2}\right|_{#3}}
\newcommand{\pd}[2]{\frac{\partial #1}{\partial #2}}

\def\be{\begin{equation}}
\def\ee{\end{equation}}
\def\bea{\begin{eqnarray}}
\def\eea{\end{eqnarray}}
\def\bal{\begin{align}}
\def\ealn{\end{align}}
\def\bean{\begin{mathletters}\begin{eqnarray}}
\def\eean{\end{eqnarray}\end{mathletters}}

\newcommand{\tbox}[1]{\mbox{\tiny #1}}

\def\text{\tbox}

\def\Es2{E_{0,{\rm sat}}^2}

\newcounter{eqletter}
\def\mathletters{%
\setcounter{eqletter}{0}%
\addtocounter{equation}{1}
\edef\curreqno{\arabic{equation}}
\edef\@currentlabel{\theequation}
\def\theequation{%
\addtocounter{eqletter}{1}\thechapter.\curreqno\alph{eqletter}%
}%
}

\begin{document}


\dsp



\title{Point Vortices: Finding Periodic Orbits and their Topological Classification}
\author{Spencer Ambrose Smith}
\degreemonth{August} 
\degreeyear{2012}
\degree{Doctor of Philosophy}
\field{Physics}
\department{Physics}
\advisor{Prof. Bruce Boghosian} 

\maketitle

\dedication
\begin{quote}
\hsp
\em
\raggedleft
Dedicated to my parents Jeff and Sandy,\\
who sparked my interest in science and beauty.\\
And to Liza for her love, encouragement, and support.
\end{quote}

\begin{figure}[htbp]
 \centering
\large
\resizebox{\linewidth}{!}{\includegraphics{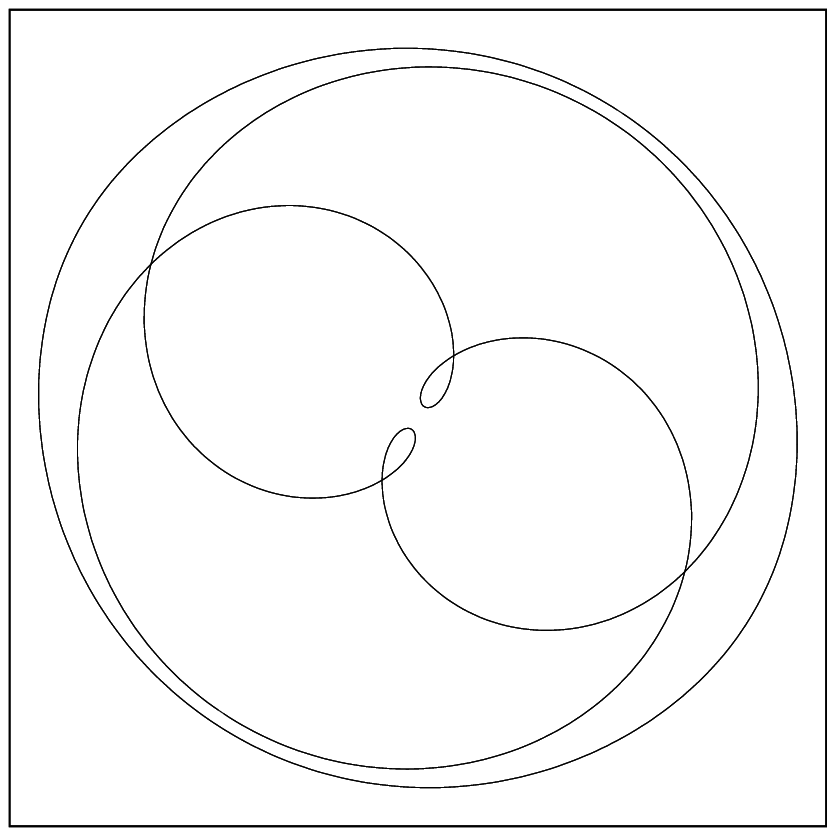}}
\end{figure}
\newpage

\thispagestyle{plain}
\begin{center}
    \textbf{Point Vortices: Finding Periodic Orbits and their Topological Classification}
    
    \vspace{0.5cm}
    \LARGE
    \textbf{Abstract}
\end{center}
The motion of point vortices constitutes an especially simple class of solutions to Euler's equation for two dimensional, inviscid, incompressible, and irrotational fluids.  In addition to  their intrinsic mathematical importance, these solutions are also physically relevant.  Rotating superfluid helium can support rectilinear quantized line vortices, which in certain regimes are accurately modeled by point vortices.  Depending on the number of vortices, it is possible to have either regular integrable motion or chaotic motion.  Thus, the point vortex model is one of the simplest and most tractable fluid models which exhibits some of the attributes of weak turbulence.

The primary aim of this work is to find and classify periodic orbits, a special class of solutions to the point vortex problem.  To achieve this goal, we introduce a number of algorithms: Lie transforms which ensure that the equations of motion are accurately solved; constrained optimization which reduces close return orbits to true periodic orbits; object-oriented representations of the braid group which allow for the topological comparison of periodic orbits.  By applying these ideas, we accumulate a large data set of periodic orbits and their associated attributes.  To render this set tractable, we introduce a topological classification scheme based on a natural decomposition of mapping classes.  Finally, we consider some of the intriguing patterns which emerge in the distribution of periodic orbits in phase space.  Perhaps the most enduring theme which arises from this investigation is the interplay between topology and geometry.  The topological properties of a periodic orbit will often force it to have certain geometric properties.


\newpage
\addcontentsline{toc}{section}{Table of Contents}
\tableofcontents

\listoftables
\listoffigures

\startarabicpagination

\chapter{Introduction and Summary}
\label{ch:intro}

\section{Point of View and Main Ideas}

		Despite its relative simplicity, the 2D point-vortex model has appeared time and again in the description of a wide variety of physical systems.  Part of this breadth of application is attributable to its long history starting with Helmholtz in 1858~\cite{bib:Aref2}, who considered point-like distributions of vorticity imbedded in a 2D ideal, incompressible fluid.  Indeed, the velocity field generated by the Hamiltonian motion of a collection of point vortices satisfies the Euler fluid equation~\cite{bib:saffman}.  This represents a remarkable conceptual simplification: trading the infinite dimensional ideal fluid field equations for the finite dimensional coupled ordinary differential equations of the point-vortex model.  It is completely natural then to ask: to what extent are various general phenomena in 2D hydrodynamics present in the point-vortex model?

	Onsager was one of the first to tackle this question when he used ideas from equilibrium statistical mechanics to show the existence of negative temperature states~\cite{bib:EyinkSpohn}; states that correspond to large-scale, long-lived vortex structures, not unlike those that form in Earth's atmosphere.  Further research in the statistical vein has resulted in a kinetic theory of point vortices~\cite{bib:Chavanis}, while consideration of structure formation dovetails nicely with the concept of Lagrangian coherent structures~\cite{bib:HallerYuan}~\cite{bib:Provenzale}.  There is also the view that point vortices are a useful toy model of 2D inviscid turbulence~\cite{bib:Babiano}~\cite{bib:Tabeling}~\cite{bib:Benzi}, an idea motivated by the non-integrable, i.e., chaotic, motion of four or more vortices.  Closely related to this is the more rigorous concept of chaotic advection~\cite{bib:Aref1}, which can help explain the transport properties of a vortex dominated ideal fluid~\cite{bib:Leoncini}.  Clearly a wide variety of physical phenomena falls under the aegis of point-vortex dynamics.
	
	In addition to a diverse spectrum of phenomena in ideal fluids, the point-vortex model is also applicable to other, more exotic, physical systems.   This includes 2D electron plasmas~\cite{bib:Driscoll1} and Bose-Einstein Condensates (BECs).  Indeed, the Gross-Pitaevskii equation, which governs the evolution of the BEC wave-function, can be re-expressed, via the Madelung transformation, as Euler's equations~\cite{bib:JerrardSpirn}.  Therefore, quantized vortex defects in a rotating BEC~\cite{bib:Fetter}~\cite{bib:WeilerNeely} interact, to first approximation, as if they are point vortices.  This also applies to other superfluids such as He-II, the main physical motivation of this work (see chapter~\ref{ch:phys}), where the vortex circulations are still quantized and the size of the vortex core is small enough to really warrant approximation by point vortices.  Since superfluid turbulence is dictated by 3D quantized line vortices~\cite{bib:Vinen}, the point-vortex model can be considered a toy model of quantum turbulence as well.
	
	Aside from physical instantiations, the point-vortex model should engender some intrinsic interest simply as a mathematical entity.  It has been described as a ``mathematical playground"~\cite{bib:Aref2}, touching upon areas such as the theory of dynamical systems, ordinary differential equations (ODEs), and Hamiltonian dynamics, whose appearance might be expected, as well as some ideas that at first seem to have no connection.  For example, it is neither readily apparent that equilibrium configurations of point vortices can be connected to the roots of certain polynomials~\cite{bib:Aref3}, nor is it immediate that one can apply topology in the guise of Thurston-Nielsen and braid theory to describe fluid mixing~\cite{bib:BoylandStremler}.  A plethora of physical and mathematical considerations give weight to the notion that the point-vortex system is an important item of study despite its relative simplicity.

	In light of these myriad physical and mathematical justifications for exploring point vortices, we aim to answer some very basic questions about their dynamics.  What types of solutions are allowed by the equations of motion?  In particular, we only consider periodic orbits, solutions which repeat after a finite time.  From the point of view of dynamic systems theory, periodic orbits form a very special set of solutions~\cite{bib:Katok}.  For hyperbolic systems, all trajectories can be decomposed and approximated by a set of periodic orbits.  Indeed, for such systems all times averages along trajectories can be replaced by weighted averages over certain attributes of periodic orbits.  In this view, periodic orbits form the skeletal structure of phase space and dictate the chaotic behavior of trajectories.  While point vortex dynamics are far from being hyperbolic, periodic orbits still order phase space and represent the most concrete way in which we can explore the qualitative types of point vortex motion.  In the specific context of point vortex dynamics, periodic orbits gain an additional useful attribute.  Loops in the phase space of point vortices can be classified topologically.  We will develop a particularly useful topological invariant related to the complexity of the braided structure of each periodic orbit.  With these ideas in mind, we consider the main goal of this work to be finding periodic orbits of the point vortex system and classifying them.


\section{Structure of Dissertation}

Due to the complexity of some of the ideas in this work, the first three chapters are devoted primarily to background material.  These are chosen to provide the physical and mathematical motivation for exploring solutions to the point vortex model, as well as introduce the prerequisites for framing and understanding later discussions.  The only possibly novel result from this part of the dissertation comes in section~(\ref{sec:TNbraidtree}), where we apply Thurston-Nielsen classification ideas to create the TN braid tree conjugacy invariant.  This idea and the related $T_1$ and $T_2$ classification schemes will constitute the most fruitful central concepts of this work, and will pave the way for finding, storing, and classifying periodic orbits as well as explaining their qualitative features.  The next chapter deals with the concepts behind the algorithms which enable us to accurately solve the equations of motion, extract periodic orbits, and compare them.  Many of these ideas are new and certainly constitute the bulk of the work in this dissertation.  At this point we have achieved our two main goals of finding and classifying periodic orbits of point vortex dynamics.  However, the large set of accrued periodic orbits necessitates some analysis, which we present in the subsequent chapter.  In addition to validating the general classification scheme, we present some interesting additional ideas which characterize individual periodic orbits and we point out various intriguing patterns in the distribution of periodic orbits in phase space.  This section also makes it apparent that there are plenty of further directions to pursue.  In the final chapter we point out the main ideas that could constitute future work, as well as summarize the conclusions previously addressed.

\subsection{Chapter~\ref{ch:phys}: The Physical Motivation}

There is one particular physical system which should be kept in mind when considering the properties of solutions to the point vortex model: quantized rectilinear vortices in rotating superfluid Helium $\lrp{\leftexp{4}{He}}$.  At first glance, the point vortex model might seem to be too simple with too many potentially unphysical attributes.  For instance, the total energy of the system remains constant as there is no viscosity, the motion is purely in two dimensions, the vortices are infinitesimally small, and they are all identical.  In this chapter we will consider the properties of superfluid $\leftexp{4}{He}$ vortices, and why their dynamics are approximated nicely by the point vortex model.

Briefly, superfluids have zero viscosity by virtue of the $U\lrp{1}$ symmetry of the Bose Einstein condensate wave-function.  This symmetry also leads to super-flow about regions of physical space which are topologically similar to a torus.  For rotating superfluids this topological defect exists as a thin straight vortex filament.  Due to this rectilinear geometry, all of the pertinent motion for vortices occurs in a plane.  The topology also guarantees that the single characteristic attribute of line vortices, the strength of the superfluid circulation about their cores, can only attain specific quantized values.  Indeed, energy arguments indicate that vortices with a single quantum of circulation are much more likely to exist than are multiply quantized vortices.  The lack of physical extent of a single point vortex also requires justification.  Empirical evidence shows that the width of vortex cores are many orders of magnitude smaller than the typical inter-vortex spacing.  This indicates that no geometric characteristic of the vortex core can appreciably affect the larger scale motion of the ensemble of vortices.  In aggregate, these considerations justify the approximation of superfluid vortices by identical point particles.  

What about the motion of point vortices?  Symmetry considerations dictate the velocity field of each vortex, and indicate that a single vortex simply moves with the local velocity field due to all other vortices.  One additional constraint arises when we account for the effect of mutual friction between the superfluid and the normal fluid in the two fluid model.  While this does change the equations of motion to be driven and dissipative, we recover the conservative point vortex dynamics at absolute zero temperature.  Thus, within the bounds of some reasonable assumptions, we can associate the motion of identical point vortices with the motion of quantized vortices in rotating superfluid $\leftexp{4}{He}$ at absolute zero.

\subsection{Chapter~\ref{ch:pvm}: Mathematical Background}

	The point vortex model has been around for a long time, and consequently has connections to a large number of disparate areas of mathematics.  In this chapter we introduce the point vortex model and consider the mathematical tools and points of view which are particularly useful for describing and finding periodic orbits.  First, we will show that the point vortex model is derivable from Euler's equation of inviscid fluid dynamics.  This recasts our search for periodic orbits as a search for special solutions to Euler's equation.  Indeed, many of the periodic orbits we have found have not been described before, and may be considered novel solutions to Euler's equation.
	
	Next, we consider the point vortex model as a key example of a Hamiltonian system, albeit an unusual one.  In standard Hamiltonian systems we have generalized coordinates paired with their conjugate momenta.  In point vortex dynamics, the x-position of a vortex is conjugate to its y-position.  There is no momentum as the vortices are mass-less and move purely by being advected in the velocity field due to all other vortices.  This means that the configuration space of this system is also its phase space; the positions of each vortex alone determine the future evolution of the whole system.  In a more mathematical light, the phase space is an example of a symplectic manifold that is not naturally the cotangent bundle of a lower dimensional manifold, as say the phase space of gravitating point particles is.
	
	The geometric view of periodic orbits tracing out loops in this phase space plays a crucial role in our subsequent understanding of point vortex motion.  To develop this point of view, we introduce mathematical ideas such as Poisson brackets, Lie derivatives, symplectic manifolds, as well as differential forms.  The utility of many of these ideas lies in a common and concise language to describe the geometric structure of phase space.  However, perhaps the most important reason for introducing these concepts is that they allow us to define the connection between chaotic or integrable motion and the existence of conserved quantities.  In particular, there are two independent conserved quantities in addition to the energy, which loosely correspond to linear and angular momentum.  They show that the motion of three vortices is integrable and that of four can be chaotic.  They also show that rigidly rotating solutions, relative equilibria, play a large role in partitioning phase space into qualitatively different regions.  We will end this chapter with a discussion of the motion of two, three, and four vortices.  The relative equilibria solutions for each case are described, and qualitative motion of general periodic orbits is discussed.

\subsection{Chapter~\ref{ch:braid}: Braid Theory and other useful Topological Ideas}

While the previous chapter dealt with periodic orbits from the geometric point of view, this chapter considers them from the topological point of view.  In phase space, periodic orbits form loops which can be topologically distinguished from each other.  A more direct way to view the topological differences lies in the representation of a periodic orbit as the direct product of the vortex configuration in the plane with an orthogonal direction corresponding to the time flow.  Each vortex forms a strand in time, and movement of vortices about each other corresponds to the twisting of these strands.  Thus, each periodic orbit corresponds to a geometric braid, which we can represent algebraically in terms of Artin braid generators.

Next, we consider the solution to two classic problems in braid theory, the word and conjugacy problems.  The first addresses when two braids, written as a string of algebraic generators, can be considered to be the same.  This is solved using a natural ordering of braids along with the usual braid relations.  More important is the conjugacy problem, which addresses when two braids are related by conjugation with a third braid.  Geometrically this corresponds to rotating the projection plane used to translate geometric braids to algebraic braids, and to a time translation.  Since neither of these actions should change the topological properties of phase space loops, the real topological quantity we want to investigate is the conjugacy classes of braids, referred to as braid types.

There are a couple of full solutions to the braid conjugacy problem, however we will choose to use a partial solution which has some properties that are useful in the context of vortices.  The basis for our classification of braid types starts with a classification of mapping class groups due to Thurston.  Since the mapping class group of the punctured disk corresponds to braid types, we can directly use this classification.  Essentially, this result divides braids into finite order braids which are particularly simple, pseudo-Anosov braids which are particularly complex, or a composite braids which are a well defined combination of the two.  We then draw some natural conclusions about how these classification ideas apply to our specific situation.  This results in the Thurston-Nielsen braid tree conjugacy invariant and two related classification schemes, which together form the backbone of the topological aspects of this work.  In particular, this classification will allow us to find, store, and distinguish periodic orbits, as well as draw some conclusions about the qualitative character of these solutions to the point vortex problem.

\subsection{Chapter~\ref{ch:alg}: Algorithmic Ideas}

In this chapter we discuss the ideas behind the algorithms which enable the extraction and classification of periodic orbits.  Before finding periodic orbits we must be able to accurately evolve the point vortex equations of motion.  The first section deals with our solution to one particular problem which arises when implementing such a numerical evolution.  When two vortices become very close to one another, we encounter a fundamental tradeoff between accuracy and run time of the numerical method.  We solve this by first transforming the vortex configuration to a related configuration where the close pair of vortices has been replaced by a single vortex of larger strength.  Then we use a near identity canonical transformation, generated using Lie perturbation theory, to rigorously equate the dynamics of the previous two systems.  Not only does this method work well, but it also turns out to be qualitatively related to many of the topological classification ideas of the previous chapter.

Next we consider our method for extracting periodic orbits.  This starts with a long regular orbit, from which we identify close return orbit segments.  These segments are then given to a relaxation algorithm, which converts them to true periodic orbits.  For this, we consider a cost functional on the space of all loops, which attains a global minimum only when the loop is a solution to the equations of motion.  The minimal values of the functional are then found via a version of Newtonian descent which is modified to account for the symmetries tied to the constants of motion.

The relaxation algorithm works well, and we have found millions of periodic orbits.  Unfortunately, with the question of existence firmly out of the way, this abundance of solutions begs the question of uniqueness.  We deal with this problem in two different ways.  First, we use the topological invariants of the previous chapter to compare orbits and decide whether they are of different braid types.  If two periodic orbits have the same topological properties, we can then continue by comparing them geometrically.  If we take rigid rotations, time translations, and vortex label permutations into account, we can then uniquely specify whether two periodic orbits are geometrically the same.  This reduces the number of unique periodic orbits to tens of thousands, a more manageable amount.  

\subsection{Chapter~\ref{ch:class}: The Classification of Periodic Orbits and Patterns in Parameter Space}

With this sizable set of periodic orbits in hand, we will turn our attention in this chapter to the different classes into which these orbits fall.  The first section outlines the various types of data that we collect from each periodic orbit.  Generally, the data is either geometric in origin such as period, angular momentum, and Floquet stability eigenvalues, or the data is topological like braid type and pseudo-Anosov expansion factor.  Since the data from a full periodic orbit, the vortex positions for each discrete time, is quite large, we instead use the geometric and topological data to compare and categorize orbits as well as describe interesting patterns which emerge in the distribution of orbits in parameter space.

Next we use this data to describe which types of braids are allowed by the point vortex dynamics.  Despite the fact that all our braids are pure (each vortex constitutes one strand), we will see that some orbits are built up of permutation orbits.  Since we are considering identical vortices, a change in vortex labeling (permutation) should not change a vortex configuration.  Permutation orbits simply return to the same vortex configuration, albeit one that permutes the vortex indices.  These periodic orbits are particularly symmetric and have some interesting properties.  Next, we consider the observed fact that all of our orbits are composed of positive braids.  These are braids which are exclusively composed of positive generators.  Perhaps more interesting is the division between braids with pseudo-Anosov (pA) components and those with just finite order components.  Surprisingly, there are no pA braids for the three vortex case.  Indeed pA braids are subtly linked to the divide between integrability and chaos, as their existence forces an orbit to have non-zero maximum Floquet exponents.   The final binary division between periodic orbits which we will consider concerns the potential chirality of orbits.  By performing a parity inversion and a time reversal, a periodic orbit is mapped to another orbit which is also a solution to the equations of motion.  However, this new orbit is either essentially the same as the old orbit and therefore non-chiral, or it is essentially distinct from the old orbit and therefore chiral.  We will see that non-chiral orbits always have a certain amount of symmetry.  While chiral orbits do not necessarily possess the same degree of symmetry, their existence will help to explain some features of the topology of phase space.

The next section is devoted to classification ideas which are not necessarily binary.  This includes the $T_2$ braid type classification scheme and the divisions imposed by the locations of relative equilibria.  The $T_2$ braid type classification played an important role in our ability to generate and store periodic orbits, however it really shows its utility by separating orbits into different classes which reflect qualitatively different vortex motions.  In the case of three vortices, there are two classes which evenly bisect a graph of angular momentum $\lrp{L}$ vs. period $\lrp{T}$.  All orbits in one class are below a certain angular momentum value, while all those in the second class are above this value.  In the case of four vortices, the $L$ vs. $T$ graph exhibits many different patterns which in aggregate form a confusing picture.  Only by considering these patterns for each of the eight $T_2$ classes can we make any sense of them.  Another classification idea involves the relative equilibria.  These are solutions where the vortex configuration rotates rigidly.  Interestingly, these orbits lie in phase space exactly at the boundary between two topologically distinct regions.  Thus, orbits with angular momentum values on either side of the angular momentum of one of these relative equilibria will have different behavior.  In the case of four vortices, the three relative equilibria split $L$ vs. $T$ graphs into three regions.  In two of these regions the periodic orbits exist in discrete families, while in the remaining region they exist in continuous families.  The overall picture that these observations and those of the other sections paint is that the topological properties of periodic orbits will often determine their geometric properties.

\subsection{Chapters~\ref{ch:conc}: Conclusions and Future Work}

This final chapter summarizes the ideas presented in this dissertation and the conclusions we can draw from finding and classifying periodic orbits.  It also briefly covers some of the areas in which we can make progress in the future.  In particular, we are interested in extending these results to the more realistic driven-dissipative case.  Which orbits will survive at finite temperatures, if any?  We would also like to expand the TN braid tree classification to be a full conjugacy invariant capable of answering the conjugacy problem.  This would also enable us to distinguish topologically chiral braids.  It would be interesting to tease out the connection between topological and geometric chirality.  Next, we would like to find periodic orbits for larger sets of vortices, and see what conclusions might hold in the limit of large numbers.  Finally, we would like to extend the analytical results related to the Lie transform algorithm to include the integrable motion of three or more vortices treated as a single vortex.

\chapter{Physical Background}
\label{ch:phys}


In this chapter we will focus exclusively on the connection between point vortices and a particular physical system, rotating superfluid Helium-4 (He-II).  An overview of this connection is as follows:  If we ignore for a moment the fact that the Helium atoms in this system are interacting, then this superfluid can be loosely described as a Bose-Einstein condensate, where a finite number of particles are in the same quantum ground state with a common macroscopic wavefunction.  Through some symmetry arguments and ideas from mean field theory, we can determine that the dynamics of this wave-function are dictated by the Gross-Pitaevskii equation.  This equation can then be interpreted hydrodynamically, with a velocity field that follows the Euler equation.  In a rotating superfluid, we find topological defects (rectilinear vortices) which are quantized in strength.  These vortices advect in the velocity field of their compatriots, and their motion can be described by the point vortex model.  However, we must take the interaction of Helium atoms into consideration.  This changes the picture slightly, and necessitates use of the two fluid model with mutual friction between the vortices and normal fluid.  Fortunately, we will recover conservative point vortex dynamics in the limit of small temperatures.

\section{Superfluid Helium}
\subsection{Bose-Einstein Condensates}	

Probably the most important attribute of Helium atoms is their spin-statistics.  In this case $\leftexp{4}{He}$ atoms have integer spin, and are therefore bosons.  This means that, unlike fermions, quantum mechanics does not prohibit a group of these atoms from inhabiting the same quantum state.  Indeed at a finite temperature, $T_c = 2.17 K$, a second order phase transition occurs, separating the normal liquid Helium-I phase from the superfluid Helium-II phase.  To first approximation, the superfluid state can be thought of as a Bose-Einstein condensate (BEC).  To be sure, unlike a true BEC gas of non-interacting particles, the $\leftexp{4}{He}$ atoms do have interactions that must be taken into account for an accurate picture of superfluid phenomena.  Fortunately, at low temperatures we can treat most of the effects due to interacting particles by only slightly modifying the phenomenological picture of rotating superfluids.  Much of this story can be found in~\cite{bib:Pismen}\cite{bib:Pethick}.

In this section, we will develop the dynamics of the superfluid state.  Our starting point will be the superfluid wavefunction, $\psi$.  The fundamental assumption that we are making is that there are an appreciable number of atoms in the ground state.  To justify this, consider the following outline of a simple analysis, found in most statistical mechanics texts~\cite{bib:Huang}\cite{bib:Feynman}\cite{bib:LandauSP}, of a number of bosons trapped in a box.  The boundary conditions necessitate quantized energy levels, from which one can define a grand canonical ensemble and all the thermodynamic variables.  The stipulation of a fixed total number of bosons allows the partition function to be written in a particularly simple form.  If one then tried to calculate any thermodynamic variable by approximating the energy spectrum as continuous, he will find that this is only applicable at temperatures above $T_c = \frac{2\pi \hbar^2}{m k}\lrp{\frac{\rho/ s}{\zeta_{3/2}\lrp{1}}}^{2/3}$.  Here $m$ is the mass of one $\leftexp{4}{He}$ atom, $k$ is Boltzmann's constant, $\rho$ is the number density, $s$ is the number of spin states per atom ($s = 1$ since $\leftexp{4}{He}$ has spin 0), and $\zeta_{3/2}\lrp{1} = 2.612$ where $\zeta_r\lrp{\alpha} = \sum^{\infty}_{n=1}\frac{\alpha^n}{n^r}$.  Below this temperature, the continuum approximation breaks down, which implies that there are an appreciable number of particles occupying a single energy level.  If we split off the term corresponding to the lowest energy level, we can then redo the analysis and find that the fraction of atoms in the lowest energy state is given by, $\langle n_0\rangle / \langle N \rangle = \lrp{1-\lrp{T/T_c}^{3/2}}$.   Thus, the number of $\leftexp{4}{He}$ atoms participating in the condensate are vanishingly small at the transition temperature, but grow to be a totality of atoms at near absolute zero.  This simple analysis does not account for the proper nature of the phase transition.  Phenomena, such as the divergence of the specific heat, are captured only in a very crude and qualitative manner by the BEC assumptions.  However, as we shall see, this is a good starting point on which to base the more accurate, though phenomenological, two fluid model of Landau.  We will now concentrate on what we can explain about the dynamics of the superfluid based on the macroscopic BEC wave-function.

\subsection{Mean Field Theory and the Gross-Pitaevskii Equation}

All of the $\leftexp{4}{He}$ atoms in the ground state share a single macroscopic wave function.  We will treat this condensate wave function $\psi$ as an order-parameter in a Landau-Onsager type mean field theory.  Order parameter is perhaps a misleading term, as $\psi$ is really a complex valued scalar field over our three dimensional fluid domain.  The essence of mean field theory is to specify an action functional
\be
S = \int  \mathcal{L} d\bfx dt,
\label{eq:actionfunctional}
\ee 


which has a Lagrangian density, $\mathcal{L}$, that is invariant to all the symmetries of our system.  This Lagrangian density itself is a function of the order parameter $\psi$, its complex conjugate $\overline{\psi}$, time derivatives $\partial_t \psi$,  and gradients of the of the order parameter $\nabla \psi$.  The symmetries that we want to take into consideration are Galilean coordinate transformations, $\bfx' = \bfx - \bfv t$ \& $t' = t$, and the $U\lrp{1}$ symmetry of the order parameter itself, $\psi' = e^{i \phi} \psi$.  The appropriate invariant kinetic and potential terms are: $\frac{i}{2}\lrp{\psi\partial_t \overline{\psi} - \overline{\psi}\partial_t \psi}$ and $\frac{1}{2}\lrp{1+\left| \psi \right|^2}^2$ respectively.  We will additionally add in a gradient term that will discourage spatial inhomogeneities, giving a Lagrangian density
\be
\mathcal{L} = \frac{i}{2}\lrp{\psi\partial_t \overline{\psi} - \overline{\psi}\partial_t \psi} + \nabla \psi \cdot \nabla \overline{\psi} + \frac{1}{2}\lrp{1-\left| \psi \right|^2}^2.
\label{eq:LagrangianDensity}
\ee

Varying $S$, i.e. requiring $\delta S = 0$, gives the dynamic equation for the order parameter field: $-i \partial_t \psi = \nabla^2\psi +\lrp{1 - \left| \psi \right|^2} \psi$.  A more rigorous derivation, using second quantized boson particle fields and the Hartree-Fock approximation, results in the same equation with physical constants reinstated
\be
i 2 \kappa^{-1} \pd{ \psi}{t} = \lrp{\nabla^2 + \ell^{-2}\lrp{1- \left| \psi \right|^2}} \psi.
\label{eq:GrossPitaevskii}
\ee
Here we have $\kappa = \hbar / m$, $\ell$ is the healing length, i.e. the characteristic distance over which perturbations in $\left| \psi \right|$ return to the bulk value, and it is assumed that any external potential is zero or constant.  This is the Gross-Pitaevskii equation, often referred to as the Nonlinear Schr\"{o}dinger equation in optics, which determines the evolution of the condensate wave function for a BEC.  

\subsection{Hydrodynamic Interpretation}	
	
The Gross-Pitaevskii equation, Eq.~(\ref{eq:GrossPitaevskii}), has a number of interesting solutions, including soliton solutions.  We are interested in those that have a hydrodynamic flavor.  To advance this view, consider a transformation that expresses the complex order parameter in terms of its modulus and angle, $\psi = \left| \psi \right| e^{i \theta}$.  Or to cut out much of the interpretational work later on, consider the related transformation
\be
\psi = \sqrt{\rho} e^{i\phi/\kappa}.
\label{eq:MadelungTransformation}
\ee
This is often called the Madelung transformation.  It will facilitate a natural interpretation of $\rho$ and $\phi$ in fluid dynamic terms.  If we write Eq.~(\ref{eq:GrossPitaevskii}) in terms of Eq.~(\ref{eq:MadelungTransformation}), divide out the common exponential term, and then group real and imaginary terms together, we get the equations
\be
\partial_t\rho + \nabla \cdot \lrp{\rho\nabla \phi} = 0,
\label{eq:GPcontinuity}
\ee
\be
\partial_t\phi + \frac{1}{2}\left|\nabla \phi\right|^2 + P = 0,
\label{eq:GPbernoulli}
\ee
where $P = c^2\lrp{\lrp{\rho-1} - \ell^2\rho^{-\frac{1}{2}}\nabla^2\rho^{\frac{1}{2}}}$ is considered a pressure term.  Here $c = \kappa/\lrp{\ell \sqrt{2}}$ is the speed of sound, and the non-standard term in the pressure, $- c^2\ell^2\rho^{-\frac{1}{2}}\nabla^2\rho^{\frac{1}{2}}$, is called the quantum pressure.  Notice that if we identify the gradient of the angular variable as a velocity, $\bfv = \nabla\phi = \kappa \nabla \theta$, and the square of the modulus as the fluid density, $\rho$, then the above two equations have a familiar interpretation.  The first, Eq.~(\ref{eq:GPcontinuity}), is simply the continuity equation, which expresses the conservation of mass flowing into and out of arbitrary closed volumes in the fluid domain.  The second, Eq.~(\ref{eq:GPbernoulli}), is Bernoulli's equation, which describes potential fluid flow.  We can obtain a more familiar form for this equation by taking the gradient, which gives
\be
\partial_t \bfv + \lrp{\bfv\cdot \bfnabla} \bfv + \nabla P = 0.
\label{eq:GPeuler}
\ee

This is simply Euler's equation for ideal fluids.  In particular this tells us that the fluid represented by the velocity field $\bfv = \kappa \nabla \theta$ and density $\rho = \left| \psi \right|^2$ is inviscid (viscosity free) and irrotational ($\nabla \times \bfv = \nabla \times \nabla \phi \equiv 0$).  Furthermore, notice that $\ell$, the healing length is very small.  Near the transition temperature it can be modeled as $\ell = \ell_0\lrp{1 - T/T_c}^{-0.67}$, where $\ell_0 \simeq 4.0 \; \AA$~\cite{bib:PhysRevLett.23.1276}.  Thus, close to the transition temperature the healing length diverges, and the quantum pressure becomes increasingly important.  However, for the lower temperatures that we are interested in, the healing length is small enough, a few angstroms, to warrant ignoring the quantum pressure.  Another important simplification arrises when considering the density to be constant, $\rho \simeq 1$.  This is tantamount to saying that the fluid is incompressible, and therefore that the velocity is divergence free, $\bfnabla \cdot \bfv = 0$.  Energy considerations are the main justification for this approximation.  The terms in the Lagrangian density, Eq.~(\ref{eq:LagrangianDensity}), which involve spatial derivatives discourage steady state solutions with varying order parameter magnitude, $\left| \psi\right|$, or equivalently with varying density, $\rho$.  Indeed, the solution to Eq.~(\ref{eq:GPcontinuity} and \ref{eq:GPbernoulli}) which globally minimizes the action, Eq.~(\ref{eq:actionfunctional}), are simply $\rho = 1$, $\phi = constant$.  This corresponds to a completely quiescent superfluid.  How then do vortices and superfluidity arise?
	
\section{Rotating Superfluids and Quantized Vortices}

Since the motionless, quiescent, solution to the Gross-Pitaevskii equation, Eq.~(\ref{eq:GrossPitaevskii}), minimizes energy, all solutions that involve a moving superfluid must be meta-stable.  Soliton solutions gain this stability dynamically, through the interplay between dissipative and nonlinear terms in the Gross-Pitaevskii equation.  However, by far the most prevalent solutions are those that involve vortices, which do not gain meta-stability in this manner.  Instead, they are stable due to topology.

Consider superfluid filling the inside of a torus.  If this is put in continuous motion around the torus, then at every point the gradient of the phase is non-zero.  Pick out any closed path that circles the torus once, and consider the phase along this path at any one moment in time.  The phase will be increasing in the direction of fluid flow along this path, and after one circuit the phase will be larger than its starting value.  This should seem odd, since the wave function at this point hasn't changed.  In particular, because the wave function is single valued, the phase must have changed by an integer multiple, $n$, of $2\pi$.  More precisely
\be
\oint \bfv \cdot d\ell = \oint \kappa \nabla \theta \cdot d\ell = \kappa \Delta \theta = \kappa n 2 \pi.
\label{eq:QuantizedCirculation}
\ee

This quantity, the velocity integrated about a closed loop, is called the circulation, $\Gamma$.  This equation effectively says that the circulation for any closed loop is quantized for a superfluid, $\Gamma = n 2 \pi \kappa$.  For a quiescent superfluid, this circulation is zero for any loop.  For the torus, or indeed any topology of the fluid manifold that is not simply connected, any integer is possible.  The most important aspect of this idea is that a solution with a non-zero circulation can not be continuously deformed into a solution with zero circulation.  Therefore super-flow can not easily decay, and there is an attendant energy penalty which discourages such phase-slips.

Like the torus example, superfluid vortices have quantized circulation for any loop that encircles the vortex core.  The vortex core is an area where the superfluid density drops to zero, creating the necessary topology.  Another way to justify the existence of a vortex core in an area that contains circulation is to apply Stokes law to Eq.~(\ref{eq:QuantizedCirculation}).
\be
\oint_{\partial S} \bfv \cdot d\ell = \int_S \lrp{\bfnabla \times \bfv} \cdot \hat{n} \; da = \int_S \lrp{\bfnabla \times \nabla \phi} \cdot \hat{n} \; da = 0
\label{eq:HoleExistence}
\ee
This is in contradiction with a non-zero circulation, and implies that there must be some area within $S$ where all the vorticity, $\bfnabla \times \bfv$, resides.  Since the superfluid is vorticity free, this region must be devoid of superfluid and therefore have zero superfluid density.  This is what constitutes the vortex core.  In cold superfluid helium, these vortex cores are very small, on the order of a few helium atoms in diameter.  What do these vortices look like?  The easiest attribute to calculate is the velocity field for a single straight-line vortex.  By symmetry it is everywhere tangent to a circle centered on the vortex.  Using Eq.~(\ref{eq:QuantizedCirculation}), we can find the magnitude of this velocity to be
\be
\left| v \right| = \frac{\Gamma}{2\pi r} =  \frac{ n \kappa}{r}.
\label{eq:VortexVelocity}
\ee

Thus, the superfluid velocity decays away from the vortex as $\frac{1}{r}$.  This also means that the velocity blows up as the distance to the core gets smaller, indicating that the vortex core must have a finite size.  Indeed, a rough calculation of the vortex core diameter determined by the point at which the velocity grows larger than the critical Landau velocity gives the correct order of magnitude.  To find how the superfluid density behaves near the core, consider the solution ansatz: $\psi = \varrho\lrp{r}e^{i n \theta} $.  That is, the magnitude of the order parameter, or density squared, is dependent only on the radius from the vortex, and the phase is dependent only on the azimuthal angle $\theta$.  Consider a stationary solution to the Gross-Pitaevskii equation, Eq.~(\ref{eq:GrossPitaevskii}), where $\psi_t = 0$.  Use of the ansatz gives the following differential equation
\be
\varrho'' + \frac{1}{r}\varrho' + \lrp{\ell^{-2}-\lrp{\frac{n}{r}}^2-\ell^{-2}\varrho^2}\varrho = 0.
\label{eq:CoreStructure}
\ee

This is not analytically solvable, however it behaves in an expected fashion.  For $r \ll \ell$, the solution is approximately $\varrho \propto r^{\left| n \right|}$, and for $r \gg \ell$ the regular density is quickly recovered, $\varrho = 1$.  The characteristic vortex core size is related to both the healing length, $\ell$, as well as the quantum number, $n$.  The actual core structure is not very well understood, and different models give different characteristic vortex core sizes, $a_0$.  As this is usually a very small length for temperatures much lower than the transition temperature, We will assume that $\rho = 1$ outside of $a_0$, and $\rho = 0$ inside this radius.  The motion of a set of vortices will not be affected by the precise value of $a_0$, and we will not worry about it from now on.

One thing that can be easily calculated is the kinetic energy of the superfluid around a single vortex.  Assuming that this vortex lies along the center line of a cylindrical container of radius $R$ and height $h$, the kinetic energy is
\be
\mathcal{E} = \int^h_0 \int^{2\pi}_0\int^R_{0} \frac{1}{2} \rho v^2\; r\; dr d\theta \; dz = h \pi \rho \int^R_{a_0} \lrp{\frac{\kappa n}{r}}^2 \; r\; dr = h \pi \rho \kappa^2 n^2 \ln{\lrp{R/a_0}}.
\label{eq:LineVortexKE}
\ee
The most important aspect of this equation is that the energy is proportional to $n^2$.  Thus the energy of a single vortex with circulation $\Gamma = 2 \cdot \lrp{2 \pi \kappa}$ is two times larger than that of two vortices, each with circulation $\Gamma = 2 \pi \kappa$.  Because of this, the formation of vortices with a single quantum of circulation are energetically preferred over fewer vortices with larger circulations.  This is one of the main reasons, besides simplicity, that we will be focusing on vortices with identical strengths (circulations).

How do quantum vortices form?  Consider a rotating drum of helium-I, that is liquid helium above the transition temperature.  The finite viscosity will cause the fluid to eventually rotate rigidly, in step with the walls of the cylinder, and the fluid particles will have some angular momentum about the center axis.  If the temperature is decreased below the lambda transition, a finite number of particles will now be in the superfluid condensate.  Since the angular momentum is conserved, these superfluid particles will need to somehow rotate, in aggregate, about the center.  This is only possible if a number of quantized vortices form.

A related argument, due to Feynman~\cite{bib:Feynman}, points out that the presence of line vortices minimizes the free energy of the rotating superfluid.  In a cylinder rotating with angular frequency $\omega$, the free energy of the fluid in the rotating frame is 
\be
\mathcal{E} = \mathcal{E}_0 - M\omega,
\label{eq:RotFreeEnergy}
\ee
where $\mathcal{E}_0$ is the energy and $M$ is the angular momentum of the fluid, both in the fixed lab frame.  For a regular fluid, the motion that minimizes the free energy is solid body rotation.  For this type of motion, there is circulation about every loop, or equivalently, the vorticity ($\bfnabla \times \bfv$) is nowhere zero.  This is certainly not compatible with the motion of a superfluid.  A plausible free-energy-minimizing state might be that of the quiescent superfluid, which has $M = 0$ and $\mathcal{E}_0 = 0$.  However, under the right conditions, a superfluid with line vortices has an even lower free energy.  As seen in Eq.~(\ref{eq:LineVortexKE}), the addition of a line vortex will have a larger lab frame energy, but the momentum will also be larger.  Thus, the free energy of this system with a single vortex will decrease as the angular frequency is increased.  Indeed, there is a critical angular frequency above which the existence of a single line vortex is thermodynamically preferred over quiescent superfluid.  This angular frequency is very low, and it is very difficult to produce a superfluid sample without quantized vortex lines.

A set of line vortices will also be energetically favored over no super-flow, given a large enough angular frequency.  However, the free energy is now dependent on the relative location of the vortices, in addition to the number of vortices.  The energy minimizing configuration turns out to be a triangular lattice.  In the context of super-conducting vortices, such a configuration is called an Abrikosov lattice.  In a sense, the velocity field of the Abrikosov lattice most closely mimics that of a rigidly rotating fluid.  Indeed, we can find the inter-vortex spacing, $d$, in terms of the angular frequency.  For a triangular, Abrikosov vortex array, the vortex number density is $\rho_n = \frac{3}{2\sqrt{2}}\frac{1}{d^2}$.  Therefore, the circulation about a circle of radius $r$ is
\be
\Gamma = n \kappa = \rho_n \pi r^2 \kappa = \frac{3 \pi r^2}{2\sqrt{2} d^2} \kappa.
\ee
Notice that this is proportional to $r^2$, just as rigid rotation is $\Gamma = \omega r \lrp{2 \pi r} = 2 \pi r^2 \omega$.  This shows that a constant number density array of vortices does in fact mimic rigid body rotation.  Equating the two circulations gives
\be
d = \sqrt{\frac{3}{4\sqrt{2}}\frac{\kappa}{\omega}}.
\ee

\begin{figure}[htb]
 \centering
\large
\resizebox{\linewidth}{!}{\includegraphics{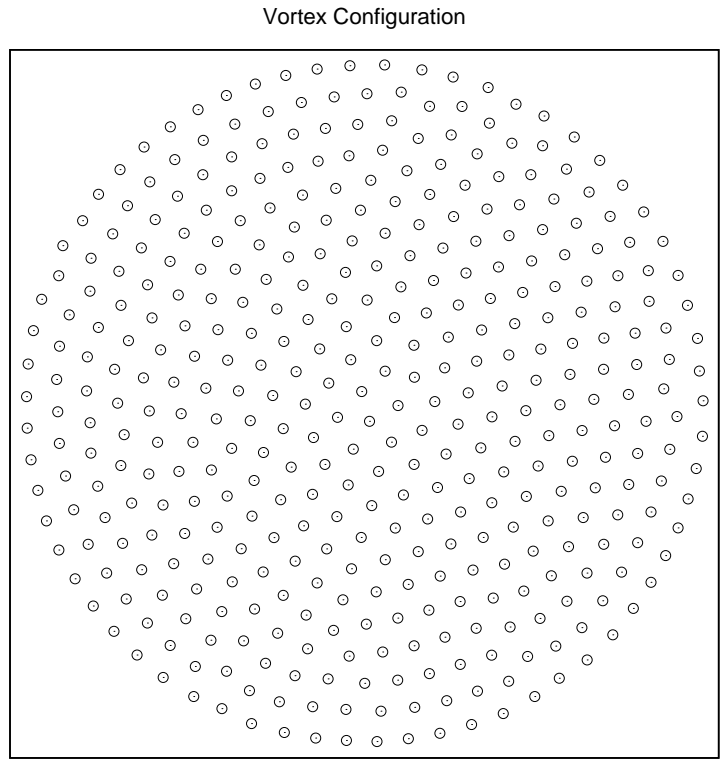}}
\normalsize
    \caption{This is a configuration of 400 point vortices, evolved under the driven dissipative dynamics of Eq.~{\ref{eq:FiniteTemperatureVortexMotion}}.  Notice that the center closely approximates an Abrikosov triangular lattice, while the outer vortices are grouped into concentric circles.  There are also lattice dislocations, which are longer time scale transients of the approach to the asymptotic solution which minimizes free energy.}
    \label{figure:VortexCrystal}
\end{figure}

Thus, an increase in rotational frequency results in a more compressed vortex lattice.  These lattices have been observed in superfluid helium \cite{Hall:1956eh}, though they expectedly deviate from this ideal near the edges of the lattice, see Fig.~(\ref{figure:VortexCrystal}).  Because this solution minimizes the energy, it is expected to be ubiquitous wherever there is a mechanism for dissipating energy.  For point-vortex dynamics, this dissipative mechanism arises when we have mutual friction between the normal component of the two-fluid model and the vortex cores.  This will be the subject of the next section, though it should be noted that this mechanism disappears at zero temperature, and the conservative dynamics, which are the main focus of this work, are recovered.

Finally, before we move on to the two-fluid model, we should mention how line vortices move in the absence of dissipative forces.  we will spend much more time on this point in chapter~(\ref{ch:pvm}), and will only mention the essentials.  Galilean invariance of the Gross-Pitaevskii equation implies that a quantized vortex solution remains symmetric when placed in a constant phase gradient and viewed from the moving frame.  This means that vortices move as if they were fixed in the ambient superfluid.  Thus, each vortex simply advects in the velocity field of all other vortices, and can be considered to have no mass or inertia.  Additionally, the velocity field at any point is the linear superposition of the velocity field due to all vortices.  This leads to particularly simple equations of motion, Eq.~(\ref{eq:PVevolution}), which constitute the point-vortex model.

\section{Two Fluid Model}

In three dimensions, the motion and geometry of superfluid vortex lines can be very complicated.  First of all, they must either terminate at the boundaries of the superfluid or form closed loops.  Those vortex lines that do extend to the boundaries often experience an attractive potential due to the impurities of the confining surface.  This ``pinning" force can often be the determining factor in the motion of vortices.  Furthermore, line vortices move in the velocity field due to all other vortices, and in the case where the vortex line has some curvature, there will be a locally induced velocity field contribution.  In the rest of this work, we will be assuming that the line vortices are all rectilinear, which is a good approximation for a rotating superfluid.  We will also ignore pinning completely.  These assumptions lead to the relatively simple point vortex dynamics of Eq.~(\ref{eq:PVevolution}), which is the focus of this work.  A slightly harder complication to ignore is the effect of strong interactions between helium atoms, which modifies the simple BEC picture.  In this section, we will explain how the point vortex model is generalized under these changes, and how this is not a problem in the zero temperature limit.

At absolute zero, all of the fluid is in the superfluid state.  As the temperature increases, there arise quantized excited states with particle-like properties.  The lowest energy quasiparticles are phonons, quantized collective sound vibrations.  These all move with the same velocity, $c$, and have a linear dispersion relation, $\epsilon = c p$.  A little higher on the energy spectrum are rotons, which have a quadratic dispersion relation and an energy gap, $\epsilon = \epsilon_0 + \lrp{p - p_0}^2/2\mu$.  Overall, both quasiparticles are part of a continuous excitation spectrum, $\epsilon\lrp{p}$.  Following Landau, the critical velocity of the superfluid can be defined as $v_c = \min_p \epsilon\lrp{p}/p$~\cite{bib:LandauSP}.  The existence of a critical velocity is one of the defining features of superfluid helium, a feature that is noticeably absent in the simple BEC description.  At absolute zero quasiparticles are not thermally produced, but still may be produced by the relative motion of the superfluid and any fixed obstacle, such as the wall of the container.  However, this may only happen if the relative speed is greater than the critical velocity, $v_c$.  Otherwise, no phonons or rotons will form, no momentum can be transfered from the superfluid, and there will be no mechanism for the superfluid to slow down.  Thus super-flow can occur for relative velocities below that of the critical velocity.

This association of the quasiparticles with momentum transfer to obstacles, naturally leads one to consider the gas of thermal phonons and rotons as a viscous fluid.  These quasiparticles interact enough to reach a thermal equilibrium, and as a fluid carry all of the entropy of the overall fluid.  Note that this ``normal" fluid is comprised of the collective modes of the superfluid state, and can not be thought of in exactly the same way at higher temperatures near the lambda transition, where the superfluid has vanishing density.  None the less, one can construct a consistent phenomenological model, valid at temperatures below the critical temperature, that treats liquid helium-II as two interpenetrating fluids.  One fluid, the superfluid, has density $\rho_s$ and velocity field $\bfv_s$, while the other, the normal fluid, has density $\rho_n = \rho - \rho_s$ and velocity field $\bfv_n$.  This two fluid model, proposed by Tisza and then Landau, behaves according to the hydrodynamic equations
\begin{subequations}
\begin{align}
\rho_s \pd{\bfv_s}{t} + \lrp{\bfv_s \cdot \bfnabla}\bfv_s &= - \frac{\rho_s}{\rho}\nabla P + \rho_s S \nabla T - F_{ns}	\\
\rho_n \pd{\bfv_n}{t} + \lrp{\bfv_n \cdot \bfnabla}\bfv_n  &= - \frac{\rho_n}{\rho}\nabla P - \rho_s S \nabla T + F_{ns} + \nu \nabla^2 \bfv_n, 
\label{eq:twofluidmodel}
\end{align}
\end{subequations}
where $P$ is the pressure, $S$ is the entropy density, and $\nu$ is the normal fluid viscosity \cite{BARENGHI:1983uy}.  Aside from the two terms $\rho_s S \nabla T$ and $F_{ns}$ in each, these two equations are essentially the Euler equation for a perfect fluid and the Navier-Stokes equation for a viscous fluid, respectively.  Both of the normal fluid and superfluid inhabit the same space, and their interactions lead to some interesting phenomena.  First of all, there are the expected sound waves due to pressure gradients, where the two fluids oscillate in phase with each other.  This is, however, not the only sound mode that is supported by the two fluid equations.  The temperature gradient terms lead to so-called ``second sound", where the two fluids oscillate out of phase with each other.  These entropy waves are only one of the many phenomena, from the fountain effect to the sudden cessation of boiling upon crossing the lambda transition, that result from the normal fluid carrying all of the entropy.  We are mostly interested in the low-temperature dynamics of superfluid vortices, where the entropy is negligibly small and we can consider the temperature to be constant.  We will therefore ignore the temperature gradient induced effects on vortex dynamics.  

The main change that two-fluid hydrodynamics introduces to vortex dynamics is mutual friction.  Mutual friction, $F_{ns}$, was originally observed in rotating cylinders, where the whole fluid was unexpectedly observed to participate in rotation.  It was postulated that there was some bulk friction, with which the rotating normal fluid acted on the superfluid.  It is now known that the mutual friction only acts on the cores of superfluid vortices, which in turn constrains the overall superfluid motion.  Importantly, this provides a dissipative mechanism to realize the Abrikosov lattice of rectilinear vortices.  In the absence of this mutual friction, a rectilinear vortex, or point vortex, moves as if frozen in with the local superfluid flow.  We will now try to motivate the equations of motion for a point vortex at finite temperature, where mutual friction is important.  Much of the background material can be found in \cite{BARENGHI:1983uy}.

Consider a line vortex which is moving with velocity $\bfv_L$.  In the area of the vortex core, excitations have a drift velocity $\bfv_R$.  If these two velocities are the same then the quasiparticles do not act on the vortex with any overall force.  However, if there is relative motion, then two forces arise: a drag force in the direction of the difference, and a Magnus-like lift force which is perpendicular to the difference.  Together they constitute the force due to excitations
\be
F_{ex} = D\lrp{\bfv_R-\bfv_L} + D'\hat{\bfomega} \times \lrp{\bfv_R-\bfv_L},
\label{eq:excitationForce}
\ee
where $D$ and $D'$ are temperature dependent constants, and $\hat{\omega}$ is a unit vector out of plane.  We have that $D = \rho_n v_g \sigma_{\parallel}$ and $D' = \rho_n v_g \sigma_{\perp}$, where $v_g$ is the roton group velocity, and $\sigma_{\parallel}$, $\sigma_{\perp}$ are the excitation cross-sections parallel and transverse to the relative velocity.  The most important thing to note is that they are both proportional to the normal density, which exponentially decays to zero at absolute zero.  There is another subtle force, the Iordanskii force, that arises due due to an accumulated geometric phase of phonons about the vortex core, $F_I = -\rho_n \Gamma\hat{\omega} \times \lrp{\bfv_R-\bfv_L}$, where $\Gamma$ is the quantum of circulation.  This is in the opposite direction of transverse force due to rotons, and we can group them together by replacing $D'$ with $D_t = D' - \rho_n \Gamma$.  This is positive for temperatures above $1 K$, where rotons dominate, negative at lower temperatures where the excitation gas is mostly phonons, and goes to zero with the decreasing normal fluid density.  The final force on the vortex core is due to a difference in velocity between the vortex itself and the local superfluid velocity, $\bfv_s$.  This is also a transverse force, and we will call it the superfluid Magnus force
\be
F_{M} = - \rho_s \Gamma \hat{\bfomega} \times \lrp{\bfv_s-\bfv_L}.
\label{eq:MagnusForce}
\ee
The vortex has insignificant inertia, and therefore all of the previous forces balance out, $F_M + F_I + F_{ex} = 0$
\be
\rho_s \Gamma \hat{\bfomega} \times \lrp{\bfv_s-\bfv_L} = D\lrp{\bfv_R-\bfv_L} + D_t\hat{\bfomega} \times \lrp{\bfv_R-\bfv_L}.
\label{eq:ForceBalance}
\ee

Before we solve for the vortex velocity, there is one more wrinkle.  The drift velocity of the excitations $\bfv_R$ is not exactly equal to the averaged normal fluid velocity $\bfv_n$, since the force with which the excitations act on the vortex has a reaction force acting back on the excitation cloud by Newton's third law.  This locally slows down the excitation velocity from the averaged normal fluid velocity.  Indeed the difference between the two is proportional to the total force on the normal fluid, $\bfv_R-\bfv_n = F_n/E$, for some constant $E$.  Furthermore, the total force on the normal fluid, $F_n$, must balance out the total force on the superfluid, which is the Magnus force, $F_n = F_M$.  Putting these two relations together with the force balance relation, Eq.~(\ref{eq:ForceBalance}), gives, after some work, the following expression for the velocity of the vortex
\be
\bfv_L = \bfv_s + \alpha \lrp{\bfv_n - \bfv_s} + \alpha' \hat{\omega} \times \lrp{\bfv_n - \bfv_s}.
\label{eq:FiniteTemperatureVortexMotion}
\ee
Both $\alpha$ and $\alpha'$ are positive for low temperatures, and disappear at absolute zero.  Notice that the conservative dynamics of point vortices are what remain in this limit.  That is, the vortex moves with the local superfluid velocity due to all other vortices, $\bfv_L = \bfv_s$.  This is yet another justification for focusing on the dynamics of point vortices in the conservative regime.  

For a rotating drum at non-zero temperatures, Eq.~(\ref{eq:FiniteTemperatureVortexMotion}) has the rotating Abrikosov lattice as an asymptotic solution.   The normal fluid is rotating rigidly, $\bfv_n = \bfomega \times \bfr$, and any vortex that does not follow this motion will have extra components to its velocity, in addition to the local superfluid velocity.  In particular, if at a vortex the local superfluid velocity is larger than the normal velocity, then the $\alpha$ term of Eq.~(\ref{eq:FiniteTemperatureVortexMotion}) will slow the vortex down, while the $\alpha'$ term will push it outward from the center of the drum, to where the normal fluid is faster.  Conversely, a relatively slow vortex will be sped up and pushed inward, toward the slower center of the drum.  This dissipative motion will eventually result in a lattice like that in Fig.~(\ref{figure:VortexCrystal}).

The creation of this vortex lattice highlights one of the roadblocks to realizing the full range of conservative vortex motions.  Notice that the vortex lattice is also a solution to the purely conservative dynamics at absolute zero.  Thus, as one decreases the temperature, the naturally forming vortex lattice will remain, even at absolute zero.  To see any of the types of motion described in the rest of this work, one must somehow perturb the vortex lattice.  If this and the pinning problem could be overcome, then the point vortex model would give accurate predictions about the motion of quantized superfluid vortices.  Indeed, as we have seen, all the important aspects of superfluid vortices are captured by the point vortex model.  First of all, they are very straight in the case of a rotating drum.  The rectilinear vortices have their motion confined to a plane, and are therefore well modeled as objects in two dimensions.  Furthermore, the superfluid vortex core is on the order of a few Angstroms in diameter, whereas the usual inter-vortex distance is on the order of tenths of millimeters.  Thus, we can reasonably ignore any internal structure of the vortices and consider them to be point objects.  The nature of superfluid vortices also constrains the free parameters in the point vortex model.  For superfluid vortices in a rotating drum, all circulations, or vortex strengths, are of the same sign.  That is, all the vortices have velocity fields that circulate fluid about them in the same direction.  More interestingly, all of the strengths have the same magnitude, one quantum of circulation.  Thus, our task in this work is to understand the motion of a collection of identical point-like particles that have motions in two dimensions described by the point-vortex model.


\chapter{Point Vortex Model}
\label{ch:pvm}

We have demonstrated in the previous chapter that point vortices exist in nature.  Now we turn our attention to the mathematical model that describes their motion.  We will first highlight where the point vortex model comes from with a rough derivation starting with the Euler equations of fluid motion.  Next we will explain the Hamiltonian nature of this dynamical system.  After this we will rewrite the model in the more insightful language of differential geometry, specifically highlighting the symplectic nature of Hamiltonian dynamics.  This will allow us then to talk about symmetries and conserved attributes of the system.  In particular this will elucidate the geometric nature of the divide between integrable and chaotic motion.  Finally, as a break from the abstraction, we will review what these ideas imply about the motion of two, three, and four vortices.  Throughout this section, we rely on ideas and notation from~\cite{bib:NewtonTNVP}\cite{bib:saffman}.

\section{Whence Point Vortices}

The point-vortex model is a highly idealized, though very useful, model of fluid dynamics.  It is therefore important to understand where it comes from, and in particular, what physical assumptions are made when assuming that a system is well described by point vortex dynamics.  In chapter~\ref{ch:phys}, we argued that rotating liquid He-II is one such system, partially based on the Euler equation interpretation of the wave function evolution.  In this section we hope to show how the point vortex model naturally arises from the Euler equation.

Before turning to the Euler equation, briefly consider the Navier-Stokes equation in 3 dimensions.  In this case, the homogeneous and Newtonian version:
\be
\rho\left(\pd{\bfv}{t}+ \left(\bfv \cdot \bfnabla\right) \bfv \right) =  -\nabla p + \mu \nabla^2 \bfv +\lrp{\frac{1}{3}\mu+\mu^{v}}\bfnabla\lrp{\bfnabla\cdot\bfv},
\label{eq:compNavStokes}
\ee
along with the compressible continuity equation
\be
\pd{\rho}{t}+ \nabla \cdot \left(\rho\bfv\right) =  0.
\label{eq:compCont}
\ee
Here, $\bfv = \left(v_x,v_y,v_z\right)$, $\rho$ is the density, $p$ is the pressure, $\mu$ is the dynamic viscosity, and $\mu^{v}$ is the volume viscosity.  We have assumed no external forcing.  Additionally, we are assuming that our fluid is inviscid $\left(\mu = 0\right)$, and therefore energy is not dissipated away on any scale.  This is tantamount to completely decoupling the macroscopic (fluid) degrees of freedom from the microscopic (molecular or thermal) degrees of freedom.  Since our fluid is incompressible, $\rho$ is constant, and we might as well set it equal to one.  This stipulation of incompressibility simplifies the continuity equation to be $\bfnabla \cdot \bfv = 0$, i.e. our vector fields are divergence free, and gets rid of the last term in Eq.~(\ref{eq:compNavStokes}).  Now our equation of fluid motion has been reduced to Euler's equation:
\be
\left(\pd{\bfv}{t}+ \left(\bfv \cdot \bfnabla\right) \bfv \right) =  -\nabla p.
\label{eq:EulerFluid}
\ee

It should be mentioned that both the incompressible and inviscid limits to Eq.~(\ref{eq:compNavStokes}) are very subtle~\cite{bib:LandauFluid}.  We discuss the Navier-Stokes equation mainly to highlight the assumptions which are implicit in Euler's equation.  For our purposes, it is especially fruitful to rewrite this equation in terms of vorticity $\left( \bfomega = \bfnabla \times \bfv \right)$.  Taking the curl of both sides of Eq.~(\ref{eq:EulerFluid}), we obtain:
\be
\left(\pd{\bfomega}{t}+ \left(\bfv \cdot \bfnabla\right) \bfomega \right) =  \left(\bfomega \cdot \nabla \right) \bfv.
\label{eq:EulerVort3D}
\ee
One advantage of this equation is that the pressure term has dropped out.  Notice that we made use of the vector identity $\bfnabla \times \bfnabla p = 0$.  However, the real reason for thinking in terms of vorticity comes to light when we restrict the fluid motion to 2 dimensions.  In this case, vorticity is out of plane, and we may think of it as a scalar field over $\mathbb{R}^2$.  We will denote two dimensional position as $\bfr \equiv \langle x, y\rangle$. Since the velocity has no component aligned with the vorticity, the right hand side of Eq.~(\ref{eq:EulerVort3D}) vanishes, and we get:
\be
\left(\pd{\omega}{t}+ \left(\bfv \cdot \bfnabla\right) \omega \right) =  0.
\label{eq:EulerVort2D}
\ee

The left hand side of Eq.~(\ref{eq:EulerVort2D}) is often referred to as the convective derivative $\left(\frac{D}{Dt} \equiv \frac{\partial}{\partial t}+ \left(\bfv \cdot \bfnabla\right)\right)$ of vorticity.  The convective derivative has a nice interpretation as the rate of change of a function with time, in the co-moving frame of the fluid.  Because $\frac{D\omega}{Dt} = 0$, we can say that the vorticity of a fluid particle is conserved.  Furthermore the total vorticity, or first vorticity moment, $\int_{\mathbb{R}^2}\omega\left(\bfr\right)d\bfr$ is also conserved in time, and therefore, in two dimensions,  vorticity is neither created nor destroyed.\\

If the domain in which we are considering our fluid motion is simply connected, i.e., all loops can be deformed into any other loop (the fundamental group is trivial), then $\bfnabla \cdot \bfv = 0$ implies the existence of a vector potential for the velocity field.  We are considering $\mathbb{R}^2$ and therefore can write $\bfv = \bfnabla \times \bfA$.  Furthermore, if we choose $\bfA$ such that it is divergence free, we get a simple Poisson equation relating the vector potential and the vorticity
\be
\nabla^2 A = - \omega.
\label{eq:PoissonVorticity}
\ee
Notice that, much like the vorticity, the vector potential is effectively a scalar, only having an extent out of the x-y plane.  Now, the existence of a Poisson equation allows us to use the appropriate Greens function to solve for the potential, and therefore the velocity field, given a vorticity field.  This effectively inverts the original relation between vorticity and velocity.  For a fluid domain in $\mathbb{R}^2$ without boundary the appropriate Green's function is
\be
G\left(\bfr,\bfr'\right) = \frac{\ln\left| \bfr - \bfr' \right|}{2 \pi}
\label{eq:2DGreens}
\ee

From this we can express the potential A as
\be
A\left(\bfr\right) = -\frac{1}{2 \pi}\int_{\mathbb{R}^2} \ln\left| \bfr - \bfr' \right| \omega\left(\bfr'\right)\, d\bfr'.
\label{eq:PotentialOmega}
\ee
Taking the curl, we can then recover the velocity field in terms of the vorticity
\be
\bfv\left(\bfr\right) = -\frac{1}{2 \pi}\int_{\mathbb{R}^2} \frac{\left( \bfr - \bfr' \right)}{\left| \bfr - \bfr' \right|^2}\times \bfomega\left(\bfr'\right) \, d\bfr'.
\label{eq:VelocityFromOmega}
\ee
This is very similar in form to the Biot-Savart law, and indeed in $\mathbb{R}^3$ they are mathematically identical in form.

At this point we must make our assumptions about the vorticity field explicit to go any further.  We can think of point vortices as a discrete number of points where all the vorticity is concentrated.  We saw earlier in chapter~\ref{ch:phys} that the effective diameter of a superfluid vortex in superfluid He-II is on the order of a few angstroms.  This justifies our ignoring any physical extent to the vortices, as long as the distance between the vortices remains large when compared to this effective diameter.  We also saw that a path integral $\int_{C} \bfv \cdot dl$ is equal to the minimal, quantized, strength if the path encloses a single vortex and zero if if does not.  This tells us that the vorticity is zero outside of the vortex diameter.  We can model this by assuming discrete Dirac delta function like support
\be
\omega\left(\bfr\right) = \sum^N_{i =1}\Gamma_i\delta\left(\bfr_i - \bfr\right),
\label{eq:DiscreteVorticity}
\ee
where $\Gamma_i$ is the strength of the i-th vortex and $\left\{\bfr_i\right\}$ is the set of positions for the N vortices.  In much of what follows, we will assume that $\Gamma_i = 1$, reflecting our focus on a system of positive identical vortices.  However, we will often leave in the strengths to preserve the generality of a particular result.  Now that we have a vorticity distribution, we can find explicit formula for the potential and the velocity at any point for a given set of vortex strengths and positions.  Using Eq.~(\ref{eq:PotentialOmega}), we get
\be
A\left(\bfr\right) = -\frac{1}{2\pi}\sum^N_{i =1}\Gamma_i\ln\left|\bfr_i - \bfr\right|.
\label{eq:NvortexA}
\ee
Now using Eq.~(\ref{eq:VelocityFromOmega}), we find
\begin{align}
v_x\left(\bfr\right) &= -\frac{1}{2\pi}\sum^N_{i =1}\Gamma_i\frac{\left( y - y_i \right)}{\left| \bfr - \bfr_i \right|^2} \nonumber\\
v_y\left(\bfr\right) &= +\frac{1}{2\pi}\sum^N_{i =1}\Gamma_i\frac{\left( x - x_i \right)}{\left| \bfr - \bfr_i \right|^2}.
\label{eq:PVvelocityfield}
\end{align}
This is the velocity field for a configuration of vortices.  Notice that the velocity field is basically the superposition of velocity fields induced by each point vortex.  Importantly, the velocity falls off to zero with increasing distance from the vortices as $\frac{1}{r}$, and diverges in the same manner as we approach one of the vortices.  This latter trend is unphysical and serves to remind us that this model is only valid when considering distances larger that the vortex core diameter.

Our goal now is to find the equations of motion for the vortices themselves.  Since the vorticity is simply advected with the local velocity field $\left(\frac{D\omega}{Dt}\right) = 0$, we can use the above equation at the site of each vortex to find its velocity vector.  However, this is a divergent quantity.  Physically, we can remove this obstacle by noting that the divergent quantity is a self-interaction term.  It is the velocity at a vortex due to its own field.  We can safely exclude this term from the sums.  Note that a single vortex in $\mathbb{R}^2$ is stationary and therefore does not feel its own velocity field.  The exclusion of self interaction terms leads us to the fundamental equations for vortex motion
\begin{align}
\frac{d x_i}{dt} &= -\frac{1}{2\pi}\sum^N_{j \neq i}\Gamma_j\frac{\left( y_i - y_j \right)}{\left| \bfr_i - \bfr_j \right|^2}, \nonumber\\
\frac{d y_i}{dt} &= +\frac{1}{2\pi}\sum^N_{j \neq i}\Gamma_j\frac{\left( x_i - x_j \right)}{\left| \bfr_i - \bfr_j \right|^2}.
\label{eq:PVevolution}
\end{align}

It is worth noting that in considering point vortices, we have gone from a potentially hard-to-solve partial differential equation, Eq.~(\ref{eq:EulerFluid}), to a set of relatively simple coupled ordinary differential equations, Eq.~(\ref{eq:PVevolution}).  This expresses the idea that all of the fluid motion, excepting that of the vortices, is completely passive and determined by the vortices alone.  It is a very powerful idea, and indeed the main reason for the popularity of more general ``vortex methods" in various fluid dynamic situations.   For our purposes, this means that point vortex dynamics are easily explored computationally.

This equation is the heart of this work, and as such, we will have a number of ways of expressing it, each giving us some different mathematical insight into its nature.  The most natural starting point is to express Eq.~(\ref{eq:PVevolution}) in Hamiltonian form.

\section{Hamiltonian Structure}

It is interesting and quite beautiful that these particle-like point vortices should follow Hamiltonian mechanics, much like massive point particles.  Indeed, Arnold showed in~\cite{bib:ArnoldTop} that even the Euler equation itself could be thought of in terms of Hamiltonian dynamics on the infinite dimensional space of divergence free vector fields.  This attribute of the point vortex system allows one to formulate statistical mechanics on a large number of vortices and, more importantly for this work, allows one to deal coherently with symmetries.

Instead of simply stating the Hamiltonian that gives rise to the point vortex motion, we will try to motivate it to a certain extent by deriving it from energy considerations.  Consider the total energy of the fluid, which, since there is no potential energy, is just the kinetic energy 
\be
E = \frac{1}{2} \int_{\mathbb{R}^2} \bfv \cdot \bfv \, d\bfr.
\label{eq:totalKE}
\ee
Since $\bfv = \bfnabla \times \bfA$, we can use the vector identity $\left(\bfnabla \times \bfA\right)\cdot \left(\bfnabla \times \bfA\right) = \bfnabla \cdot \left( \bfA \times \left(\bfnabla \times \bfA\right)\right) + \bfA \cdot \left(\bfnabla \times \left(\bfnabla \times \bfA\right)\right)$ and the definition of vorticity to break Eq.~(\ref{eq:totalKE}) into two terms.
\be
E = \frac{1}{2} \int_{\mathbb{R}^2} \left(\bfA \cdot \bfomega \right) d\bfr + \frac{1}{2} \int_{\mathbb{R}^2} \bfnabla \cdot \left( \bfA \times \bfv \right) d\bfr
\label{eq:totalKEsub}
\ee

If the domain were in $\mathbb{R}^3$, then the second term would be a boundary term that vanishes.  We, however, are not so lucky.  In two dimensions, this term is logarithmically divergent, which means that our kinetic energy is formally infinite.  Fortunately, this term is effectively a constant in terms of the positions of the vortices.  It is controlled by how the velocity field decays far from the vortices, where the vortices appear as if they are simply a single vortex of larger strength $\sum \Gamma_i$.  Because of this, we can treat this term as an additive constant whose derivatives with respect to vortex coordinates are zero, and therefore do not affect the vortex motion in the Hamiltonian perspective.  Note that if we have a neutral vortex mixture where $\sum \Gamma_i = 0$, the velocity field falls off much more quickly with distance from the vortices, much like the electric field of a dipole.  If we also set a minimal distance from each vortex in the integration, we could expect a finite value for this term.  However, since we are only considering positive vortices, and this term does not affect the dynamics,  we will be ignoring it from this point on.\\

Now, turning our attention to the first term, we can get an explicit form by using Eqs.~(\ref{eq:NvortexA} and \ref{eq:DiscreteVorticity})
\be
E = -\frac{1}{4\pi} \sum^N_{i=1}\sum^N_{j=1}\Gamma_i\Gamma_j\ln\left| \bfr_i - \bfr_j \right|.
\label{eq:totalKE2}
\ee
Once again, we must contend with infinite terms in Eqs.~(\ref{eq:totalKE2}), this time coming from terms where $i = j$.  We can view a generic term in this sum as the kinetic energy due to the interaction between two vortices.  Thus, the infinite terms are due to self interaction, and can be neglected much as we did in setting up Eq.~(\ref{eq:PVevolution}).  We will associate what remains of the sum with the Hamiltonian of the point vortex system.  While this is not the total energy of the fluid, we can still salvage it physically, by thinking of it as the kinetic energy of the fluid due to the relative motions of the point vortices.  For future reference this Hamiltonian is given as
\be
H = -\frac{1}{4\pi} \sum^N_{i}\sum^N_{j\neq i}\Gamma_i\Gamma_j\ln\left| \bfr_i - \bfr_j \right|.
\label{eq:VortexHamiltonian}
\ee
To get the equations of motion, Eq.~(\ref{eq:PVevolution}), from this Hamiltonian, we must take the partial derivatives with respect to the positions and divide by the strengths
\begin{align}
\frac{d x_i}{dt} &= +\frac{1}{\Gamma_i}\pd{H}{y_i} \nonumber\\
\frac{d y_i}{dt} &= -\frac{1}{\Gamma_i}\pd{H}{x_i}.
\label{eq:PVhamEv}
\end{align}

These equations are slightly odd for Hamiltonian dynamics for two reasons.  First of all, there is the factor of $\frac{1}{\Gamma_i}$ in each, which makes the equations non-canonical if the $\Gamma_i \neq 1$.  Indeed some people prefer to make the some what awkward variable transformations $q_i = \sqrt{\left|\Gamma_i\right|}\sgn{\left(\Gamma_i\right)}\, x_i$ and $p_i = \sqrt{\left|\Gamma_i\right|}\sgn{\left(\Gamma_i\right)}\, y_i$ to force the equations to be canonical.  However, since the extra factor arose naturally in the analysis of this system we shall keep it.  Indeed, we will see later on that this does not adversely impact any of our mathematical machinery.

In most usual Hamiltonian systems the conjugate variables are generalized position and momentum, whereas here they are the $x$ and $y$ coordinates of each vortex!  This reflects the fact that the velocity vector of each vortex is uniquely determined by the positions of all of the other vortices.  Indeed, one can think of the vortex particles as having no mass and therefore simply moving with the local fluid flow.  This has a number of enormous ramifications, which will become clearer in later sections.  First of all, the configuration space of the point vortex system, with coordinates $\left\{ x_i,y_i\right\}$, is the phase space as well.  Each point in the configuration space uniquely determines the state of the system, and the Hamiltonian equations describe a vector field over this space which generates the phase space flow.  Additionally, this means that, unlike most regular Hamiltonian systems, the phase space is not the cotangent bundle of some other manifold.  Because of this, there is no natural way to view this problem in terms of Lagrangian dynamics on a tangent bundle; the Legendre transformation is not defined.  So this curious system is an example of how Hamiltonian dynamics encompass more general problems than Lagrangian dynamics.

\section{Geometric View}

V. I. Arnold said~\cite{bib:Arnold1} that ``Hamiltonian Mechanics cannot be understood without differential forms."  Some physicists might take issue with this sentiment, having little need to think of symplectic structure, differential forms, and Lie derivatives to know how a harmonic oscillator works.  However, geometry is indeed fundamental in this case, and a little nomenclature plus a few ideas from modern geometry will go a long way toward elucidating the structure of solutions to the point vortex problem.  Our goal in this chapter is to introduce all of the modern geometric ideas needed throughout this work specifically from the viewpoint of our particular Hamiltonian system.  Due to the breadth of this background material we will handle many of the topics superficially.  Fortunately there are many fine introductions to differential geometry, algebraic topology, and symplectic geometry out there.  In this chapter we have relied heavily on~\cite{bib:Arnold1} and have adopted much of his notation.  For an overview of differential geometry see~\cite{bib:TuMan}.  For algebraic topology, see~\cite{bib:Hatcher}.   Many of the ideas from these areas are put to use in physical situations in the book~\cite{bib:Frankel}.  For Symplectic geometry see~\cite{bib:Arnold1}\cite{bib:Silva}.

\subsection{Phase Space Manifold}
\label{subsec:PSmanifold}

From the geometric perspective, the dynamics of a system occur on its phase space, where each point encodes the entire state of the system at that time.  Our phase space is also identical to its configuration space, the space of vortex positions.  We will denote a point in this space as $\bfz \equiv \left\{x_1, \ldots, x_N, y_1, \ldots, y_N \right\}$, where a single vortex position in the plane is $\bfr_i \equiv \lrbk{x_i,y_i}$.  Since we are considering all positive strength vortices, no two vortices will ever collide (see comments relating to integrals of motion in section \ref{sec:IntMotSymmChaos}).  Thus we must exclude the collision set
\be
\mathcal{C} = \left\{ \bfz \in \left(\mathbb{R}^2\right)^N \mid \bfr_i = \bfr_j \text{ for at least one pair } i \neq j \right\}
\label{eq:CollisionSet}
\ee
from our phase space ($\myman = \left(\mathbb{R}^2\right)^N - \mathcal{C}$).  Hence, the manifold on which we consider the dynamics to take place is
\be
\myman = \left\{ \bfz \in \left(\mathbb{R}^2\right)^N \mid \bfr_i \neq \bfr_j \text{ for } i \neq j \right\}.
\label{eq:PVmanifold}
\ee
Locally, $\myman$ looks just like regular Euclidean space $\mathbb{R}^{2N}$.  Globally, however, the two are very different due to the collision set.  In fact, one can show that this set forms enough of an obstruction to prohibit one dimensional loops from being deformed into one another.  Since the periodic solutions we are looking to find are just one dimensional loops, the topological structure of all possible loops on this manifold is particularly interesting.  Consider two loops to be similar if one can be deformed into the other.  The set of all such similarity classes forms what we refer to as the first homotopy group, or fundamental group $\pi_1$.  As the name implies, these classes have a group structure, which in the case of $\myman$ is the pure braid group $\pi_1\left(\myman\right) = PB_N$ on N strands.  As this topological feature is such a large aspect of the categorization of periodic orbits,  we have devoted a whole chapter to it, ch.~(\ref{ch:braid}), and will hold off on any further discussion here.

\subsection{Differential Geometry}

Continuing with the geometric point of view, we can consider the time evolution of our vortex configuration to be the flow on $\myman$ due to a vector field.  To coherently deal with vector fields, we must introduce a few concepts from differential geometry.  For a review of this material see~\cite{bib:Arnold1}.  We call the space of vector fields the tangent bundle $T\myman$, which reflects the notion that it is a collection of the tangent vector spaces at each point, with some notion of continuity from point to point.  In this view a vector field picks out an element of the tangent space at each point of the manifold, and we say that a vector field is a section of the tangent bundle.  Consider  for a moment a single vector field $\bfX$ on our manifold.  In local coordinates, which in our case are valid everywhere on the manifold, we can express this arbitrary vector field as
\be
\bfX\left(\bfz\right) = \sum^N_i a_i\left(\bfz\right)\pd{}{x_i}+b_i\left(\bfz\right)\pd{}{y_i}  = \sum^N_i a_i\left(\bfz\right)\partial_{x_i}+b_i\left(\bfz\right)\partial_{y_i},
\ee
where $\bfz \in \lrp{\mathbb{R}^2}^N$, and $\left\{ a_i\left(\bfz\right),b_i\left(\bfz\right)\right\} $ are real valued functions.  Notice that the set of differential operators $\left\{ \partial_{x_i},\partial_{y_i}\right\}$ act as the basis for the tangent space at any point on the manifold.  There is a dual space to this tangent space, whose basis we will denote as $\left\{ dx_i,dy_i\right\}$.  Thus we have $\intprod{\partial_{x_i}}{dx_j} = \intprod{\partial_{y_i}}{dy_j} = \delta_{ij}$ and $\intprod{\partial_{x_i}}{dy_j} = \intprod{\partial_{y_i}}{dx_j} = 0$, where $\bfimath$ is the basis contraction or interior product and $\delta_{ij}$ is the usual Kronecker delta.  This dual space is the space of 1-forms.  We can generically write a 1-form $\bfalpha$ as
\be
\bfalpha\left(\bfz\right) = \sum^N_i \alpha_i\left(\bfz\right) dx_i+\beta_i\left(\bfz\right) dy_i.
\ee

There are a number of equivalent ways to think about 1-forms.  Just as a single vector field is a section of the tangent bundle over our manifold $T\myman$, a 1-form is a section of the cotangent bundle $T^{\ast}\myman$.  From a more functional point of view, a 1-form acts as a linear map that takes vector fields to $\mathbb{R}$-valued scalar fields.  Restricted to a point on the manifold, a 1-form takes vectors to real numbers.  In particular, we can view the above $\bfalpha$ as acting on $\bfX$ via the interior product
\be
\bfalpha\left(\bfX\right)\left(\bfz\right) \equiv \intprod{\bfXs}{\bfalpha}\left(\bfz\right) =  \sum^N_i \left(\alpha_i a_i+\beta_i b_i\right)\left(\bfz\right).
\ee
If we are simply dealing with a Euclidean vector space then there is no difference between vector fields and 1-forms, and this contraction is locally the dot product.  However, as we shall see, Hamiltonian dynamics necessitates the use of Symplectic manifolds, where this distinction is important.  

In a similar functional vein, we can think of 2-forms as bilinear, skew-symmetric functions that take two vector fields and return a scalar field
\begin{align}
\bfomega^2\left(a\bfX_1+b\bfX_2, \bfX_3\right) &= a\bfomega^2\left(\bfX_1, \bfX_3\right) + b\bfomega^2\left(\bfX_2, \bfX_3\right), \nonumber\\
\bfomega^2\left(\bfX_1, \bfX_2\right) &= -\bfomega^2\left(\bfX_2, \bfX_1\right).
\end{align}

Analogously, $k$-forms can be thought of as taking $k$ vector fields to a scalar field $\bfomega^k\left(\bfX_1,\cdots, \bfX_k\right) \in C^\infty\left(\myman\right)$.  The interior product of a $k$-form with a vector field is a $k-1$ form, which can be given by $\intprod{\bfXs}{\bfomega^k} \left(\bfX_1,\cdots, \bfX_{k-1}\right) = \bfomega^k\left(\bfX,\bfX_1,\cdots, \bfX_{k-1}\right)$.  Because of this step down from $k$-forms to $k-1$ forms due to the interior product, we can think of scalar fields as 0-forms.  In the opposite direction, there is a way of taking two lower order forms and creating a higher order form.  We call this the exterior, or wedge product between two forms $\extprod{\bfalpha}{\bfbeta}$.  If given a $k$-form and a $j$-form, the wedge product gives us a $k+j$ form.  This product is skew-symmetric, meaning that $\extprod{\bfalpha}{\bfbeta} = - \extprod{\bfbeta}{\bfalpha}$.  In particular this means that the exterior product of any form with itself is zero.  With this in mind we can express any $k$-form as the wedge product of the $k$ basis 1-forms
\be
\bfomega^k = \sum_{i_1 < \cdots < i_k} a_{i_1 ,\cdots , i_k}\left(\bfz\right) dz_{i_1}\wedge \cdots \wedge dz_{i_k},
\label{eq:kformdecomp}
\ee
where $dz_{i_j}$ are one of the $2N$ elements of the basis for 1-forms $\left\{ dx_1, \cdots, dx_N, dy_1, \cdots, dy_N\right\}$, and $\left\{i_1 ,\cdots , i_k\right\}$ is just an ordered (increasing) set of integers between 1 and $2N$.  Notice that there are $\binom{2N}{k}$ basis k-forms.  In particular there is only one basis element for $2N$-forms, the top form, and it is a differential volume element for the manifold.  There are no k-forms for $k > 2N$.  To do actual calculations we must know the rule for evaluating a k-form on k vector fields.  Due to Eq.~(\ref{eq:kformdecomp}), it is sufficient to evaluate $dz_{i_1}\wedge \cdots \wedge dz_{i_k}$ on k vector fields $\bfX_1, \cdots,\bfX_k$
\be
 dz_{i_1}\wedge \cdots \wedge dz_{i_k} \left(\bfX_1, \cdots,\bfX_k \right) =
 \begin{vmatrix}
 dz_{i_1}\left(\bfX_1\right) & \cdots & dz_{i_k}\left(\bfX_1\right) \\
  \vdots  & \ddots & \vdots  \\
 dz_{i_1}\left(\bfX_k\right) & \cdots & dz_{i_k}\left(\bfX_k\right) 
 \end{vmatrix}
\ee

Since the coefficients in a basis expansion of a k-form are just functions over $\myman$, we can think of differentiating this k-form.  In particular, we can define the exterior derivative, $d$, as acting on a k-form and giving a k+1 form.  If $\bfomega^k = \sum a_{i_1 ,\cdots , i_k} dz_{i_1}\wedge \cdots \wedge dz_{i_k}$, then 
\be
d\bfomega^k = \sum_{i_1 < \cdots < i_k} da_{i_1 ,\cdots , i_k} \wedge dz_{i_1}\wedge \cdots \wedge dz_{i_k},
\ee
where
\be
da =  \sum^{2N}_i \pd{a}{z_i} dz_i.
\label{eq:extderfunc}
\ee
Importantly $d^2 = 0$, or differentiating any k-form twice gives us zero.  This exterior derivative generalizes the vector calculus notions of divergence, grad, and curl.  With it we can state all variations of Stokes', Green's, or Gauss' integral relations rather succinctly
\be
\int_{\partial c}\bfomega^k = \int_{c}d \bfomega^k,
\label{eq:Stokes}
\ee
where, very loosely, $c$ is a k+1 dimensional subset of our manifold, and $\partial c$ is its boundary.  Our main interest in the external derivative is that we can associate the Hamiltonian vector field with the exterior derivative of the Hamiltonian function.  The exact nature of this association is explored in the next section.

\subsection{Symplectic Structure}

The relationship between vector fields and differential 1-forms is essentially the same thing as the relationship between covariant and contravariant vectors in Riemannian geometry.  And much as Riemannian geometry has a special structure, the metric $g_{\mu\nu}$, which allows one to associate covariant vectors with contravariant vectors and vice versa, our manifold is equipped with an analogous structure, the symplectic 2-form $\bfomega^2$.  It is therefore a symplectic manifold $\left( \myman, \omega^2\right)$, as are all manifolds on which Hamiltonian dynamics occur.  In this section, we will explain some of the implications of having our dynamics occur on a symplectic manifold.

To be a symplectic 2-form, $\bfomega^2$ must be closed and non-degenerate.  Closed simply means that $d\bfomega^2 = 0$.  The non-degeneracy condition says that if we take any non-zero vector field $\bfX_1$, then we can find another vector field $\bfX_2$ such that $\bfomega^2\left(\bfX_1,\bfX_2\right) \neq 0$.  In our global coordinates the symplectic 2-form is
\be
\bfomega^2 = \sum^N_i \Gamma_i dx_i\wedge dy_i.
\label{eq:sym2form}
\ee

We can use this 2-form to create the connection between vector fields and 1-forms in the following way.  Consider two vector fields $\bfX_1 = \sum^N_i \left(a_i\partial_{x_i} + b_i\partial_{y_i}\right)$ and a generic one $\bfX_2$.  Using some of the ideas in the last section we can calculate
\be
\bfomega^2\left(\bfX_1,\bfX_2\right) = \sum^N_i \Gamma_i dx_i\wedge dy_i \left(\bfX_1,\bfX_2\right) = \sum^N_i \Gamma_i \left(a_idy_i -b_i dx_i\right)\left(\bfX_2\right).
\ee
Therefore $\intprod{\bfXs_1}{\bfomega^2} \equiv \bfomega^2\left(\bfX_1, \ast \right)$ is a 1-form.  We will abuse notation slightly, and refer to 1-forms and vector fields both as $\langle a_1, \cdots, a_N, b_1, \cdots, b_N \rangle$, where the difference in respective basis is neglected.  This isomorphism induced by $\bfomega^2$ can be understood as taking a vector field $\langle a_1, \cdots, a_N, b_1, \cdots, b_N \rangle$ to a 1-form $\langle -\Gamma_1 b_1, \cdots, -\Gamma_N b_N, \Gamma_1 a_1, \cdots, \Gamma_N a_N \rangle$.  For our purposes, the inverse of this transformation is more useful.  It is easily verified that this transformation exists, and takes a 1-form $\langle a_1, \cdots, a_N, b_1, \cdots, b_N \rangle$ to a vector field $\langle \frac{1}{\Gamma_1} b_1, \cdots, \frac{1}{\Gamma_N} b_N, -\frac{1}{\Gamma_1} a_1, \cdots, -\frac{1}{\Gamma_N} a_N \rangle$.  If we consider the 1-forms and vector fields to be column vectors, we can then express the inverse mapping, which we will denote $\pten$, as a matrix
\be
 \pten = 
 \begin{pmatrix}  0 & \mathbb{A} \\
  -\mathbb{A} & 0
 \end{pmatrix},  \; \; \text{where} \; \;
 \mathbb{A} = 
 \begin{pmatrix} 
  \frac{1}{\Gamma_1} &   & \text{\huge{0}}  \\
   &  \ddots & \\
   \text{\huge{0}} &  & \frac{1}{\Gamma_N}
 \end{pmatrix}.
 \label{eq:PoissonTensor}
 \ee

We will sometimes refer to this as the Poisson tensor.  So, for a given 1-form $\bfxi$ we have an associated vector field $\bfX = \pten \bfxi$.  In particular, we will be using this formalism to define Hamiltonian vector fields $\bfX_{H}$ generated by a function H
\be
\bfomega^2\left(\bfX_H,\ast \right) = dH \;\; \text{OR} \;\; \bfX_H = \HamVec{H}.
\label{eq:HamVecField}
\ee
Notice that we can write the equations of motion as $\dot{\bfz} = \HamVec{H}$.  So use of our Hamiltonian Eq.~(\ref{eq:VortexHamiltonian}), the exterior derivative of a function Eq.~(\ref{eq:extderfunc}), and the Poisson tensor Eq.~(\ref{eq:PoissonTensor}) will give us the vortex equations of motion Eq.~(\ref{eq:PVevolution}).  We will be using this notation henceforth.

From the geometric perspective, we can formally define our Hamiltonian system as the triple $\left\{ \myman,\bfomega^2, H\right\}$.  We have defined on our symplectic manifold $\left\{ \myman,\bfomega^2\right\}$ a function H.  The exterior derivative of H, a 1-form $dH$, can be associated with a vector field due to the symplectic structure.  The vector field then induces a flow of points on the manifold.  Since the manifold is the phase space, this flow is the evolution of our point vortex system.

We can think of this flow as a differentiable map, $g^t_H\lrp{\bfz}$, of the manifold back to itself, a family of diffeomorphisms, parameterized by time.  This flow is implicitly defined by 
\be
\derivc{}{t}{t=0} g^t_H\lrp{\bfz} = \HamVec{H}\lrp{\bfz}.
\label{eq:HamFlow}
\ee
Knowing $g^t_H\lrp{\bfz}$ is tantamount to solving the point vortex problem.  If it is integrable, we can find an analytical form for it, if not, which is the more general case, we must solve for individual trajectories numerically.  The forward trajectory of a point in time is called the orbit of this point.  We are interested in finding orbits that close in on themselves, $g^t_H\lrp{\bfz^{\ast}} = \bfz^{\ast}$, or periodic orbits.  We will have more to say about such orbits in section~\ref{sec:IntMotSymmChaos} of this chapter.  For now, we will explore some important attributes of Hamiltonian flows and vector fields.

First of all, Hamiltonian flows preserve symplectic structure; both $\bfomega^2$ and $\pten$ do not change under the flow diffeomorphism $g^t_H$.  Maps with this property are called symplectomorphisms.  From a slightly different view, we can think of such maps as a simple change of coordinates $\bfz \rightarrow \bfz^{\ast} = g^t_H\lrp{\bfz}$.  The invariance of symplectic structure simply implies that the Hamiltonian equations remain unchanged under this coordinate transformation.  Symplectomorphisms are simply canonical coordinate transformations, and for each scalar field defined on our manifold, we can associate a one-parameter family of coordinate transformations that are canonical.  We will be using this and other similar methods of generating canonical transformations in section~\ref{sec:LieTrans}.  One last thing to note is that all powers of $\bfomega^2$ are invariant.  In particular the top form $\bfomega^{2N}$ is preserved, which leads to the conservation of phase space volume under Hamiltonian flows.

\subsection{Lie Derivatives and Poisson Brackets}

One of the more powerful reasons for using all of this geometric machinery is to easily and coherently deal with symmetries and conserved quantities of our dynamical system.  Consider for a moment the conservation of energy.  The quantity $dH\lrp{\bfX}$ is essentially the directional derivative of H in the direction of $\bfX$.  If we set $\bfX = \HamVec{H}$ then $dH\lrp{\HamVec{H}} = \bfomega^2\lrp{\HamVec{H},\HamVec{H}} = 0$.  This tells us that H is not changing along its flow direction, and therefore energy is a conserved quantity.

We will consider the other conserved quantities of the point vortex system in a moment, but first we need some additional structure.  Consider a general vector field, which is not necessarily Hamiltonian
\be
\derivc{}{t}{t=0} X^t\lrp{\bfz} = \bfX\lrp{\bfz}.
\ee
Define the Lie derivative of a function $\phi$ in the direction of the flow due to $\bfX$ as the directional derivative
\be
\lrp{\Lieder{\bfXs}{\phi}}\lrp{\bfz} = \derivc{}{t}{t=0} \phi \lrp{X^t\lrp{\bfz}}.
\label{eq:LieDerFunc}
\ee
So $\Lieder{\bfXs}{}$ is an operator that takes scalar fields to scalar fields.  We can expand this definition to include the action of the Lie derivative on a k-form $\bfalpha$
\be
\lrp{\Lieder{\bfXs}{\bfalpha}} = \intprod{\bfXs}{d\bfalpha} + d\lrp{\intprod{\bfXs}{\bfalpha}}
\label{eq:LieDerForm}
\ee

The action of a Lie derivative on vector and tensor fields is also defined, and again measures the change of these items in the direction of the flow.  We can use the Lie derivative to measure the degree to which two vector fields do not commute.  Consider the commutator of two Lie derivatives, which surprisingly turns out to be the Lie derivative with respect to some other vector field
\be
\lrbc{\Lieder{\bfXs_1}{}, \Lieder{\bfXs_2}{} } = \Lieder{\bfXs_1}{}\Lieder{\bfXs_2}{} - \Lieder{\bfXs_2}{}\Lieder{\bfXs_1}{} = \Lieder{\bfXs_3}{}.
\ee

The relationship between this new vector field and the two others defines the Poisson bracket $\lrbc{\bfX_1,\bfX_2} = \bfX_3$.   If the flows locally commute, $X^t_1X^s_2X^{-t}_1X^{-s}_2\lrp{\bfz} = \bfz$, then the Poisson bracket is zero $\lrbc{\bfX_1,\bfX_2} = 0$.  If not, then the resulting flow is generated by the vector field $\bfX_3$.  This bracket is bilinear, skew-symmetric, and obeys the Jacobi identity, and is therefore a Lie bracket.  This makes the space of vector fields into a Lie algebra.

The Lie bracket is particularly useful if we restrict our attention to Hamiltonian vector fields.  In this case, the resulting vector field is also Hamiltonian $\lrbc{\bfX_P,\bfX_F} = \bfX_H$.  So Hamiltonian vector fields form a subalgebra of the Lie algebra of general vector fields.  Furthermore, $H = \lrbk{P,F}$, also called the Poisson bracket (of functions), where
\be
\lrbk{P,F} = dP\lrp{\HamVec{F}} = -dF\lrp{\HamVec{P}} = \bfomega^2\lrp{\HamVec{F},\HamVec{P}}.
\ee
In our coordinates we can write out the Poisson bracket as
\be
\lrbk{P,F} =
\sum_i^N\frac{1}{\Gamma_i}\left(
\pd{P}{x_i}\pd{F}{y_i} -\pd{P}{y_i}\pd{F}{x_i}
\right).
\label{eq:pb}
\ee

Notice that Hamiltonian functions form a Lie algebra, with this Poisson bracket as the Lie bracket.  We are interested in certain elements of this Lie algebra, specifically those that Poisson commute with the primary Hamiltonian function H:  $\lrbk{F_i} = \lrbk{F_i \in C^\infty\lrp{\myman} \mid \lrbk{F_i,H} = 0}$.  These functions are first integrals of motion and are constant along orbits $F_i\lrp{g^t_H\lrp{\bfz}} = F_i\lrp{\bfz}$.  To see this more clearly, note that $\lrbk{F_i,H} = dF_i\lrp{\HamVec{H}} = \Lieder{\HamVecm{H}}{F_i}$, i.e. the derivative of $F_i$ along the flow direction $\HamVec{H}$ is zero if $F_i$ and $H$ Poisson commute.  Another, perhaps more familiar way of writing this is $\dot{F_i} =\lrbk{F_i,H}$.

With this formalism, we can once again establish that H itself is a first integral of motion and energy is conserved by calculating $\deriv{H}{t} = \lrbk{H,H} = 0$.  That this is the case is due to the skew-symmetry of the bracket.  We can even write the canonical Hamiltonian equations of motion in this fashion
\begin{align}
\frac{dx_j}{dt} &= \left\{x_j, H\right\} \nonumber \\
\frac{dy_j}{dt} &= \left\{y_j, H\right\}.
\end{align}

We also can easily calculate the Poisson bracket of vector fields as $\lrbc{\HamVec{F_i},\HamVec{H}} = \HamVec{\lrbk{F_i,H}}$.  This immediately tells us that the flow of a first integral commutes with the system flow.  Lastly, we can consider the Poisson bracket of two first integrals of motion $\lrbk{F_1,F_2}$.  Due to the Jacobi identity
\be
\lrbk{\lrbk{F_1,F_2},H} + \lrbk{\lrbk{H,F_1},F_2} + \lrbk{\lrbk{F_2,H},F_1} = 0.
\label{eq:jacobi}
\ee
Due to the definition of first integrals, the right two terms disappear and we have that $\lrbk{\lrbk{F_1,F_2},H} = 0$.  This means that first integrals are closed under the Poisson bracket, and they form a subalgebra of the Lie algebra of scalar fields.

\section{Integrals of Motion,  Symmetry, and Chaos}
\label{sec:IntMotSymmChaos}

Imagine that we have found a set of integrals of motion for our point vortex system: $\bfF = \lrbk{F_0 = H, F_1, \cdots}$.  A single one of these first integrals defines a set of submanifolds, each of co-dimension 1, and each taking a constant value of this first integral on itself.  These submanifolds are referred to as level sets of this first integral $\lrbk{\bfz} = \lrbk{\bfz \in \myman \mid F_i\lrp{\bfz} = f_i \in \mathbb{R}}$.  For instance, we already have conservation of energy, and therefore constant energy submanifolds.  The flow of a phase space point is constrained to lie on the intersection of the level sets for each integral of motion.  We will refer to this submainifold as $\mymans$, where $\mymans = \lrbk{\bfz \in \myman \mid \bfF\lrp{\bfz} = \lrbk{E,f_1,\cdots}}$.  If we have a sufficient number of first integrals, meeting some criteria to be defined shortly, then this geometric constraint is enough to ensure that the phase space motion is integrable (Liouville Theorem).  One of the criteria is that all of the first integrals Poisson commute with one another $\lrp{\lrbk{F_i,F_j} = 0 \;\text{for all}\; i,j}$.  This extra constraint on the first integrals ensures that the Hamiltonian vector fields associated with each first integral will be tangent to the submanifold $\mymans$.  Such a set of first integrals, that pair-wise Poisson commute, is said to be in involution, or an involute set.

We would certainly like to find a maximal set of first integrals in involution for our point vortex system.  To aid in identifying these first integrals, consider the result from last section: $\Lieder{\HamVecm{F_i}}{H} = 0$.  This says that the Hamiltonian is an invariant of the flow due to a first integral.  This is just another way to state N\"{o}ther's theorem.  For every constant of the motion, $F_i$, there is an associated continuous symmetry of the Hamiltonian function and vice versa.  The continuous symmetry can be associated with the flow, $g^t_{F_i}\lrp{\bfz}$, which is generated by the first integral Hamiltonian vector field, $\HamVec{F_i}$.  If you look at the Hamiltonian, Eq.~(\ref{eq:VortexHamiltonian}), there are a few obvious continuous symmetries that it admits when thinking about the vortices lying on the plane.  Since only the differences in vortex positions appear, a translation of the whole vortex configuration in the $x$ or $y$ direction is one such symmetry.  Furthermore, since only the magnitude of the differences appears, rigid rotations of the vortex configuration about any point in the plane are another symmetry.  The first of these gives the linear momentum in each direction, or linear impulse
\begin{subequations}
\begin{align}
Q &= \sum^N_i \Gamma_i x_i, \\
P &= \sum^N_i \Gamma_i y_i.
\label{eq:LinearImpulse}
\end{align}
\end{subequations}
From these we can define a point on the plane called the center of vorticity, which is also constant in time, and about which the vortices will rotate
\be
\langle X, Y \rangle = \frac{1}{\sum \Gamma_i}\langle Q, P \rangle.
\label{eq:CenterOfVorticity}
\ee
For an equal number of positive and negative vortices of equal strength, the center of vorticity is not defined.  This means that such a configuration is unbounded, and pairs of opposite signed vortices can pair up and translate off to infinity.  This is one reason why we will only be considering like signed vortices, in addition to the fact that such configurations arise as the result of a rotating superfluid.

The vector fields due to these first integrals are particularly simple
\begin{subequations}
\begin{align}
\HamVec{Q} &= \langle 0, \cdots, 0, 1, \cdots, 1 \rangle, \\
\HamVec{P} &= \langle -1, \cdots, -1, 0, \cdots, 0 \rangle.
\label{eq:MomentumVectorField}
\end{align}
\end{subequations}
Indeed this is just bulk linear motion of the vortex configuration in the plane.  Next is the rotational symmetry.  Since the center of vorticity is invariant, we will consider this to be the point about which the rigid body rotations take place.  This gives a first integral, the angular momentum or angular impulse
\be
L = \sum^N_i\Gamma_i\lrp{\lrp{x_i-X}^2+\lrp{y_i-Y}^2}.
\label{eq:AngularMomentum}
\ee
The associated vector field is
\be
\HamVec{L} = \langle -\lrp{y_1-Y}, \cdots, -\lrp{y_N-Y}, \lrp{x_1-X}, \cdots, \lrp{x_N-X} \rangle.
\label{eq:AngularMomentumVector}
\ee
This is, as promised, a rigid clockwise rotation about the center of vorticity.  However, this set of $\lrp{H,Q,P,L}$ is not in involution
\be
\lrbk{Q,P} = \sum \Gamma_i, \;\; \lrbk{Q,L} = 2P, \;\; \lrbk{P,L} = -2Q.
\label{eq:PQdonotPcommute}
\ee

Instead of using P and Q, consider the single first integral 
\be
Q^2+P^2 = \lrp{\sum \Gamma_i x_i}^2 + \lrp{\sum \Gamma_i y_i}^2,
\label{eq:P2Q2integral}
\ee
which commutes with L: $\lrbk{Q^2+P^2,L} = \lrbk{Q^2,L} + \lrbk{P^2,L} = 2Q\lrbk{Q,L} + 2P\lrbk{P,L} = 4QP - 4PQ = 0$.  Now we have a set of first integrals of motion in involution $\lrp{H,Q^2+P^2,L}$.  It can be shown that this is a maximal set, and there are no more integrals of motion that we could add to this set \cite{Aref:1983up}.  This has important implications for integrability as we shall see in a moment.  However, let us first look at what continuous symmetry the first integral $Q^2+P^2$ corresponds to.  Its Hamiltonian vector field is
\begin{align}
\HamVec{\lrp{Q^2+P^2}} &= 2\langle-P,\cdots,-P,Q,\cdots,Q\rangle \nonumber \\
&= 2\lrp{\sum \Gamma_i}\langle-Y,\cdots,-Y,X,\cdots,X\rangle.
\end{align}
\label{eq:P2Q2HamVec}

We can think of the flow of this vector field as rotating the center of vorticity, and therefore the vortex configuration, about the origin without the configuration itself rotating about its center of vorticity.  Since an increase in $Q^2+P^2$ translates the center of vorticity further away from the origin, we can think of  $Q^2+P^2$ and the angular parameterization of its flow as polar coordinates for the linear impulses P and Q.  

Returning to the issue of integrability, consider a fundamental result by Liouville.  If a Hamiltonian system with N degrees of freedom (a 2N dimensional phase space) has N independent first integrals of motion in involution (including H), then the system is integrable.  Since we have three such integrals, the motion of three vortices is integrable, while that of four vortices is generically chaotic.  This is the primary justification for our focus on the periodic orbits of three and four vortices.  We would like to better understand the nature of this transition to chaos via adding additional vortices.

Consider for a moment the three vortex case in more detail.  We have established that a level set of our manifold is $\mymans = \lrbk{\bfz \in \myman \mid H\lrp{\bfz} = E, \lrp{Q^2+P^2}\lrp{\bfz} = c_1, L\lrp{\bfz} = c_2}$.  Notice that on its own the level set of L forms a $\lrp{2N-1}$-sphere centered on the center of vorticity.  This is compact, and therefore $\mymans$ is also compact (for any number of vortices).  Because of this compactness, we can say that $\mymans$ is topologically equivalent to a 3-torus.  In our case, with equal strength and sign vortices, we can go a little further by noting that the vector field $\HamVec{\lrp{Q^2+P^2}}$ not only Poisson commutes with the other two vector fields $\HamVec{H}$ and $\HamVec{L}$, but also is everywhere orthogonal to them using the Euclidean metric
\be
\HamVec{\lrp{Q^2+P^2}}\cdot \HamVec{H} = \HamVec{\lrp{Q^2+P^2}}\cdot \HamVec{L} = 0.
\ee
This tells us that none of the phase space flow in time is in the direction of the flow due to $\lrp{Q^2+P^2}$.  Effectively, this means that the motion of phase space points is constrained even further to a 2-torus.  

Another result that follows from the Liouville theorem is that this invariant torus can be described by a set of canonical variables called action-angle coordinates.  In the N = 3 case, they consist of three angle variables $\lrbk{\phi_1,\phi_2,\phi_3}$ which are defined mod $2\pi$.  Their conjugate variables, the action variables $\lrbk{I_1,I_2,I_3}$, are only dependent on the involute first integrals of motion, and are therefore constants of the motion themselves.  Because of this, we have that $\dot{I_i} = 0 = -\partial_{\phi_i}H$, and the Hamiltonian is not a function of any of the angle variables.  Looking at the conjugate equations of motion gives: $\dot{\phi_i} = \partial_{I_i}H = \omega_i\lrp{I_1,I_2,I_3} = const.$, and therefore, the angle variables increase linearly in time: $\phi_i\lrp{t} = \omega_i t + \phi_i\lrp{0}$.

Another way to say that the phase space flow occurs on a 2-torus, is to have a set of canonical action angle variables where $\omega_i = 0$ for one of them.  Then the 2-torus is defined by the remaining angle variables.  Since $Q,P$ and $Q^2+P^2$ are always conserved and play no significant part in the topology of phase space, We will set them all to zero from now on.  This is equivalent to setting the center of vorticity to coincide with the origin. We will also be setting the energy of the point vortex system to zero.  To justify this, consider a coordinate transformation that radially expands the vortices outward from the center of vorticity (which is now at the origin) by a factor a:  $\bfz \rightarrow a\bfz$.  If we assume that time is also dilated, $t \rightarrow a^2 t$, then the equations of motion are invariant.  Under this transformation the energy changes as
\be
E \rightarrow E - \frac{1}{4\pi}\lrp{\sum_{i}\sum_{j\neq i} \Gamma_i\Gamma_j}\ln{a}.
\label{eq:EnergyDilation}
\ee
Thus, if we have a periodic orbit, we can map it to another periodic orbit of any energy.  The only real difference between periodic orbits in this one parameter family is that those on higher energy level sets will have smaller periods.  The angular momentum also changes under this transformation as: $L \rightarrow a^2L$.  Notice that under this transformation the energy is a monotonically decreasing function of the angular momentum
\be
E\lrp{L} = E_0 - \frac{1}{8\pi}\lrp{\sum_{i}\sum_{j\neq i} \Gamma_i\Gamma_j}\ln{\frac{L}{L_0}}.
\ee
Thus, we would be justified in setting either H or L to a specific value, though not both.  Since L is always positive and H can be either positive or negative, the only ``natural" choice is to set the energy equal to zero.  We will do this for the remainder of the work.  One last point about this transformation:  it effectively tells us that there is no absolute preferred length scale in the system, only relative length scales.  This is an artifact of the point vortices having no physical extent.  Realistically this is only valid when the inter-vortex lengths are much larger than any diameter of the vortex core.  This provides another justification for only considering positive vortices, because the inter-vortex lengths for positive vortices can be shown to be bounded from above and below.  We can then restrict our attention to values of L whose associated submanifold does not have a large disparity between the largest and smallest inter-vortex scales.  This turns out to be equivalent to restricting our attention to smaller values of L.

Now, the goal of this work is to find and categorize the periodic orbits of the point vortex system.  Since we have ``modded" or quotiented  out the first integrals H, Q, and P, changes in the periodic orbits can only be parameterized by the remaining integral of motion, the angular momentum L.  In general, the value of the angular momentum will help determine the type of periodic orbit that is allowed to exist on the invariant submanifolds $\mymans$.  We will refer to these level sets of the energy momentum mapping as $\mymanL = \lrbk{\bfz \in \myman \mid H\lrp{\bfz} = Q\lrp{\bfz} = P\lrp{\bfz} = 0, L\lrp{\bfz} = L}$.  One major organizing principle in this vein originates again with the Liouville theorem.  One of the requirements on our involute integrals of motion is that they are independent of each other.  This is equivalent to requiring that the vector fields generated by these first integrals be linearly independent.  When they are not, the topology of the invariant manifold changes.  For us, the vector field due to $\lrp{Q^2+P^2}$ is always linearly independent of the other two, and therefore provides no problems.  The vector fields due to H and L are for the most part linearly independent of each other, except at a finite number of submanifolds with particular values of L.  At these values of L, the vector fields $\HamVec{L}$ and $\HamVec{H}$ are antiparallel.  Since $\HamVec{L}$ generates rigid body rotations about the center of vorticity, the vortex motion for these values of L are rigid rotations as well.  We call these periodic orbits relative equilibria, because they are equilibrium points of the manifold when it has been reduced by identifying all phase space points that are rotationally equivalent to each other as a single reduced phase space point.

Many people have made a cottage industry of finding these relative equilibria \cite{CAMPBELL:1979un}\cite{LaurentPolz:2004je}.  The Abrikosov lattice mentioned in chapter~\ref{ch:phys} is one such solution.  Another is the set of N collinear vortices, whose locations along the axis have been shown~\cite{Aref:2011ce}\cite{bib:Aref3} to correspond intriguingly to the zeros of the Nth Hermite polynomial.  For the $N = 3$ and $N = 4$ case, there are respectively two and three such relative equilibria.  

In addition to the continuous symmetries corresponding to the integrals of motion and the overall scale invariance, there are a number of discrete symmetries.  Under permutation of vortex indices the Hamiltonian remains the same.  This is due to the pairwise interactions, and the fact that each pairwise term is present in the Hamiltonian.  If we have that all vortex strengths are the same, then the vortices are effectively identical.  In this case, some people choose to picture the periodic solutions as existing on the phase space with the symmetric group quotiented out $\myman / S_n$, rather than the full phase space.  Here the fundamental group is the full braid group $\pi_1\lrp{\myman \/ S_n} = B_n$ rather than the pure braid group.  Even though we are considering equal strength vortices, we choose to think of the periodic orbit in the full phase space as a personal preference.

Consider the three operations: ``charge" conjugation $C: \lrbk{\Gamma_i} \rightarrow \lrbk{-\Gamma_i}$, time reversal $T: t \rightarrow -t$, and parity reversal $P: \lrbk{x,y} \rightarrow \lrbk{-x,y}$.  All three are non-canonical in that they transform the equations of motion from $\dot{\bfz} = \HamVec{H}$ to $\dot{\bfz} = -\HamVec{H}$.  Therefore any combination of two of these operations will be a discrete symmetry of the system.  For our purposes, the most interesting is the PT symmetry.  It generically maps a point in phase space to a different point.  The flow of these two points will mirror each other.  The periodic orbits will fall into two classes:  periodic orbits which are mapped, via PT, back onto themselves, or at least to another part of the submanifold $\mymanL$ that is reachable from the original by time translation and a rigid body rotation;  and periodic orbits whose PT image is distinct.  We will refer to these as non-chiral and chiral pairs respectively, and will have much more to say about this topic in chapter~\ref{ch:class}.

\section{Point Vortex Motion}
\label{sec:PtVM}

Now that we have outlined a number of abstract results, we will consider the specific solutions to the point vortex problem when there are $N = 2, 3$, and $4$ vortices.  In the case of a single vortex, there are no other vortices to be advected by, so this vortex does not move.  There are sufficiently many integrals of motion to constrain the submanifold $\mymanL$ to a single point.

\subsection{Two Vortices}
\label{sec:pvm2}

The solution to the two vortex problem is particularly simple.  Every solution is a relative equilibrium, where the two vortices rotate about their center of vorticity.  In the case of two opposite signed vortices of equal strength, this center of vorticity is at infinity and the two translate along the line that bisects them.  This is analogous to a propagating smoke ring in three dimensions.  For positive vortices, the center of vorticity lies in between the two.  For later use, in chapter~\ref{ch:alg}, we will solve the general case of two positive vortices of strengths $\lrbk{\Gamma_1,\Gamma_2}$ and locations $\bfr_1 = \lrang{x_1,y_1}$ and $\bfr_2 = \lrang{x_2,y_2}$.  The Hamiltonian is
\be
H = -\frac{\Gamma_1\Gamma_2}{2\pi}\ln{\left| \bfr_1-\bfr_2 \right|}.
\ee
Since the linear moments are conserved, we can transform to new coordinates $\lrbk{\bfr,\bfR}$, where one pair is the center of vorticity or circulation
\begin{subequations}
\begin{equation}
\bfR\equiv\frac{\Gamma_1\bfr_1+\Gamma_2\bfr_2}{\Gamma_1+\Gamma_2},
\label{eq:CenterOfCirculation}
\end{equation}
and the other pair is the relative displacement
\begin{equation}
\bfr\equiv\bfr_1 - \bfr_2.
\label{eq:RelativeDisplacement}
\end{equation}
\end{subequations}
We can eliminate $\bfr_1$ and $\bfr_2$ in favor of $\bfR=\langle X,Y\rangle$ and $\bfr=\langle x,y\rangle$ by using the inverse transformation
\begin{subequations}
\begin{eqnarray}
\bfr_2 &=& \bfR - \frac{\Gamma_1}{\Gamma_2 + \Gamma_1}\bfr\label{eq:inversem}\\
\bfr_1 &=& \bfR + \frac{\Gamma_2}{\Gamma_2 + \Gamma_1}\bfr.\label{eq:inversen}
\end{eqnarray}
\end{subequations}

The Poisson bracket relations among the new coordinates may be readily calculated to be:
\begin{subequations}
\begin{equation}
\left\{ X, Y \right\} = \frac{1}{\Gamma_R},
\label{eq:PBR1}
\end{equation}
\begin{equation}
\left\{ x, y \right\} = \frac{1}{\Gamma_r},
\label{eq:PBR2}
\end{equation}
and
\begin{equation}
\{x, Y\} = \{y, X\} = 0,
\label{eq:PBR3}
\end{equation}
\end{subequations}
where the total circulation of the vortex pair is
\begin{subequations}
\begin{equation}
\Gamma_R\equiv\Gamma_2 + \Gamma_1,
\end{equation}
and the reduced circulation of the vortex pair is
\begin{equation}
\Gamma_r\equiv\frac{\Gamma_2\Gamma_1}{\Gamma_2 + \Gamma_1}.
\end{equation}
\end{subequations}
It follows that $X$ and $Y$ are a canonically conjugate pair, as are $x$ and $y$.  The Hamiltonian is now
\be
H = -\frac{\Gamma_R\Gamma_r}{2\pi}\ln{\left|\bfr \right|}.
\ee
Notice that neither $X$ or $Y$ appear in this, and are therefore constants of the motion (as we designed them to be).  Next, consider the conservation of angular momentum $L$, which we can express as
\be
L = \Gamma_r\lrp{x^2+y^2}.
\ee
This implies that the distance between the two vortices is constant.  We define the action variable $J$ and the angle variable $\theta$,
\begin{equation}
J \equiv \frac{1}{2}\left|\bfr\right|^2 = \frac{1}{2}\left(x^2 + y^2\right), \; \; \;
\theta\equiv\arg\left(y + ix\right).
\label{eq:DimerActionAngle}
\end{equation}
We transform coordinates again, this time to eliminate $x$ and $y$ in favor of $\theta$ and $J$.  The Poisson bracket relations among the new coordinates may be calculated as,
\begin{eqnarray}
\left\{\theta, J\right\} &=&
\frac{1}{\Gamma_r}\left(
\frac{\partial\theta}{\partial x}\frac{\partial J}{\partial y} -
\frac{\partial\theta}{\partial y}\frac{\partial J}{\partial x}
\right)\nonumber\\ 
&=&
\frac{1}{\Gamma_r}\left(
\frac{y}{x^2+y^2}\;y -
\frac{-x}{x^2+y^2}\;x
\right)\nonumber\\
&=&
\frac{1}{\Gamma_r}.
\label{eq:Jthpbracket}
\end{eqnarray}
It follows that $\theta$ and $J$ comprise a conjugate pair of coordinates.  Overall, our coordinates are now $\bfz = \lrang{\theta,X, J, Y}$, and the Hamiltonian can be written
\be
H = -\frac{\Gamma_R\Gamma_r}{4\pi}\ln{\lrp{2J}}.
\ee
Since there is no dependance on $\theta$, we have that $\dot{J} = 0$ as expected.  The angle variable $\theta$ is linear in time
\begin{equation}
\dot{\theta} = \HamVec{H} = \frac{1}{\Gamma_r}\frac{\partial H}{\partial J} = -\frac{\Gamma_R}{4\pi J}.
\end{equation}
It follows that
\begin{equation}
\theta(t) = \theta_0-\Omega t,
\label{eq:unperturbed}
\end{equation}
where $\theta_0$ is a constant of integration, and where we have defined the rotation frequency
\begin{equation}
\Omega\equiv\frac{\Gamma_R}{4\pi J} = \frac{\Gamma_R\Gamma_r}{2\pi L} = \frac{\Gamma_1\Gamma_2}{2\pi L}.
\label{eq:RotationRate}
\end{equation}
In our case where $\Gamma_1 = \Gamma_2 =1$ and $E = 0$, we have that $L = \frac{1}{2}$ and $\Omega = 1/\pi$.  In summary, two positive vortices will rotate about their center of vorticity with a rotational frequency $\Omega$.

\subsection{Three Vortices}
\label{sec:pvm3}

The motion of three vortices is somewhat more difficult to solve.  The motion is integrable as there are three independent integrals of motion $\lrbk{H,L, Q^2+P^2}$ in involution.  In closed form, it is expressed in terms of Jacobi elliptic functions~\cite{Aref:1992ur}.  We will limit our explanation of the qualitative nature of the solutions in this section.

The orbits exist on a family of 2-tori, defined by the angle variables $\lrbk{\phi_1,\phi_2}$ and parameterized by the action variables $\lrbk{I_1,I_2}$.  We will be ignoring the other two, inconsequential directions corresponding to rigid body translations.  The angle variable are linear (mod $2\pi$) in time $\phi_i = \omega_i t$, where $\omega_i = \pd{H}{I_i}$.  Because the Hamiltonian does not depend on the angle variables, the angular frequencies only depend on the integrals of motion.  Since we have set $H=0$, these frequencies are functions of the angular momentum only, $\omega_i\lrp{L}$.

\begin{figure}[htb]
  \centering
  \large
  \scalebox{0.7}{\includegraphics{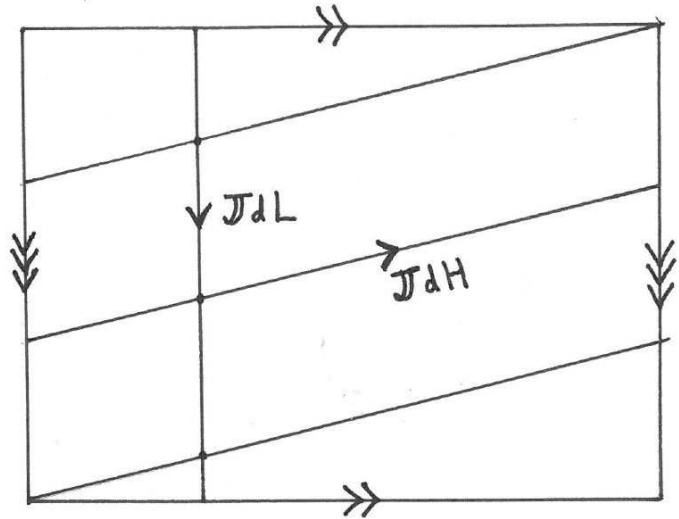}}
  \normalsize
    \caption{The fundamental domain of a torus is shown.  To reconstruct the torus, just connect the top and bottom edges and then the right and left edges.  They should be glued together such that the orientation of the arrows is consistent.  On the torus is shown an idealized periodic orbit, identified as the lines following the flow direction $\HamVec{H}$.  The rigid rotation symmetry direction is identified by $\HamVec{L}$.}
    \label{graph:TorusFdomain}
\end{figure}

General solutions are conditionally periodic and the orbit is dense on the torus.  This indicates that the ratio of the angular frequencies is irrational.  If this ratio is a rational number on the other hand, $\omega_1 m = \omega_2 n$ for integers m and n, then the orbit will close back on itself.  This periodic orbit will rotate around the $\phi_1$ direction of the torus n times for every m times it rotates about the $\phi_2$ direction.  We can view a periodic orbit as a straight line in the following cartoon, Fig.~(\ref{graph:TorusFdomain}), of the fundamental domain of the torus.

The ratio of angular frequencies gives the winding angle of the above plot.  Since this angle is functionally dependent on the angular momentum, it is reasonable to ask how it behaves as we increase L.  While a more detailed answer to this question must wait until chapter~\ref{ch:class} where we analyze the multitude of found periodic orbits, we will give a partial answer here.  Remember that $\HamVec{H}$ and $\HamVec{L}$ will be antiparallel for certain values of L, which we will denote as $L^{\ast}_i$, where there is a relative equilibrium solution.  At these values of L, the topology of $\mymanL$ changes, and the 2-torus collapses to a simple circle.  For two values of L that do not have an $L^{\ast}_i$ between them, we can continuously deform the action variables from one submanifold $\mymanL$ to the other.  Where there is $L_1 < L^{\ast}_i < L_2$, the action variables for the submanifold on either side of $L^{\ast}_i$ correspond to qualitatively different types of motion.

There are two relative equilibria for the $N = 3$ case.  The first occurs at $L^{\ast}_1 = 1$ and consists of a rigidly rotating equilateral triangle.  It is an easy exercise to show that this configuration has a period of $T = \frac{4}{3}\pi^2 = 13.1595$ and a distance of $r = 3^{-\frac{1}{2}} = 0.577$ from the center of vorticity to each vortex.  The next relative equilibria occurs at $L^{\ast}_2 = 2^{\frac{1}{3}} = 1.25992$ and consists of collinear vortices.  It has a period of $T = \frac{4}{3}\pi^2 2^{\frac{1}{3}} = 16.5799$ and a distance of $r = 2^{-\frac{1}{3}} = 0.7937$ from the center of vorticity, where the central vortex lies, to the outer two vortices.  Now we can consider there to be three classes of submanifold $\mymanL$ partitioned by the two relative equilibria.
\begin{itemize}
\item $\lrbc{L < L^{\ast}_1}$:	This region is actually quite trivial, in that the submanifold is empty $\mymanL = \lrbk{\emptyset}$.  This simply means that the respective level sets for L and $H = 0$ do not intersect.

\item $L = L^{\ast}_1$:	As we increase L, this is the first time that the energy and angular momentum level sets intersect.  Topologically $\mymanL$ is a circle, which is parameterized by the angle $\phi_1$ of the solid body rotation.

\item $\lrbc{L^{\ast}_1 < L < L^{\ast}_2}$:	Increasing L even more, we get a new angular variable $\phi_2$, corresponding to periodic orientation-preserving oscillations in the area of the triangle between the three vortices.  $\mymanL$ is now a 2-torus, parameterized by $\phi_1$ and $\phi_2$.

\item $L = L^{\ast}_2$:	The three vortices are collinear, and $\mymanL$ is once again a circle.  This is a very unstable periodic orbit, and small perturbations that change L slightly will lead to two qualitatively different motions.  A slight decrease in L will result in large oscillations in the area that preserve the orientation, while a slight increase will lead to equally large oscillations of the area that switch orientation twice per period.

\item $\lrbc{L^{\ast}_2 < L}$:	In this regime, two vortices are always closer together than the other two pairings.  These two rotate about each other, and then they as a single unit rotate with the remaining vortex.  Thus, there are two rotational motions, the ``inner" and ``outer" rotations.  These now correspond to $\phi_1$ and $\phi_2$, and $\mymanL$ is once again a 2-torus.  Notice how in going through $L^{\ast}_2$, the two motions that make up the torus rotational directions have qualitatively changed.  As L increases even further, the inner vortex pair gets closer together when compared to their distance from the remaining vortex.  
\end{itemize}

\subsection{Four Vortices}
\label{sec:pvm4}

The case of four vortices is where interesting behavior really starts to occur.  The motion is, in general, chaotic since there are not enough integrals of motion to reduce the phase space.  There are, however, a few exceptions.  If the total strength of the vortices is zero, $\sum \Gamma_i = 0$, then we have that Q and P Poisson commute with each other, Eq.~(\ref{eq:PQdonotPcommute}).  Furthermore, we can generically set $Q =P = 0$, which allows Q and P to Poisson commute with L.  The are now four involute integrals of motion, which is enough to force this special four vortex problem to be integrable.  This is the case when you have two ``leap-frogging" pairs of vortices.  Additional special cases occur when the vortex configuration has a high degree of symmetry.  The general idea that we can glean from this is that even in the identical vortex case, there will be regions of phase space where the motion is integrable, interspersed amongst the chaotic sea.

Fortunately there is a wealth of topological structure to help us make sense of the regions of phase space with qualitatively different motions.  Much of this structure is wrapped up in the topology of braids associated with the periodic orbits.  This is the main theme of this dissertation, and we will develop it further in the following chapters.  As in the $N=3$ case, we have relative equilibria to help sort out the families of $\mymanL$.  In our case $\mymanL$ is now four dimensional.  We could reduce it to three dimensions by modding out the angular momentum flow direction $\HamVec{L}$, which completes the symplectic reduction~\cite{bib:Abraham}, however we will not go this far.

There are now three relative equilibria.  The first, with an angular momentum of $L = 2^{\frac{2}{3}} = 1.5874$, is the square configuration.  It has a period of $T = \frac{2}{3}\pi^2 2^{\frac{2}{3}} = 10.4447$ and all four vortices are a distance $r = 2^{-\frac{2}{3}} = 0.62996$ from the center of vorticity.  Next is the equilateral triangle with one vortex at the center, at $L = 3^{\frac{1}{2}} = 1.73$.  It has a period of $T = \frac{2\pi^2}{\sqrt{3}} = 11.3964$ and three of the vortices are a distance $r = 3^{-\frac{1}{4}} = 0.7598$ from the center.  Finally we get to the collinear relative equilibria at $L = \frac{\lrp{\sqrt{6}+3}2^{\frac{1}{3}}}{\lrp{9\sqrt{3}+11\sqrt{2}}^{\frac{1}{3}}} = 2.18225$.  It has a period of $T = T = \frac{2\pi^2}{\sqrt{3}}2^{\frac{1}{3}} = 14.3586$.  Since it is symmetric, there are two lengths which characterize it.

These three relative equilibria partition the submanifolds $\mymanL$ into three different groups.  Within each group, the manifolds $\mymanL$ must be homeomorphic to each other.  Ideally, we would like to be able to describe the topology of each of these classes of manifolds, however there are a couple of complications to achieving this.  First of all, classifying 4-manifolds is notoriously hard, even in our restricted case of closed 4-manifolds which admit smooth structure ($\HamVec{H}$ is defined everywhere on $\mymanL$).  Second, our manifold could be comprised of multiple path connected components, each of which we would need to classify.  We will largely ignore this issue of exact classification, though we will remark on it further after we have developed the necessary machinery of braid classification.  It will turn out that the existence of periodic orbits with corresponding braids of specific type will restrict the type of manifold topology possible for that range of angular momentum values.

\chapter{Braids}
\label{ch:braid}

We briefly mentioned in section~(\ref{subsec:PSmanifold}) that the first homotopy group of our manifold is equivalent to the pure braid group: $\pi_1\lrp{\myman} = PB_N$.  This roughly says that if we pick a distinguished base-point, then loops which begin and end at this point and can be continuously deformed into each other are topologically equivalent to an element of the pure braid group.  Because periodic orbits are loops in $\myman$, they can be topologically distinguished from one another by virtue of this association with braids.  Additionally, we will be able to categorize the periodic orbits via their braid type, and extract a whole host of interesting information about them which is encoded in their topology.  In this chapter we will introduce basic braid notation and ideas~\cite{bib:BirmanBLMCG}\cite{bib:Kassel}\cite{bib:Dehornoy}, as well as develop more sophisticated topological classification ideas based on mapping class groups.  

\section{Basic Braid Theory}
	
\begin{figure}[htb]
  \centering
  \large
  \scalebox{0.6}{\includegraphics{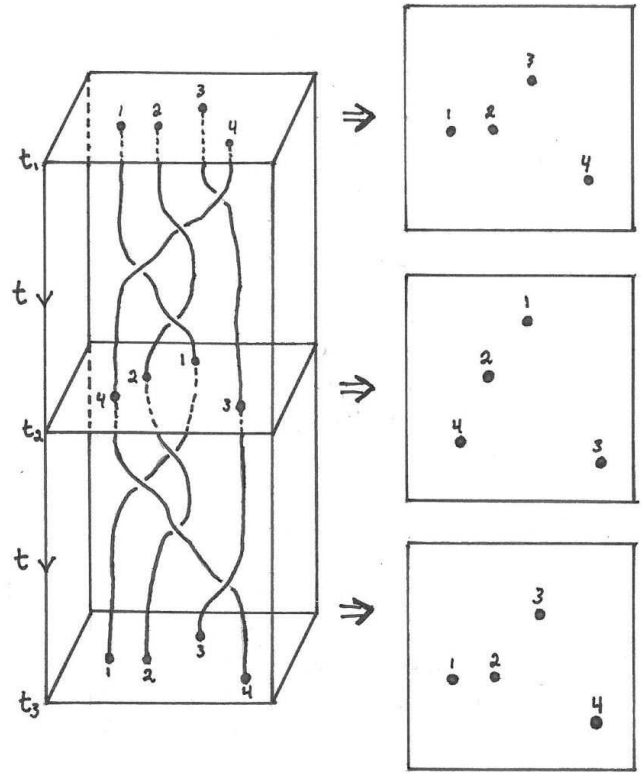}}
  \normalsize
    \caption{The motion of vortices on a plane can be viewed as a geometric braid.  Time advances downward, and each horizontal slice is a vortex configuration at a certain time.  If we additionally require that at $t_1$ and $t_3$ the vortex configurations are not only identical, but also without permutations, then the braid is a pure braid $PB_4$.}
    \label{graph:BraidPTconfig}
\end{figure}

	As mentioned above, the periodic orbits of the point vortex system are loops in phase space.  Though it is certainly hard to visualize loops in this potentially high dimensional space, there is fortunately a much more immediate way to see the relationship between braids and periodic orbits.  Consider a configuration of vortices on the plane $\mathbb{R}^2$ and their time evolution in a third orthogonal direction, say $-\hat{z}$ downward.  In this view, each horizontal slice is a snapshot of the vortex configuration at a given time, with time advancing downward, see Fig.~(\ref{graph:BraidPTconfig}).  If we consider a periodic orbit, then the time-slice at $t=0$ and $t = T$ are the same, and we can identify the top and bottom of this space.  

\begin{figure}[htb]
 \centering
\large
\resizebox{0.2\linewidth}{!}{\includegraphics{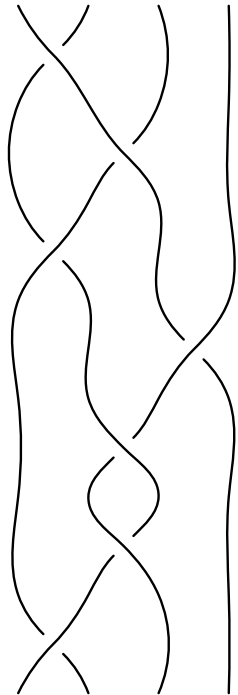}}
\normalsize
    \caption{A braid diagram on four strands.  The corresponding braid word is $\beta = \sigma_{1}\sigma_{2}\sigma_{1}^{-1}\sigma_{3}^{-1}\sigma_{2}\sigma_{2}\sigma_{1}^{-1}$}
    \label{figure:InitialBraidEx}
\end{figure}

	To think about this geometric picture from the perspective of braids, we will first ease up on the restriction that the vortices start and end at the same positions.  Instead, we will just assume that the set of original and final positions $\lrbk{\bfr_i}$ are the same.  That is, we will allow a permutation of the positions, which will allow us to talk about braids in general, as opposed to just pure braids.  We will refer to the trajectory of a single vortex as a strand of the braid.  It is convenient to view the strands projected onto an arbitrary axis, say the x-axis, with the over/under crossings identified.  This view is referred to as a braid diagram, Fig.~(\ref{figure:InitialBraidEx}).
	
	So how do we know if two braid diagrams represent the same braid?  If one braid diagram can be continuously deformed into another, then we say the two are isotopic, and they represent the same braid.  Since the vortices are not allowed to collide, the strands can not pass through one another.  Thus, when two braids cross over each other, the over/under-ness of the crossing is preserved by isotopy.  These crossings are fundamental to the braid structure, and we'll see that they form generators for the braid group.  It is always possible to push, isotopically, crossings up or downward on a braid diagram so that no two crossings occur at the same time.  Thus we can equate a braid diagram with a succession of braid crossings in time.  To make this structure more precise and algebraic, consider a set of N strands, which at a given time are ordered in the x-direction.  We will refer to the $i$-th strand crossing over the $\lrp{i+1}$-th strand as $\sigma_i$.  If it instead crosses under the $\lrp{i+1}$-th strand, then the element is $\sigma_i^{-1}$.  Notice that these two elements are inverses of each other, in that $\sigma_i\sigma_i^{-1}$ is isotopic to the identity $\lrp{id}$.  The positive $\lrp{N-1}$ elements and their inverses acting on N strands generate the braid group $B_n$.  That is, we can write out any element $\beta$ of $B_n$ as a product of these generators
\be
\beta = \sigma_{i_1}^{\pm 1}\sigma_{i_2}^{\pm 1} \cdots \sigma_{i_k}^{\pm 1}.
\label{eq:GenericArtinBraid}
\ee

	This is referred to as the Artin representation of the braid group.  It is read left to right, meaning that $\sigma_{i_1}^{\pm 1}$ is the first generator in time, and the top crossing on the corresponding braid diagram.  Now this group is certainly not the free group on these generators.  There are a couple of relations on a series of generators that quantify when two braid diagrams are equivalent.  First off, consider a pair of strands which switch places, and then a separate pair also switch places.  It does not matter which of the two go first in time.  In terms of the generators, this is expressed as $\sigma_i\sigma_j = \sigma_j\sigma_i$ for $\left| i-j \right| \leq 2$, where the generators could be either positive or negative.  The next restriction is given by $\sigma_i\sigma_{i+1}\sigma_i = \sigma_{i+1}\sigma_{i}\sigma_{i+1}$, where the generators are either all positive or all negative, and is best justified by looking at its corresponding braid diagram in the following figure.
\begin{figure}[htb]
 \centering
\large

\resizebox{\linewidth}{!}{\subfigure{\includegraphics{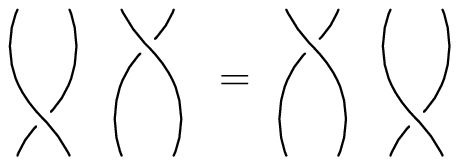}} \quad \text{and} \quad
\subfigure{\includegraphics{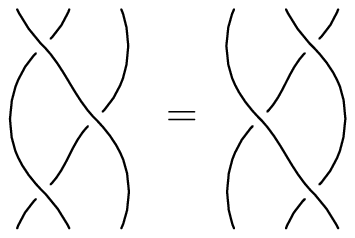}}}
\normalsize
    \caption{The braid relations 1) $\sigma_1\sigma_3 = \sigma_3\sigma_1$ and 2) $\sigma_1\sigma_{2}\sigma_1 = \sigma_{2}\sigma_{1}\sigma_{2}$}
    \label{figure:BraidRelations}
\end{figure}

These two restrictions are called the braid relations on $B_n$
\begin{align}
\sigma_i\sigma_j &= \sigma_j\sigma_i \;\; \text{for} \;\; \left| i-j \right| \leq 2 \notag \\
\sigma_i\sigma_{i+1}\sigma_i &= \sigma_{i+1}\sigma_{i}\sigma_{i+1}.
\label{eq:BraidRelations}
\end{align}
These relations and the generators $\lrbk{\sigma_i}$ define a group structure.  To see this consider two braids $\beta_1 = \sigma_{i_1}^{\epsilon_{1}}\sigma_{i_2}^{\epsilon_{2}} \cdots \sigma_{i_k}^{\epsilon_{k}}$ and $\beta_2 = \sigma_{j_1}^{\epsilon'_{1}}\sigma_{j_2}^{\epsilon'_{2}} \cdots \sigma_{j_m}^{\epsilon'_{m}}$, where the $\epsilon$ are either positive or negative integers corresponding to the number of successive times a generator or its inverse is applied.  The product of these two braids $\beta_1\beta_2$ corresponds to the braid diagram of $\beta_1$ followed in time by $\beta_2$.  It is given algebraically by $\beta_1\beta_2 = \sigma_{i_1}^{\epsilon_{1}}\sigma_{i_2}^{\epsilon_{2}} \cdots \sigma_{i_k}^{\epsilon_{k}}\sigma_{j_1}^{\epsilon'_{1}}\sigma_{j_2}^{\epsilon'_{2}} \cdots \sigma_{j_m}^{\epsilon'_{m}}$.  So the product of two braids is well defined and gives another braid in $B_N$.  Furthermore, each element has an inverse.  For $\beta_1$ this is easily seen to be $\beta_1^{-1} = \sigma_{i_k}^{-\epsilon_{k}}\sigma_{i_{k-1}}^{-\epsilon_{k-1}} \cdots \sigma_{i_1}^{-\epsilon_{1}}$.

For $N=2$ this is a simple infinite cyclic group, where every element is a simple power of $\sigma_1$ or its inverse.  There are no braid relations since there is only one generator, and the group is Abelian.  $B_3$, on the other hand, is non-Abelian due to the second of the braid relations, Eq.~(\ref{eq:BraidRelations}).  This is the case for all $B_N$ for $N \geq 3$.

The pure braid group, $PB_N$, has a simple relationship to the full braid group.  Consider the map on $B_N$ which simply ``forgets" the over/under relationship encoded in $\sigma$ and $\sigma^{-1}$.  This is a map from the braid group to the symmetric group, $B_N \rightarrow S_N$, i.e. it assigns a permutation of the original strand indices to each braid.  The pure braid group is simply the kernel of this map, the set of braids that are mapped to the identity permutation.  There is a different representation for the pure braid group, where the generators themselves are elements of the pure braid group.  However, these generators do not correspond to natural motions of the vortex dynamics, and we will not use them.  

There are a few attributes of $B_N$ and $PB_N$ that we will need to use in what follows.  First of all, consider the center of $PB_N$, the subgroup of $PB_N$ whose elements commute with every other element in $PB_N$.  It is the set generated by all powers of the full twist $\Delta^2$.  The half twist and full twist are given by
\begin{subequations}
\begin{align}
\Delta &= \lrp{\sigma_{1}\sigma_{2}\cdots\sigma_{N-1}}\lrp{\sigma_{1}\sigma_{2}\cdots\sigma_{N-2}}\cdots\lrp{\sigma_1\sigma_2}\sigma_1, \\
\Delta^2 &= \lrp{\sigma_{1}\sigma_{2}\cdots\sigma_{N-1}}^N
\label{eq:FullAndHalfTwist}
\end{align}
\end{subequations}

So we have that $\beta \lrp{\Delta^2}^k = \lrp{\Delta^2}^k \beta$ for any braid $\beta$ in $PB_N$.  This also happens to be the center for the full braid group.  Another result, which may be more intuitive, is that both $B_N$ and $PB_N$ are torsion free.  This means that there are no elements, $\beta$, of either group, such that some power of this element is equal to the identity.  A braid can only be undone by its inverse.

Finally, we will want to consider quantities that are the same for all braid words that represent the same braid.  One such quantity, called the exponent sum, is given, not surprisingly, by the sum of the exponents $\epsilon$:
\be
es\lrp{\beta} = \sum \epsilon_i.
\label{eq:ExponentSum}
\ee
It is easily checked that this quantity remans unchanged for all of the braid relations.

\section{Word and Conjugacy Problem}
	
The Artin representation of a braid element is useful because it renders a topological concept into algebraic terms.  In particular, we can easily deal with braids, and therefore the topological aspects of periodic orbits, computationally.  There are, however, two impediments to comparing the topology of two periodic orbits.  The first, termed the word problem, can be dealt with in a rather straight-forward fashion, while the second, called the conjugacy problem, is somewhat more difficult.

Consider two braid words $\beta_1$ and $\beta_2$ written in terms of the Artin generators.  How can we determine whether these two braid words represent the same braid?  This basically asks if we can use the braid relations, Eq.~(\ref{eq:BraidRelations}), repeatedly to algebraically transform $\beta_1$ into $\beta_2$,  or equivalently whether the braid diagrams corresponding to each braid word are isotopic.  We already showed that there are certain quantities, like the exponent sum, which are the the same for all equivalent braid words.  There are, however, many braid words with the same exponent sum that do not represent equivalent braids, so the exponent sum is not sufficient for distinguishing braids.  There are a couple of different ways to solve this problem.  One popular method involves rewriting any given braid word in a ``normal form" that is the same for all braid words corresponding to the same braid.  An example is the left greedy normal form~\cite{bib:Kassel}.  There are a number of advantages to singling out a particular representative for each braid, however we will be doing this in a slightly different way, using an order on braids.

It was proved in the early 1990s that the braid groups are left orderable~\cite{bib:Dehornoy}.  A left order is a relation $<$ between any two elements $\beta_1, \beta_2$ of $B_N$ such that either $\alpha \beta_1 < \alpha \beta_2$, $\alpha \beta_2 < \alpha \beta_1$, or $\alpha \beta_1 = \alpha \beta_2$ for all $\alpha \in B_N$.  This provides a rigorous sense in which one element of the braid group can be considered larger than another.  The existence of an order is dependent on finding a subset of braids $S$ such that the whole braid group is the disjoint union of $S$, the identity, and $S^{-1}$.  The way that one defines this subset will then determine the character of the order.  Now we can define $\beta_1 < \beta_2$ if and only if $\beta_1^{-1}\beta_2 \in S$.  One immediate result is that $\beta_1 = \beta_2$ iff $\beta_1^{-1}\beta_2 \notin S$ and $\beta_2^{-1}\beta_1 \notin S$.  This solves the word problem for $B_N$.  We have built up a whole object oriented class to computationally deal with braids, and this class checks braid word equivalency in this manner, using the Dehornoy left ordering~\cite{bib:Dehornoy}.

Next, consider three braids $\beta_1, \beta_2$ and $\alpha$ in $B_N$.  We say that $\beta_1$ is conjugate to $\beta_2$ if there exists some $\alpha$ such that $\alpha\beta_1\alpha^{-1} = \beta_2$.  Of particular interest are the conjugacy classes of $PB_N$.  The conjugacy class $\lrbc{\beta_1}$ for $\beta_1 \in PB_N$ is defined as the set of braids that are conjugate to $\beta_1$: $\lrbc{\beta_1} = \lrbk{\beta_2 \in PB_N \mid \alpha\beta_1\alpha^{-1} = \beta_2 \; \text{for some} \; \alpha \in B_N}$.  Note that even though $\alpha$ is in $B_N$, the conjugate braid remains in $PB_N$.  Since conjugacy is reflexive, every element of a conjugacy class is conjugate to every other element, and it does not matter which element is chosen as a representative.  Particularly simple conjugacy classes include $\lrbc{\Delta^{2k}}$, each of which has only itself as a member since the full twist commutes with all braids.  Importantly, any two conjugacy classes are disjoint (do not have any common elements), and the full pure braid group can be broken up into the disjoint union of conjugacy classes.

Why are conjugacy classes important?  Think about a braid associated with a periodic orbit.  In choosing this braid, we have implicitly chosen a distinguished point on the periodic orbit.  If we were to choose another point, close enough to the first, while keeping the same projection axis, the new braid would be related to the old by a cyclic rotation of the generators: $\beta = \sigma_{i_1}\sigma_{i_2}\sigma_{i_3}\cdots\sigma_{i_k} \; \rightarrow \; \beta' = \sigma_{i_2}\sigma_{i_3}\cdots\sigma_{i_k}\sigma_{i_1}$.  Notice that we can write this transformation as a conjugation, $\beta' = \sigma_{i_1}^{-1}\beta\sigma_{i_1}$.  Such a braid is said to be a closed braid, $\hat{\beta}$.  Additionally, the braid representation of the periodic orbit is dependent on the choice of projection axis.  If instead of projecting the vortex motion onto the $x-t$ plane as viewed from the $-y$ axis, we chose the $x-t$ plane as viewed from the $+y$ axis, the new braid would be rotated by $\pi$.  The new and old braids would once again be related by a conjugation: $\beta' = \Delta^{-1}\beta\Delta$.  In general, if we took a periodic orbit and deformed it in any manner that did not have a section of the orbit passing through the collision set, Eq.~(\ref{eq:CollisionSet}), the braid representative would remain within the conjugacy class of the original braid.  Thus, it is readily apparent that the topological object that describes the periodic orbit is not the pure braid itself, but the conjugacy class of the pure braid.  We will refer to braid conjugacy classes as ``braid types."
	
How can we tell if two braids are of the same braid type?  Or more generally, how can we distinguish or characterize different braid types?  The first of these questions is what we call the conjugacy problem, and has been solved in a number of interesting ways.  One way is to use the ``Garside" algorithm, or some of its more recent variants~\cite{Birman:1998tc}.  This algorithm is fairly complicated, and we will only give the roughest idea of how it works.  Given a braid $\beta$, we can write it in the Garside normal form, $\beta = \Delta^k\alpha$.  Here $\Delta$ is the half twist, and $k$ is the largest integer such that the ``tail," $\alpha$, is composed of all positive generators.  Additionally, $\alpha$ can be written in a unique way, making this a word invariant.  Now consider all of the elements of a conjugacy class.  There is an iterative procedure for obtaining a representative finite set of such elements, each written in Garside normal form.  All of these elements will have the same exponent, and there will be a single element with the smallest tail.  This unique braid word is taken to be the normal form for the whole conjugacy class.  Thus you can determine whether two braids are conjugate by comparing their normal forms.  While this algorithm works well, especially some of the more modern variations, the normal form it produces does not contain as much topological information as we would like.  We will be using a separate method that we developed (though we are sure it has been thought of before) based on some beautiful ideas of Thurston.
		
\section{Thurston-Nielsen Classification}	

Before we get into the classification theorem of Thurston~\cite{THURSTON:1988ty}\cite{Boyland:1994ud}, we must explain the link between mapping class groups and braids.  Consider a disk $D$ with a set of N points, $Q_N = \lrbk{\bfr_1,\bfr_2,\cdots,\bfr_N}$, identified on its surface.  We can think of the group, $\mathcal{F}= Aut\lrp{D,Q_N}$, of all possible orientation preserving homeomorphisms, or maps, which fix the boundary $\partial D$ point-wise and fix each identified point.  That is, $f: D \rightarrow D$, such that $f\lrp{\bfr} = \bfr$ if $\bfr \in \partial D, Q_N$.  This group, $\mathcal{F}$, can be given a topological structure, and we can define the mapping class group, $\mathcal{MCG}\lrp{D,\partial D,Q_N}$, as the group of path connected components in $\mathcal{F}$.  That is to say that two maps of $\mathcal{F}$ are considered equal, and therefore in the same class, if you can continuously deform one map into another, and therefore consider them to be isotopic.  Interestingly, we can associate this mapping class group with the pure braid group, $\mathcal{MCG}\lrp{D,\partial D,Q_N}\simeq PB_N$.  One can think of this as associating a whole class of surface maps with each element of the pure braid group, where each map is consistent with the movement of the braid strands.

	There is one concrete example that should be kept in mind when thinking of these types of mappings.  Consider a periodic orbit of the point vortex system.  It describes the motion of the vortices as they trace out strands of a pure braid.  The fluid around the vortices is advected according to Eq.~(\ref{eq:PVvelocityfield}).  If we are given a periodic orbit, then we can talk about the time-T maps, where T is the period, that it induces on all non-vortex points of the plane: $F_T^t\lrp{\bfr} = \bfr'$.  This map is parameterized by time $t$, which just serves to remind us that starting at different points on the periodic orbit will give different maps.  $F_T^t$ fixes the vortex positions, and since the velocity field falls off inversely with distance, we can always choose a disk large enough such that the boundary is fixed to an arbitrary precision.  Therefore, the above mapping class result tells us that two periodic orbits with the same braid type will have time-T maps in the same class.  Thus the topological aspects of a periodic orbit will constrain the dynamical aspects! 
	
	Since there is this correspondence between the mapping class group and the pure braid group, we can associate results in one arena with results in the other.  In particular, the Thurston-Nielsen classification of mapping classes will give us not only valuable information about the dynamics of time-T maps, but will also point the way to a classification of braid types.  Before we state the content of the Thurston-Nielsen classification result, we must introduce a few concepts.  A map $f  \in \mathcal{F}$ is said to be reducible if there exists a family of closed curves $R_c = \lrbk{\mathcal{C}_i}$, which are non-intersecting with each other ($\mathcal{C}_i \cap \mathcal{C}_j = \emptyset$ for $i\neq j$), do not self-intersect,  and are mapped back onto themselves ($f\lrp{\lrbk{\mathcal{C}_i}} = \lrbk{\mathcal{C}_i}$).  For the case of maps associated with pure braids, we restrict our attention to closed curves that are individually mapped back to themselves ($f\lrp{\mathcal{C}_i} = \mathcal{C}_i$).  Furthermore, our set of reduction curves $R_c$ excludes those curves that are deformable, while not passing over any distinguished points $Q_N$, to other curves in the set or to a point.  This set of reduction curves breaks up the disk into components, $M_i \in \lrp{D,Q_N}\setminus R_c$, each containing two or more of the identified points.  The restriction of the map $f$ to these components $M_i$ are themselves called components, and denoted $f_i$.  
\begin{figure}[htb]
  \centering
  \large
  \scalebox{0.7}{\includegraphics{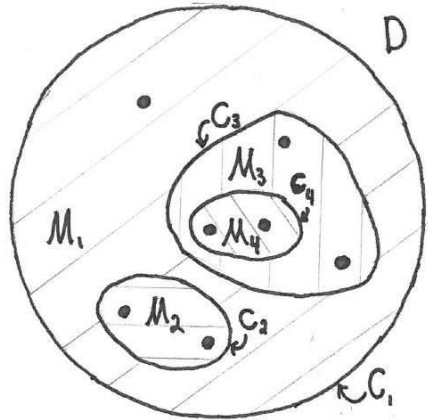}}
  \normalsize
    \caption{This is an example of a potential set of reduction curves, $R_c = \lrbk{\mathcal{C}_i}$, on the 7 punctured disk.  The components of the disk, $M_i \in \lrp{D,Q_N}\setminus R_c$, are also labeled.}
    \label{graph:ReductionCurve}
\end{figure}	

The Thurston-Nielsen classification theorem tell us that in each mapping class of a mapping class group there exists a representative map, $f^{\ast}$, which is either reducible or irreducible:	
\begin{itemize}

\item Irreducible:  If $f^{\ast}$ is irreducible, then it is one of the following two types
	\begin{itemize}
		\item Finite Order (fo): This is the simplest type of map.  Some positive power of $f^{\ast}$ is equal to the identity, $\lrp{f^{\ast}}^k = id$.  In the case of the pure braid induced maps, these are just the maps associated with some power of the full twist$\lrp{\Delta^2}^j$. \footnote{This definition of finite order assumes that the elements of the mapping class group we are considering fix the boundary $\partial D$ as a set.  That is, twists are allowed.  This allows an iteration of the map to be equivalent to the identity.  On the other hand, we have defined our mapping class group on maps which fix the boundary point-wise.  This was needed to associate it with the pure braid group, which has no torsion, or members that iterate to the identity.  To make these two definitions jive, consider two boundaries.  One far away enough to be fixed by the time-T maps under consideration, and the other close enough to allow twists.  A finite order map is now just a map $f^{\ast}$, such that $\lrp{f^{\ast}}^k = id$ inside the inner boundary, and $\lrp{f^{\ast}}^k$ is just a twist map on the annulus outside the inner boundary.} 
		
		\item Pseudo Anosov (pA):  $f^{\ast}$ is a pseudo Anosov diffeormorphism.  These maps are locally hyperbolic with stretching and contracting directions.  We will have much more to say about pA maps.
		
	\end{itemize} 

\item Reducible: There exists a set of reduction curves $R_c = \lrbk{\mathcal{C}_i}$, such that the components, $f^{\ast}_i$ of $f^{\ast}$ are irreducible maps.

\end{itemize}

There are few things to point out about this result.  First, note that it pertains to a single map, $f^{\ast}$, the TN representative, within a mapping class.  How are the properties of other maps within this mapping class related to those of the TN representative?  If the TN representative is reducible, then all other maps in the same class are also reducible with reduction curves that are isotopic to the reduction curves of the TN representative.  Thus the sets of isotopic reduction curves help to characterize mapping classes.  In particular, these isotopy classes are conjugacy invariants, and will be useful in categorizing pure braids.  We will follow up on this idea in the next section.  If the TN representative is irreducible, then so are the other maps in the mapping class.  These other maps will also be ``more complicated," in a sense to be defined later, than their TN representative.  In a sense a pA TN representative sets a lower bound for the topological complexity of all other maps in its mapping class.  Finally, there is no way of knowing what sort of support a general map in a mapping class with a pA TN representative will have.  The behavior ascribed to the TN representative could possibly be evident only in a vanishingly small area (range) of a general map.  We will try to give some evidence in the results section to support the idea that pA time-T maps have finite, non-zero support.

\begin{figure}[htb]
  \centering
  \large
  \scalebox{0.5}{\includegraphics{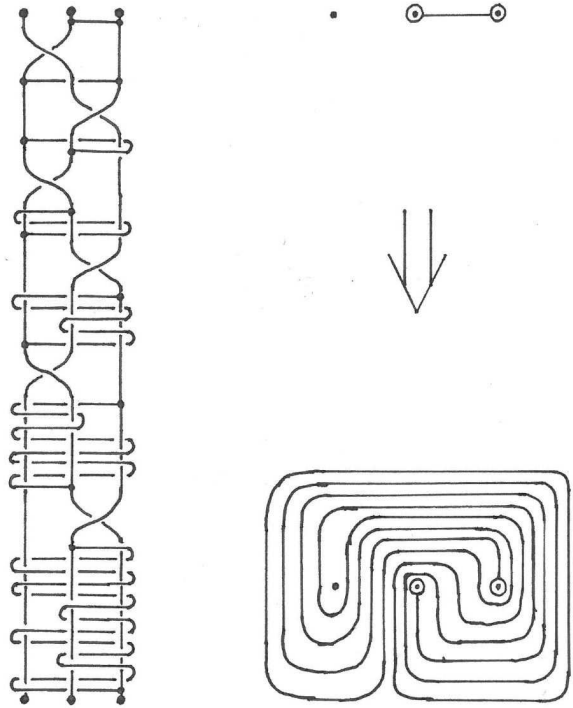}}
  \normalsize
    \caption{This shows the action of the pseudo-Anosov (pA) braid, $\beta = \lrp{\sigma_1\sigma_2^{-1}}^3$, in $PB_3$ on a rubber band that encircles two strands.  Notice that the length of the band after each subsequent generator follows the Fibonacci sequence.  Hence this braid is often referred to as the golden ratio braid.}
    \label{graph:PhiBraid}
\end{figure}

We would also like to point out that the TN categorization of our mapping classes automatically extends to a categorization of pure braids.  Pure braids can be finite order, pseudo Anosov, or reducible, in which case they can be broken down into sets of irreducible pure braids.  There is a natural geometric way of thinking about a reducible pure braid, by considering the reduction curves to be rubber bands.  These rubber bands are such that one can slide them down the braid and end up with an un-streatched rubber band at the end.  This action traces out a cylinder around each subset of strands encircled by each rubber band.  For a pA braid, say the traditional hair braid on three strands $\beta = \lrp{\sigma_1\sigma_2^{-1}}^3$, see Fig.~(\ref{graph:PhiBraid}), a rubber band around any two strands which excludes the third will exponentially increase in length after pulling it to the bottom of the braid.  In fact, if the length of the rubber band is initially 1, then the final length after having the rubber band travel through the braid $k$ times is, $L_k \simeq e^{k 6\ln{\phi}}$, where $\phi = \frac{1}{2}\lrp{1+\sqrt{5}}$ is the golden ratio.

What in particular characterizes a finite order or pseudo Anosov map?  Finite order maps are very simple, and in our case correspond directly to pure braids that are powers of the full twist $\Delta^2$.  When $N=2$ all braids are finite order braids.  This is not to say that all the maps in a finite order mapping class are simple.  This result simply says that simple finite order maps exist within this class.  We will certainly see examples of time-T maps of finite order periodic orbits that have considerable complexity.  The main use of this finite order concept will be to help in the classification of braid types.

Pseudo Anosov maps, on the other hand, imply a wealth of dynamic information.  pA maps are a generalization of linear Anosov maps on the torus.  These maps are everywhere hyperbolic, with an expansion factor $\lambda > 1$.  This means that at each point on the torus there are two directions, the unstable and stable directions.  One can think of integral curves that follow one of these two directions at each point.  A family of these curves fills up the whole torus, and is said to foliate the torus.  Thus we have a stable $F^s$ and an unstable $F^u$ foliation of the torus.  An Anosov map preserves these foliations, in that an integral curve will be mapped to an integral curve.  If two close points, $\left|\bfr_1-\bfr_2 \right| = \epsilon$, are separated along a leaf of the unstable foliation $F^u$, then there is an unstable integral curve that connects them, and their distance under the mapping will be larger by a multiplicative factor lambda: $\left|f\lrp{\bfr_1}-f\lrp{\bfr_2} \right| = \epsilon \lambda$.  If these two points are on a leaf of the stable foliation $F^s$, then they will be mapped closer together by an inverse factor of lambda: $\left|f\lrp{\bfr_1}-f\lrp{\bfr_2} \right| = \epsilon \frac{1}{\lambda}$.  Pseudo Anosov maps, which in our case live on the punctured disk, have all of these properties and also have a finite number of singularities of the foliations.  At these singularities there can be three or more expanding and contracting directions, as opposed to the two directions along the unstable integral curves and two directions along the stable integral curves of a regular non-singular point.  More intuitively, if each regular point looks locally like a saddle point with a stable and unstable direction, then an $n$-pronged singularity looks like a saddle for someone with $n$ legs.  The expansion factor also tells us something about the number of periodic points of the mapping.  Consider $P\lrp{f^n}$ to be the number of points that are periodic under the $n$-th iterate of the map $f$, that is $P\lrp{f^n} = \lrbk{x\mid f^n\lrp{x}=x}$.  We can define, roughly, the topological entropy to be
\be
h_{top}\lrp{f} = \lim_{n \to \infty} \frac{1}{n}\ln{\lrp{P\lrp{f^n}}}.
\label{eq:TopEntropy}
\ee

It turns out that for pA maps, the topological entropy is equal to the log of the expansion factor, $h_{top}\lrp{f} = \ln{\lrp{\lambda}}$.  Thus, there are an exponentially increasing number of periodic points with increasing period: $P\lrp{f^n} \simeq e^{n \ln{\lrp{\lambda}}} = \lambda^n$.  Notice how this is in a similar form to the dilation of the rubber band along the golden ratio braid.  Indeed the topological entropy can be thought of as a measure of how much stretching there is in a map.  If the support of a pA map is compact, then there is not only stretching, but also folding as well, and we say that this map has the mixing property~\cite{Boyland:2000uc}\cite{Gouillart:2006hi}\cite{Thiffeault:2006jp}\cite{Thiffeault:2008er}.  In our fluid advection terms, this means that if a time-T map is associated with a periodic orbit that is pA, then this map effectively mixes the fluid particles.  Some people have even used this idea to find the most efficient stirring devices~\cite{Finn:2010wg}\cite{Finn:2003hd}.	
	
\section{TN Braid Trees}
\label{sec:TNbraidtree}

Since our main focus is to categorize periodic orbits via their braid type, we will now investigate what structure is imbued upon braid types by the TN classification.  The first step is to follow through on a point that we noted in the last section, that the set of reduction curves for a braid is, up to isotopy, an invariant of conjugacy classes.  That is to say, two braids in the same conjugacy class, i.e. of the same braid type, will have reduction curves that can be smoothly deformed into one another.  For a given number of braid strands, it turns out that there are a finite number of these reduction curve isotopy classes.  This classification is rather coarse, and many different braid types will be in the same reduction curve isotopy class.  For example, pure braids with two strands, which are just powers of $\Delta^2 = \sigma_1^2$, are all in the same reduction curve isotopy class.  This class consists of one single curve encircling both strands.  Despite this degeneracy these reduction curve isotopy classes do a good job of lumping braid types together that have similar periodic orbit related properties.  As a small preview, we will mention that the two regions of qualitatively different motion for the $N = 3$ vortex problem, see section~\ref{sec:pvm3}, are distinguished by two different reduction curve isotopy classes.  For the sake of brevity, we will also refer to reduction curve isotopy classes as $T_1$ classes, and the $T_1$ classes for a specific number of strands, $N$, as $T^N_1$ classes.

\begin{figure}[htbp]
  \centering
  \large
  \scalebox{0.7}{\includegraphics{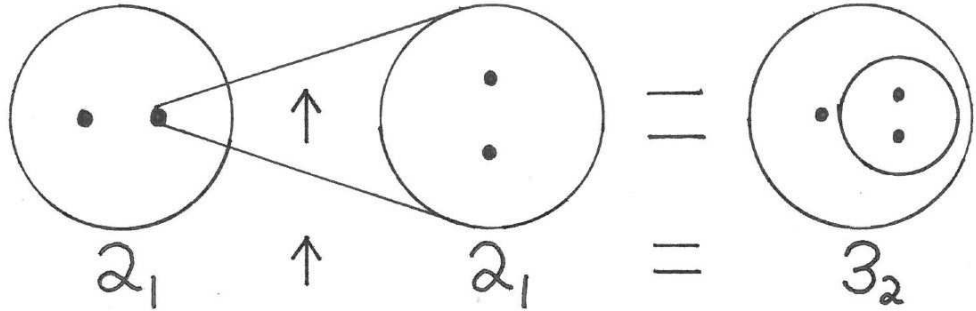}}
  \normalsize
    \caption{This diagram shows the injective process that builds up $3_2$ out of two copies of $2_1$.}
    \label{graph:Injective}
\end{figure}

As a reminder, a reduction curve must encircle at least two points of $Q_N$ and can not be deformed to another curve in the reduction curve set by passing over one of the points.  There will always be one curve, the fundamental curve, that encircles all the points, and this is the only curve in the case where the braid is pA or finite order.  So, for a given number of points, how can we enumerate all possible $T^N_1$ classes?  This is difficult if you simply draw a bunch of reduction curves and try to see which ones can be deformed into each other.  Fortunately these classes have some algebraic and combinatorial structure.  To see this, consider the $T^3_1$ classes.  There are only two of them, corresponding to a single loop encircling all three points, as well as this loop plus one encircling two of the three points.  The first reduction curve set exists for any N, and we will refer to this class as $N_1$, or in this case $3_1$.  The second reduction curve class, $3_2$, can be thought of as being built up from $T^2_1$ classes, in this case $2_1$ is the only choice.  This building-up action is essentially replacing one of the two points that the reduction curve in $2_1$ encircles by a copy of $2_1$.  We will refer to this algebraically as, $3_2 = 2_1\uparrow2_1$, where $X\uparrow Y$ should be read as class $Y$ replaces one of the available points in class $X$.  An available point is one that is only encircled by the outer-most curve.  Refer to figure~(\ref{graph:Injective}) to see this injective or replacement process.

For $N = 4$ this injective process gives us 5 $T^4_1$ classes, $T^4_1 = \lrbk{4_1,4_2,4_3,4_4,4_5}$.  The first of these is the usual fundamental reduction curve, but the rest are built up of all possible combinations of elements of $T^3_1$ and $T^2_1$ that leave us with four points.  We can write $4_2 = 3_1\uparrow2_1$, $4_3 = 3_2\uparrow2_1$, $4_4 = 2_1\uparrow3_1$, and $4_5 = 2_1\uparrow3_2$.  Since $3_2$ can be expressed as $2_1\uparrow 2_1$, we can further expand $4_3$ and $4_5$ as $4_3 = \lrp{2_1\uparrow 2_1}\uparrow2_1$ and $4_5 = 2_1\uparrow \lrp{2_1\uparrow 2_1}$, where the parentheses determine order of operation.  Notice that in general, we can express any element of any class as a combination of elements from lower classes that only have the fundamental reduction curve, $\lrbk{2_1,3_1,4_1,\cdots}$.  This is a direct reflection of the fact that all reducible braids and reducible mapping classes can be built up from strictly irreducible components.  One might then assume that the number of classes in $T^N_1$, denoted $a_n$, increases combinatorially as
\be
a_n = a_{n-1}a_2 + a_{n-2}a_3 + \cdots + a_2 a_{n-1} +1.
\label{eq:CombinatoricalGrowth}
\ee

However, there are two reasons why the growth in number of classes is smaller than this.  First of all, notice that $4_3 = \lrp{2_1\uparrow 2_1}\uparrow2_1$, which is composed of the fundamental reduction curve as well as two other curves which each encircle a different pair of points, has no available points in the sense mentioned above.  We will call this type of element of $T_1$, one that has no available points, a ``closed" element.   This means that in $T^N_1$ classes of higher $N$, closed elements can not be used as a starting point for the injective process.  That is $4_3\uparrow X$ is not a valid algebraic representation for any reduction curve isotopy class.  However, $X\uparrow 4_3$ is valid as long as  $X$ is not closed.  For each $N \geq 3$ there is at least one closed element of $T^N_1$.  The second reason for smaller growth than Eq.~(\ref{eq:CombinatoricalGrowth}) implies, is that there is a slight degeneracy in the algebraic representation of elements of $T^N_1$.  The first examples arise in the $N=5$ case, where it can be shown that $\lrp{2_1\uparrow 2_1}\uparrow3_1 = \lrp{2_1\uparrow 3_1}\uparrow2_1$ and $\lrp{2_1\uparrow 2_1}\uparrow\lrp{2_1\uparrow 2_1} = \lrp{2_1\uparrow \lrp{2_1\uparrow 2_1}}\uparrow 2_1$.  The first of these implies that it does not matter which of $3_1$ or $2_1$ we inject into the two available points of $2_1$.  This degeneracy can be encapsulated in the following relation on arbitrary elements $X, Y$, and $Z$ of $T_1$, where $X$ has at least two available points.
\be
\lrp{X\uparrow Y}\uparrow Z = \lrp{X\uparrow Z}\uparrow Y.
\label{eq:T1restriction}
\ee
Even taking into consideration closed elements and the restrictions due to degeneracy Eq.~(\ref{eq:T1restriction}), the number of classes in $T^N_1$, $a_n$, still grows quickly in a combinatorial manner.  It would be interesting to come up with a closed or recursive expression for $a_n$, but for this work, we only need to consider the 2 classes for $T^3_1$ and the 5 classes for $T^4_1$.  We have summarized much of the previous discussion in table~(\ref{table:RCclassification}).

\begin{table}[htbp]
  \centering
  \footnotesize
  \begin{center}
  \be
 \begin{array}[c]{|c|c|c||c|c|c|}
 \hline
RC & T_1 & decomposition & RC & T_1 & decomposition\\
\hline
\multirow{2}{*}{\scalebox{0.3}{\includegraphics{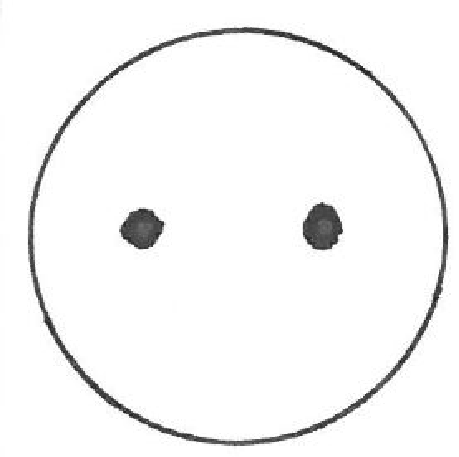}}}
& \multirow{2}{*}{$2_1$} & \multirow{2}{*}{$2_1$} &
\multirow{2}{*}{\scalebox{0.3}{\includegraphics{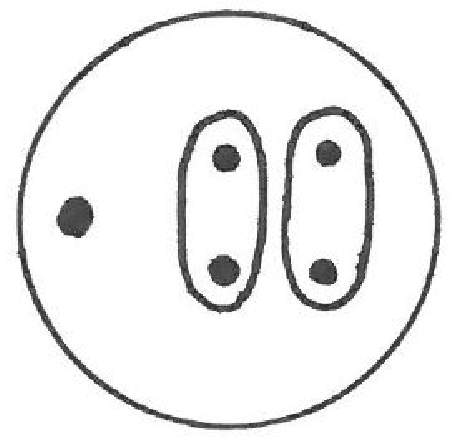}}}
& \multirow{2}{*}{$5_3$} & \multirow{2}{*}{$4_2 \uparrow 2_1 = \lrp{3_1 \uparrow 2_1} \uparrow 2_1$} \\
& & & & & \\
\hline
\multirow{2}{*}{\scalebox{0.3}{\includegraphics{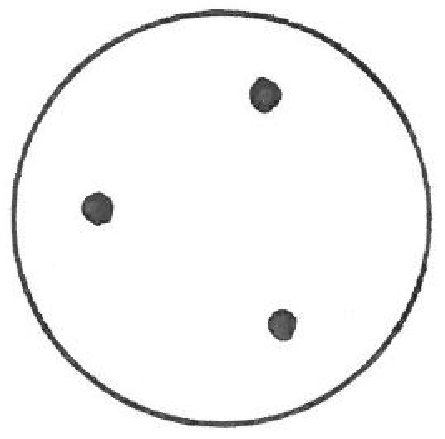}}}
& \multirow{2}{*}{$3_1$} & \multirow{2}{*}{$3_1$} &
\multirow{2}{*}{\scalebox{0.3}{\includegraphics{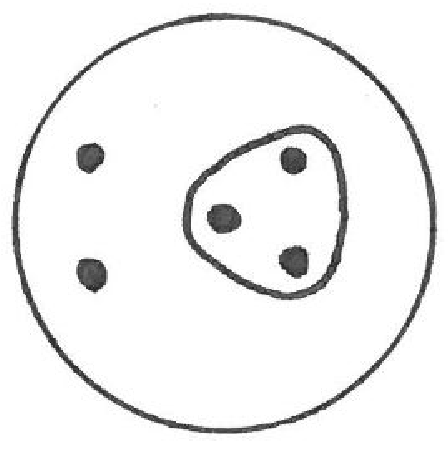}}}
& \multirow{2}{*}{$5_4$} & \multirow{2}{*}{$3_1 \uparrow 3_1$} \\
& & & & & \\
\hline
\multirow{2}{*}{\scalebox{0.3}{\includegraphics{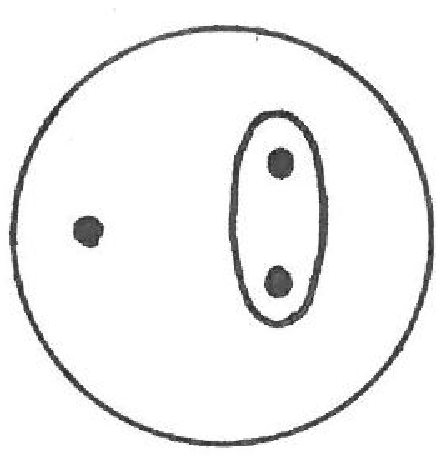}}}
& \multirow{2}{*}{$3_2$} & \multirow{2}{*}{$2_1 \uparrow 2_1$} &
\multirow{2}{*}{\scalebox{0.3}{\includegraphics{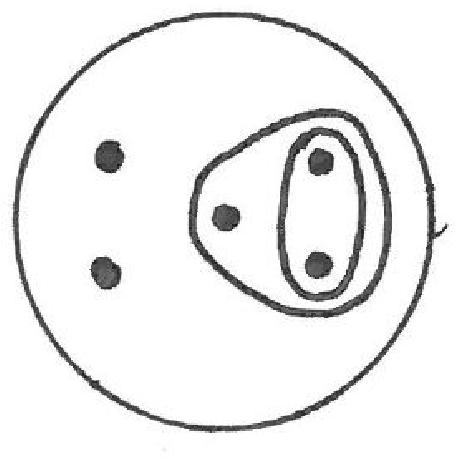}}}
& \multirow{2}{*}{$5_5$} & \multirow{2}{*}{$3_1 \uparrow 3_2 = 3_1 \uparrow \lrp{2_1\uparrow 2_1}$} \\
& & & & & \\
\hline
\multirow{2}{*}{\scalebox{0.3}{\includegraphics{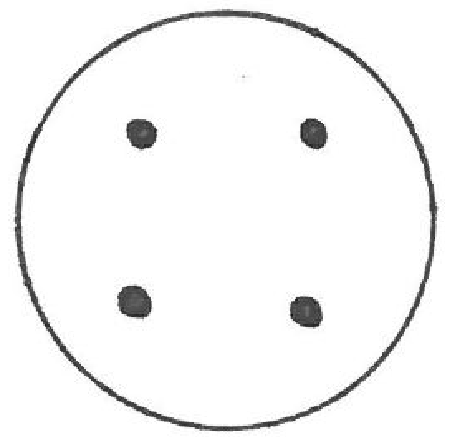}}}
& \multirow{2}{*}{$4_1$} & \multirow{2}{*}{$4_1$} &
\multirow{2}{*}{\scalebox{0.3}{\includegraphics{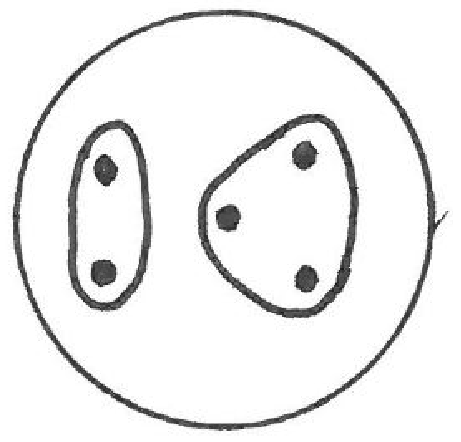}}}
& \multirow{2}{*}{$5_6$} & 4_4 \uparrow 2_1 = \lrp{2_1 \uparrow 3_1}\uparrow 2_1\\
& & & & & 3_2 \uparrow 3_1 = \lrp{2_1 \uparrow 2_1}\uparrow 3_1 \\
\hline
\multirow{2}{*}{\scalebox{0.3}{\includegraphics{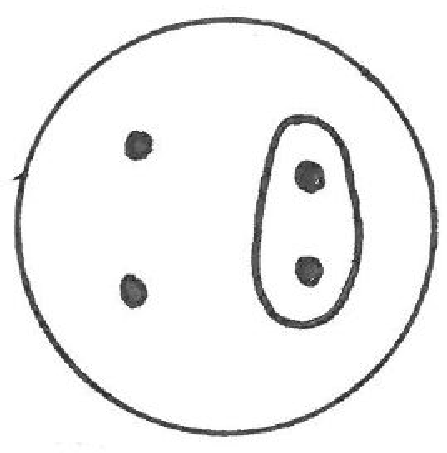}}}
& \multirow{2}{*}{$4_2$} & \multirow{2}{*}{$3_1 \uparrow 2_1$} &
\multirow{2}{*}{\scalebox{0.3}{\includegraphics{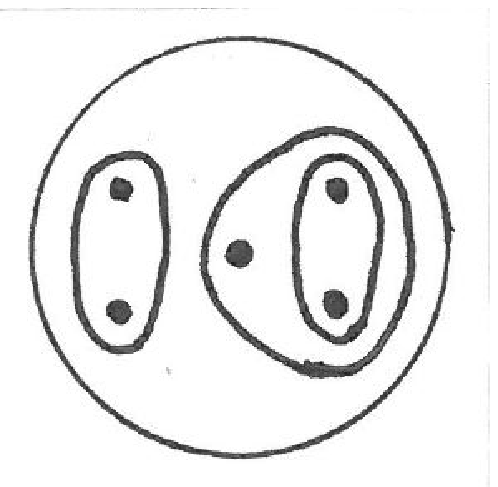}}}
& \multirow{2}{*}{$5_7$} & 4_5 \uparrow 2_1 = \lrp{2_1 \uparrow \lrp{2_1\uparrow 2_1}}\uparrow 2_1\\
& & & & & 3_2 \uparrow 3_2 = \lrp{2_1 \uparrow 2_1}\uparrow \lrp{2_1 \uparrow 2_1} \\
\hline
\multirow{2}{*}{\scalebox{0.3}{\includegraphics{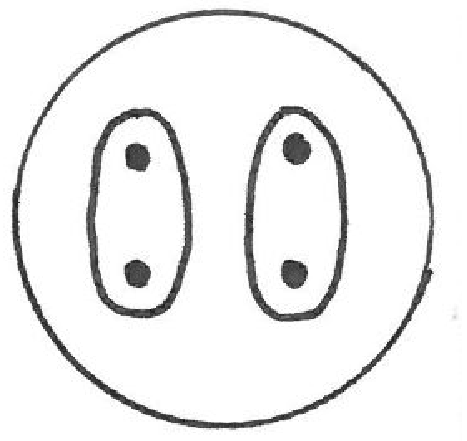}}}
& \multirow{2}{*}{$4_3$} & \multirow{2}{*}{$3_2 \uparrow 2_1 = \lrp{2_1 \uparrow 2_1} \uparrow 2_1$} &
\multirow{2}{*}{\scalebox{0.3}{\includegraphics{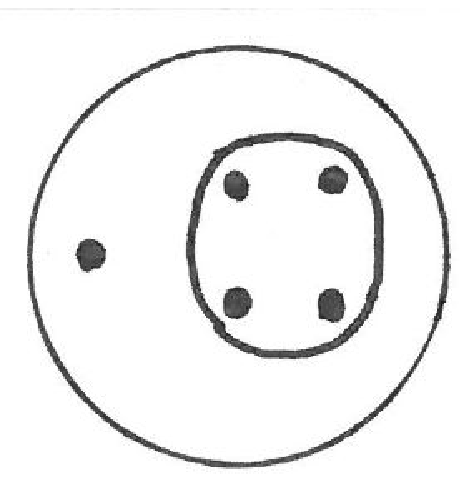}}}
& \multirow{2}{*}{$5_8$} & \multirow{2}{*}{$2_1\uparrow 4_1$} \\
& & & & & \\
\hline
\multirow{2}{*}{\scalebox{0.3}{\includegraphics{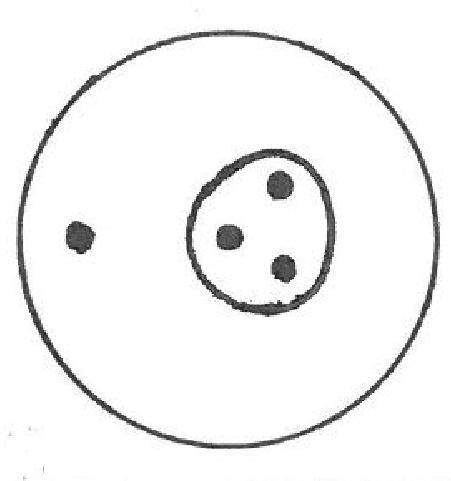}}}
& \multirow{2}{*}{$4_4$} & \multirow{2}{*}{$2_1 \uparrow 3_1$} &
\multirow{2}{*}{\scalebox{0.3}{\includegraphics{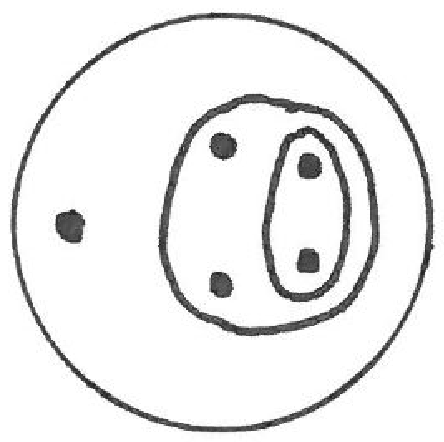}}}
& \multirow{2}{*}{$5_9$} & \multirow{2}{*}{$2_1\uparrow 4_2 = 2_1 \uparrow \lrp{3_1 \uparrow 2_1}$} \\
& & & & & \\
\hline
\multirow{2}{*}{\scalebox{0.3}{\includegraphics{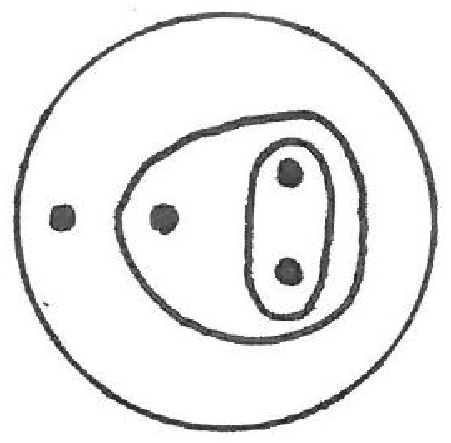}}}
& \multirow{2}{*}{$4_5$} & \multirow{2}{*}{$2_1 \uparrow 3_2 = 2_1 \uparrow \lrp{2_1 \uparrow 2_1}$} &
\multirow{2}{*}{\scalebox{0.3}{\includegraphics{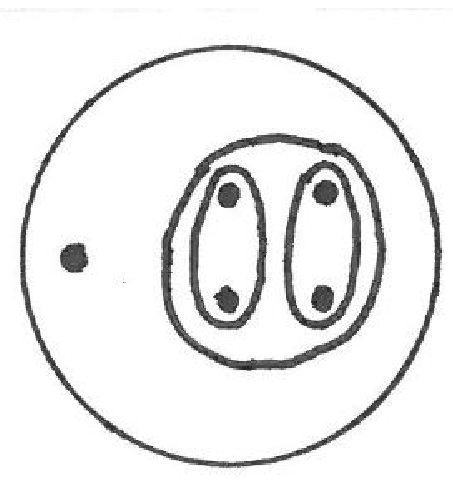}}}
& \multirow{2}{*}{$5_{10}$} & \multirow{2}{*}{$2_1 \uparrow 4_3 = 2_1 \uparrow \lrp{\lrp{2_1 \uparrow 2_1} \uparrow 2_1}$} \\
& & & & & \\
\hline
\multirow{2}{*}{\scalebox{0.3}{\includegraphics{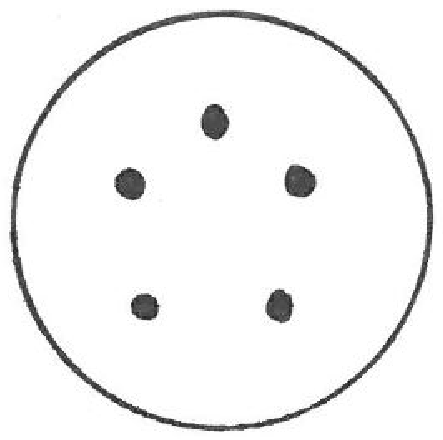}}}
& \multirow{2}{*}{$5_1$} & \multirow{2}{*}{$5_1$} &
\multirow{2}{*}{\scalebox{0.3}{\includegraphics{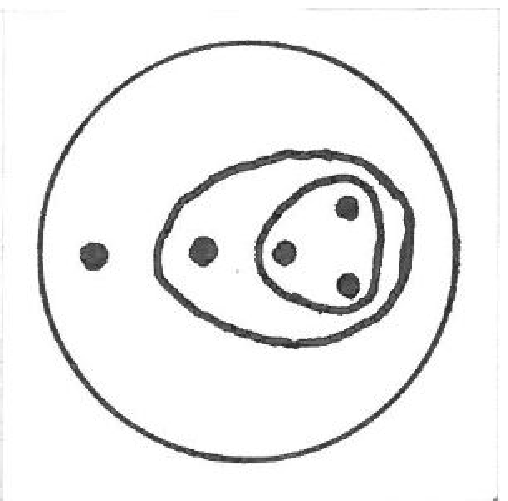}}}
& \multirow{2}{*}{$5_{11}$} & \multirow{2}{*}{$2_1 \uparrow 4_4  = 2_1 \uparrow \lrp{2_1 \uparrow 3_1}$} \\
& & & & & \\
\hline
\multirow{2}{*}{\scalebox{0.3}{\includegraphics{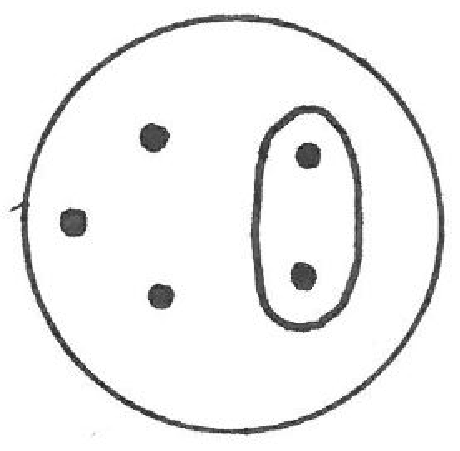}}}
& \multirow{2}{*}{$5_2$} & \multirow{2}{*}{$4_1 \uparrow 2_1$} & 
\multirow{2}{*}{\scalebox{0.3}{\includegraphics{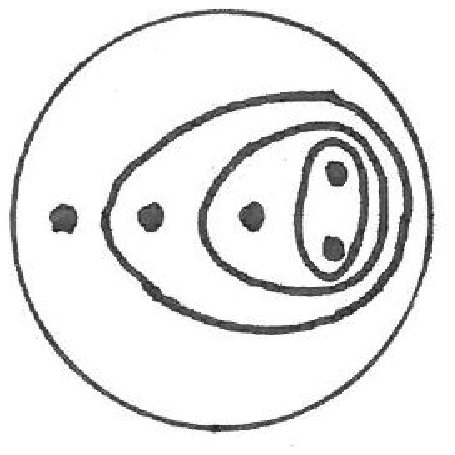}}}
& \multirow{2}{*}{$5_{12}$} & \multirow{2}{*}{$2_1 \uparrow 4_5 = 2_1 \uparrow \lrp{2_1 \uparrow \lrp{2_1 \uparrow 2_1}}$} \\
& & & & & \\ 
\hline 
 \end{array} \nonumber
\ee
  \end{center}
\normalsize
 \caption{This table shows a representative of each reduction curve ($RC$) isotopy class for $N = \lrbk{3,4,5}$, as well as its $T_1$ label and decomposition.}
 \label{table:RCclassification}
\end{table}
Notice that in the $T_1$ classification, there is no mention of the type of irreducible component.  That is, we have not taken the difference between finite order and pseudo Anosov braids into account.  We can augment the $T_1$ classification by distinguishing fo and pA braids, which gives a new, though related, classification $T_2$.  In a certain sense $T_2$ is finer than $T_1$, and we will often refer to the two as the fine and gross classifications respectively.  For enumeration purposes, remember that only components with three or more points can be pseudo Anosov.  That is for each element of $T_1$ whose reduced algebraic expression contains one factor of the form $N_1$, for $N \geq 3$, we have two elements of $T_2$.  There is a bit of subtlety involved in accounting for $T_1$ elements with more than one $N_1$ factor.  If they are like $3_1\uparrow 3_1$, which is three strands inserted into one of three other strands, we can have the inner and outer groupings be any of the four combinations: $\lrp{fo,fo}$, $\lrp{fo,pA}$, $\lrp{pA,fo}$, or $\lrp{pA,pA}$.  Therefore there are four elements in $T_2$ corresponding to the one element in $T_1$.  On the other-hand, consider $\lrp{2_1\uparrow 3_1}\uparrow 3_1$, which is two separate groups of three points, each injected into one of the two available points of $2_1$.  Since there is nothing to distinguish the two subgroups of three points from each other at the level of reduction curve isotopy, there are only three unique combinations: $\lrp{fo,fo}$, $\lrp{fo,pA}$, or $\lrp{pA,pA}$.  There are only three elements of $T_2$ that correspond to $\lrp{2_1\uparrow 3_1}\uparrow 3_1$ in $T_1$.  As you can see, accounting for the growth of classes in $T_2$ is even more difficult than in the $T_1$ case.  Fortunately the $N = 3$ and $N = 4$ cases are straight forward, and we have listed them in table~{\ref{table:RCclassification2}}.

\begin{table}[htbp]
  \centering
  \footnotesize
  \begin{center}
  \be
 \begin{array}[c]{|c|c|c|c||c|c|c|c|}
 \hline
RC & T_1 & fo/pA & T_2 & RC & T_1 & fo/pA & T_2 \\
\hline
\multirow{2}{*}{\scalebox{0.3}{\includegraphics{RC3_1}}}
& \multirow{2}{*}{$3_1$} & fo & \overline{3}_1 &
\multirow{2}{*}{\scalebox{0.3}{\includegraphics{RC4_3}}}
& \multirow{2}{*}{$4_3$} & \multirow{2}{*}{$fo$} & \multirow{2}{*}{$\overline{4}_5$} \\ \cline{3-4}
 & & pA & \overline{3}_2 & & & & \\
\hline
\multirow{2}{*}{\scalebox{0.3}{\includegraphics{RC3_2}}}
& \multirow{2}{*}{$3_2$} & \multirow{2}{*}{$fo$} & \multirow{2}{*}{$\overline{3}_3$} &
\multirow{2}{*}{\scalebox{0.3}{\includegraphics{RC4_4}}}
& \multirow{2}{*}{$4_4$} & fo & \overline{4}_6 \\ \cline{7-8}
 & & & & & & pA & \overline{4}_7 \\
 \hline
 \multirow{2}{*}{\scalebox{0.3}{\includegraphics{RC4_1}}}
& \multirow{2}{*}{$4_1$} & fo & \overline{4}_1 &
\multirow{2}{*}{\scalebox{0.3}{\includegraphics{RC4_5}}}
& \multirow{2}{*}{$4_5$} & \multirow{2}{*}{$fo$} & \multirow{2}{*}{$\overline{4}_8$} \\ \cline{3-4}
 & & pA & \overline{4}_2 & & & &  \\
 \hline
  \multirow{2}{*}{\scalebox{0.3}{\includegraphics{RC4_2}}}
& \multirow{2}{*}{$4_2$} & fo & \overline{4}_3 & & & & \\ \cline{3-4}
 & & pA & \overline{4}_4 & & & &  \\
 \hline
 \end{array} \nonumber
\ee
  \end{center}
\normalsize
 \caption{The $T_2$, ``Fine", classification for $N = 3$ and $N = 4$.}
 \label{table:RCclassification2}
\end{table}
This classification gives an arbitrary braid type a natural rooted tree structure.  First consider a graph, which is just a set of vertices connected by edges.  A tree is a particular type of graph, where any two vertices are connected by only one path (collection of edges).  A rooted tree is a tree with a specific vertex designated as the root.  This creates a direction on each of the edges, either toward the root or away from the root.  At each vertex there is only one edge that leads toward the root, while possibly many that lead away from it.  Consider the subgraph associated with any given vertex, which includes itself, those vertices that are further away from the root, and all edges that connect these vertices and our designated vertex.  Each of these subgraphs are themselves rooted trees.  Thus any rooted tree can be described in a recursive manner, which will be important when we represent these ideas in C++ code.

\begin{figure}[htb]
  \centering
  \large
  \scalebox{0.7}{\includegraphics{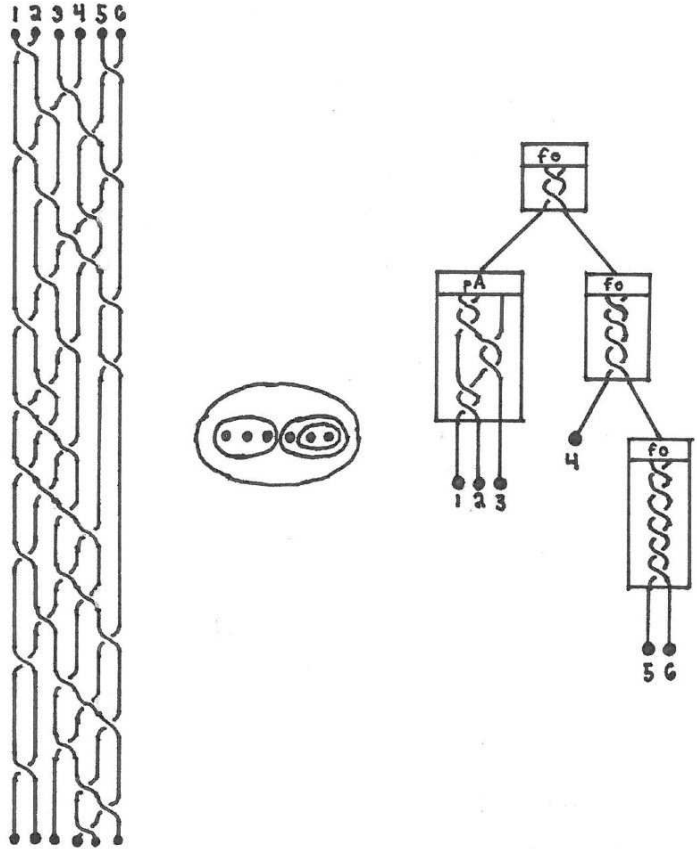}}
  \normalsize
    \caption{This diagram shows a pure braid on the left, a representative reduction curve $\lrp{2_1\uparrow 3_1}\uparrow \lrp{2_1 \uparrow 2_1}$ in the center, and the TN braid tree structure on the right.}
    \label{graph:BraidTree}
\end{figure}

In what sense does a pure braid have this structure?  First of all, we will identify each vertex as a pure braid of irreducible type, and each edge as a strand from one of the vertex braids.  The single edge associated with a given vertex which is closer to the root vertex is the strand into which this vertex's braid is ``injected."  This reflects the replacement or insertion process inherent in the $T_1$ classification.  The root vertex is just the braid corresponding to the points and curves immediately within the fundamental, or outer-most reduction curve.  Each vertex is either a finite order braid, a pseudo Anosov braid, or a single strand called a leaf vertex.  This whole picture is much more understandable with an an actual diagram, Fig.~(\ref{graph:BraidTree}).

We can picture the original braid by repeatedly inserting each vertex braid into the strand above it, until there is only one braid left.  This view of a pure braid has a number of nice properties.  First of all, the tree structure is itself a conjugacy invariant, and directly reflects not only the $T_1$ classification, but also the $T_2$ classification, due to identified fo and pA vertices.  Furthermore, each finite order pure braid is uniquely identified just by the power, $k$, of the full twist, $\beta_{fo}^k = \lrp{\Delta^2}^k$.  Different pA braids can be roughly sorted by their stretching factor $\lambda$, which is certainly a conjugacy invariant.  However, this does not uniquely determine a pA braid type, since there can be two non-conjugate pA braids with the same stretching factor.  Because of this, the tree structure does not, on its own, solve the conjugacy problem for pure braids.\footnote{Different pA braid types can be differentiated by some version of the Garside algorithm.  Even with this included, the tree structure does not act as a perfect solution to the conjugacy problem.  For a set of vertex braids under a vertex that is finite order, we must find a standard order to label these ``children" vertices by.  Notice that we can permute the order with out changing the ``parent" braid, since the permutation amounts to a conjugation, and powers of the full twist do not change upon conjugation with any braid.  We could create any coherent order we like, such as: fo braids first and by ascending order, then pA braids sorted first by stretching factor, second by twist power of the Garside algorithm, and finally by the Garside "tail" length.  On the other-hand, the order of ``children" vertices under a pA ``parent" vertex would have a fixed ordering determined by the Garside normal form of the parent vertex.  Even after all this there is some additional, unexplored, subtlety involved in making the TN tree structure a normal form for pure braids.}  This does, however, solve the conjugacy problem for pure braids that have only fo components.  It also reduces the the computational complexity of the Garside algorithm, since it is necessary to use it only on the possibly smaller pA components.  Essentially, by using the tree structure (without the ordering mentioned in the footnote) we have given up on a complete solution to the conjugacy problem.  However this is more than made up for by the natural fit of this description for vortex motion.  Vortex fluid motion naturally occurs on many different length and time scales, which are represented by the different levels, or degrees away from the tree vertex, of the tree structure.  The motions of a group of vortices represented by a braid vertex must necessarily be on a smaller length scale, and therefore a faster timescale, than those vortex motions represented by any vertex braid that includes the previous braid in its subtree.  The notion of reduction curves that predicate the whole tree structure is a natural way of describing Lagrangian coherent structures.  Lagrangian coherent structures are regions of fluid that do not mix with other regions.  That is that they are bounded by material lines, which are analogous to our reduction curves.  One final advantage of using the tree structure is that there are completely analogous data structures that are well studied in computer science.  For these we simply change some nomenclature, calling vertices nodes.  Each node is its own object, with pointers, corresponding to edges, to its ``parent" object and all of its ``child" objects.  Because of this structure, a whole tree can be created recursively.

All of this classification is based on one simple idea that applies to multiple related areas.  There are often groups of elements of a system that we can treat as a single element.  For mapping classes, these elements were the maps restricted to a component of the plane broken up by reduction curves.  The map within the fundamental reduction curve treats other reduction curves as if they were just identified points.  Each reduction curve does the same thing for other reduction curves that are internal to it.  For braids, these elements are groups of strands that can have a rubber band wrapped around them slide to the bottom of the braid without distortion.  These subsets of braids interact with other braid strands as if they were just one single strand.  For point vortices, these elements are groups of vortices that are close to one another, and appear from far away to be just single vortices of larger strength.  This idea relates all three of these areas in a topological sense.  Vortex motions trace out braids, which constrain the class of map that the advected fluid particles will experience due to periodically moving vortices.  In certain regimes this idea is rigorous, not only up to topology, but also analytically.  In the next chapter on Algorithms, we will present an analytical method for replacing a pair of vortices with a single larger strength vortex, while keeping the dynamics the same.

\chapter{Algorithms}
\label{ch:alg}

At times it seemed that the writing and development of algorithms was the sole component to this project, and if judged by hours worked and lines of code written (about 50,000), this might not be too far from the truth.  However, it would have been far more intractable if we had not made the choice to encode the ideas in an object-oriented fashion.  In each piece of code written, vortices, collections of vortices, orbits, periodic orbits, dynamics, and much more are treated as objects in their own right, with their own attributes and methods.  This facilitated code reuse, allowed for a much more legible coding style, and freed us up to focus on the ideas behind the code.  This chapter will focus, to a large extent, on three broad categories of ideas that form the backbone of the code.

First, is the concept of using Lie Transform perturbation theory to help the accuracy when numerically evolving the equations of motion.  When two vortices are very close together, a number of problems arise that can be ameliorated by treating the two vortices as if they were one single vortex of a larger strength.  Perhaps more importantly, this has the feel of the general classification scheme in ch.~\ref{ch:braid}, which treats periodic orbits as being built up from the motion of collections of fewer vortices.  However, this is the rare case where this idea can be treated analytically as opposed to just topologically.  This section closely follows from our paper~\cite{Smith:2011ey}.

Second, are the ideas that allowed us to find the periodic orbits in phase space, many of which are unstable, and thus very hard to extract.  We will detail how to find orbits that return close to their starting point, and are therefore good candidates for reduction.  Then we recount how to build a representative ensemble of such candidates to ensure that the set of periodic orbits below a certain period are themselves representative.  Next, and most importantly, we outline the algorithm which takes a candidate orbit and reduces it to a true periodic orbit using a modified Newton's method on the entire orbit.

There are many more algorithmic ideas that were integral to finding, categorizing, and investigating the set of periodic orbits, which we will not have the space to recount in full.  A final subsection will give a brief accounting of some of these additional implemented ideas.  These mainly concern the concepts necessary to classify periodic orbits such as the symmetry induced equivalence of periodic orbits, orbit chirality, and geometric orbit to braid conversion.  The very important braid related concepts of braid ordering, braid type tree structure, and braid group manipulations were covered in chapter~\ref{ch:braid}, and will only elicit a brief mention here.

\section{Lie Transform Regularization}
\label{sec:LieTrans}

To find the orbit trajectories of the point vortex system, one must simply integrate the equations~(\ref{eq:PVevolution}) numerically, using an appropriate adaptive time stepping method.  As the separation between a pair of vortices relative to all other inter-vortex length scales decreases, however, the computational time required diverges.  Accuracy is usually the most discouraging casualty when trying to account for such vortex motion, though the varying energy of this ostensibly Hamiltonian system is a potentially more serious problem.  We solve these problems by a series of coordinate transformations:  we first transform to action-angle coordinates, which, to lowest order, treat the close pair as a single vortex amongst all others with an internal degree of freedom.  We next, and most importantly, apply Lie transform perturbation theory to remove the higher-order correction terms in succession.  The overall transformation drastically increases the numerical efficiency and ensures that the total energy remains constant to high accuracy.  Perhaps more importantly, this transformation is a rigorous analytical example of the idea at the heart of the $T_1$ periodic orbit classification:  that groups of vortices will often act as if they are one single vortex.
		
\subsection{Posing the Problem}

	Consider the motion of a system of point vortices, where there are two like-signed vortices that are much closer to one another than to any other vortex.  These two, to first approximation, will simply rotate about their center of circulation with an angular frequency that is inversely dependent on the separation length, squared.  Therefore, the closer this vortex pair is, the faster their angular movement will be compared to that of other vortices.  To accommodate this motion, the adaptive integration method will reduce the time-steps between integrator function-calls to maintain the prescribed tolerance.  This prohibitively and unnecessarily slows down the integration of the system as a whole.  One naive way of circumventing this problem is to adaptively change the tolerance, though this has the unpalatable consequence of decreasing the accuracy of the computed orbit.  While one could possibly ignore this with appeals to the chaotic, non-integrable nature of the dynamics, there is another serious problem to consider.  In most integrators the decreased accuracy will cause the vortices to systematically overshoot their ideal near-circular orbit.  This has the effect of increasing the vortex-pair separation, and therefore decreasing the energy in this ostensibly Hamiltonian system.  It turns out that one does not have to choose either lowered accuracy and changing energy or long integration times.  We develop a transformation in this section which largely alleviates both of these problems, and does so in a physically intriguing way.

\subsection{Outline of the Solution}

	Intuitively, a pair of like-signed vortices that are considered close when compared to all other inter-vortex distances -- what we shall call a vortex ``dimer" -- will rotate about its center in a manner that appears unaffected by the presence of all other vortices.  That is, the internal motion of the dimer, that of the constituent vortex pairs with respect to their center of circulation, is approximately the integrable motion of the pair on their own.  Similarly, other vortices will interact with the dimer roughly as if it were a single vortex.  This crude approximation has the useful feature that the now integrable dimer motion ceases to be the limiting factor in choosing the step size for numerical integration.
	
	Of course, we can not perfectly shoe-horn our problem into this picture.  After some initial transformations to express the dimer in action-angle coordinates, we find correction terms which modify this picture.  Fortunately these terms can be expanded in powers of a small factor, and therefore our system becomes amenable to perturbation methods.  In particular, we use Lie transform perturbation theory, see ~\cite{bib:DragtFinn}~\cite{bib:Deprit}~\cite{bib:Cary1}~\cite{bib:Littlejohn2}, because the invariance of the symplectic structure of our phase space is manifest with these transforms.  From here, we choose generating functions for the Lie transforms that will get rid of the correction terms successively at each order, while maintaining, as best we can, the integrable nature of the dimer's internal motion.  When a correction term which we can not transform away does arise, we will see that it does not present any real computational problem.  Indeed it will even have the interesting physical interpretation of an additional field that couples only to vortices which have an internal ``spin" degree of freedom, i.e., the dimers.  Overall we will have succeeded in converting the original N-body problem into an (N-1)-body problem.
	
	We will show that the overall transformation alleviates the small time-step issue, maintains a constant energy, and results in very accurate orbits.  It should be noted that KAM theory\cite{bib:Arnold1} and other super-convergent methods are not applicable here, seeing that they can not deal with the non-integrable system which arises after our first Lie transformation.  Likewise it should be clear that our method is not connected with fast multipole methods, which increase the speed of integrating an N-body system by including only the most important pairwise interactions for each particle.  We consider all $N(N-1)/2$ pairwise calculations, and therefore do not attempt to change how the algorithmic complexity scales with point-vortex number.  We are more concerned with accuracy and how we can preserve it, while solving the specific small time-step issue.

\subsection{A Vortex Dimer}
\subsubsection{Hamiltonian for a Vortex Dimer}

Returning to Eq.~(\ref{eq:VortexHamiltonian}) for the Hamiltonian of $N$ vortices in an infinite domain, we suppose that two like-signed vortices are very close to one another.  We further suppose that these two are the ones labeled $m$ and $n$, and we refer to this pair as a ``vortex dimer.''  We rewrite the Hamiltonian, $H\left(\bfr_1,\ldots,\bfr_m,\ldots,\bfr_n,\ldots,\bfr_N\right)$, as follows:
\begin{eqnarray}
H &\equiv&
-\frac{\Gamma_m\Gamma_n}{2\pi}\ln \left|\bfr_m-\bfr_n\right|
-\sum_{j\neq m,n}^N\sum_{k\neq j,m,n}^N\frac{\Gamma_j\Gamma_k}{4\pi}\ln \left|\bfr_j-\bfr_k\right|
\nonumber\\
& &
-\sum_{j\neq m,n}^N\frac{\Gamma_m\Gamma_j}{2\pi}\ln \left|\bfr_m-\bfr_j\right|
-\sum_{j\neq m,n}^N\frac{\Gamma_n\Gamma_j}{2\pi}\ln \left|\bfr_n-\bfr_j\right|.
\end{eqnarray}
The first term on the right is the Hamiltonian of interaction between vortices $m$ and $n$ within the dimer.  The second term is the Hamiltonian of interaction amongst all of the other vortices in the system.  The third and fourth terms describe the interaction of the dimer with all the other vortices in the system; in particular, the third term describes the interaction between all of the other vortices and vortex $m$, and the fourth term describes the interaction between all of the other vortices and vortex $n$.

\subsubsection{Transformation to Center of Circulation and Relative Coordinates}

To facilitate the eventual approximation of the vortex dimer as a single vortex, it helps to transform the dimer coordinates to center of circulation, Eq.~(\ref{eq:CenterOfCirculation}), and relative displacement, Eq.~(\ref{eq:RelativeDisplacement}), coordinates.  These coordinates, $\bfR=\langle X,Y\rangle$ and $\bfr=\langle x,y\rangle$, have Poisson bracket relations given by Eq.~(\ref{eq:PBR1} - \ref{eq:PBR3}).  To refresh our memory, they are
\be
\left\{ X, Y \right\} = \frac{1}{\Gamma_R}, \; \left\{ x, y \right\} = \frac{1}{\Gamma_r}, \; \text{and}\; \{x, Y\} = \{y, X\} = 0, \nonumber
\ee
where the total circulation, $\Gamma_R$, and reduced circulation, $\Gamma_r$, of the vortex dimer are
\be
\Gamma_R\equiv\Gamma_m + \Gamma_n \; \text{and} \; \Gamma_r\equiv\frac{\Gamma_m\Gamma_n}{\Gamma_m + \Gamma_n}. \nonumber
\ee
Thus, $X$ and $Y$ comprise a canonically conjugate pair, as do $x$ and $y$.  It follows that the Poisson bracket for the full set of coordinate is
\begin{eqnarray}
\left\{ A, B\right\} &\equiv&
\frac{1}{\Gamma_R}\left(\frac{\partial A}{\partial X}\frac{\partial B}{\partial Y} - 
\frac{\partial A}{\partial Y}\frac{\partial B}{\partial X}\right)\nonumber\\ 
&+&
\frac{1}{\Gamma_r}\left(\frac{\partial A}{\partial x}\frac{\partial B}{\partial y} - 
\frac{\partial A}{\partial y}\frac{\partial B}{\partial x}\right)\nonumber\\ 
&+&
\sum_{j\neq m, n}\frac{1}{\Gamma_j}
\left(\frac{\partial A}{\partial x_j}\frac{\partial B}{\partial y_j} - 
\frac{\partial A}{\partial y_j}\frac{\partial B}{\partial x_j}\right).
\end{eqnarray}

It remains to write the Hamiltonian in the new coordinates.  Using Eqs.~(\ref{eq:inversem}) and (\ref{eq:inversen}), this is straightforward, and we find
\begin{eqnarray}
H
&=&
-\frac{\Gamma_R\Gamma_r}{2\pi}\ln\left|\bfr\right|
-\sum_{j\neq m,n}^N\sum_{k\neq j,m,n}^N\frac{\Gamma_j\Gamma_k}{4\pi}\ln \left|\bfr_j-\bfr_k\right|+
\nonumber\\
& &
-\sum_{j\neq m,n}^N\frac{\Gamma_m\Gamma_j}{2\pi}
\ln \left|\bfR-\bfr_j - \frac{\Gamma_r}{\Gamma_m}\bfr\right|
-\sum_{j\neq m,n}^N\frac{\Gamma_n\Gamma_j}{2\pi}
\ln \left|\bfR-\bfr_j + \frac{\Gamma_r}{\Gamma_n}\bfr\right|
\nonumber\\
&=&
-\frac{\Gamma_R\Gamma_r}{2\pi}\ln\left|\bfr\right|
-\sum_{j\neq m,n}^N\sum_{k\neq j,m,n}^N\frac{\Gamma_j\Gamma_k}{4\pi}\ln \left|\bfr_j-\bfr_k\right| + 
\nonumber\\
& &
-\sum_{j\neq m,n}^N\frac{\Gamma_m\Gamma_j}{4\pi}
\ln \left[
\left|\bfR-\bfr_j\right|^2 - 2\frac{\Gamma_r}{\Gamma_m}\left(\bfR-\bfr_j\right)\cdot\bfr + \frac{\Gamma_r^2}{\Gamma_m^2}r^2\right] +
\nonumber\\
& &
-\sum_{j\neq m,n}^N\frac{\Gamma_n\Gamma_j}{4\pi}
\ln \left[
\left|\bfR-\bfr_j\right|^2 + 2\frac{\Gamma_r}{\Gamma_n}\left(\bfR-\bfr_j\right)\cdot\bfr + \frac{\Gamma_r^2}{\Gamma_n^2}r^2\right]
\nonumber\\
&=&
-\frac{\Gamma_R\Gamma_r}{2\pi}\ln\left|\bfr\right|
-\sum_{j\neq m,n}^N\sum_{k\neq j,m,n}^N\frac{\Gamma_j\Gamma_k}{4\pi}\ln \left|\bfr_j-\bfr_k\right|
-\sum_{j\neq m,n}^N\frac{\Gamma_R\Gamma_j}{2\pi}\ln\left|\bfR-\bfr_j\right|+
\nonumber\\
& &
-\sum_{j\neq m,n}^N\frac{\Gamma_m\Gamma_j}{4\pi}
\ln \left(1 - 2\frac{\Gamma_r}{\Gamma_m}\frac{\left(\bfR-\bfr_j\right)\cdot\bfr}{\left|\bfR-\bfr_j\right|^2} + 
\frac{\Gamma_r^2}{\Gamma_m^2}\frac{|\bfr|^2}{\left|\bfR-\bfr_j\right|^2}\right)+
\nonumber\\
& &
-\sum_{j\neq m,n}^N\frac{\Gamma_n\Gamma_j}{4\pi}
\ln \left(1 + 2\frac{\Gamma_r}{\Gamma_n}\frac{\left(\bfR-\bfr_j\right)\cdot\bfr}{\left|\bfR-\bfr_j\right|^2} + 
\frac{\Gamma_r^2}{\Gamma_n^2}\frac{|\bfr|^2}{\left|\bfR-\bfr_j\right|^2}\right).
\label{eq:comhamiltonian}
\end{eqnarray}

As we will be referring to the terms of this Hamiltonian often, it makes sense to label them.  The first three terms are collectively denoted $H_0$.  The first term on its own is $H_{01}$, while $H_{02}$ corresponds to the second and third terms together.  The Hamiltonian of interaction between the two vortices within the dimer is $H_{01}$.  The first term of $H_{02}$ is the interaction Hamiltonian of all non-dimer vortices in the system, and the second term of $H_{02}$ is the principal interaction Hamiltonian between the dimer and all of the other vortices.  Altogether $H_0$ makes the crude approximation that the dimer, while having internal degrees of freedom, is merely a vortex of circulation $\Gamma_R$ at position $\bfR$, interacting normally with the N-2 other vortices.  The rest of the Hamiltonian constitutes correction terms to this picture.

\subsection{Ordering}
\subsubsection{Ordering the Hamiltonian}

We would like to develop an ordering scheme whereby the fastest and most important motion is the oscillation of the relative displacement vector of the vortex dimer, described by the first term $\left(H_{01}\right)$ in Eq.~(\ref{eq:comhamiltonian}).   We introduce a formal ordering parameter, $\epsilon$, where the small quantities, $|\bfr|/|\bfR-\bfr_j|$ and $|\bfr|/|\bfr_i-\bfr_j|$, are of order $\epsilon$ or smaller.  Eq.~(\ref{eq:comhamiltonian}) is thus rewritten
\begin{eqnarray}
H
&=&
-\frac{\Gamma_R\Gamma_r}{2\pi}\ln\left|\bfr\right|
-\sum_{j\neq m,n}^N\sum_{k\neq j,m,n}^N\frac{\Gamma_j\Gamma_k}{4\pi}\ln \left|\bfr_j-\bfr_k\right|
-\sum_{j\neq m,n}^N\frac{\Gamma_R\Gamma_j}{2\pi}\ln\left|\bfR-\bfr_j\right|+
\nonumber\\
& &
-\sum_{j\neq m,n}^N\frac{\Gamma_m\Gamma_j}{4\pi}
\ln \left(1 - 2\epsilon\frac{\Gamma_r}{\Gamma_m}\frac{\left(\bfR-\bfr_j\right)\cdot\bfr}{\left|\bfR-\bfr_j\right|^2} + 
\epsilon^2\frac{\Gamma_r^2}{\Gamma_m^2}\frac{|\bfr|^2}{\left|\bfR-\bfr_j\right|^2}\right)+
\nonumber\\
& &
-\sum_{j\neq m,n}^N\frac{\Gamma_n\Gamma_j}{4\pi}
\ln \left(1 + 2\epsilon\frac{\Gamma_r}{\Gamma_n}\frac{\left(\bfR-\bfr_j\right)\cdot\bfr}{\left|\bfR-\bfr_j\right|^2} + 
\epsilon^2\frac{\Gamma_r^2}{\Gamma_n^2}\frac{|\bfr|^2}{\left|\bfR-\bfr_j\right|^2}\right)
\label{eq:orderedhamiltonian}
\end{eqnarray}
One should immediately note that in this scheme, all of $H_0$ is of order zero in $\epsilon,$ and therefore the internal dimer motion alone does not yet enjoy the distinction of being the unperturbed motion.  We will soon remedy this, when we also order the Poisson bracket, or equivalently the Poisson tensor, Eq.~(\ref{eq:PoissonTensor}), in $\epsilon$.  The Hamiltonian ordering does, however, indicate that the correction terms are of order $\epsilon^2$ and higher, though it is not yet manifest.

\subsubsection{Ordering the Poisson Tensor}

Lie transform perturbation theory will require that we use the Lie derivative, see~Eq.(\ref{eq:LieDerFunc}).  To ensure that the Lie derivative will deal with the $\epsilon$ ordering correctly, we will need to order the Poisson bracket as well as the Hamiltonian.  For the sake of transparency we adopt the Poisson tensor framework of symplectic geometry instead of the Poisson bracket formalism.  These are both interchangeable, but ordering, and indeed perturbing, the Poisson tensor, $\pten$, is conceptually much simpler.  

We have two goals that we wish to accomplish with the ordering of the Poisson tensor.  The first goal, alluded to in the previous section, deals with how the Lie derivative handles the ordering.  This primary concern will be explained later, after we have introduced the Lie derivative machinery.  However, we will state here the ordering of the Poisson tensor that arises from this consideration:
\begin{equation}
\pten = \pten_0 + \epsilon^2 \pten_2
\label{eq:PoissonOrdering}
\end{equation}
where the ${2\choose 0}$-tensors $\pten_0$ and $\pten_2,$ expressed as $2N \times 2N$ matrices, are:
\begin{equation}
 \pten_0 = 
 \begin{pmatrix}  0 & \mathbb{A} \\
  -\mathbb{A} & 0
 \end{pmatrix}, \;
 \pten_2 =  
 \begin{pmatrix}  0 & \mathbb{B} \\
  -\mathbb{B} & 0
 \end{pmatrix}, \; \; \text{where} \; \;\nonumber
 \end{equation}
 \begin{equation}
 \mathbb{A} = 
 \begin{pmatrix} 
  \frac{1}{\Gamma_r} &  & \text{\huge{0}} \\
   & 0 &  \\
  \text{\huge{0}} & & \ddots
 \end{pmatrix}, \; \;
 \mathbb{B} = 
 \begin{pmatrix} 
  0 &  &  & \text{\huge{0}}  \\
   & \frac{1}{\Gamma_R} & & \\
   & & \frac{1}{\Gamma_{r_j}} & \\
   \text{\huge{0}} & &  & \ddots 
 \end{pmatrix}.
 \label{eq:PoissonTensor2}
 \end{equation}
 
Our second goal is to recognize that the fast internal movement of the dimer is the most important motion, and therefore elevate it to the status of unperturbed motion in our eventual Lie transform perturbation theory.  The Poisson tensor ordering of~Eq.(\ref{eq:PoissonOrdering}) achieves this.  That is, the zero-order term of $\dot{z} = \pten dH,$ when expanded in $\epsilon,$ is the motion of the dimer alone:
\begin{align}
\dot{z} &= \left(\pten_0 + \epsilon^2 \pten_2\right)d\left(H_0 + \mathcal{O}(\epsilon^{2})\right)\nonumber\\  
&= \pten_0 dH_{01} + \epsilon^2 \pten_2 dH_{02} +  \mathcal{O}(\epsilon^{2}), \;  \; or \nonumber\\
\dot{z}^j &= \pten_0^{jk} \frac{\partial H_{01}}{\partial z^k} + \epsilon^2 \pten_2^{jk} \frac{\partial H_{02}}{\partial z^k} +  \mathcal{O}(\epsilon^{2}).
\end{align}

While this justifies our consideration of the internal dimer motion as the unperturbed motion, one should take care in interpreting the rest of the above dynamic expansion.  In particular, the ordering is absolutely correct only when comparing the relative size of different contributions to the evolution of an individual coordinate, and not when comparing a single contribution to the evolution of one coordinate with that of another coordinate.  This is a minor issue, and certainly will not affect the development of our overall Lie transform method.

\subsection{The Unperturbed Problem}

The motion of two vortices, that of the unperturbed problem, was solved in section~\ref{sec:pvm2}.  We have already transformed the dimer coordinates to center of circulation and relative displacement coordinates.  We can further transform these coordinates to action angle coordinates, $\lrbk{J,\theta}$, see Eq.~(\ref{eq:DimerActionAngle}), which explicitly reflect that the distance between the two dimer vortices is a constant of the unperturbed motion.

The Hamiltonian of the unperturbed problem, $\left(H_{01}\right)$ of Eq.~(\ref{eq:orderedhamiltonian}) is
\begin{equation}
H_{01} = H = -\frac{\Gamma_R\Gamma_r}{4\pi}\ln{\lrp{2J}}. \nonumber
\end{equation}
Notice that the angle coordinate, $\theta$, is absent, which tells us that $\partial_t J = 0$.  Due to their Poisson bracket relations, Eq.~(\ref{eq:Jthpbracket}), these two coordinates comprise a conjugate pair.  In particular this tells us  that the Poisson tensor and its ordering remain unchanged if the coordinates are written,
\begin{equation}
\bfz = \langle\theta,X,x_j, \cdots, J, Y, y_j, \cdots\rangle.
\end{equation}

We next aim to write the Hamiltonian in the new coordinates.  Toward this end, we define the angles
\begin{equation}
\theta_j \equiv \arg\left[\left(Y-y_j\right) + i\left(X-x_j\right)\right],
\end{equation}
so that
\begin{equation}
\left(\bfR - \bfr_j\right)\cdot\bfr =
\sqrt{2J}\; \left|\bfR - \bfr_j\right|\; \cos\left(\theta - \theta_j\right).
\end{equation}
We also make use of the straightforwardly derived Fourier expansion
\begin{equation}
-\ln\left(1 \mp 2z\cos\theta+z^2\right) =
\sum_{\ell=1}^\infty\frac{2(\pm 1)^\ell}{\ell}\; \cos\left(\ell\theta\right) z^\ell.
\end{equation}
Then, after some algebra, the last two terms of Eq.~(\ref{eq:orderedhamiltonian}) transform to give us
\begin{eqnarray}
H
&=&
-\frac{\Gamma_R\Gamma_r}{4\pi}\ln\left(2J\right)
-\sum_{j\neq m,n}^N\sum_{k\neq j,m,n}^N\frac{\Gamma_j\Gamma_k}{4\pi}\ln \left|\bfr_j-\bfr_k\right|
-\sum_{j\neq m,n}^N\frac{\Gamma_R\Gamma_j}{2\pi}\ln\left|\bfR-\bfr_j\right|
\nonumber\\
& &
+
\sum_{\ell=2}^\infty\epsilon^\ell\;\frac{2}{\ell}
\left[
\frac{\Gamma_n^{\ell-1}+(-1)^\ell\Gamma_m^{\ell-1}}{\Gamma_R^{\ell-1}}
\right]
\sum_{j\neq m,n}^N\frac{\Gamma_r\Gamma_j}{4\pi}
\left(\frac{\sqrt{2J}}{\left|\bfR-\bfr_j\right|}\right)^\ell
\cos\left[\ell\left(\theta-\theta_j\right)\right].
\label{eq:finalhamiltonian}
\end{eqnarray}
We have thus succeeded in writing the Hamiltonian as
\begin{equation}
H = H_0 + \sum_{\ell=2}^\infty \epsilon^\ell H_\ell,
\end{equation}
where $H_0 = H_{01}+H_{02}$,
\begin{equation}
H_{01} = -\frac{\Gamma_R\Gamma_r}{4\pi}\ln\left(2J\right),\label{eq:o0}
\end{equation}
\begin{eqnarray}
H_{02} &=&
-\sum_{j\neq m,n}^N\sum_{k\neq j,m,n}^N\frac{\Gamma_j\Gamma_k}{4\pi}\ln \left|\bfr_j-\bfr_k\right|+\nonumber\\
& &-\sum_{j\neq m,n}^N\frac{\Gamma_R\Gamma_j}{2\pi}\ln\left|\bfR-\bfr_j\right|,\label{eq:o1}
\end{eqnarray}
and for $\ell\geq 2$,
\begin{equation}
H_\ell =
\frac{2}{\ell}
\left[
\frac{\Gamma_n^{\ell-1}+(-1)^\ell\Gamma_m^{\ell-1}}{\Gamma_R^{\ell-1}}
\right]
\sum_{j\neq m,n}^N\frac{\Gamma_r\Gamma_j}{4\pi}
\left(\frac{\sqrt{2J}}{\left|\bfR-\bfr_j\right|}\right)^\ell
\cos\left[\ell\left(\theta-\theta_j\right)\right].\label{eq:ol}
\end{equation}
The rather simple form of the higher-order terms $H_\ell$ is encouraging.  To the extent that the unperturbed motion of $\bfr$ is oscillatory, these terms will be straightforward to average and integrate over unperturbed orbits.

It is very important to note that $\theta$ and $J$ are action-angle variables for the unperturbed problem with Hamiltonian $H_{01}$.  They are not action-angle variables for the full Hamiltonian $H$.  Our strategy now will be to use perturbation theory to find a near-identity canonical transformation so that $\theta$ and $J$ are action-angle variables to higher order in $\epsilon$.

\subsubsection{The Unperturbed Motion}

The Hamiltonian $H_{01}$ is independent of all coordinates other than $J$, and therefore the coordinates $\bfR$, $J$, and $\bfr_j$ for $j\neq m,n$ are constant along unperturbed orbits.  Only $\theta$ varies along unperturbed orbits according to, Eq.~(\ref{eq:unperturbed}).  That is $\theta$ increases linearly as $\theta(t) = \theta_0-\Omega t$, with angular frequency $\Omega\equiv\frac{\Gamma_R}{4\pi J}$.

This means that averages of a phase function $A$ along unperturbed orbits are simply averages over the angle variable $\theta$.  These are accomplished with the operator
\begin{equation}
\langle A\rangle\equiv\frac{1}{2\pi}\int_0^{2\pi}A(\theta)\; d\theta. \label{eq:avgphasefn}
\end{equation}
The oscillatory part of a phase function $A$ is then denoted
\begin{equation}
\widetilde{A}\equiv A-\left\langle A\right\rangle.
\end{equation}

\subsubsection{Averaging the Hamiltonian Over Unperturbed Orbits}

We are soon going to need the average of our Hamiltonian over unperturbed orbits, so we compute it here.  Because $H_{01}$ and $H_{02}$ are independent of $\theta$, they are unchanged by averaging over $\theta$; that is
\begin{subequations}
\begin{eqnarray}
\langle H_{01}\rangle &=& H_{01},\\
\langle H_{02}\rangle &=& H_{02}.\\
\noalign{\noindent \mbox{
For $\ell\geq 2$, since $\left\langle\cos\left[\ell\left(\theta-\theta_j\right)\right]\right\rangle = 0$, it follows that
}}\nonumber\\
\langle H_\ell\rangle &=& 0. \label{eq:avHi}
\end{eqnarray}
\end{subequations}
The oscillatory parts of the Hamiltonian are then
\begin{subequations}
\begin{eqnarray}
\widetilde{H}_{01} &=& 0\\
\widetilde{H}_{02} &=& 0\\
\widetilde{H}_\ell &=& H_\ell.
\end{eqnarray}
\end{subequations}

\subsection{Lie Transform Perturbation Theory}
\subsubsection{Overview}
At this point, our phase-space coordinates are $\bfz = \langle\theta,X,x_j, \cdots, J, Y, y_j, \cdots\rangle,$ where $j\neq m, n$.  We introduced the Lie derivative in chapter~\ref{ch:pvm}, as acting on scalar fields, Eq.~(\ref{eq:LieDerFunc}), as well as on on arbitrary k-forms, Eq.~(\ref{eq:LieDerForm}).  Here, we will need to use the Lie derivative acting on vector fields and the Poisson tensor as well.  To refresh our memory, the Lie derivative, $\Lieder{\bfv}{}$, with respect to the vector field $\bfv$ acts on scalar fields $f\left(\bfz\right)$ and on other vector fields $\bfw\left(\bfz\right)$ in the following manner:
\begin{equation}
\Lieder{\bfv}{f} = v^if_{,i} = v^i\frac{\partial f}{\partial z^i},
\label{eq:LieDerivativeScalar}
\end{equation}
\begin{equation}
\left(\Lieder{\bfv}{\bfw}\right)^k = v^iw^k_{,i} - w^iv^k_{,i}.
\end{equation}

We can take the Lie derivative of any tensor by contracting this tensor with the requisite number of arbitrary 1-forms and vector fields, applying the Lie derivative to the resultant scalar, and then applying the Leibnitz rule.  We will need the Lie derivative of a Poisson tensor
\begin{equation}
\left(\Lieder{\bfv}{\pten}\right)^{jk} = \pten^{jk}_{,i}v^i - \pten^{ik}v^j_{,i} - \pten^{ji}v^k_{,i},
\label{eq:LieTensorDerivative}
\end{equation}
but the first term in Eq.~(\ref{eq:LieTensorDerivative}) is equal to zero, since our Poisson tensor has constant components.

The method of Lie transformations usually proceeds by picking a generating function, g, which gives a Hamiltonian vector field, $\HamVec{g}$, or
\begin{equation}
v^j = \pten^{jk}g_{,k}  \label{eq:scalargenerator}
\end{equation}
and noting that the following coordinate transformation is always canonical
\begin{equation}
\mybar{z} = \exp\left(+\Lieder{\bfv}{}\right) z.
\end{equation}
That is, the Lie derivative of $\pten$ disappears under such a generating function, thereby preserving the symplectic structure.  While this choice, Eq.~(\ref{eq:scalargenerator}),  overdetermines the generating vector field, we will use it until a little more subtlety is needed to address fourth-order corrections.  This transformation results in a new Hamiltonian
\begin{equation}
\mybar{H} = \exp\left(-\Lieder{\bfv}{}\right) H.
\end{equation}
If $\bfv$ is of order $\epsilon^j$, the transformation will affect the Hamiltonian only at order $\epsilon^j$ and higher.  

Our strategy will be to determine successive vector generators $\bfv_i$, starting at second order, that preserve the Poisson tensor and eliminate the Hamiltonian at all higher orders.  To leave the largest variety of near identity transformations at our disposal, we choose an infinite product form, in the manner of Dragt and Finn~\cite{bib:DragtFinn}:  
\begin{equation}
\mybar{\bfz} = 
\exp\left(+\epsilon^2\Lieder{2}{}\right)\exp\left(+\epsilon^3\Lieder{3}{}\right)\exp\left(+\epsilon^4\Lieder{4}{}\right)\cdots
\bfz.
\label{eq:oalltranform}
\end{equation}
Here, $\Lieder{n}{} \equiv \Lieder{\bfv_n}{}$.  This results in a new Hamiltonian and Poisson tensor:
\begin{eqnarray}
\mybar{H} &=&
\cdots\exp\left(-\epsilon^4\Lieder{4}{}\right)\exp\left(-\epsilon^3\Lieder{3}{}\right)\exp\left(-\epsilon^2\Lieder{2}{}\right)
H, \; \; \; \;\label{eq:HamTrans} \\
\mybar{\pten} &=&
\cdots\exp\left(-\epsilon^4\Lieder{4}{}\right)\exp\left(-\epsilon^3\Lieder{3}{}\right)\exp\left(-\epsilon^2\Lieder{2}{}\right)
\pten. \label{eq:PoisTrans}
\end{eqnarray}

\subsubsection{Ordering: Poisson Tensor Redux}

Now that we have the Lie derivative machinery at our disposal, we can justify our use of~Eq.(\ref{eq:PoissonOrdering}) for the initial Poisson ordering.  Consider the Lie derivative of a scalar $f$,~Eq.(\ref{eq:LieDerivativeScalar}), with respect to a vector field,~Eq.(\ref{eq:scalargenerator}), generated by the scalar $g$: 
\begin{equation}
\Lieder{\bfv}{f} = \pten^{jk}\frac{\partial g}{\partial z^k}\frac{\partial f}{\partial z^j}.
\end{equation}
Because of the symplectic structure of the Poisson tensor, we can partition the terms on the right hand side into two groups: those that have derivatives with respect to variables $\langle\theta,J\rangle$, and those that have derivatives with respect to variables $\langle X,x_j, \cdots, Y, y_j, \cdots\rangle$.  For the Hamiltonians and scalar generators we are considering, the magnitude of the $\langle\theta,J\rangle$ terms in the Lie derivative will be reduced by roughly a factor of $J$ when compared to the magnitude of $g$ and $f$ together.  Similarly, the magnitude of the $\langle X,x_j, \cdots, Y, y_j, \cdots\rangle$ terms in the Lie derivative will be reduced by roughly a factor of $\left|\bfR-\bfr_j\right|^2$ or $\left|\bfr_i-\bfr_j\right|^2$.  Therefore the ratio of the $\langle X,x_j, \cdots, Y, y_j, \cdots\rangle$ terms in the Lie derivative to the  $\langle\theta,J\rangle$ terms will be of order $\epsilon^2$.  This relative ordering can be formally achieved by ordering the initial Poisson tensor in the manner given by~Eq.(\ref{eq:PoissonOrdering}):
\begin{eqnarray}
\Lieder{\bfv}{f} &=& \left(\pten_0^{jk} + \epsilon^2 \pten_2^{jk}\right)\frac{\partial g}{\partial z^k}\frac{\partial f}{\partial z^j}\nonumber\\ 
&=& \pten_0^{jk}\frac{\partial g}{\partial z^k}\frac{\partial f}{\partial z^j} + \epsilon^2 \pten_2^{jk}\frac{\partial g}{\partial z^k}\frac{\partial f}{\partial z^j}.
\end{eqnarray}

\subsubsection{Ordering: Lie Transform}

Expanding each of the exponentials of~Eqs.(\ref{eq:HamTrans} \& \ref{eq:PoisTrans}) to fourth order and putting in the explicit initial ordering of the Hamiltonian and Poisson tensor gives:
\begin{eqnarray}
\mybar{H} &=& \left(1-\epsilon^4\Lieder{4}{} \right)\left(1-\epsilon^3\Lieder{3}{} \right)\left(1-\epsilon^2\Lieder{2}{} + \epsilon^4\frac{1}{2}\Lieder{2}{}^2\right)\left(H_0 + \sum_{n=2}\epsilon^n H_n\right) \label{eq:hexpansion}\\
\mybar{\pten} &=& \left(1-\epsilon^4\Lieder{4}{} \right)\left(1-\epsilon^3\Lieder{3}{} \right)\left(1-\epsilon^2\Lieder{2}{} + \epsilon^4\frac{1}{2}\Lieder{2}{}^2\right)\left(\pten_0 + \epsilon^2 \pten_2\right) \label{eq:hbarexpansion}
\end{eqnarray}
We suppose that both the final Hamiltonian and Poisson tensor are also ordered in $\epsilon$.  Then, expansion of the above in powers of $\epsilon$ and matching terms yields the sequence of equations
\begin{subequations}
\begin{eqnarray}
\mybar{H}_0 &=& H_0\label{eq:order0}\\
\mybar{H}_1 &=& 0\label{eq:order1}\\
\mybar{H}_2 &=& H_2 - \Lieder{2}{} H_0\label{eq:order2}\\
\mybar{H}_3 &=& H_3 - \Lieder{3}{} H_0\label{eq:order3}\\
\mybar{H}_4 &=& H_4 - \Lieder{4}{} H_0  - \Lieder{2}{} H_2 + \frac{1}{2} \Lieder{2}{}^2 H_0  \label{eq:order4}\\
&\vdots&\nonumber
\end{eqnarray}
\end{subequations}
and
\begin{subequations}
\begin{eqnarray}
\mybar{\pten}_0 &=& \pten_0\label{eq:porder0}\\
\mybar{\pten}_1 &=& 0\label{eq:porder1}\\
\mybar{\pten}_2 &=& \pten_2 - \Lieder{2}{} \pten_0\label{eq:porder2}\\
\mybar{\pten}_3 &=&  - \Lieder{3}{} \pten_0\label{eq:porder3}\\
\mybar{\pten}_4 &=& - \Lieder{4}{} \pten_0  - \Lieder{2}{} \pten_2 + \frac{1}{2} \Lieder{2}{}^2 \pten_0  \label{eq:porder4}\\
&\vdots&\nonumber
\end{eqnarray}
\end{subequations}

After applying the Lie transforms, we would like the equations of motion to consist of the motion of the dimer and the motion of all the other vortices with the dimer.  That is, $ \dot{\mybar{\bfz}} = \pten_0 dH_0+\pten_2 dH_0 $.  We can ensure this by requiring that order by order, starting at second order, the transformed Poisson tensor remains unchanged and the transformed Hamiltonian disappears.  Furthermore, secular perturbation theory forbids us from including any averaged, Eq.~(\ref{eq:avgphasefn}), contributions to our transformed Hamiltonians in the calculation of the vector generators.  This means that we will need to check that  $\langle \mybar{H}_\ell\rangle = 0$ at each order.  These considerations lead to the following conditions:
\begin{subequations}
\begin{eqnarray}
\widetilde{H_2} - \widetilde{\Lieder{2}{} H_0}   &=& 0 \label{eq:CondOrder2}\\
\widetilde{H_3} - \widetilde{\Lieder{3}{} H_0}   &=& 0 \label{eq:CondOrder3}\\
\widetilde{H_4} - \widetilde{\Lieder{4}{} H_0}  - \widetilde{\Lieder{2}{} H_2} + \frac{1}{2} \widetilde{\Lieder{2}{}^2 H_0} &=&  0 \label{eq:CondOrder4}
\end{eqnarray}
\end{subequations}
and
\begin{subequations}
\begin{eqnarray}
\Lieder{2}{} \pten_0 &=& 0 \label{eq:PCondOrder2}\\
\Lieder{3}{} \pten_0 &=& 0 \label{eq:PCondOrder3}\\
\Lieder{4}{} \pten_0  + \Lieder{2}{} \pten_2 - \frac{1}{2} \Lieder{2}{}^2 \pten_0  &=& 0. \label{eq:PCondOrder4}
\end{eqnarray}
\end{subequations}

\subsubsection{Ordering: Orders Two and Three}

To maintain the second-order motion that results from the interaction of the non-dimer vortices, as well as their interaction with the dimer as though it were a single particle of circulation $\Gamma_R$, we must demand that the conditions in Eq.~(\ref{eq:CondOrder2}) \& (\ref{eq:PCondOrder2}) are met.  The second of these conditions is trivially fulfilled if
\begin{equation}
v_2^i = \pten_0^{ik}g_{2,k}.
\end{equation}
Thus the vector generator at second order, which is now completely determined by a scalar generator $g_2$, has only two non-zero components, 
\begin{equation}
\bfv_2 = \langle \frac{1}{\Gamma_r}\frac{\partial g_2}{\partial J}, 0, \cdots, -\frac{1}{\Gamma_r}\frac{\partial g_2}{\partial \theta}, 0, \cdots \rangle.  
\end{equation}
This allows us to write the condition, Eq.~(\ref{eq:CondOrder2}), on our Hamiltonian as:
\begin{equation}
\widetilde{H_2}-\frac{\Gamma_R}{4\pi J}\widetilde{\frac{\partial g_2}{\partial\theta}} = 0. \label{eq:H2trans}
\end{equation}
If we demand that $g_2$ be single-valued to preclude secular terms and consider Eq.~(\ref{eq:avHi}), it is immediately apparent that $\langle \mybar{H}_2\rangle = \langle H_2\rangle - \langle \frac{\Gamma_R}{4\pi J}\frac{\partial g_2}{\partial\theta} \rangle = 0$.  Now we must solve Eq.~(\ref{eq:H2trans}) for the generator
\begin{equation}
g_2 = \frac{4\pi J}{\Gamma_R} \int^{\theta} \widetilde{H}_2\left(\theta'\right)\; d\theta'.
\end{equation}
From Eq.~(\ref{eq:ol}), this yields the generator
\begin{equation}
g_2 = \frac{\Gamma_r}{\Gamma_R}\;J^2
\sum_{j\neq m,n}^N\Gamma_j\;
\frac{\sin\left(2\theta-2\theta_j\right)}{\left|\bfR-\bfr_j\right|^2}.
\end{equation}

At this point it would be ideal to be able to renormalize the unperturbed orbits by treating all of $H_0$ as if it were the new unperturbed problem, in the manner of a superconvergent Lie transformation.  However, this problem is already non-integrable and therefore not amenable to this treatment.  So, we proceed to third order with the same unperturbed motion. 
Identical considerations lead to the following third-order generator
\begin{equation}
g_3 = \left(\frac{2}{3}\right)^2\frac{\sqrt{2}\Gamma_r\left(\Gamma_n-\Gamma_m\right)}{\Gamma_R^2}\;J^{\frac{5}{2}}
\sum_{j\neq m,n}^N\Gamma_j\;
\frac{\sin\left(3\theta-3\theta_j\right)}{\left|\bfR-\bfr_j\right|^3}.
\end{equation}

The generator $g_2$ provides the most important correction term present.  With it, the transformation given by Eq.~(\ref{eq:oalltranform}) takes us to a set of coordinates in which the dimer may be approximated as a single vortex of circulation $\Gamma_R$ and position $\bfR$.  Whereas this approximation was valid only to order $\epsilon$ in our original coordinates, in the new coordinates it is accurate to order $\epsilon^2$ -- that is, the first corrections to it are order $\epsilon^3$.  When the generator $g_3$ is included, the validity of this approximation is extended to order $\epsilon^3$. We next refine this transformation to push the corrections to still higher order.

\subsubsection{Ordering: Order Four}

Using $\Lieder{2}{}\pten_0 = 0$ and $H_2-\Lieder{2}{}H_0 = 0$, the conditions at fourth order, Eq.~(\ref{eq:CondOrder4}) \& (\ref{eq:PCondOrder4}), can be written
\begin{equation}
\widetilde{H_4} - \widetilde{\Lieder{4}{} H_0}  - \frac{1}{2}\widetilde{\Lieder{2}{} H_2} =  0 \label{eq:smallCondOrder4}
\end{equation}
and
\begin{equation}
\Lieder{4}{}\pten_0 + \Lieder{2}{}\pten_2 = 0.
\end{equation}
This latter equation is satisfied with the choice
\begin{equation}
v_4^j = \pten_0^{jk}g_{4,k} + \pten_2^{jk}g_{2,k}.
\end{equation}
Our vector generator at fourth order now has the form
\begin{eqnarray}
\bfv_4 = \langle &\frac{1}{\Gamma_r}&\frac{\partial g_4}{\partial J}, \frac{1}{\Gamma_R}\frac{\partial g_2}{\partial Y}, \frac{1}{\Gamma_j}\frac{\partial g_2}{\partial y_j}, \cdots, \nonumber\\
&-\frac{1}{\Gamma_r}&\frac{\partial g_4}{\partial \theta}, -\frac{1}{\Gamma_R}\frac{\partial g_2}{\partial X}, -\frac{1}{\Gamma_j}\frac{\partial g_2}{\partial x_j}, \cdots \rangle.  
\end{eqnarray}
All but two of these terms are already determined from $g_2$.  We will fix $g_4$ by considering Eq.~(\ref{eq:smallCondOrder4}).  Once again we demand that $g_4$ is single-valued and check for any averaged contributions to $\mybar{H}_4$.  Unlike at second and third order, we encounter an averaged contribution to $\mybar{H}_4$, due to $\Lieder{2}{} H_2$, that secular perturbation theory forbids us from transforming away.  Before we consider how this averaged term affects our perturbation theory, we can derive $g_4$ from the purely oscillatory parts of each term:
\begin{equation}
g_4 = \frac{4\pi J}{\Gamma_R} \int^{\theta} \left(\widetilde{H}_4  - \widetilde{\pten_2^{jk}g_{2,k}H_{0,j}} - \frac{1}{2}\widetilde{\pten_0^{jk}g_{2,k}H_{2,j}}\right)\; d\theta'.
\end{equation}
From this the fourth-order generator is straightforwardly, though laboriously, derived:
\begin{equation}
\begin{split}
g_4 \; = \;
& \frac{\Gamma_r}{4\Gamma_R^2}\;J^3
\sum_{j\neq m,n}^N\Gamma_j\; 
\left[
\left(2\left( \frac{\Gamma_n^3+\Gamma_m^3}{\Gamma_R^2} \right)-\Gamma_j \right) \frac{\sin\left(4\theta-4\theta_j\right)}{\left|\bfR-\bfr_j\right|^4} 
\; + \; 8\left(\Gamma_R+\Gamma_j\right) \frac{\sin\left(2\theta-2\theta_j\right)}{\left|\bfR-\bfr_j\right|^4}\;+
\bracketnewln{+}\sum_{k\neq j,m,n}^N\Gamma_k\; \left(
8\frac{\sin\left(2\theta-3\theta_j+\theta_k\right)}{\left|\bfR-\bfr_j\right|^3\left|\bfR-\bfr_k\right|} 
\;-\; 8\frac{\sin\left(2\theta-3\theta_j+\theta_{jk}\right)}{\left|\bfR-\bfr_j\right|^3\left|\bfr_j-\bfr_k\right|} 
-\frac{\sin\left(4\theta-2\theta_j-2\theta_k\right)}{\left|\bfR-\bfr_j\right|^2\left|\bfR-\bfr_k\right|^2} \right) \right],
\end{split}
\end{equation}
where $\theta_{jk} \equiv \arg\left[\left(y_j-y_k\right) + i\left(x_j-x_k\right)\right]$.

\subsubsection{Ordering: Averaged Contribution to $\mybar{H}_4$}

At fourth order, we must consider the effect of the non-oscillatory term due to $\Lieder{2}{} H_2$.  This gives us an averaged contribution to the Hamilton, 
\begin{equation}
\left\langle \mybar{H}_4\right\rangle = \frac{3\Gamma_r J^2}{4 \pi \Gamma_R}\;
\sum_{j\neq m,n}^N \sum_{k\neq m,n}^N\Gamma_j\Gamma_k\;
\frac{\cos\left(2\theta_j-2\theta_k\right)}{\left|\bfR-\bfr_j\right|^2\left|\bfR-\bfr_k\right|^2},
\label{eq:h4contrib}
\end{equation}
which changes the dynamics
\begin{eqnarray}
\dot{\mybar{z}^j} &=& \pten_0^{jk} \frac{\partial H_{01}}{\partial z^k} + \epsilon^2 \pten_2^{jk} \frac{\partial H_{02}}{\partial z^k} + \epsilon^4\pten_0^{jk} \frac{\partial \left\langle \mybar{H}_4\right\rangle}{\partial z^k}\nonumber\\ 
&=&\left(\pten_0+ \epsilon^2 \pten_2 \right) d \left( H_{01} +  H_{02} + \epsilon^4 \left\langle \mybar{H}_4\right\rangle \right).
\label{eq:FinalDynamics}
\end{eqnarray}
Only the angle variable of the dimer, $\theta$, is affected at fourth order.  In particular, $J$ remains a constant of the motion and 
 \begin{equation}
\dot{\theta} = -\Omega + \epsilon^4\frac{3J}{2 \pi \Gamma_R}\;
\sum_{j\neq m,n}^N \sum_{k\neq m,n}^N\Gamma_j\Gamma_k\;
\frac{\cos\left(2\theta_j-2\theta_k\right)}{\left|\bfR-\bfr_j\right|^2\left|\bfR-\bfr_k\right|^2}. \label{eq:freqcorrection}
\end{equation}

	Thus the rotational frequency of the dimer decreases at fourth order, though the extent to which it decreases depends on the configuration of other vortices.  To help interpret Eq.~(\ref{eq:freqcorrection}), we can write the unit vector from vortex $k$ to the dimer as $\hat{\bfR}_k = \frac{\bfR - \bfr_k}{\left|\bfR-\bfr_k\right|}$ and denote rotation of a vector by $\frac{\pi}{2}$, $\hat{z}\times\bfv = \hodge{\bfv}$, by the $\hodge{}$ operator.  This gives the following for the important part of the frequency correction:
 \begin{eqnarray}
& &\frac{\cos\left(2\theta_j-2\theta_k\right)}{\left|\bfR-\bfr_j\right|^2\left|\bfR-\bfr_k\right|^2} = \nonumber\\
& &\qquad\frac{\left[ \hat{\bfR}_j \cdot \left(\hat{\bfR}_k+\hodge{\hat{\bfR}_k} \right)\right]\left[ \hat{\bfR}_k \cdot \left(\hat{\bfR}_j+\hodge{\hat{\bfR}_j} \right)\right] }{\left|\bfR-\bfr_j\right|^2\left|\bfR-\bfr_k\right|^2}. \label{eq:freqcorrection2}
\end{eqnarray}

	Though it is not obvious, any configuration of any number of extra vortices will always decrease or maintain the dimer's rotational frequency.  For a single extra vortex, with either positive or negative circulation, this is apparent.  The complications arise when considering the three-body contributions to~Eq.(\ref{eq:freqcorrection}).  We can see, from~Eq.(\ref{eq:freqcorrection2}), that two extra, like-signed vortices which are co-linear with the dimer will further decrease the rotation rate.  If, however, they are both an equal distance from the dimer and are at right angles to each other with respect to the dimer, there will be zero net change in the dimer's rotation rate.  The results of these two scenarios are interchanged when the two extra vortices have opposite signed circulations.  When there are more than two extra vortices, it becomes more difficult to argue that the change in the dimer's rotation rate is negative-definite, though we have not seen a counter example in numerical simulations.

	As a physics analogy, we consider $\theta$ to be an internal variable of the dimer, much like spin.  Up until now the rest of the system has been transparent to the spin of the dimer.  With the addition of the fourth-order averaged Hamiltonian, we can view the spin as coupling to a new field; a field generated by the vortices, but coupled to only by vortices with non-zero spin, i.e., the vortex dimers.  This new field gives the dimer a position-dependent effective spin, which is smaller than the bare spin ($\Omega$).  This interesting view remains valid through at least fifth order in our calculations.

\subsubsection{Transformed Variables}

To fourth order, the transformation given in Eq.~{\ref{eq:oalltranform}} is written as
\begin{equation}
{\mybar{\bfz} \brace \bfz} = 
\left(1\pm \epsilon^2 \Lieder{2}{} \pm \epsilon^3 \Lieder{3}{}  + \epsilon^4 \left(\pm\Lieder{4}{}+\frac{1}{2}\Lieder{2}{}^2 \right)\right)
{\bfz \brace \mybar{\bfz}},
\end{equation}
and results in the following specific transformations, where $\alpha_j \equiv \left(2\frac{\Gamma_n^3+\Gamma_m^3}{\Gamma_R^2} - \Gamma_j \right)$ and $\beta_j \equiv  \left(\Gamma_R+\Gamma_j\right)$.  To get the forward transformation, simply use the upper operator of $\pm$ or $\mp$; to get the backward transformation, use the lower of $\pm$ or $\mp$ and replace all transformed coordinates $\mybar{\bfz}$ with the original coordinates $\bfz$ and vice-versa.

\begin{equation}
\begin{split}
\mybar{\theta} = \; \theta \; & \pm \; \epsilon^2\frac{2}{\Gamma_R}J \sum_{j\neq m,n}^N\Gamma_j   \frac{\sin\left(2\theta-2\theta_j\right)}{\left|\bfR-\bfr_j\right|^2}
 \; \pm \; \epsilon^3\frac{10 \sqrt{2}}{9}\frac{\left(\Gamma_n-\Gamma_m\right)}{\Gamma_R^2} J^{\frac{3}{2}} \sum_{j\neq m,n}^N\Gamma_j\frac{\sin\left(3\theta-3\theta_j\right)}{\left|\bfR-\bfr_j\right|^3} \;+\; \\
\; &\pm \; \epsilon^4\frac{1}{\Gamma_R^2}J^2 \sum_{j\neq m,n}^N\Gamma_j\; 
\left[  \left(\pm\Gamma_j+\frac{3}{4}\alpha_j \right)\frac{\sin\left(4\theta-4\theta_j\right)}{\left|\bfR-\bfr_j\right|^4} 
\;+\;6\beta_j \frac{\sin\left(2\theta-2\theta_j\right)}{\left|\bfR-\bfr_j\right|^4} \;+\; \right. \\
& \left. \quad+\;\sum_{k\neq j,m,n}^N\Gamma_k \left(
6\frac{\sin\left(2\theta-3\theta_j+\theta_k\right)}{\left|\bfR-\bfr_j\right|^3\left|\bfR-\bfr_k\right|}
\;-\; 6\frac{\sin\left(2\theta-3\theta_j+\theta_{jk}\right)}{\left|\bfR-\bfr_j\right|^3\left|\bfr_j-\bfr_k\right|}
\;+\;\left(\pm1-\frac{3}{4}\right)\frac{\sin\left(4\theta-2\theta_j-2\theta_k\right)}{\left|\bfR-\bfr_j\right|^2\left|\bfR-\bfr_k\right|^2} \right) \right]	\label{eq:Thetatransform}
\end{split}
\end{equation}

\begin{equation}
\begin{split}
\mybar{J} = \; J \: & \mp \; \epsilon^2\frac{2}{\Gamma_R}J^2 \sum_{j\neq m,n}^N\Gamma_j   \frac{\cos\left(2\theta-2\theta_j\right)}{\left|\bfR-\bfr_j\right|^2}
\;\mp\; \epsilon^3\frac{4 \sqrt{2}}{3}\frac{\left(\Gamma_n-\Gamma_m\right)}{\Gamma_R^2} J^{\frac{5}{2}} \sum_{j\neq m,n}^N\Gamma_j\frac{\cos\left(3\theta-3\theta_j\right)}{\left|\bfR-\bfr_j\right|^3} \;+\;  \\
\; & \mp\; \epsilon^4\frac{1}{\Gamma_R^2}J^3 \sum_{j\neq m,n}^N\Gamma_j\; 
\left[  \alpha_j\frac{\cos\left(4\theta-4\theta_j\right)}{\left|\bfR-\bfr_j\right|^4} 
\;+\;4\beta_j \frac{\cos\left(2\theta-2\theta_j\right)}{\left|\bfR-\bfr_j\right|^4}
\;\mp \frac{4 \Gamma_j}{\left|\bfR-\bfr_j\right|^4} \;+\;\right. \\
& \left.  \quad+\;\sum_{k\neq j,m,n}^N\Gamma_k \left(
4\frac{\cos\left(2\theta-3\theta_j+\theta_k\right)}{\left|\bfR-\bfr_j\right|^3\left|\bfR-\bfr_k\right|} 
\;-\;4\frac{\cos\left(2\theta-3\theta_j+\theta_{jk}\right)}{\left|\bfR-\bfr_j\right|^3\left|\bfr_j-\bfr_k\right|}\;+\; \right. \right. \\
& \left. \left. \qquad \mp4\frac{\cos\left(2\theta_j-2\theta_k\right)}{\left|\bfR-\bfr_j\right|^2\left|\bfR-\bfr_k\right|^2}
\;-\; \frac{\cos\left(4\theta-2\theta_j-2\theta_k\right)}{\left|\bfR-\bfr_j\right|^2\left|\bfR-\bfr_k\right|^2} \right) \right]	\label{eq:Jtransform}
\end{split}	
\end{equation}

\begin{equation}
\begin{split}
\mybar{X} =  \; X \; & \mp \; \epsilon^4\frac{2\Gamma_r}{\Gamma_R^2}J^2 \sum_{j\neq m,n}^N\Gamma_j\;  \frac{\sin\left(2\theta-3\theta_j\right)}{\left|\bfR-\bfr_j\right|^3} \\
\mybar{Y} =  \; Y\; & \pm \; \epsilon^4\frac{2\Gamma_r}{\Gamma_R^2}J^2 \sum_{j\neq m,n}^N\Gamma_j\;   \frac{\cos\left(2\theta-3\theta_j\right)}{\left|\bfR-\bfr_j\right|^3}\\
\mybar{x}_i =  \;  x_i \;&  \pm\; \epsilon^4\frac{2\Gamma_r}{\Gamma_R}J^2  \frac{\sin\left(2\theta-3\theta_i\right)}{\left|\bfR-\bfr_i\right|^3} \\
 \mybar{y}_i = \;  y_i\;& \mp \; \epsilon^4\frac{2\Gamma_r}{\Gamma_R}J^2  \frac{\cos\left(2\theta-3\theta_i\right)}{\left|\bfR-\bfr_i\right|^3}	\label{eq:XYtransform}
\end{split}
\end{equation}

It is important to explicitly note that our new, transformed system, given by Eq.~(\ref{eq:FinalDynamics}) and transformations Eq.~(\ref{eq:Thetatransform} - \ref{eq:XYtransform}), is in Hamiltonian form and therefore still conservative.  Furthermore, despite the fact that we perturbed the Poisson tensor, the symplectic structure is unchanged.

\subsection{Numerical Results}
We propose to use the above-described transformation as the basis of a numerical method, whereby we make this transformation, integrate the reduced system, and invert the transformation.  The overall efficacy of this Lie transform method will depend to a large degree on various factors such as the number, configuration, and circulation of vortices in the system.  To probe the essential behavior of this method, we will consider the simplest case, that of three vortices of equal circulation, two of which are much closer to one another.  This enables us to unambiguously state that our small perturbation parameter is  $\epsilon = |\bfr|/|\bfR-\bfr_j|$.  For reference, we used a Runge-Kutta-Fehlberg (4, 5) adaptive time-stepping numerical integrator.  It should be noted that there are numerical integrators which, when tailored, are better suited for use with the 2D point vortex system.  In particular, an appropriate member of the large family of symplectic integrators, which can preserve symplectic structure and conserve first integrals of motion, would be ideal~\cite{bib:Hairer} and~\cite{bib:Leimkuhler}.  

\subsubsection{Efficiency}

\begin{figure}[htbp]
  \centering
  \large
  \resizebox{\linewidth}{!}{\includegraphics{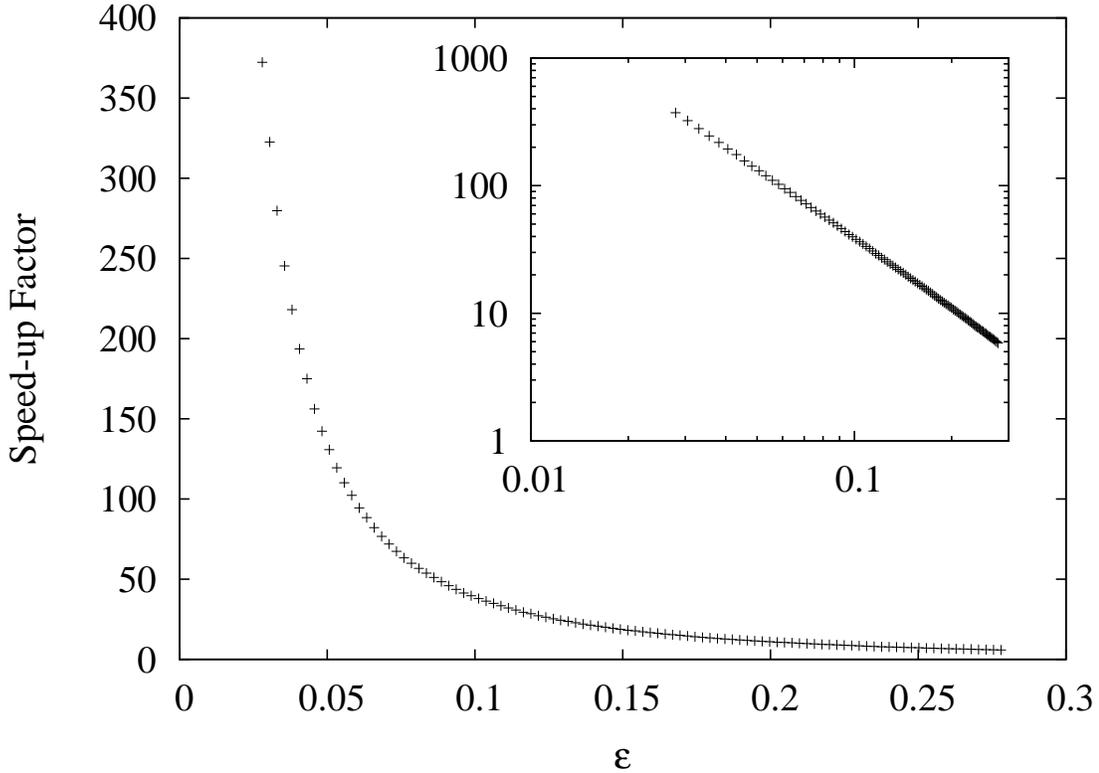}}
  \normalsize
    \caption{Speedup and Efficiency: The vertical axis shows the ratio of the number of calls to the adaptive time-stepper for the Regular Method $\left(N_R\right)$ to the Lie Method $\left(N_L\right)$, i.e. the speed-up factor.  This is plotted against  $\epsilon,$ the vortex dimer separation $\left|\bfr_n-\bfr_m \right|$ divided by the separation between the remaining vortex and the dimer $\left|\bfR-\bfr_j\right|.$  \; A log-log version of this graph is inset.  This corresponds to a power law $\frac{N_R}{N_L} \propto \epsilon^{\alpha}$ where $\alpha \approx -1.8$}
    \label{graph:graph1}
\end{figure}

We would like to measure how much faster an algorithm based on the dynamics Eq.~(\ref{eq:FinalDynamics}) and transformations Eq.~(\ref{eq:Thetatransform} - \ref{eq:XYtransform}) would run when compared to one based on the original point-vortex dynamics Eq.~(\ref{eq:PVevolution}).  The pertinent quantities we want to track are the number of calls each method makes to the base adaptive timer stepper.  In particular, we will consider the ratio of these two numbers, i.e., the speedup factor, and how it behaves when $\epsilon$ becomes small and our transformations therefore become more applicable.  Since the time that it takes a vortex in the dimer to rotate through a given angle scales as the dimer separation squared Eq.~(\ref{eq:unperturbed}, \ref{eq:RotationRate}), we would roughly expect the speedup factor to scale as $\epsilon^{-2}$ for all orders of Lie transforms.  As one can see in Figure \ref{graph:graph1}, this is very nearly the case.  The magnitude of the speedup factor will certainly depend on the tolerance that one sets.  However, increasing or decreasing the tolerance will not affect how the speedup factor scales with $\epsilon$.  Neither will this scaling be modified by the contributions to the Lie transform algorithm run-time from the transformations themselves.  While these do involve pairwise sums, they are calculated only when saving the state of the system is needed, and therefore do not constitute anything worse than a small multiplicative contribution to the overall run-time.  Indeed their inclusion is only an additive contribution to the run-time if one is interested only in initial and final vortex configurations.

\subsubsection{Accuracy}

\begin{figure}[htbp]
  \centering
  \large
  \resizebox{\linewidth}{!}{\includegraphics{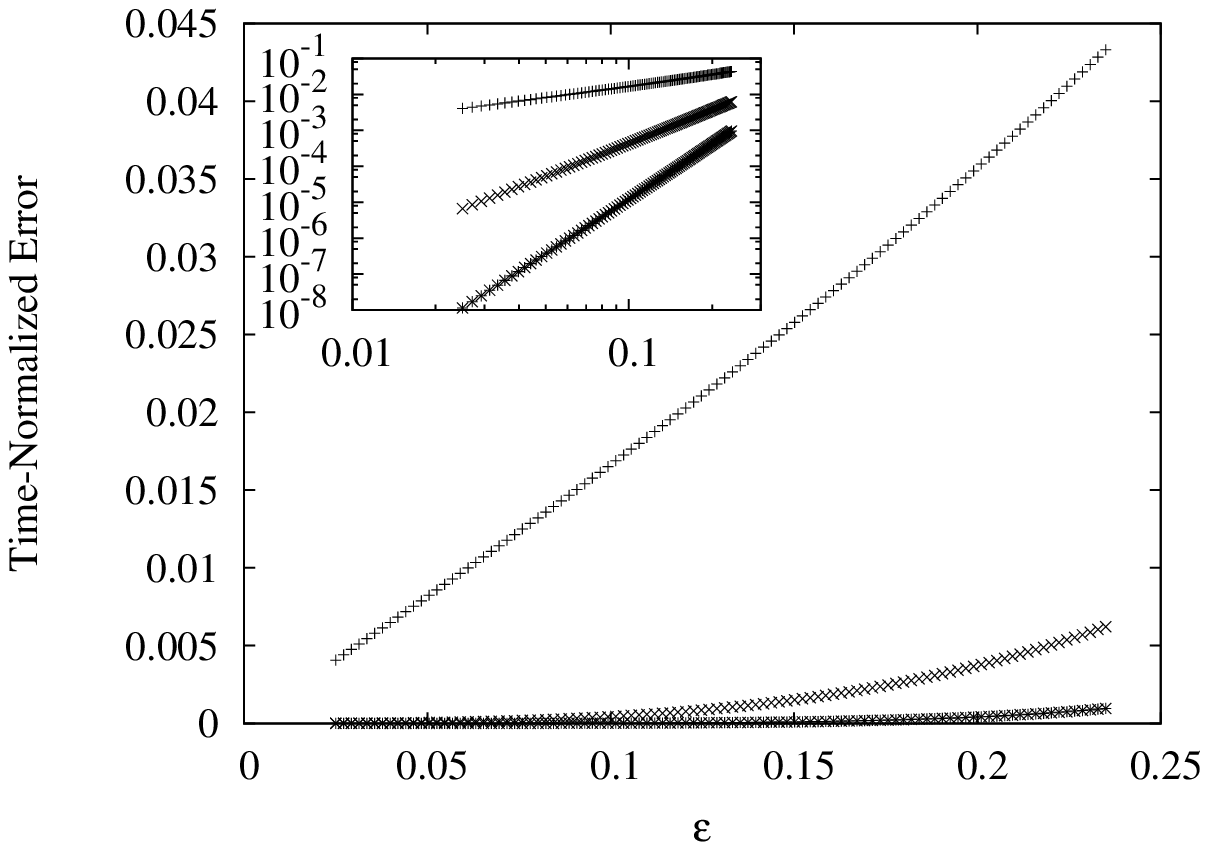}}
  \normalsize
    \caption{Accuracy of the Dimer Internal Motion: The vertical axis shows the natural norm of the difference between the final configuration space points evolved under the Lie and Regular methods, normalized by the time elapsed, i.e. $\frac{\left|\mbox{\bf\boldmath{z}}_{L,f}-\mbox{\bf\boldmath{z}}_{R,f}\right|}{\Delta t}$.  In particular, in this graph we consider the positions of only the dimer vortices with respect to their center of circulation.  The three lines, from top to bottom, represent the Lie transform at 0th, 2nd, and 4th orders respectively.  The inset log-log plot shows that they all follow power laws with exponents 1, 3, and 5 respectively.}
    \label{graph:graph2}
\end{figure}

\begin{figure}[htbp]
  \centering
  \large
  \resizebox{\linewidth}{!}{\includegraphics{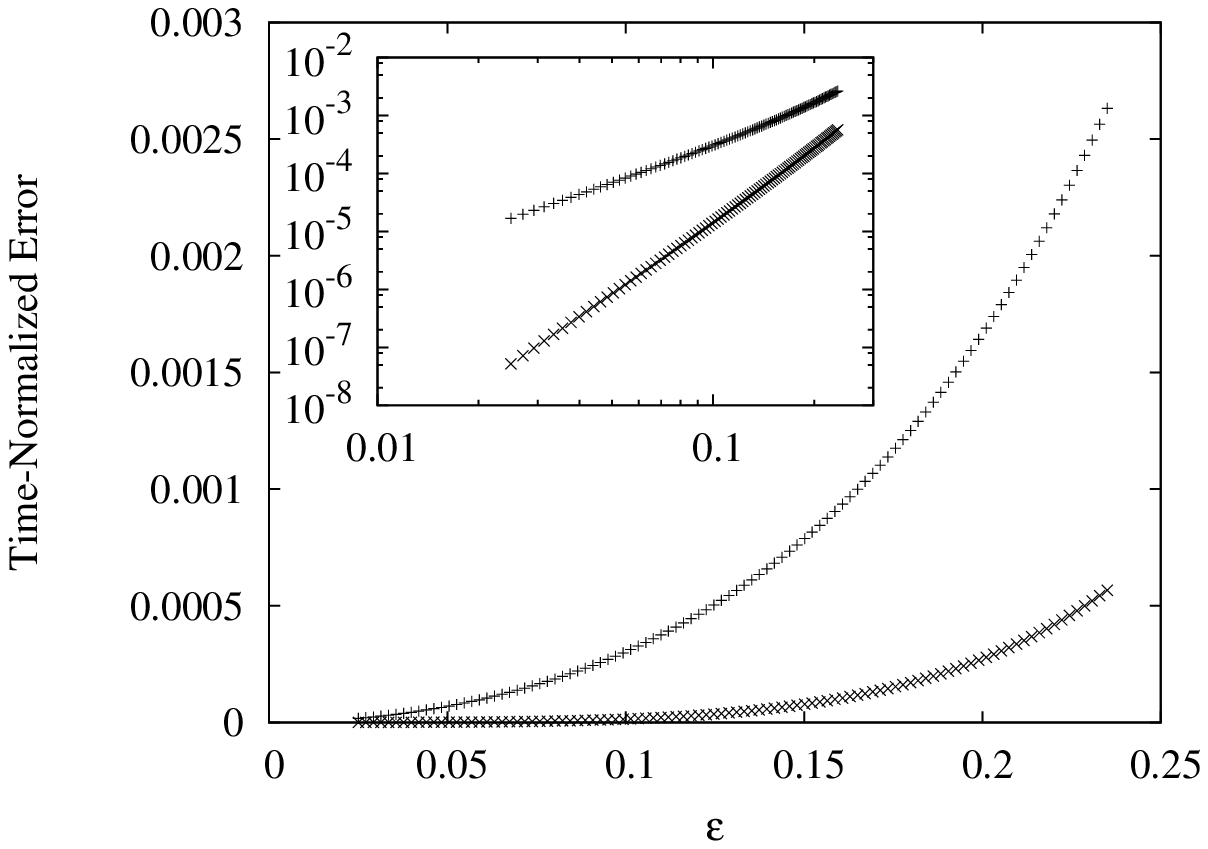}}
    \normalsize
    \caption{Accuracy of the Rest of the System: The vertical axis shows the natural norm of the difference between the final configuration space points evolved under the Lie and Regular methods, normalized by the time elapsed, i.e. $\frac{\left|\mbox{\bf\boldmath{z}}_{L,f}-\mbox{\bf\boldmath{z}}_{R,f}\right|}{\Delta t}$.  In particular, in this graph we consider the positions of only the dimer and other vortex.  The two lines, from top to bottom, represent the Lie transform at 0th - 2nd and 4th orders respectively.  The inset log-log plot shows that they both follow power laws with exponents 2 and 4 respectively.}
    \label{graph:graph3}
\end{figure}

We characterize the accuracy of the Lie transform algorithm by comparing how two identical point-vortex systems differ after evolving one with the regular method, and the other with the Lie method.  Their deviation is quantified by the natural norm of the difference of their positions in configuration space.  To have the regular method be a stand-in for the actual motion we must set the tolerance to be very small, which forces us to consider only the deviations that occur at small times.  These deviations scale linearly with time, so the quantity of interest is really the deviation divided by time.  Furthermore, there is a natural separation of the overall time-normalized deviation into that of the positions of the dimer vortices with respect to their center of circulation and that of the positions of the dimer itself and extra vortex.  The time-normalized deviation of the first group is plotted against $\epsilon$ in Fig.~\ref{graph:graph2} for three orders of the Lie transformation, while that of the second group is plotted in a similar manner in Fig.~\ref{graph:graph3}, though for only two orders of the Lie transformation.  Notice that because the constituents of the vortex pair have equal circulation, there is no 3rd order transformation to consider.  The most important thing to note about each graph is that the time-normalized deviation scales as $\epsilon^{\alpha}$, where $\alpha$ increases with the order of Lie transform used.  That is, the Lie method gets more accurate as the dimer pair separation decreases and as we include higher orders of the Lie transform.

\subsubsection{Energy Conservation}

\begin{figure}[htbp]
 \centering
\large
\resizebox{\linewidth}{!}{\includegraphics{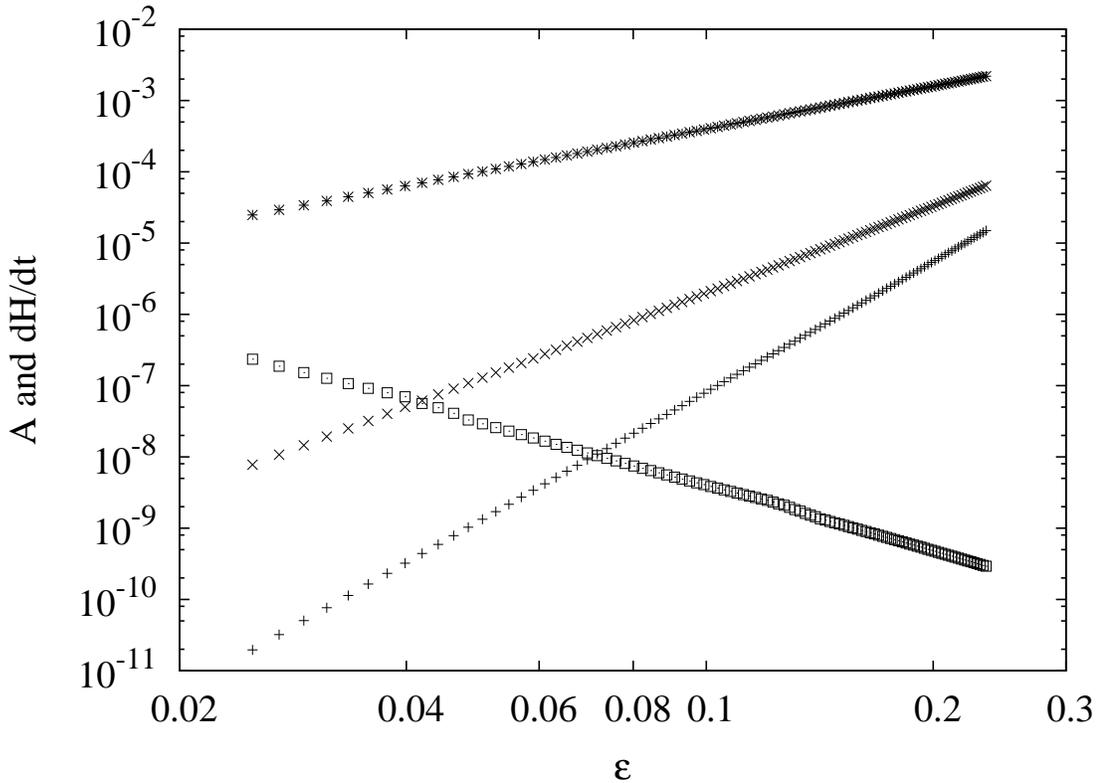}}
\normalsize
    \caption{Energy Conservation: The vertical axis of this log-log plot shows two different items.  The first is the change in energy per unit time, $\frac{dH}{dt}$, for evolution with the regular method.  This corresponds to the line which increases as $\epsilon$ decreases. Note that this is a power law, $\frac{dH}{dt} \propto \epsilon^{\alpha}$, with $\alpha$ of roughly -3.  The second item is the amplitude, $A$, of oscillations in the energy about the fixed average for evolution with the Lie method.  This corresponds to the three lines which decrease as $\epsilon$ decreases.  Note that each follows a power law $A \propto \epsilon^{\alpha}$.  From top to bottom, which corresponds to the Lie method at 0th, 2nd, and 4th orders, $\alpha$ is roughly 2, 4, and 6 respectively.}
    \label{graph:graph4}
\end{figure}

With respect to the close vortices circulating in a dimer, most numerical integration methods will systematically accrue errors that increase their separation.  This is equivalent to decreasing the total energy, which is forbidden in an autonomous Hamiltonian system.  With the regular integration method, the decrease in total energy per unit time is roughly proportional to the specified tolerance.  More importantly, we can see from Figure \ref{graph:graph4} that this change in total energy per unit time roughly scales as $\epsilon^{-3}$.  This indicates that the non-constancy of the energy will certainly become a problem at small separations, irrespective of the tolerance.  The performance of our Lie method is in marked contrast to that of the regular method.  At every order of the Lie transform, the average energy remains constant.  On top of this desired constant energy there are sinusoidal oscillations with a frequency commensurate with that of the dimer's rotation and an amplitude that depends on the order of the Lie transform and $\epsilon$.  As indicated in Figure \ref{graph:graph4} the amplitude scales as $A \propto \epsilon^{\alpha}$.  For the Lie transform at orders 0, 2, and 4, we have that $\alpha$ is nearly 2, 4, and 6 respectively.  Thus the accuracy of energy conservation under each order of the Lie transform method becomes better as $\epsilon$ decreases.

\subsubsection{Algorithmic Concerns}

In our implementation of this algorithm, the Lie transform method is triggered only when $\epsilon$ dips below a predetermined value.  One then transforms the variables, evolves the new Hamiltonian, and waits for $\epsilon$ to go above some de-triggering value before transforming back.  It is fruitful to implement this algorithm in an object oriented manner such that the collection of N vortices is itself an object.  When the method is triggered, we simply make a new (N-1)-vortex object with the transformed variables, and keep track of $J$ and $\theta$ externally.  This view needs to be modified when the transformation is used at fourth order or higher, where the modification to the Hamiltonian forces us to integrate the internal vortex dimer variables in conjunction with all the other variables.  If using the Lie transforms only to third order does not seem to be much of a sacrifice, then the N-vortex object allows us to deal with more than one dimer pair at the same time.  In particular we would just ``pop" down one more level to an (N-2)-vortex system where the integrable dimer motions are treated separately.  In particular, this would allow us to treat our simplest case of three vortices in a completely integrable fashion, where after considering the close vortices as a dimer, we could then look at the dimer and remaining vortex as a dimer itself.  In this way, our method finds subsections of the phase-space that are integrable in a particular manner and exploits that for numeric gain.  So, one must choose at third order whether one wants increased accuracy or the ability to treat other vortex pairs in a similar manner.

\subsection{Lie Transform Algorithm Conclusions}


The forward time dynamics of the point vortex model are easily obtained from a coupled nonlinear system of ODEs.  However several problems arise when using an adaptive time-stepping method to numerically integrate the system while two like-signed vortices are very close compared to other vortex separations.  If one keeps a small tolerance, the time-step will become prohibitively small and unnecessarily slow down the integration of the whole system.  On the other hand, if one increases the tolerance to alleviate this problem, not only does the accuracy suffer, but the overall energy of the supposedly Hamiltonian system tends to drift.

	To address these issues, we transformed the system such that it separated into two separate components: the motion of the vortex pair, or dimer, about its center of circulation and the motion of the dimer itself, considered as a single vortex, along with the rest of the vortices.  We viewed this as an (N-1)-vortex system with the same dynamics, where one of the vortices, the dimer, has an internal ``spin" degree of freedom.  The corrections to this appealing picture were fortunately ordered in a small parameter.  We then used Lie transform perturbation theory to develop a near-identity canonical transformation that preserved this picture order by order in the small parameter.  At fourth order, a slight alteration to the dynamics was necessary, though this affected only the dimer's spin degree of freedom by slightly lowering its rotation rate.  With this inclusion, the (N-1)-vortex picture was altered only in so far as the spin of the dimer weakly coupled to a new field produced by the (N-1) vortices.

	When quantifying the efficacy of the algorithmic implementation of our transformations, the Lie method, we chose the particularly simple example of three vortices of equal circulation with one close pair.   The first metric we considered was the speed-up factor, which measures the ratio of the number of calls to the numerical integrator made by the regular method versus those made by the Lie method.  Most importantly, we found this value to scale as $\epsilon^{-1.8}$, which indicates that the regular method run-time diverges as $\epsilon$ becomes small, since the Lie method run-time remains constant.  Next we considered the accuracy of the Lie method versus the real motion as given by the regular method with a very low tolerance.  The accuracy is quantified by the distance between two initially identical phase space points after evolving under the regular and Lie methods for a unit time.  This was further broken down into the phase space variables of the dimer relative to the center of circulation, and of the center of circulation and remaining vortex.  For both cases we found that the measure of accuracy gets smaller, i.e. more accurate, as a $\epsilon$ decreases.  As expected, the relationship is a power-law, with the integer power depending on the Lie transform order that is used.  Additionally we examined the total energy of the system.  The Lie method has zero drift in the value of the Hamiltonian as well as very small amplitude oscillations.  These oscillations drastically decrease in amplitude as $\epsilon$ decreases, and do so in a manner that gets better with inclusion of successively higher orders of the Lie transformation.  This is to be contrasted with the consistent decrease in energy that occurs with the regular method when using an insufficiently low tolerance.  Finally, we described how this transformation could be used as the basis of an object-oriented method for accurate numerical integration of point-vortex dynamics.

\section{Finding Periodic Orbits}

	Finding periodic orbits of the 2D point vortex system was the main goal of our work, and it turned out to be a particularly hard goal to achieve.  Many periodic orbits are unstable in the sense that nearby points will often diverge exponentially in distance from the periodic orbit under time evolution.  This is bothersome from the perspective of extracting periodic orbitss from orbits with close returns.  One standard method is to consider a Poincare section transverse to the direction of motion of a point in phase space.  A close return will intersect this hyper-plane, by definition, near to the original point.  If the original point is moved in a direction that will cause the final point to close the relative distance between the two on the Poincare section, then the loop can be closed and a periodic orbit extracted.  However when we have an unstable periodic orbit, there is sensitive dependance on the initial position and any such method is bound to preferentially pick out periodic orbits that are stable.  Since unstable solutions are expected to abound in a generically chaotic system, this is a problem.  Indeed, we never got this method to work very well with point vortices.
	
	Another method that we tried was based on a paper by Cvitanovic~\cite{Lan:2004uv}.  The main idea of this paper is to use an artificial periodic orbit, a loop in phase space that does not necessarily follow the local velocity vector $\HamVec{H}$, as a starting point for reduction.  At each point on this loop, there might be a discrepancy between the dynamical velocity vector and the vector tangent to the loop itself.  One can then use the length of the difference between these two vectors (suitably scaled), integrated over the loop, as a cost functional over all possible loops.  Then a discretization of the loop will allow the use of Newton's method to reduce this cost function and hopefully end up with a true periodic orbit.  This method worked sporadically, and we eventually abandoned it.
	
	The method that ended up working very well is a modification of a method outlined in~\cite{Boghosian:2011ec} and \cite{Boghosian:2011gy}.   It has much in common with~\cite{Lan:2004uv}, but treats the period as a main variable to be varied, among other differences.  We will go over this method and the modifications to it in detail, but first we will describe how to generate close returns to feed to this method. 		
		
\subsection{Generating Close Returns}

The first step in generating close returns is to create very long orbit segments.  This is easily done by picking an initial vortex configuration and numerically integrating it forward for a given amount of time.  However, in order to try and capture a representative sample of periodic orbits below a certain period, if not all periodic orbits below this period, we must start with a representative ensemble of initial vortex configurations.  We solved this by setting up a two dimensional grid on which the vortices sit.  First collect all possible configurations of vortices on the grid, eliminating redundancies like translationally or rotationally equivalent configurations, as well as permutations of vortices (since they are identical).  Then translate each configuration so that its center of vorticity coincides with the origin, and radially expand or contract the vortices about this point, as in Eq.~(\ref{eq:EnergyDilation}), so as to set all the energies to zero.  This gives an ensemble of points on the $H = P = Q = 0$ level set submanifold of the phase space which are relatively evenly distributed.  The maximum angular momentum value for elements of this ensemble is determined by the fineness of the grid, where a finer grid results in a larger maximum bounding angular momentum.  This procedure is really only possible for the $N=3$ and $N = 4$ case, where the number of configurations generated is manageable.  Past this, the combinatorics generate a prohibitive number of choices, despite the symmetry reductions, to practically enable an accurate sampling of initial positions in phase space.  One could reduce the number of points in the grid to help manage this combinatorial problem, however the coarseness of the grid would limit the variety of periodic orbits that are found.

\begin{figure}[htbp]
\centering
\large
\resizebox{\linewidth}{!}{\includegraphics{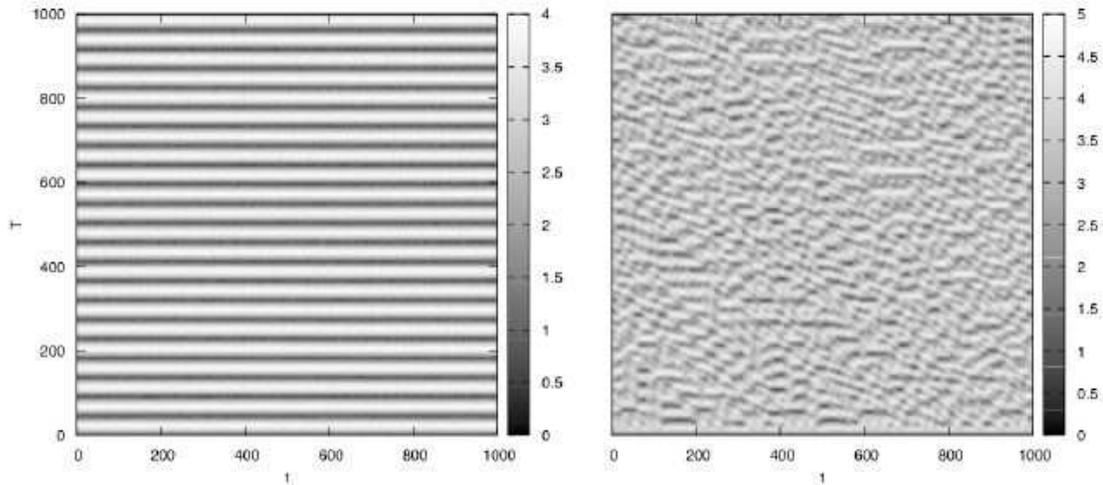}}
\normalsize
\caption{Two Tt-maps for orbit segments of four vortices.  The one on the left corresponds to an orbit segment that is periodic, while the map on the right has a chaotic generating orbit segment.  The axes measure time in multiples of a discrete time step.}
\label{figure:tTmaps}
\end{figure}

Now that an ensemble of starting vortex configurations is generated, we can take each one and generate a long orbit segment.  The Poincar$\acute{e}$ recurrence theorem, and other related ideas, will ensure that an orbit of sufficient length will return to within a specified distance of the original point.  Practically, this means that the long orbit segment will contain close returns.  We would like to pick out sub-segments of this overall orbit which have the smallest difference in initial and final orbit configurations, or distance in phase space.  To do this, we use the so-called $Tt$ method~\cite{Boghosian:2011gy}.  For an initial vortex configuration, $\bfz_0$, we have an orbit segment, $\mathcal{OS}\lrp{\bfz_0,t_f} = \lrbk{\bfz \in g^t_H\lrp{\bfz_0} \mid t\in\lrbc{0,t_f}}$, where the flow is defined by Eq.~(\ref{eq:HamFlow}).  Consider the $L^2$ norm of the difference between phase space points on the orbit segment, $f\lrp{t,T} = \| g^{t+T}_H\lrp{\bfz_0} - g^t_H\lrp{\bfz_0} \|$.  Notice that this function is parameterized by its starting time $t$ and its length or period $T$.  Thus, we could find close returns by finding minima of the map $f\lrp{t,T}$, subject to some maximum and minimum values of $T$.  A visual example of this is given in figure~(\ref{figure:tTmaps}).

For a given orbit segment, we can use the $Tt$ method to generate a large set of close return orbit segments with periods below a certain value.  We can then select amongst this set for orbit segments that, when closed, have unique braid types or other distinguishing features that make them likely candidates for a successful reduction to a periodic orbit.

\subsection{Loop Functional}

A close return orbit segment can be slightly deformed at the beginning and end to create a closed loop.  This loop is certainly not a solution to our equations of motion, Eq.~(\ref{eq:PVevolution}).  However, one hopes that it is sufficiently close to an actual periodic orbit so as to converge on this orbit after some iterative procedure.  The goal of this section is to outline just such a procedure, based on the variational ideas in~\cite{Boghosian:2011ec}.

It helps to think of a loop, $\alpha$, as a periodic function which assigns a point in phase space to a time parameter, along with the period of this function, $\alpha = \lrbk{\bfz\lrp{t},T}$.  This loop satisfies the periodicity requirement, $\bfz\lrp{t+T} = \bfz\lrp{t}$, for all $t$.  If this loop is a periodic orbit, then it additionally satisfies $\dot{\bfz}\lrp{t} = \HamVec{H}\lrp{\bfz\lrp{t}}$ for all $t$.  That is, at every point on the periodic orbit, the Hamiltonian vector field must be pointed in the same direction and have the same magnitude as the directional derivative of the parameterized loop.  This points toward an intuitive functional, which tells us, in a rough sense, how close our loop is to being a periodic orbit
\be
\mathcal{F}\lrbc{\alpha} = \mathcal{F}\lrbc{\bfz,T} \equiv \frac{1}{2}\int^T_0 \left\| \dot{\bfz}\lrp{t} - \HamVec{H}\lrp{\bfz\lrp{t}}\right\|^2 dt.
\label{eq:InitialFunctional}
\ee

Most importantly, this functional is positive-definite and has a value of 0 only for periodic orbits.  Notice that it depends not only on the geometry of the loop, $\alpha$, but also on how it is parameterized.  A change just in the period would generically change the value that the functional returns.  That is, a minimum of this functional fixes the orbit as well as the period.  To make this more explicit, we can transform the variables to the following
\be
\chi \equiv \frac{t}{T} \in \lrbc{0,1} \; \text{and} \; \bfrho\lrp{\chi}\equiv \bfz\lrp{\chi T}.
\label{eq:PeriodOneTransform}
\ee
We will also consider the Hamiltonian vector field as a function on phase space, which for brevity we will denote as $\bff \equiv \HamVec{H}$.  With these changes, the functional, Eq.~(\ref{eq:InitialFunctional}), becomes
\be
\mathcal{F}\lrbc{\bfrho,T} \equiv \frac{1}{2}\int^1_0 \left\| \frac{\bfrho'\lrp{\chi}}{\sqrt{T}} - \sqrt{T}\bff\lrp{\bfrho\lrp{\chi}}\right\|^2 d\chi.
\label{eq:Functional}
\ee

The space that our loop and periodic orbits live in is the infinite dimensional space of periodic functions with period one.  The minimization problem would then consist of finding solutions such that the functional, or Fr$\acute{e}$chet, derivative with respect to the loop variable is zero, $\delta \mathcal{F}/\delta \bfrho = 0$, the partial derivative with respect to the period is zero, $\partial \mathcal{F}/\partial T = 0$, and the functional itself is zero, $\mathcal{F} = 0$.  This leads to some interesting analytical constraints on periodic orbits in~\cite{Boghosian:2011ec}.  However, we are interested in using this functional as a cost function to numerically reduce an arbitrary loop to a periodic orbit.  As a practical concern, we can only store a finite number of points to represent any given loop.  This discretization of the loop can be given by the coordinates $\lrbk{\bfrho_0, \cdots, \bfrho_{N_d-1}}$, where $\bfrho_j = \bfrho\lrp{j/N_d}$, and $N_d$ is the number of discrete points along the loop.  Along with this, we must necessarily use a discrete representation of this functional.  In particular, the time derivative is approximated as $\lrp{\bfrho_{j+1}-\bfrho_j}/\lrp{\frac{1}{N_d}}$ and a point on the loop itself is given, for stability, as the mean of two adjacent coordinates, $\lrp{\bfrho_{j+1}+\bfrho_{j}}/2$.  Thus the functional becomes a simple function of the coordinates and period
\be
\mathcal{F}\lrbc{\bfrho_0, \cdots, \bfrho_{N_d-1},T} = \frac{1}{2} \sum_{j=0}^{N_d-1} \left| \frac{\bfrho_{j+1}-\bfrho_{j}}{\sqrt{\frac{T}{N_d}}} - \sqrt{\frac{T}{N_d}} \bff\lrp{ \frac{\bfrho_{j+1}+\bfrho_{j}}{2} } \right|^2.
\label{eq:DiscreteFunctional}
\ee

It is this function that we want to minimize over the discretized version of loop space, $\lrbk{\bfrho_0, \cdots, \bfrho_{N_d-1}, T}$.  Since each coordinate, $\bfrho_j$, has twice as many components as there are vortices, $2N_v$, the total dimensionality of the discretized loop space is $N_T = 2\times N_v \times N_d +1$.  Of course, a random point in this potentially large space is not likely to represent anything close to a loop, which is why we start with the close return loops previously mentioned.  

\subsection{Minimization Methods}

A first approach to evolving our loop, represented as a single point in the discretized loop space, toward a periodic orbit would be to move the point in the opposite direction from the gradient of the cost function.  This is the gradient descent method.  Denote a point in the discretized loop space as $\bfp = \lrbk{\bfrho_0, \cdots, \bfrho_{N_d-1}, T}$, and the $j$th point in the iterative method as $\bfp_j$.  Then the gradient descent step, $\bfs_j^G$, which determines the next iteration, $\bfp_{j+1} = \bfp_j + \bfs_j^G$, is given by $\bfs_j^G = - \tau_j \left. \nabla \mathcal{F}\right|_{\bfp_j}$.  Here $\tau_j$ is just a small number that is chosen at each iteration to ensure that $\mathcal{F}\lrp{\bfp_{j+1}} \leq \mathcal{F}\lrp{\bfp_j}$.  The gradient is given by $\nabla = \lrang{\partial_{\bfrho_0},\cdots, \partial_{\bfrho_{N_d-1}}, \partial_T}$, which can be thought of as a $N_T \times 1$ matrix operator.  The action of each $2N_v \times 1$ block, $\partial/\partial \bfrho_k$, on the cost functional is
\begin{align}
\pd{\mathcal{F}}{\bfrho_k} &= - \frac{N_d}{T}\lrp{\bfrho_{k+1} -2 \bfrho_{k}+\bfrho_{k-1}} + \bff\lrp{\bfrho_{k+\frac{1}{2}}} - \bff\lrp{\bfrho_{k-\frac{1}{2}}} + \nonumber\\
& -\frac{1}{2}\lrbc{\bfJ\bff\lrp{\bfrho_{k+\frac{1}{2}}}}^{\intercal}\lrp{\bfrho_{k+1}-\bfrho_{k}} -\frac{1}{2}\lrbc{\bfJ\bff\lrp{\bfrho_{k-\frac{1}{2}}}}^{\intercal}\lrp{\bfrho_{k}-\bfrho_{k-1}} + \nonumber\\
& +\frac{T}{2 N_d}\lrbc{\bfJ\bff\lrp{\bfrho_{k+\frac{1}{2}}}}^{\intercal}\bff\lrp{\bfrho_{k+\frac{1}{2}}} +\frac{T}{2 N_d}\lrbc{\bfJ\bff\lrp{\bfrho_{k-\frac{1}{2}}}}^{\intercal}\bff\lrp{\bfrho_{k-\frac{1}{2}}},
\label{eq:GradF}
\end{align}
where $\bfJ\bff$ is the Jacobian of $\bff$, $\lrbc{}^{\intercal}$ is the transpose, $\bfrho_{k+\frac{1}{2}} \equiv \lrp{\bfrho_{k+1}+\bfrho_{k}}/2$, and $\bfrho_{k-\frac{1}{2}} \equiv \lrp{\bfrho_{k}+\bfrho_{k-1}}/2$.  In addition to this, we have the contribution to the overall gradient due to the period
\be
\pd{\mathcal{F}}{T} = \frac{1}{2}\sum^{N_d-1}_{k = 0} \lrbc{- \frac{N_d}{T^2}\left| \bfrho_{k+1} - \bfrho_{k} \right|^2 + \frac{1}{N_d}\left| \bff\lrp{\bfrho_{k+\frac{1}{2}}}  \right|^2 }.
\label{eq:TGradF}
\ee

However, gradient descent only converges on local minima in a q-linear fashion~\cite{bib:DennisSchnabel}.  This means that if $\bfp_{\ast}$ is a solution, i.e. local minima, then for all $j$ greater than some integer there exists a number $0 \leq c < 1$, such that $\left| \bfp_{j+1} - \bfp_{\ast} \right| \leq c \left| \bfp_{j} - \bfp_{\ast} \right|$.  Furthermore, if the maximum and minimum eigenvalues of the Hessian of the cost function, $\lrbc{Hess\lrp{\mathcal{F}}}^{ij} \equiv \partial_i\partial_j \mathcal{F}$, are very different in magnitude, then $c$ can be arbitrarily close to one.  In this case the steps tightly zig-zag back and forth, while making little progress toward $\bfp_{\ast}$.  This problem of ill-conditioning of the Hessian will remain a problem, however, we can help the local convergence rate by choosing a iteration step size by way of Newton's method.  This will give q-quadratic convergence, where $\left| \bfp_{j+1} - \bfp_{\ast} \right| \leq c \left| \bfp_{j} - \bfp_{\ast} \right|^2$.  The step, $\bfs_j^N$, is given by the solution of the linear equation
\be
Hess\lrp{\mathcal{F}\lrp{\bfp_j}} \bfs_j^N = - \nabla \mathcal{F}\lrp{\bfp_j},
\label{eq:NewtonStep}
\ee
which is the needed step size and direction to land on the minimum of the local quadratic model.  While Netwon's method has nice convergence properties near local minima, it does require the computation of the $N_T\times N_T$ Hessian, as well as the solution of the above linear equation.  For four vortices and a typical loop discretization of two hundred points, this amounts to a Hessian matrix with over 2.5 million elements.  It is easy to see that the size of the matrix and therefore the computational complexity of solving the linear equation, Eq.~(\ref{eq:NewtonStep}), can quickly become prohibitive.  Fortunately, this matrix is very sparse as well as symmetric.  It has a sparsity pattern
\be
 Hess\lrp{\mathcal{F}} = 
 \begin{pmatrix} 
A_0 & B_1 & 0 & \cdots & 0 & B_0^{\intercal} & C_0 \\
B_1^{\intercal} & A_1 & \ddots & \ddots & \text{\huge{0}} & 0 & \vdots \\
0 & \ddots & \ddots & B_j & \ddots & \vdots & \vdots \\
\vdots & \ddots & B_j^{\intercal} & A_j & \ddots & 0 & C_j \\
0 & \text{\huge{0}}  & \ddots & \ddots & \ddots & B_{N_d-1} & \vdots \\
B_0 & 0 & \cdots & 0 & B_{N_d-1}^{\intercal} & A_{N_d-1} & C_{N_d-1} \\
C_0^{\intercal} & \cdots & \cdots & C_j^{\intercal} & \cdots & C_{N_d-1}^{\intercal} & D
 \end{pmatrix},
 \label{eq:SparsityPattern}
 \ee
where the $2N_v$ by $2N_v$ sub-matrices are given by $A_j = \partial^2\mathcal{F}/\partial \bfrho_j^2$ and $B_j = \partial^2\mathcal{F}/\partial \bfrho_{j+1} \partial \bfrho_j$.  Also, $C_j = \partial^2\mathcal{F}/\partial \bfrho_j \partial T$ is a $2N_v\times 1$ matrix, and $D = \partial^2\mathcal{F}/\partial T^2$ is just a single number.  Thus, the number of non-zero entries is $4N_v\lrp{3N_v+1}N_d+1$ which is only linear in $N_d$, and therefore much less than the total number of possible entries $\lrp{2N_vN_d+1}^2$.  For four vortices and two hundred loop points, only 1.6$\%$ of the entries are non-zero, and the matrix is certainly sparse.  To solve this sparse linear system, we used UMFPACK \cite{Davis:2004:CPS:992200.992205}\cite{Davis:2004:AUV:992200.992206}\cite{Davis:1999:CUM:305658.287640}, which is a sophisticated LU decomposition for unsymmetric sparse matrices.  There are faster methods for symmetric matrices, however we will soon need to make some modifications to the basic linear equation, Eq.~(\ref{eq:NewtonStep}), which will result in an unsymmetric matrix.  The sub-matrices $\lrbk{A_j,B_j,C_j,D}$ are rather complicated, though easy to compute, and we will not be including them here.  They do, however, have a lot of internal structure and symmetry that can be be exploited to minimize the number of computations required in calculating the Newton step.

\subsection{Global Convergence}

Solving Eq.~(\ref{eq:NewtonStep}) and implementing the resultant Newtonian step, $\bfp_{j+1} = \bfp_j + \bfs_j^N$, will result in a sequence of points in discretized loop space that, if near a periodic orbit, will converge in a q-quadratic fashion.  However, this desired fast convergence only exists close to the periodic orbit.  In-fact, further away from the desired solution the Newtonian step is not guaranteed to even decrease the cost function!  Thus, we must augment Newton's method in order to gain better global convergence.  There are many different ways of doing this, though they can be grouped into two basic categories, line searches and model-trust regions.  Roughly, lines searches fix the direction of the step and vary the size of the step, while model-trust regions vary both the length and the direction of the step.  Model-trust region algorithms, like the hook-step, require multiple computations of linear equations similar to Eq.~(\ref{eq:NewtonStep}), and can therefore be computationally expensive when the Hessian matrix is very large.  Whether the benefits of using such algorithms outweigh these costs is very much dependent on the minimization problem being considered.  In our case, the conceptually simpler line search works well and is less computationally expensive to implement.  We will now focus on the line search.

Given a Newtonian step, $\bfs_j^N$, we can think about the function, $\mathcal{G}_{\bfp_j}\lrp{\tau} = \mathcal{F}\lrp{\bfp_j+\tau \bfs_j^N}$, where $\tau \in \lrbc{0,1}$.  This is a one dimensional function, which we would like to minimize.  That is, we would like to find the fictitious time step, $\tau$, that would locally minimize our cost function.  Notice that if the Hessian of the cost function is positive definite, then a sufficiently small step, $\tau$, will result in a decrease in the cost function, $\mathcal{F}\lrp{\bfp_j+\tau \bfs_j^N} \approx \mathcal{F}\lrp{\bfp_j} - \tau \lrbc{\bfnabla \mathcal{F}}^{\intercal}Hess\lrp{\mathcal{F}}^{-1}\bfnabla \mathcal{F}$.  Thus, we are assured that the minimum of $\mathcal{G}_{\bfp_j}\lrp{\tau}$ exists.  We can then start with the Newtonian step of $\tau = 1$, and backtrack until an acceptable minimum is found.

Occasionally there will be regions of the phase space where repeated use of the Newtonian step will cause, $\bfp_j$, to converge on a set of points (limit cycle) which are not the local minima.  Crucially, checking to see if the step simply reduces the cost function is not sufficient to evade this problem.   In these cases, the full Newtonian step is successively overshooting the actual minima and decreasing the cost function by an increasingly small amount.  We will need to require that each step size results not just in a decrease in the cost function, but a sufficiently large decrease so as to avoid this phenomenon.  Such a condition is given by the Armijo, or Wolfe constraints.  The first of which is
\be
\mathcal{F}\lrp{\bfp_j+\tau \bfs_j^N} \leq \mathcal{F}\lrp{\bfp_j} + \alpha \tau \lrbc{\bfnabla \mathcal{F}\lrp{\bfp_j}}^{\intercal} \bfs_j^N,
\label{eq:ArmijoCond}
\ee
where $\alpha \in \lrp{0,1}$.  This is essentially saying that after our step of $\tau \bfs_j^N$ the cost function must decrease at least as much as it would under a linear decrease with a slope proportional to the initial directional derivative along the Newtonian step direction.  An intermediate value of $\alpha$ usually works the best, though the act of choosing is a dark art and depends on the cost function.  After $\alpha$ is chosen, one repeatedly reduces the value of $\tau$, usually by a constant multiplicative factor, until this condition is met.  Furthermore, this condition is guaranteed to hold for small enough $\tau$ if the step direction, $\bfs_j$, is a descent direction, $\lrbc{\bfnabla \mathcal{F}\lrp{\bfp_j}}^{\intercal} \bfs_j < 0$.  The second Armijo condition is used to avoid step sizes that are too small.  However, it is not often needed in practice, so we will forgo its description.

Thus, the line search consists of trying the full Newtonian step, $\tau = 1$, checking to see if Eq.~(\ref{eq:ArmijoCond}) holds, and then if not, reducing $\tau$ by a multiplicative factor and reevaluating.  This procedure is sufficient, in our case, to increase the basin of convergence of a given periodic orbit to include most of the close return loops.

\subsection{Constrained Optimization}

Unfortunately, for Hamiltonian systems, Newton's method on its own will fail to work well.  To see why, consider an actual periodic orbit solution, identified as the point $\bfp^{\ast}$ in discretized loop space.  By definition, the cost function is zero here, $\mathcal{F}\lrp{\bfp^{\ast}} = 0$.  However, there are marginal directions, $\bfX_i\mid_{\bfp^{\ast}}$, along which $\mathcal{F}$ continues to be zero.  That is, if we denote the flow acting on a point, $\bfp$, of the discretized loop space due to the vector field $\bfX_i$ for ``time" $\tau$ as $g^{\tau}_{\bfXs_i}\lrp{\bfp}$, then we can express the condition that $\bfX_i\mid_{\bfp^{\ast}}$ is a marginal direction as
\be
\mathcal{F}\lrp{g^{\tau}_{\bfXs_i}\lrp{\bfp^{\ast}}} = \mathcal{F}\lrp{\bfp^{\ast}} = 0.
\label{eq:MarginalDef}
\ee
These marginal directions imply that the Hessian matrix is not positive definite, and more importantly that it is singular and therefore not invertible.  As the Newtonian descent algorithm depends on the solution to the linear equation, Eq.~{\ref{eq:NewtonStep}}, this presents a fundamental problem.

Before we solve this problem, we must be able to account for all of the marginal directions.  They generically arise in Hamiltonian systems due to the existence of first integrals of motion.  For each constant of motion, $F_i$, we can define two marginal directions in discretized loop space: one corresponding to the associated continuous symmetry at each point, $\HamVec{F_i}$, and another corresponding to the gradient of the constant of motion at each point, $\nabla F_i$.  More specifically, we have
\begin{align}
\bfX_{2i}\mid_{\bfp^{\ast}} &= \lrang{\HamVec{F_i}\lrp{\bfrho_0}, \cdots, \HamVec{F_i}\lrp{\bfrho_{N_d-1}},0}, \nonumber\\
\bfX_{2i+1}\mid_{\bfp^{\ast}} &= \lrang{\nabla F_i\lrp{\bfrho_0}, \cdots, \nabla F_i\lrp{\bfrho_{N_d-1}},0},
\label{eq:marginalDirections}
\end{align}
where the last entry tells us that there is no component in the $T$ direction.  Thus, the marginal directions correspond to the combined symmetry or gradient directions of each point on the discretized loop.  

To get a clearer picture of these directions, consider the concrete cases of $F_0 = H$, $F_1 = P^2+Q^2$, and $F_2 = L$.  First, since $\HamVec{H}$ is the direction of the time flow at each point on the loop, the total vector, $\bfX_0$, in the discretized loop space would generate a flow that simply rotates the discretized coordinates, $\lrbk{\bfrho_0, \cdots, \bfrho_{N_d-1}}$, about the loop.  As this flow maps each periodic orbit back onto itself, motion in this direction would certainly not change the value of the cost function.  The associated dual marginal direction, $\bfX_1$, induces a flow which maps periodic orbits to new periodic orbits with larger energy.  In a similar vein, the flows $\bfX_2$ and $\bfX_3$, associated with $F_1 = P^2+Q^2$, correspond to rigid translations of each point on the discretized loop in the same direction.  This wholesale translation of a periodic orbit produces a two parameter family of periodic orbit solutions.  Finally, we can consider the two marginal directions associated with the angular momentum, $F_2 = L$.  The flow $g^{\tau}_{\bfXs_5}$, will deform a periodic orbit to a new periodic orbit on a different angular momentum level set of the phase space.  The symmetry induced flow $g^{\tau}_{\bfXs_4}$ acts on every point of the loop discretization by rigidly rotating the corresponding vortex configuration about its center of vorticity.  Like the flow due to $\bfX_0$, this flow results in a one-parameter family of solutions which is periodic.\footnote{The action of this symmetry, due to $L$, is particularly interesting in teasing out the structure of periodic orbits.  Some periodic orbits will be mapped identically back onto themselves after a partial rigid rotation and a partial time translation.  If the periodic orbit is an integrable solution, then the set of rotations where this occurs give the winding number about one direction of the invariant torus.}  Fortunately it is sufficient to concentrate on these six marginal directions alone.

Given the fundamental problem that these marginal directions present to solving Eq.~{\ref{eq:NewtonStep}} for the Newtonian step, we must amend some aspect of the minimization algorithm so far developed.  One way to forbid movement in these directions is to use Lagrange multipliers, $\lambda^i$, to enlarge the cost function,
\be
\mathcal{L} = \mathcal{F} + \lambda^i c_i.
\label{eq:LagrangeMult}
\ee
Here, $c_i$ are functions of the discretized loop space whose gradients are in the direction of the marginal directions we want to forbid.  Traditionally one introduces Lagrange multipliers to minimize a function while maintaining extra equality constraints, $c_i = 0$.  That is, $c_i = 0$ defines a level set that we want our solution to lie on.  However, we are much more interested in the fact that using Lagrange multipliers will forbid iterative steps from being in the directions defined by $\nabla c_i$.  That is, we are not as concerned with maintaining a particular energy or angular momentum for a loop, (since we can always transform these quantities once a PO is found), as we are with ensuring that the Hessian is positive definite and invertible.  Normally, we would not need to emphasize this point, however for the marginal directions associated with a symmetry of the system, no global function exists such that its gradient is in this direction.  That is, there is no $c_i$ such that $\nabla c_i = \HamVec{H}$ everywhere.  At best we could create a multi-valued function that does this locally.  To make the Lagrange multiplier method work, we will not attempt to create a local function for these symmetry directions, but will simply use a constant function, $c_i \equiv 0$, and the marginal symmetry direction instead of a gradient.

Consider for a moment the vector $\bfX_0$.  Using this as the gradient of some $c_i$,  even though such a function does not exist, will prohibit the marginal direction in which periodic orbit is rotationally mapped to itself along the time flow direction.  Implementing this would require $N_d$ evaluations of $\HamVec{H}$.  Fortunately, we can get the same result with only one evaluation of $\HamVec{H}$, by restricting our attention to only one point on the loop.  Define $\nabla_{\bfrhos} c_0 = \lrang{0, \cdots , 0, \lrbc{\pten \partial_{\bfrho_{k_0}} H\lrp{\bfrho_{k_0}}}^{\intercal}, 0, \cdots, 0}$.  This effectively bars a single point, at $\bfrho_{k_0}$, on the discretized loop from moving in the direction of the time flow, and therefore disallows the general bulk motion described above.  Applying this idea to the other marginal directions, we have a list of 6 integers mod $N_d$, $\lrbc{k_0,k_1, \cdots, k_5}$.  Each integer describes the position on the loop where we would like to apply the appropriate restriction.  One additional wrinkle is that the Hessians for $c_0$, $c_2$, and $c_4$ are not symmetric, (see Table~\ref{table:ConstraintAdditions} below).  This does not seem to hurt the performance of the algorithm, however it does explain our choice of sparse matrix solver.  The following table summarizes the additional structures needed to use the Lagrange multiplier approach.
\begin{table}[htbp]
  \centering
\be
 \begin{array}{c|c|c}
 F_i & F_0 = H,\; F_1 = \lrp{P^2+Q^2}, \; F_2 = L & F_0 = H,\; F_1 = \lrp{P^2+Q^2}, \; F_2 = L \\
  \hline
 j & 2i & 2i+1 \\
  \hline
c_j & 0 & F_i\lrp{\bfrho_{k_j}} -  F_i\lrp{\bfrho_{k_j}\mid_{t=0}} \\
  \hline
 \nabla c_j & \lrang{0, \cdots , 0, \lrbc{\pten \partial_{\bfrho_{k_j}} F_i\lrp{\bfrho_{k_j}}}^{\intercal}, 0, \cdots, 0} & \lrang{0, \cdots , 0, \lrbc{\partial_{\bfrho_{k_j}} F_i\lrp{\bfrho_{k_j}}}^{\intercal}, 0, \cdots, 0} \\
  \hline
Hess\lrp{c_j} &
 \begin{pmatrix} 
\text{\huge{0}} & \cdots & \text{\huge{0}} \\
\vdots & \partial_{\bfrho_{k_j}} \pten \partial_{\bfrho_{k_j}} F_i\lrp{\bfrho_{k_j}} & \vdots \\
\text{\huge{0}} & \cdots & \text{\huge{0}}
 \end{pmatrix}
 &
  \begin{pmatrix} 
\text{\huge{0}} & \cdots & \text{\huge{0}} \\
\vdots & \partial_{\bfrho_{k_j}} \partial_{\bfrho_{k_j}} F_i\lrp{\bfrho_{k_j}} & \vdots \\
\text{\huge{0}} & \cdots & \text{\huge{0}}
 \end{pmatrix} 
 \end{array}\nonumber
\ee
 \caption{The Lagrange constraint functions, their gradients, and their associated Hessian matrices.}
 \label{table:ConstraintAdditions}
\end{table}

The specific Lagrange multiplier method we will be using is referred to as the prime-dual method~\cite{bib:Bonnans}, and works as follows.  Consider the augmented phase space as composed of points of the form $\tilde{\bfp} = \lrbk{\bfrho^i, T, \lambda^j}$.  If we use $\mathcal{L} = \mathcal{F} + \lambda^i c_i$ as the new objective function, then Newton's method now appears as
\be
\begin{pmatrix} 
Hess\lrp{\mathcal{F}} + \lambda^i Hess\lrp{c_i} & \partial_T \nabla_{\bfrhos} \mathcal{F} & \nabla_{\bfrhos} c_i \\
\lrbc{\partial_T \nabla_{\bfrhos} \mathcal{F}}^{\intercal} & \partial^2_{T} \mathcal{F} & 0 \\
\lrbc{\nabla_{\bfrhos} c_i}^{\intercal} & 0 & 0
 \end{pmatrix} 
 \begin{pmatrix}
 d\bfrho \\
 d T \\
 d\lambda
 \end{pmatrix}
 =
 -
 \begin{pmatrix}
 \nabla_{\bfrhos}\mathcal{F} + \lambda^i \nabla_{\bfrhos}c_i \\
 \partial_T \mathcal{F} \\
 c_i
 \end{pmatrix},
 \label{eq:UpdatedNewton}
\ee
where we have allowed that the $c_i$ have no $T$ dependence.  Notice that the solution to this linear equation ensures that the updated loop coordinates, $\bfrho_{j+1} = \bfrho_j + d\bfrho$, do not lie in one of the restricted directions: $\lrbc{\nabla_{\bfrhos} c_i}^{\intercal} d\bfrho = - c_i$, if they remain on the $c_i = 0$ manifold.  Since $c_0 = c_2 = c_4 \equiv 0$, the symmetry flow directions are always forbidden.  The prime-dual method proceeds in the usual fashion, by solving Eq.~(\ref{eq:UpdatedNewton}) and updating the loop coordinate, the period $T$, and the Lagrange multipliers, $\lambda^i_{j+1} = \lambda^i_{j} + d \lambda^i$.  However, we must now be careful since the new cost function is unbounded from below.  Indeed, there exist pathological descent directions where $\lambda^i$ are large, negative, and grow faster than $\mathcal{F}$ increases.  To preclude this possibility, we simply calculated reasonable values for $\lambda^i$ by fiat every iteration, given by
\be
\lambda^j = - \frac{\nabla_{\bfrhos}\mathcal{F} \cdot \nabla_{\bfrhos} c_j }{\nabla_{\bfrhos} c_j \cdot \nabla_{\bfrhos} c_j}.
\label{eq:lambdaFiat}
\ee
This comes from considering the simple case of gradient descent.  It is reasonable to want the descent direction not to be in one of the marginal directions, or equivalently require that $\lrp{\nabla_{\bfrhos} \mathcal{F} + \lambda^i \nabla_{\bfrhos} c_i}\cdot \nabla_{\bfrhos}c_j = 0$.  We can choose the loop coordinate sites $\lrbc{k_0,k_1, \cdots, k_5}$ such that $\nabla_{\bfrhos}c_i \cdot \nabla_{\bfrhos}c_j = 0$ for $i \neq j$, leading to Eq.~(\ref{eq:lambdaFiat}).  One should certainly be suspicious that this applies to Newtonian descent, however numerical experimentation has shown that this is a helpful construction.

There are many small tweaks to this overall algorithm that we employed, but which would not be all that enlightening to discuss.  One exception, and the final idea in this section, involves the rotation of the constraint points.  The constraint sites, $\lrbc{k_0,k_1, \cdots, k_5}$, are crucial to prohibiting movement of the loop in marginal directions.  However, we have found that in practice these sites often hold the loop back from reaching a periodic orbit.  That is, most of the loop will be evolving in the direction of a periodic orbit, but the constraints at each constraint site will be too prohibitive and locally pin the loop down.  We solved this problem by rigidly rotating the constraint sites, $\lrbc{k_0+s,k_1+s, \cdots, k_5+s}\%N_d$, every $t$ iteration steps.  The optimal choice of $\lrp{s,t}$ is hard to determine.  On one hand, $t$ should not be too small or otherwise we would need to recompute the sparsity pattern of the matrix in Eq.~(\ref{eq:UpdatedNewton}).  On the other hand, a large value of $t$ would be tantamount to not changing the constraint sites at all.  There are a number of choices for $s$ that seem to work well.  We settled on the closest integer $s$ such that $s/N_d \simeq 1- \phi$, where $\phi = 1.61803399$ is the golden ratio.  This choice works well, because $\phi$ is the most irrational of the irrational numbers, in the sense that its continued fraction, $\lrbc{1,1,1, \cdots}$, is the slowest to converge.  This means that it will take the largest number of successive rotations of this type to rotate the original constraint sites back to themselves.  It is interesting to note that many plants use this strategy in the successive helical placement of leaves about a central stalk in order to maximize exposure to the sun.

\subsection{Summary and Numerical Results}

We have extracted over 500 periodic orbits for the case of three vortices and over 12,500 periodic orbits for the four vortex case.  Each of these periodic orbits was generated using the ideas put forward in the past few sections.  They started out as close returns of carefully selected long orbit segments found via the $Tt$ method.  We then reduced these close returns to actual periodic orbits using Newton's method, Eq.~(\ref{eq:NewtonStep}), on a cost function, Eq.~(\ref{eq:DiscreteFunctional}).  To expand the area of convergence we computed the actual iterative step taken by doing a line search in the Newtonian descent direction.  Finally, to ensure that the linear equation defining the Newtonian step is not singular due to symmetries of the Hamiltonian, we used Lagrange multipliers to prohibit movement in marginal directions.

\begin{figure}[htbp]
 \centering
\large
\resizebox{\linewidth}{!}{\includegraphics{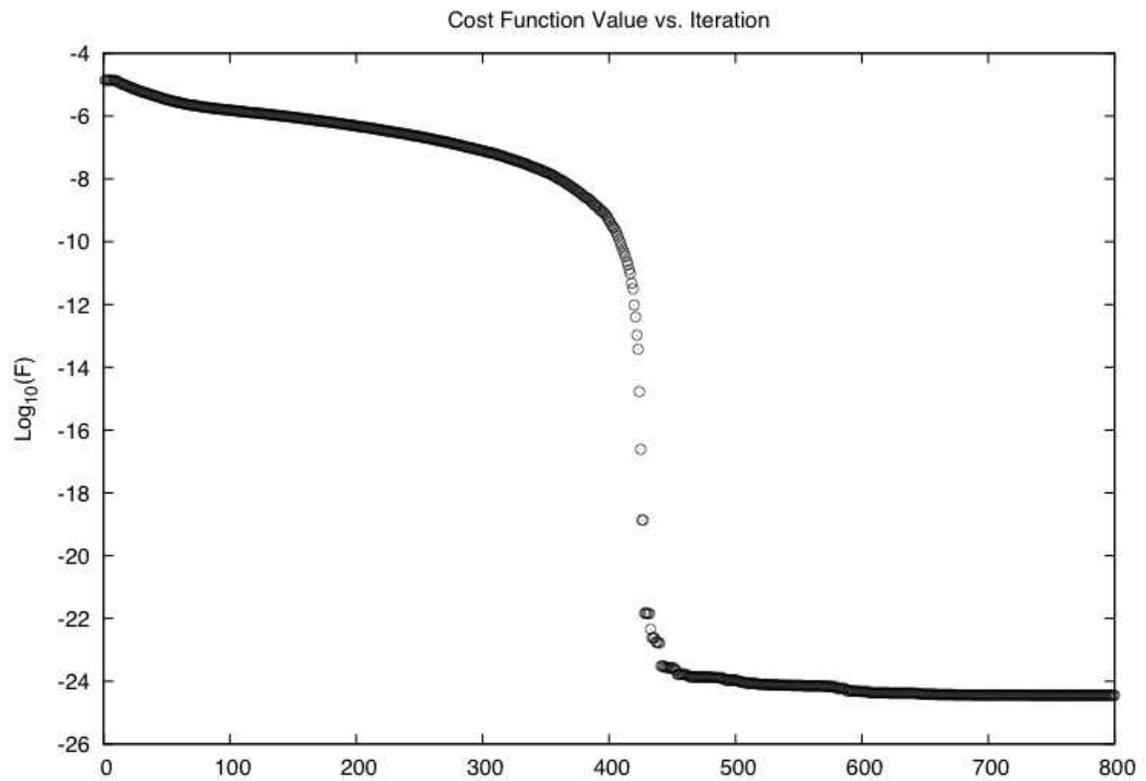}}
\normalsize
    \caption{The log base 10, $log_{10}\lrp{\mathcal{F}}$, of the cost function is plotted vs. the iteration step.}
    \label{graph:CostFunctionDecrease}
\end{figure}

We would like conclude this section by briefly highlighting the performance of the reduction algorithm.  Consider the trajectory in discretized loop space of a given close return loop being reduced to a periodic orbit.  At each iterative step we can calculate the value of the cost function, Eq.~(\ref{eq:DiscreteFunctional}), and monitor how it decreases.  Every successful reduction follows a similar pattern, shown in Fig.~(\ref{graph:CostFunctionDecrease}), whereby the cost function is initially relatively slow to decrease before the loop reaches an area of discretized loop space sufficiently close to a solution to allow the full q-quadriatic performance of Newtonian descent to kick in.

As one can see in Fig.~(\ref{graph:CostFunctionDecrease}), there are three qualitatively different sections to the reduction.  The loop starts out far away, in discretized loop space, from the eventual minimum, and therefore decreases slowly.  The rate of reduction for $\mathcal{F}$ significantly increases when the loop is close enough to the solution for the quadratic model, which Newtonian descent is based on, to accurately reflect the local geometry of the solution.  This reduces the cost function to a minimal value that is determined by the coarseness of the discretization.  That is, the larger the number of points in the discretization of the loop, $N_d$, the smaller the lower bound on $\mathcal{F}$ will be.  The third section appears to have little appreciable reduction in the cost function, and in an automated run the halting criteria would have stopped the reduction at the beginning of this section.  

\begin{figure}[htbp]
 \centering
\large
\resizebox{\linewidth}{!}{\includegraphics{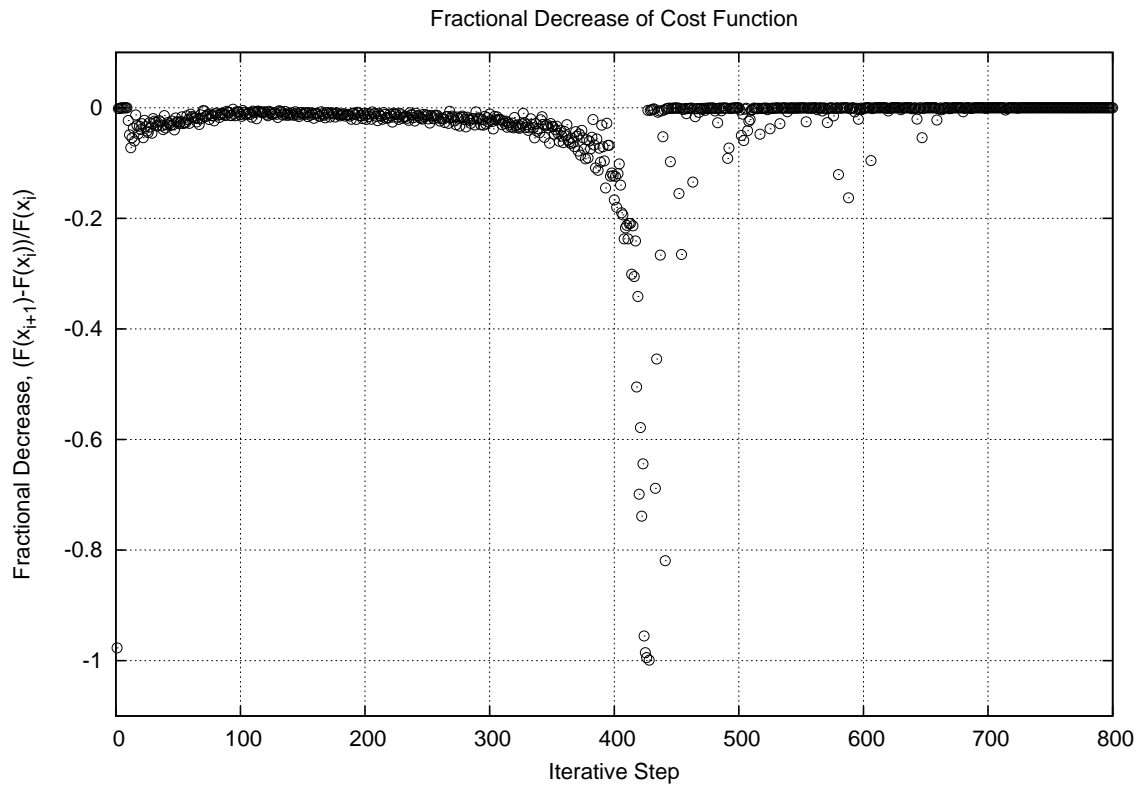}}
\normalsize
    \caption{The fractional decrease in the cost function, $\frac{\mathcal{F}\lrp{\bfp_{i+1}} - \mathcal{F}\lrp{\bfp_{i}}}{\mathcal{F}\lrp{\bfp_{i}}}$, vs. the iteration step.}
    \label{graph:FractionalDecrease}
\end{figure}
These qualitative sections can also be seen in a graph of the fractional decrease in the cost function, Fig.~(\ref{graph:FractionalDecrease}).  Notice that the main appreciable decrease happens in the middle section, where the full power of Newtonian descent finally kicks in.  Since the reduction of each periodic orbit qualitatively follow the course of the example in both graphs, we can be confident that each periodic orbit generated is a solution to Eq.~(\ref{eq:PVevolution}).  Indeed, from this point on we will assume that this is the case, and will focus on other aspects of our solutions.

\subsection{Classification and Braid Related Ideas}

Implementing the reduction ideas of the last few subsections on the Tufts cluster resulted in millions of potential periodic orbits.  Given the amount of data that one periodic orbit represents, storage of each orbit is certainly not feasible.  It is also not necessary, since there is a high degree of redundancy in the generated periodic orbits.  Many periodic orbits are copies of each other, related by a combination of time translations and rigid body rotations of the whole orbit.  We developed algorithms which compare two periodic orbits and determine whether they are really the same orbit.  However, it would still be computationally expensive to compare each potential periodic orbit with the large group of those that have already been generated and determined to be unique.  Fortunately we have also classified each periodic orbit according to its braid type, and therefore only need to compare orbits within the same class.  This highlights just how important the algorithms involving braids and their classification are.  This subsection reviews these algorithms, albeit briefly, since many of the core ideas were covered in chapter~(\ref{ch:braid}).

The periodic orbits that our reduction algorithm finds are geometric objects, in that they are described as a series of points in phase space, $\bfrho = \lrp{\bfrho_1, \bfrho_2, \cdots, \bfrho_{N_d}}$, which join to form a closed loop.  Before we associate a braid with any given periodic orbit, we must be able to manipulate these geometric orbits themselves.  In particular, we must be able to distinguish two different orbits as well as declare two similar orbits to be equal.  Declaring two single points in phase space to be equal is easy.  Simply calculate the distance that these two points are from each other in phase space using the Euclidean metric, and set a minimal distance to act as the dividing line.  The minimal distance is effectively the tolerance for declaring the points equal.  For the set of points in a discrete version of a periodic orbit, we could just use the same distance notion, but in the larger discretized loop space.  However, this fails to capture the intrinsic symmetry families, whose members should be considered equal.  For example, the time evolution flow, $g^t_{H}$, maps a periodic orbit back onto itself with the discretized points rotated or advanced along the loop.  Periodic orbits before and after this operation are certainly the same, however they would have a large distance between them in discretized loop space.  Similarly, the flow, $g^{\theta}_{L}$, rigidly rotates every point (vortex configuration in the plane) of a periodic orbit through an angle $\theta$ about the center of vorticity.  The resultant periodic orbit should certainly be considered as equivalent to the original.  These two flows, due to vector fields $\HamVec{H}$ and $\HamVec{L}$, are the only continuous symmetries that we must worry about.  The remaining integral of motion, $P^2+Q^2$, is unimportant since the center of vorticity of each vortex configuration has been set to zero by fiat.  There is, however, an additional symmetry, a discrete symmetry, which we must also take into consideration.  Since we are dealing with identical vortices, a simple permutation of vortex labels should not result in a different periodic orbit, yet this action would certainly result in a large distance between discretized loop space points.   

To determine how close two periodic orbits are to each other, while taking into consideration the symmetries that we must ``mod out," we employ the following procedure.  Pick a single point (vortex configuration), $\bfrho^1_0$, from the first periodic orbit and compare it to each of the points, $\bfrho^2_k$, of the second periodic orbit.  For any pair of vortex configurations we can assign a closeness value which is a function of how much we rigidly rotate the first configuration and what permutation, $\pi \in \bfS_{N_v}$, of the second configuration's vortex indices we choose
\be
d\lrp{\bfrho^1_0,\bfrho^2_k, \theta, \pi} = \left| g^{\theta}_{L}\lrp{\bfrho^1_0} - \pi\lrp{\bfrho^2_k} \right|.
\label{eq:rotationCloseness}
\ee
Furthermore, we can minimize this function analytically over $\theta$ for a given permutation.  Thus, if we check each permutation, we can find the minimal distance $d^{\ast}_{k}$ as well as the corresponding angle $\theta^{\ast}_k$ and permutation $\pi^{\ast}_k$.  Next, consider each point, $\bfrho^2_k$, along the second periodic orbit and its value of $d^{\ast}_k$.  There will naturally be a point $\bfrho^2_{k^{\ast}}$, for which $d^{\ast}_k$ is minimal.  This gives us a set, $\lrp{k^{\ast}, \pi^{\ast}_{k^{\ast}}, \theta^{\ast}_{k^{\ast}}}$, consisting of the discrete time translation, the vortex label permutation, and the rotation angle, which overall minimize Eq.~(\ref{eq:rotationCloseness}).  Now rotate every point of the first periodic orbit, $\bfrho^1_j \rightarrow g^{\theta^{\ast}_{k^{\ast}}}_{L}\lrp{\bfrho^1_j}$ and translate it in time, $\bfrho^1_j \rightarrow \bfrho^1_{\lrp{j+k^{\ast}}mod N_d}$ as well as permute the vortex indices of the second periodic orbit, $\bfrho^2_j \rightarrow \pi^{\ast}_{k^{\ast}}\lrp{\bfrho^2_j}$.  If the two periodic orbits are indeed identical, then they should be very close in discretized loop space after these operations.  That is, the function
\be
d\lrp{\bfrho^1,\bfrho^2} = \lrp{\frac{1}{N_d}\sum^{N_d-1}_{i = 0} \left| g^{\theta^{\ast}_{k^{\ast}}}_{L}\lrp{\bfrho^1_{i+k^{\ast}}} - \pi^{\ast}_{k^{\ast}}\lrp{\bfrho^2_i} \right|^2}^{\frac{1}{2}},
\label{eq:FullDistance}
\ee
is an effective tool for distinguishing periodic orbits.

Furthermore, we can use this ability to compare periodic orbits to define and find chiral orbits.  In chapter~(\ref{ch:pvm}) we determined that a parity transformation combined with a time reversal was a discrete symmetry of the equations of motion.  Thus, an application of this PT symmetry to any periodic orbit results in another periodic orbit.  Interestingly, this transformed orbit is not always the same orbit as the original.  We can use Eq.~(\ref{eq:FullDistance}) to tell us just how close these two orbits are.  We will call this the chiral distance, $d_c\lrp{\bfrho} = d\lrp{\bfrho,PT\lrp{\bfrho}}$, and will put it to good use in the next chapter.

Once we can distinguish different geometric orbits from one another, we would like to be able to distinguish them at the level of topology as well.  This is tantamount to computing the braid type of a periodic orbit.  An essential first step in this direction is to extract a pure braid from the geometry of the orbit.  We do this in a relatively straightforward way, by projecting the geometric braid onto a plane, observing the over/under crossing of the strands in time, and then recording the corresponding braid generators.  There are a few nuances that complicate the actual calculation.  First of all, we have the freedom to choose the projection plane, which will result in different braids.  However, as long as the plane is defined by the time direction and some line on the $x$-$y$ plane which does not intersect the vortex motion at any time, then the different possible braids will at least be related by braid conjugacy.  Next, we can encounter the situation where two or more crossings occur at essentially identical time values.  If the separate crossings do not involve any common strands, then the time ordering is unimportant, and we choose the order in some consistent manner.  In the case where there are common strands, we will solve the evolution equation with a greater time resolution, until the multiple crossings occur at different times.  Of course, there are some symmetric orbits, like the collinear configuration of three vortices, where more than one crossing does happen at exactly the same time.  In these cases a small tilt in the projection plane is sufficient to tease out the proper ordering of braid generators for this section of an orbit.  After creating the braid representation of an orbit, we check that the braid is indeed a pure braid by computing the corresponding element of the symmetric group (obtained by forgetting the over/under information), and checking to see if it is the identity.

This check for whether a braid is pure or not highlights the importance of braid manipulations.  To help facilitate the many needed operations on braids, we created a whole C++ class to represent braids.  With this structure, we can easily multiply, invert, and conjugate braids, as well as perform more complicated tasks such as the topological PT transformation, Eq.~(\ref{eq:PTbraidTrans}), and ordering braids.  This last idea is particularly important as it allows us to solve the word problem, whether two braid words represent the same braid or not, chapter~(\ref{ch:braid}).  We certainly would not have been able to accumulate, let alone classify, so many periodic orbits without a computational representation of the braid group.

The final step in distinguishing periodic orbits by braid type involves the construction of another C++ class, one which represents the TN classification of pure braids.  As mentioned in chapter~(\ref{ch:braid}), the TN braid type has a natural rooted tree structure which lends itself to a recursive representation.  Each node in this tree structure represents a finite order or pseudo-Anosov braid, which we identify using an algorithm by Bestvina and Handel~\cite{BESTVINA:1995tx} with some ideas incorporated from~\cite{BERNARDETE:1995ua}.  Each child node represents the braid which is viewed as being injected into the corresponding strand of the parent node, see Fig.~(\ref{graph:BraidTree}).  This TN braid class consists of individual nodes as objects, which are then linked together to form the rooted tree structure.  This rooted tree structure is a conjugacy invariant, and allows us to answer the conjugacy problem in most important situations.  In particular, it is the primary tool which we will use to not only distinguish different periodic orbits, but also tease out how the topology of phase space dictates some of the qualitative properties of periodic orbits.

\chapter{Periodic Orbit Classification and Results}
\label{ch:class}

By using the reduction and classification algorithms highlighted in the previous chapter, we have accumulated a large set of periodic orbits.  In this chapter we explore some of the defining individual attributes of orbits as well as the collective patterns which emerge in this data set.  To set the stage, we first discuss the types of data that we associate with each periodic orbit.  This includes such obvious items as the period, angular momentum, and braid type classification, as well as some unexpected data such as the chiral distance and Floquet exponents.  Consideration of these attributes then enable us to classify each periodic orbit into a few binary categories.  Solutions can be permutation orbits or full periodic orbits, have pseudo-Anosov components or just finite order components, and they can be chiral or non-chiral.  Beyond these binary categories, there are other general principles which help to group the set of periodic orbits into useful classes.  In particular, we will consider the effect that relative equilibria have on the topology of the level-set submanifolds of phase space, the surfaces on which the orbits exist.  We will also discuss the $T_2$ classification scheme, and how it allows us to clearly consider the patterns which emerge in parameter space.  The general picture that emerges from this analysis is that many of the geometric attributes of periodic orbit solutions to the point vortex problem are determined by the topology of the orbits.

\section{Data}

For each periodic orbit collected, we culled a significant amount of associated data.  The main data are, of course, the positions of each vortex at a discrete number of time steps.  However, to compare many different periodic orbits with each other, we must consider simpler data, such as period and angular momentum.  The idea is that these global attributes of periodic orbits will tell us something about their classification, and indeed about vortex dynamics in general.  This section details the different types of data collected, in order to make the rest of this chapter easier to understand.  In general, the data can be grouped into three categories: data which are mainly used to aid the evaluation of the algorithms and do not tell us much about the periodic orbits themselves, data which are primarily dependent on the geometry of the periodic orbits, and data which are primarily dependent on the topology of periodic orbits.

\subsection{Evaluative Data}
\begin{figure}[htbp]
 \centering
\large
\scalebox{1.0}{\includegraphics{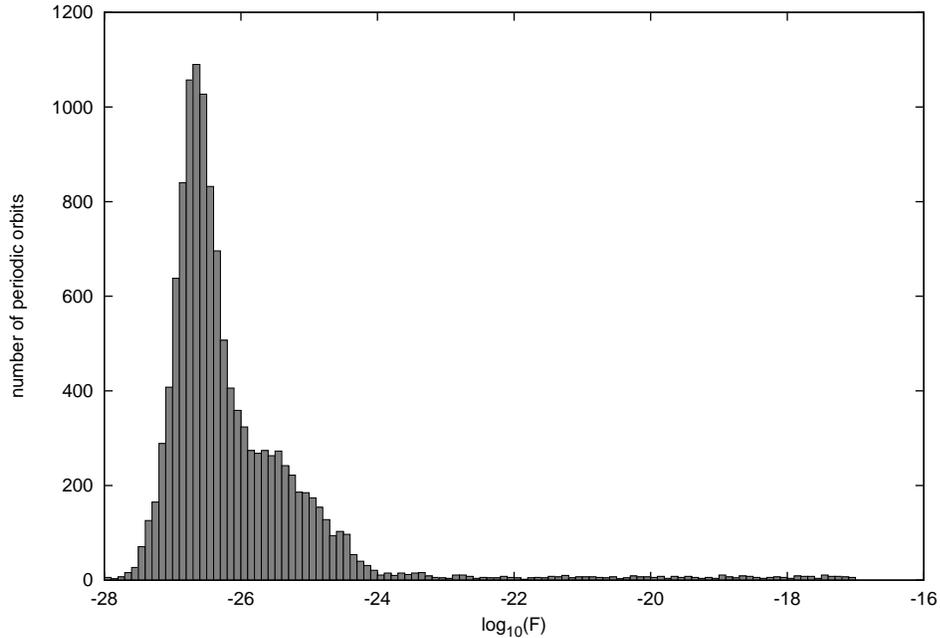}}
\normalsize
    \caption{Histogram of the number of $\lrp{N = 4}$ periodic orbits per bin (of width 0.1) vs. the order of magnitude, $log_{10}\mathcal{F}$, of the cost function evaluated on these orbits.}
    \label{graph:POhistogramF}
\end{figure}

In the course of generating periodic orbits, we must check that the algorithms from chapter~(\ref{ch:alg}) are indeed giving us solutions with the desired characteristics.  With this in mind, we catalogue three diagnostic values: the value attained by the cost function, $\mathcal{F}$ of Eq.~(\ref{eq:DiscreteFunctional}), after a periodic orbit has been reduced; the average energy of the vortex configurations along the orbit; and the average value of $P^2+Q^2$ along the orbit.  First of all, for found solutions to be actual periodic orbits, they must not only locally minimize the cost function, but also attain a value of $\mathcal{F}$ that is sufficiently small.  While the minimum possible value of $\mathcal{F}$ is determined by the coarseness of the loop discretization and the maximum curvature of the orbit, it is not a quantity that we can easily calculate.  Thus we must judge whether $\mathcal{F}$ for a given solution is sufficiently small by other means.  One way of doing this is to consider the value of the cost function for known periodic orbits, such as the rigidly rotating solutions.  These generally have cost function values ranging from $10^{-19}$ to $10^{-28}$.  With this in mind, consider the following histogram, Fig.~(\ref{graph:POhistogramF}), which shows the number of periodic orbits for the four vortex case against the $log_{10}$ of their cost function.

The overwhelming majority of periodic orbits have cost function values that lie in the range of $10^{-24}$ to $10^{-28}$, which are much smaller than the largest cost function value of a known periodic orbit.  This, coupled with the discussion concerning Fig.~(\ref{graph:CostFunctionDecrease}) should serve as ample justification that the solutions we consider in this chapter are indeed true periodic orbits of the point vortex system.  The other two values collected for diagnostic purposes, the average energy and $P^2+Q^2$, were used to double check that the energy of each periodic orbit was identical to zero and that the center of vorticity was also set to zero.

\subsection{Geometric Data}
Of the geometric data, the most important values are perhaps the period $T$ and the angular momentum $L$.  Both values are invariants of a periodic orbit, in that two orbits can not be considered equivalent through time translations and rigid vortex configuration rotations, if their values of $T$ and $L$ are incompatible.  This provides the first and easiest check for inequality amongst orbits, and reduces the number of periodic orbits that we must compare through Eq.~(\ref{eq:FullDistance}).  While necessary, equality in these two values is not a sufficient criteria to ensure that two periodic orbits are equal.  An important example of this degeneracy occurs with chiral pairs of orbits, which are related through parity and time symmetry inversions.  None the less, considering the period and angular momentum of the entire set of found periodic orbits will be the first and perhaps most important touchstone in addressing the structure and classification of periodic orbits.  We will often show $L$ vs. $T$ graphs.

\begin{figure}[htb]
 \centering
\large
\scalebox{1.0}{\includegraphics{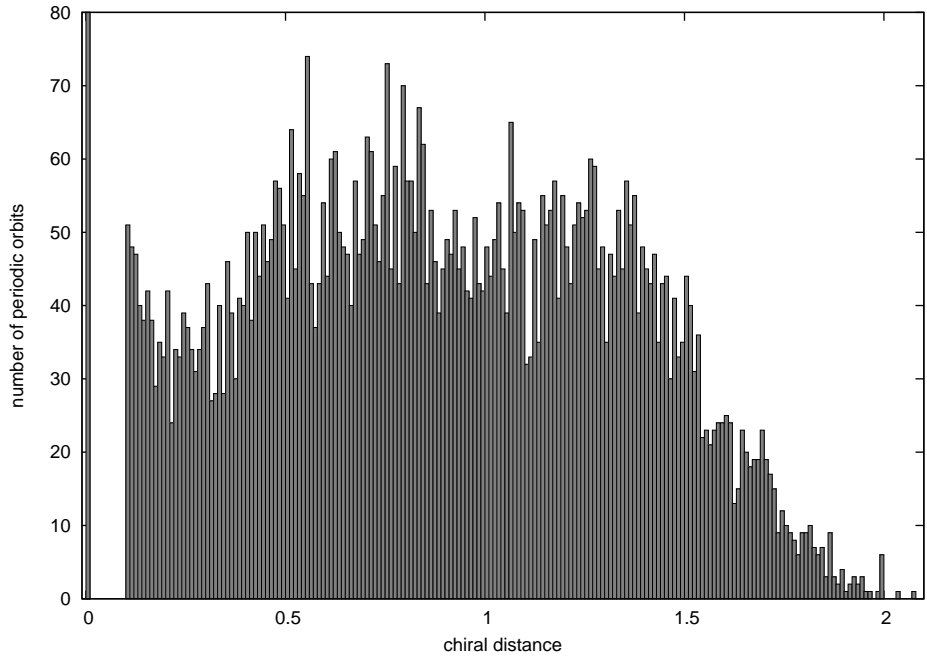}}
\normalsize
    \caption{Histogram of the number of $\lrp{N = 4}$ periodic orbits per bin (of width 0.01) vs. the chiral distance.  There are thousands of periodic orbits with a chiral distance less than 0.01.  The y-range has been lowered to more clearly show the gap between the chiral and non-chiral orbits.}
    \label{graph:POhistogramChiral}
\end{figure}

Whether or not a periodic orbit is chiral is also a geometric attribute.  We consider a periodic orbit to be chiral if its image after a time and parity inversion is greater than a certain distance away using Eq.~(\ref{eq:FullDistance}), $d\lrp{\bfrho, PT\lrp{\bfrho}} \geq d_{cutoff}$.  Fortunately there is a discrete jump in these ``chiral distance" values that allow us to unambiguously define the cutoff distance, shown in Fig.~(\ref{graph:POhistogramChiral}).

There are two salient points to note about this graph.  The first is the existence of the gap, which tells us that there is a minimum value of the chiral distance of chiral orbits (about $0.1$).  Next, the chiral distance value for each non-chiral orbit is consistent with zero, which indicates that each pair of values compared in the definition of chiral distance was similar to within the limits of double precision.  This coupled with the gap provide sufficient evidence that the chirality of periodic orbits is not an artifact of our definition of chiral distance.  Therefore, the concept of a chiral periodic orbit is a valid and interesting concept to explore.

The next piece of geometric data helps to quantify how close a given periodic orbit is to one of the relative equilibria or rigid rotations of the vortex configuration.  Recall, from section~(\ref{sec:IntMotSymmChaos}), that a relative equilibrium occurs when the Hamiltonian vector fields $\HamVec{H}$ and $\HamVec{L}$ are completely antiparallel at each point along the periodic orbit.  Thus, we could consider the quantity
\be
S_j = 1+\left.\overline{\HamVec{H}}\right|_{\bfrho_j}\cdot\left.\overline{\HamVec{L}}\right|_{\bfrho_j},
\label{eq:DiscreteLHprod}
\ee
where the over-bar indicates that the vectors are unit magnitude, and they are evaluated at point $\bfrho_j$ along the periodic orbit.  We could also imagine a continuous version of this quantity that varies with time along the orbit.  For most periodic orbits $S_j$ will oscillate in a complicated, though periodic, fashion, while for relative equilibria, $S_j$ is constant and zero.  For each periodic orbit, we store the maximum, $S_{max}$, and minimum, $S_{min}$, values of $S_j$.  This not only easily picks out relative equilibria, but also gives us extra information about the behavior of general periodic orbits.

The final bit of geometric information that we will extract from the periodic orbits concerns their stability.  Consider a point  on a periodic orbit and a sphere of small radius centered on this point.  The image of this sphere under the time flow for time $T$, the period of the orbit, will in general resemble an ellipsoid if we start with a sphere of sufficiently small radius.  In some directions the distance to the edge of the ellipsoid will have exponentially increased, some exponentially decreased, and still others will not have changed appreciably.  We call these eigen-directions unstable, stable, and marginal respectively.  In a Hamiltonian system, the volume of the sphere is preserved under symplectomorphisms, and therefore equal to that of the ellipsoid.  This gives us a relationship between the amount of stretching and contraction of the ellipsoid.  To follow this idea further, we must define the above concepts more precisely, and will be borrowing ideas and notation from~\cite{bib:ChaosBook}.

Consider what happens to a point $\bfrho_0 + \delta \bfrho_0$ under the flow $g^t_H$, for any sufficiently small $\delta \bfrho_0$.  This point is mapped to $g^t_H\lrp{\bfrho_0 + \delta \bfrho_0} = g^t_H\lrp{\bfrho_0} + \delta \bfrho\lrp{t}$.  We can expand the left hand side as a Taylor series expansion about $\bfrho_0$ and neglect terms of second order and higher (hence the sufficiently small $\delta \bfrho_0$).  This results in the association $\delta \bfrho\lrp{t} = \bfJ^t\lrp{\bfrho_0} \delta \bfrho_0$, where the Jacobian matrix is given by $\bfJ^t_{ij}\lrp{\bfrho_0} = \left. \partial \bfrho^i\lrp{t}/ \partial \bfrho^j \right|_{\bfrho = \bfrho_0}$.  This matrix measures the change in one component of a point on an orbit after time $t$, given a change in a possibly different component of the original starting point of the orbit.  We would like to consider this Jacobian matrix after one period, $T$, of the orbit has elapsed.  This matrix has some special properties, and accordingly we call it the fundamental matrix, $\bfJ \equiv \bfJ^T\lrp{\bfrho_0}$.  In particular, its eigenvalues, $\lrbc{\Lambda_1, \cdots, \Lambda_{2N_v}}$, do not depend on the starting point, $\bfrho_0$, of the periodic orbit.  Thus, these eigenvalues, which in the case of a periodic orbit are called Floquet multipliers, are important invariants of each periodic orbit.  We say that an eigenvalue and its associated eigen-direction are either unstable, marginal, or stable if the absolute value of the eigenvalue is respectively greater than, equal to, or less than one.  Since the flow $g^T_H$ is a symplectomorphism, these Floquet multipliers come in sets of four $\lrbc{\Lambda, 1/\Lambda, \Lambda^{\ast}, 1/\Lambda^{\ast}}$.  This means that the set of unstable multipliers are in $1-1$ correspondence with the set of stable multipliers.  While there is much interesting information in the structure of a periodic orbit's set of Floquet multipliers, we will be primarily interested in the largest multiplier, which is a good single measure of the instability of this orbit.  Since this is often a very large number, it is much more fruitful to deal instead with Floquet exponents $\lambda_k$, defined by $\Lambda_k = \exp\lrp{\lambda_k T}$.  While these are complex valued, we will abuse notation slightly and use $\lambda_k$ to refer to the real part of the Floquet exponents as well.  In a more general setting these are often referred to as Lyapunov exponents in the literature.

To extract these Floquet exponents for a periodic orbit, we must first compute the fundamental matrix, $\bfJ^T$.  This is achieved by integrating Eq.~(\ref{eq:FindFundMat}) along the full orbit, with $\bfJ^0 \equiv \mathbb{I}$ (the identity matrix).
\be
\frac{d\bfJ^t\lrp{\bfrho\lrp{t}}}{dt} = \bfA\lrp{\bfrho\lrp{t}} \bfJ^t \lrp{\bfrho\lrp{t}}.
\label{eq:FindFundMat}
\ee
Here $\bfA$ is a matrix which quantifies the variation of the velocity with a change in position.  That is, if $\bfv\lrp{\bfrho} = \dot{\bfrho}$, then $\bfv\lrp{\bfrho_0+\delta\bfrho_0} = \bfv\lrp{\bfrho_0} + \bfA \delta\bfrho_0$, and we can express $\bfA$ as $\bfA_{ij} = \partial \bfv^i\lrp{\bfrho}/\partial \bfrho^j$.  For us, $\bfv = \HamVec{H}$, and we can calculate the Floquet exponents in a relatively straightforward manner.

\subsection{Topological Data}

Perhaps the most important data associated with each periodic orbit are those which depend only on its topology.  All of our examples of topological data come from the braid type (braid conjugacy class) of an orbit.  At the most basic level, we have the number of vortices, either $N = 3$ or $N = 4$.  This distinction is fundamental, since it forms the dividing line between integrable and possibly chaotic motion.  Next we have the exponent sum, a conjugacy invariant of braids that counts all positive generators and subtracts off the number of negative generators.  Since all our braids are composed of positive generators only (more on this later), this measures the algebraic length of a braid.  A more important topological invariant is the $T_1$ and $T_2$ class that a braid type belongs to.  These classes will play a starring role in organizing the structure of periodic orbits in phase space.  Lastly, we record the growth factor if a braid type has a pseudo-Anosov component.

\section{Allowed Braid and Orbit Types}

Of the many different braid types that our periodic orbits could form, which are actually allowed by the dynamics?  Which are disallowed?  These interesting questions are evocative of the fundamental question in symbolic dynamics: given an alphabet of letters, and a dynamic system which generates an infinite sequence of these letters, which words are prohibited from occurring?  In this context, a complete accounting of the prohibited words is tantamount to describing the dynamic system.  In our case, one could hope that an understanding of the allowed and disallowed braid types yields a deeper understanding of the dynamics of point vortices.  

We will first consider the constraint that our periodic orbits are represented by pure braids.  Though true by fiat, consideration of this fact will naturally bring up the concept of permutation orbits.  These solutions are particularly symmetric and simple examples of periodic orbits.  Next we will reflect on the simple observation that all the orbits currently found have associated braids which are exclusively constructed out of positive braid generators.  Another defining characteristic of braids types is whether they have a pseudo-Anosov component or not.  We will show that this topological feature of periodic orbits does indeed dictate some aspects of the stability of orbits, a geometric feature.  Finally, we will consider the differences between chiral and non-chiral orbits, which dictate the amount of symmetry that can be present in an orbit.  While this is actually a geometric property, we will speculate about its connection to a topological notion of chirality.  

These concepts serve to sort our periodic orbits into several binary categories: permutation orbits or full orbits, pA components or just finite order components, and chiral or non-chiral orbits.  To better illustrate these concepts and provide a more visual connection to the individual periodic orbits, we will provide figures which show the, quite beautiful, vortex motion of specific orbits throughout this section.

\subsection{Pure Braids}

\begin{figure}[htb]
 \centering
\large
\resizebox{\linewidth}{!}{\includegraphics{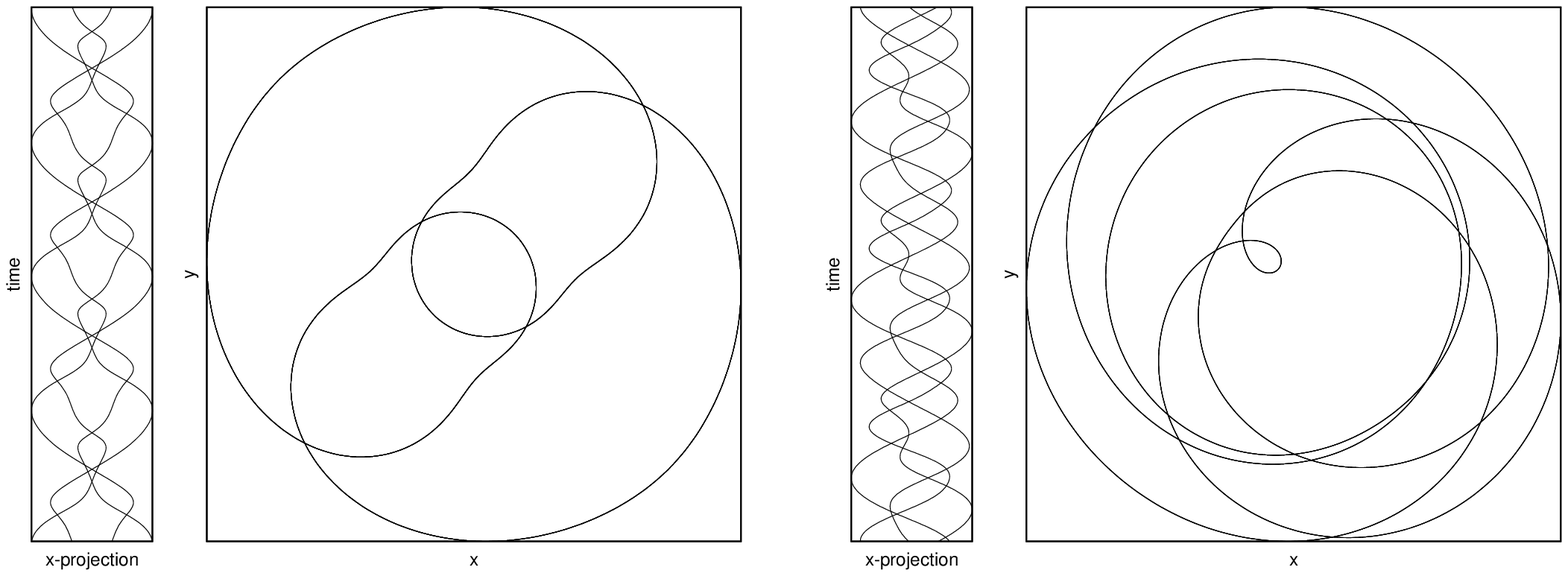}}
\normalsize
    \caption{The above shows two periodic orbits, each as a projection on the $x$-$t$ plane as well as the $x$-$y$ plane.  Both have a $T_2$ braid type of $\overline{4}_1$.  The left orbit consists of three full twists, $\lrp{\Delta^{2}}^3$, and can be constructed by gluing together four identical permutation orbits.  The periodic orbit on the right consists of four full twists, $\lrp{\Delta^2}^4$, and can be constructed out of three identical permutation orbits}
    \label{graph:permutationOrbit1}
\end{figure}

\begin{figure}[htb]
 \centering
\large
\resizebox{\linewidth}{!}{\includegraphics{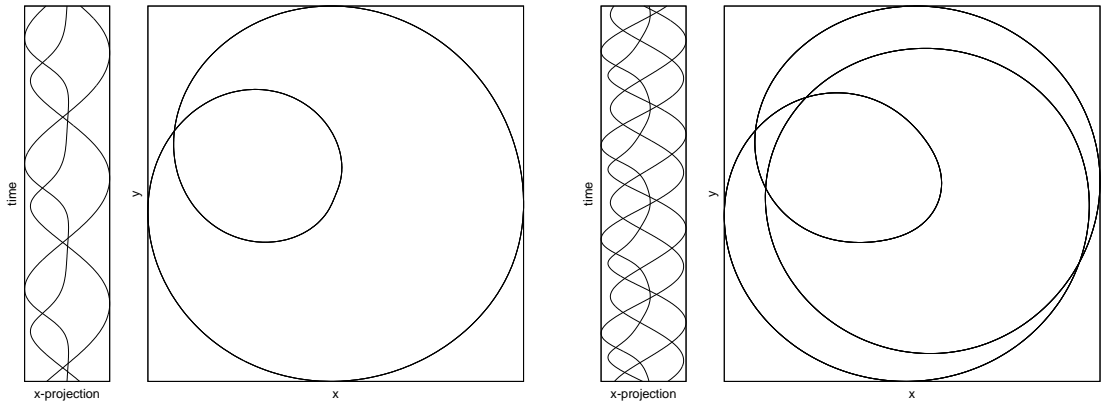}}
\normalsize
    \caption{The left side shows a periodic orbit of three vortices, which is composed of three identical permutation orbits.  The corresponding braid, $\lrp{\sigma_2\sigma_1\sigma_2\sigma_1}^3$, is conjugate to the braid with two full twists, $\lrp{\Delta^2}^2$.  The right side shows a periodic orbit on four vortices.  Similar to figure~(\ref{graph:permutationOrbit1}), this periodic orbit is composed of four identical permutation orbits, and corresponds to a braid type of three full twists, $\lrp{\Delta^2}^3$.}
    \label{graph:permutationOrbitComb}
\end{figure}

That all of our periodic orbits form pure braids is a relatively uninteresting question, as this is a condition which defines whether an orbit is periodic in phase space or not.  However, since all of our vortices have identical strengths, we could have additionally considered as periodic those orbits which take a given vortex configuration to an identical configuration, except with a permutation of the vortex indices.  Notice that these permutation orbits can be iterated a number of times, the order of the permutation element, to form a full periodic orbit with a braid in the pure braid group.  So, despite searching for pure braid periodic orbits exclusively, we find these permutation orbits as factors of full periodic orbits.  Examples of periodic orbits composed of copies of permutation orbits are shown in figures~(\ref{graph:permutationOrbit1}) and (\ref{graph:permutationOrbitComb}).  A more geometric picture of periodic orbits composed of permutation orbits emerges when we think of the action of a permutation on an arbitrary periodic orbit.  A permutation will send a point of phase space to another point in phase space, and therefore will generically send a periodic orbit to another periodic orbit in a different region of phase space.  We can now view periodic orbits composed of permutation orbits as those that are mapped back onto themselves (modulo the usual rigid vortex configuration rotations and time translations) under the action of a permutation.

The vast majority of periodic orbits in our collection can not be factored into multiples of permutation orbits in this manner.  This is to be expected for several reasons.  First of all, braids in the corresponding braid type must be expressible as a factor, $\beta = \alpha^k$, which is a very restrictive condition.  From a combinatorial perspective, there are many more ways to form pure braids of a given exponent sum which do not fit this restriction than those that do.  Second, even within a given braid type which satisfies this condition, most periodic orbits will require the full orbit to return to the original vortex configuration.  As an example, consider the $N=4$ braid type which consists of three full twists, $\Delta^6$.  We found 50 unique periodic orbits which have this braid type, but only five of these are decomposable into factor orbits (one of these five is shown on the left of Fig.~(\ref{graph:permutationOrbit1}) and another on the right side of Fig.~(\ref{graph:permutationOrbitComb})), and three of these orbits are repeats of the three relative equilibria.  Despite this relative scarcity, or perhaps because of it, periodic orbits that are multiples of permutation orbits form a special set.  They tend to be particularly symmetric.  Indeed the most symmetric orbits, the rigidly rotating relative equilibria, are all composed of permutation orbits.

There are a few natural questions which we could ask about the existence of permutation orbits.  Of the finite set of permutations (elements of the symmetric group $S_3$ or $S_4$), which are represented in permutation orbits?   For the braid type consisting of braids conjugate to $\lrp{\Delta^2}^k$, in both $N=3$ and $N=4$, which permutation braids arise as factors, $\alpha^j \simeq \lrp{\Delta^2}^k$?  In this case, we can consider permutation braids to be rational fractions, $\frac{k}{j}$, of the full twist.  What are the permissible fractions, and are there multiple permutation orbits that correspond to the same fraction?  These are difficult questions to answer in their entirety, so we will simply give an accounting of the permutation braids which have been found, in the context of these questions.  Much of the relevant information is summarized in table~(\ref{table:PermutationOrbits}).

\begin{table}[htbp]
  \centering
  \begin{center}
  \small
  \be
 \begin{array}[c]{|c|c|c|c|c|c|c|}
 \hline
\# & N & \frac{k}{j} & relative \; equilibrium & permutation & conj. \; class & braid \\
\hline
1 & 3 & \frac{1}{3} & yes & \lrp{132} & 2 & \sigma_1\sigma_2 \\
\hline
2 & 3 & \frac{1}{2} & yes & \lrp{13} & 1 & \sigma_1\sigma_2\sigma_1 \\
\hline
3 & 3 & \frac{2}{3} & no & \lrp{123} & 2 & \sigma_1\sigma_2\sigma_1\sigma_2\\
\hline
4 & 3 & \frac{5}{3} & no & \lrp{132} & 2 & \lrp{\sigma_1\sigma_2}^5 \\
\hline
5 & 3 & \frac{7}{3} & no & \lrp{123} & 2 & \lrp{\sigma_1\sigma_2}^7\\
\hline
6 & 3 & \frac{8}{3} & no & \lrp{132} & 2 & \lrp{\sigma_1\sigma_2}^8 \\
\hline
7 & 3 & \frac{11}{3} & no & \lrp{132} & 2 & \lrp{\sigma_1\sigma_2}^{11}\\
\hline
\hline
8 & 4 & \frac{1}{4} & yes & \lrp{1243} & 4 & \sigma_1\sigma_3\sigma_2 \\
\hline
9 & 4 & \frac{1}{3} & yes & \lrp{124} & 2 & \sigma_1\sigma_2\sigma_3\sigma_2 \\
\hline
10 & 4 & \frac{1}{2} & yes & \lrp{14}\lrp{32} & 3 & \sigma_1\sigma_2\sigma_3\sigma_1\sigma_2\sigma_1 \\
\hline
11 & 4 & \frac{3}{4} & no & \lrp{1342} & 4 & \sigma_1\sigma_3\sigma_2\sigma_1\sigma_3\sigma_2\sigma_1\sigma_3\sigma_2 \\
\hline
12 & 4 & \frac{3}{4} & no & \lrp{1432} & 4 & \sigma_1\sigma_2\sigma_1\sigma_3\sigma_2\sigma_1\sigma_2\sigma_3\sigma_2 \\
\hline
13 & 4 & \frac{4}{3} & no & \lrp{132} & 2 & \sigma_2\sigma_1\sigma_2\sigma_3\lrp{\sigma_1\sigma_2\sigma_1\sigma_3\sigma_2\sigma_3}^2 \\
\hline
14 & 4 & \frac{5}{3} & no & \lrp{132} & 2 & \lrp{\sigma_1\sigma_3\sigma_2}^4\lrp{\sigma_3\sigma_2\sigma_1\sigma_2}^2 \\
 \hline
 \end{array} \nonumber
\ee
\end{center}
\normalsize
\caption{This table summarizes the information concerning the set of found permutation orbits which are rational fractions of the full twist.  Orbit 11 and 13 are shown on the left and right of Fig.~(\ref{graph:permutationOrbit1}), while orbit 3 and 12 are shown on the left and right sides of Fig.~(\ref{graph:permutationOrbitComb}) respectively.  The permutations are given in cycle notation and the conjugacy classes of each permutation are listed.  The relative equilibria are given in their natural order in terms of increasing angular momentum.  So, orbit 1 is the configuration with vortices on the corners of an equilateral triangle, orbit 2 corresponds to three collinear vortices, orbit 8 vortices on the corners of a square, orbit 9 vortices on the corners and center of an equilateral triangle, and orbit 10 corresponds to four collinear vortices.}
\label{table:PermutationOrbits}
\end{table}
First notice that permutations of all possible orders are represented.  In particular, for each $N$, there is one relative equilibrium for each possible permutation order ($j$-value).  Of the $N!-1$ permutations possible (excluding the identity), only 3 in the $N=3$ case, and 6 in the $N=4$ case are represented.  However, a more appropriate question would consider how many conjugacy classes of the permutations are represented.  Since we consider braids to be equivalent in each braid type if they are conjugate to each other, we should also treat their corresponding permutations as being equal.  For example, the two $N=3$ permutations $\lrp{123}$ and $\lrp{132}$ are equivalent if we rotate our projection plane for the braids through 180 degrees.  With this in mind, there are only two conjugacy classes for $N=3$ and four classes for $N=4$.  In the former case, each permutation class is represented by a permutation orbit, while in the latter case, each permutation class is represented except for that consisting of permutations conjugate to $\lrp{12}$.  This class most likely has a representative permutation orbit outside of the $4_1$ braid type that is being considered.  

Next, there are a few things we can say about the allowed fractions $\frac{k}{j}$ of the full twist.  The $j$-value is confined to the permissible orders of a permutation, $\lrp{2,\cdots,N}$, while the $k$-value is confined to numbers which are relatively prime with respect to the order of the permutation.  Any other restrictions on the possible $k$-values would probably require a larger set of found permutation orbits.  As a final observation, we can point out that it is possible to have multiple unique permutation orbits which are the same fraction of a full twist.  Both orbits 11 and 12 require four multiples to be conjugate to three full twists.  This suggests that there are many more permutation orbits of the vortex dynamics which we simply have not yet found.

\subsection{Positive Braids}

	In the most general case, braids are composed of both positive generators, $\sigma_i$, and their inverse (negative) generators $\sigma_i^{-1}$.  Since we are dealing with identical vortices of positive strength, vortices tend to rotate about each other in the same fashion, counter clockwise when viewed on the plane from the positive z-axis.  This might lead one to predict that the braids corresponding to our found periodic orbits would be primarily composed of one type of generator, in this case positive generators.  It turns out that a much stronger restriction on the generators seems to hold.  Of the tens of thousands of periodic orbits found, none have braids that are even partially composed of negative generators.  That is to say, all these braids are expressible in terms of positive generators exclusively.  
	
	This is indeed a surprising result, and it requires some further explanation.  The braid that results from directly reading off the over and under crossings of a periodic orbit projection will generally have a few negative generators in addition to the preponderance of positive generators.  However, for each of our periodic orbits two things happen to remove these negative generators.  More often than not, a negative generator is directly adjacent to its inverse, and the pair, being equal to the identity, can be removed.  More generally, a few applications of the braid relations, Eq.~(\ref{eq:BraidRelations}), will often transform braid words into this case.  If a negative generator persists, it seems that it inevitably disappears after converting the braid to the TN braid type.  That is, we have not found any examples where the braids in TN braid tree nodes have negative generators.
	
	Due to the large number of periodic orbits in our collection, it seems unlikely that we are simply missing braid types with negative generators due to issues of sample size.  However, it is possible that negative generators will be necessary to describe braid types of periodic orbits which are longer than the cutoff we currently impose on the period.  Regardless, we would ideally want a proof that this drastic pruning indeed holds for all braid types generated by the point vortex system consisting of three or four identical positive strength vortices.  We certainly do not know if this result holds for more than four identical vortices.  Though if it does, then it should also hold for vortices of any positive integer strength, due to the essential equivalence of a group of close identical vortices and a single vortex of equal total strength (see the arguments in section~(\ref{sec:LieTrans})).	

\subsection{Pseudo-Anosov Braids}	

\begin{figure}[htb]
 \centering
\large
\resizebox{\linewidth}{!}{\includegraphics{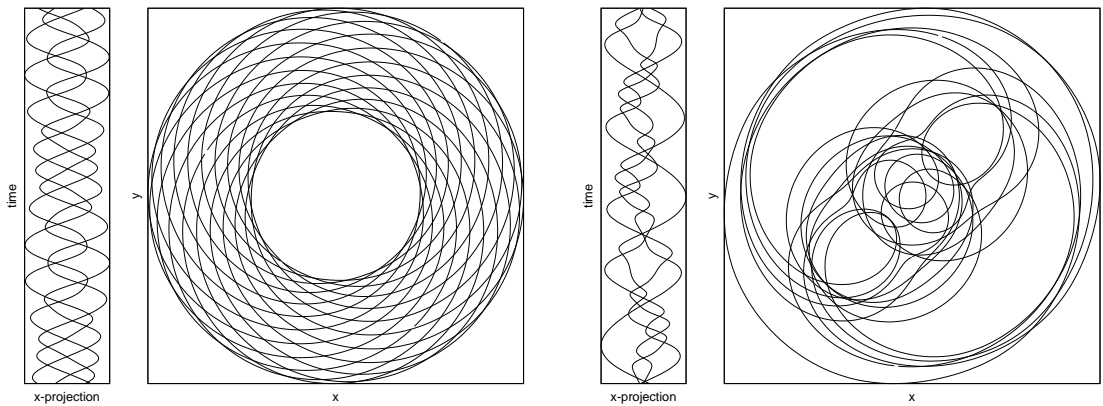}}
\normalsize
    \caption{The periodic orbit on the left is finite order, with a $T_2$ type of $\overline{4}_1$ and a braid conjugate to $\lrp{\Delta^2}^5$.  The periodic orbit on the right is pseudo-Anosov, with a $T_2$ type of $\overline{4}_2$ and a growth factor of $\lambda_{pA} = 17.94$.  Both periodic orbits are non-chiral and their projection in the $x-y$ plane therefore has at least one axis of bilateral symmetry.}
    \label{graph:pAfo}
\end{figure}
	One of the fundamental distinctions in the overall $T_2$ classification of braid types involves the difference between finite order (fo) and pseudo-Anosov (pA) braid types.  Finite order braid types, in our case of positive pure braids, are particularly simple.  They are all uniquely identified with a positive power, $k$, of the full twist, $\lrp{\Delta^2}^k$.  Pseudo-Anosov braid types on the other hand, are quite complicated, and are only partially characterized by the growth/expansion factor, $\lambda_{pA}$, of the associated minimal map (see chapter~{\ref{ch:braid}}).  Even the pair of expansion factor and exponent sum, both braid conjugacy invariants, are not sufficient to distinguish two different braid types with pseudo-Anosov components.  A complete set of such invariants would effectively solve the conjugacy problem of pure braids.

\subsubsection{pA Braid Existence}		
	We can, however, say quite a lot about the pseudo-Anosov periodic orbits which we have accumulated.  First of all, they do indeed exist, and an example is shown in Fig.~(\ref{graph:pAfo}).  Perhaps most surprisingly, they exist only for periodic orbits of four vortices.  Despite the fact that three stranded braids can be pA, the point vortex dynamics completely exclude braids from being in the $\overline{3}_2$ braid class.  One might be tempted to associate the non-occurrence of pA braids for three vortices with the integrability of three vortex dynamics, and the occurrence of pA braids for four vortices with the non-integrability of four vortex dynamics.  However, the situation is more subtle that this.  Integrability concerns the geometry and topology of the phase space of point vortices, while the pA braid expansion factor, $\lambda_{pA}$, of a periodic orbit in this phase space tells us about the time-$T$ behavior of non-vortex points on the plane being advected by the periodic motion of the vortices.  To tease out the connection between non-integrable solutions and the existence of pseudo-Anosov braids, we must must be careful in our notions of chaos.
	
\subsubsection{Eulerian Chaos vs. Lagrangian Chaos}	
	For our point vortex system there are two separate, though related, areas in which we can formulate the dynamics.  The most straightforward picture is that of the vortices in the plane.  Here we can additionally talk about the rest of the fluid, and its advected motion due to the vortices.  We can also picture the dynamics as occurring in phase space, where only the position of the vortices is explicitly considered.  In both of these views we can talk about the existence of two types of chaos: Eulerian and Lagrangian.  These follow the two principle ways to describe the dynamics of a fluid: by considering the spatially fixed velocity field or the motion of individual fluid particles.  In a loose sense, Eulerian chaos occurs when the velocity field varies in time in a non-periodic manner, while Lagrangian chaos occurs when two nearby points diverge exponentially in time.  Eulerian chaos is closest to what we would call turbulence, though true turbulence has a spatially chaotic component as well.
	
	So, it makes sense to think of Eulerian and Lagrangian chaos in both the plane and in phase space.  Fortunately, there are relations and simplifications among these four combinations.  First of all, Eulerian chaos in phase space does not occur for us.  Our Hamiltonian is autonomous, and therefore the Hamiltonian vector field in phase space, $\HamVec{H}$, is time independent.  Furthermore, there is an intimate connection between Lagrangian chaos in phase space and Eulerian chaos in the plane.  Recall that the velocity field on the plane is uniquely determined by the positions of the vortices, Eq.~(\ref{eq:PVvelocityfield}), and therefore by a single point in phase space.  Since we can more easily quantify Lagrangian chaos through Lyapunov exponents than Eulerian chaos through some measure of irregularity in velocity time series, we will address both phase space Lagrangian chaos and Eulerian chaos in the plane as the same thing and quantify it with Lyapunov exponents.  Lyapunov exponents are defined by $\lambda_{L} = \lim_{t \to \infty} \log\lrp{\bfrho^2\lrp{t}-\bfrho^1\lrp{t}}/t$, where the initial distance between the two points in phase space has been normalized to one.  Since we are considering periodic orbits, the Lyapunov exponents reduce to Floquet exponents which we more rigorously defined in the discussion of Eq.~(\ref{eq:FindFundMat}).  Finally, we have Lagrangian chaos on the plane, also called chaotic advection~\cite{bib:Aref1}.  In this case, we can define a Lyapunov exponent  of the time-$T$ map for a particular fluid particle.  This notion depends on both the initial position of the fluid particle and the point along the periodic orbit which we consider to coincide with the initial time.  Because of these complications, we will not consider Lagrangian chaos on the plane in much detail.
	
\subsubsection{pA Braid Implications}

	Now we are better equipped to consider the significance of $N=4$ pA braids.  Each braid defines a time-$T$ map of the plane, and pA braids ensure that the Lyapunov exponents of a subset of points on the plane under this map are at least as large as the pA expansion factor $\lambda_{pA}$.  Thus, the existence of pA braids force the existence of a certain amount of chaotic advection.  Notice, however that the absence of pA braids does not preclude chaotic advection.  Indeed, chaotic advection can occur even when the velocity field is stationary and there is no Eulerian chaos.  There has been much written about chaotic advection~\cite{bib:Aref1}\cite{bib:Babiano}, but it is much more prominent in viscous fluid flows and we will not pursue it further.  
\begin{figure}[htb]
 \centering
\large
\resizebox{\linewidth}{!}{\includegraphics{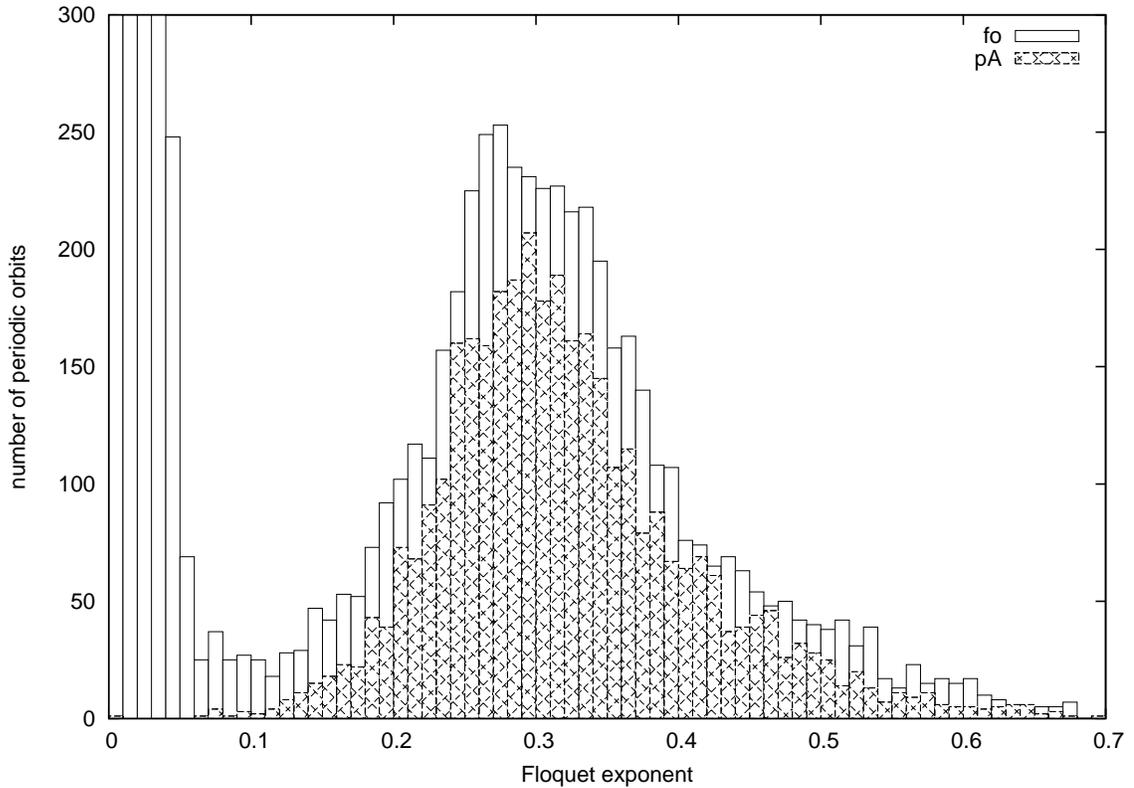}}
\normalsize
    \caption{A histogram showing the number of $N=4$ periodic orbits against their maximum Floquet exponent.  The cases of periodic orbits with pseudo-Anosov components and those without pA components are shown for comparison.  Notice that the pA periodic orbits have a lower bound to their maximum Floquet exponents.  In contrast, there are a significant number of periodic orbits with no pA component which have near zero maximum Floquet exponents (too many to show on this graph).  The boxes are of width 0.01.}
    \label{graph:POhistogrampAFloquet}
\end{figure}	

	Interestingly, the existence of pA periodic orbits correlates in a weak way with Eulerian chaos.  Consider the maximum Floquet exponents for all $N=4$ periodic orbits.  If we divide the orbits into two groups, those with pA components and those without, as in Fig.~(\ref{graph:POhistogrampAFloquet}), we can see a qualitative difference in the distribution of maximum Floquet exponents.  Notice that there are a large number of non-pA orbits which have near zero maximum Floquet exponents.  These periodic orbits are stable to perturbations and phase space trajectories in their vicinity are integrable or nearly so.  In contrast, this large bulge is completely absent in the set of pA orbits.  This suggests that the topology of the pA braids excludes their corresponding periodic orbits from being stable or near integrable.  Aside from this feature, it is interesting to note that the pA and non-pA sets mirror each other in every other respect.  This suggests that the main contributing factor to the instability of orbits with finite maximum Floquet exponent is due to dynamics and not due to the topology.  Indeed, for pA orbits there is no correlation between the log of the pA expansion factor and the maximum Floquet exponent.  As a final note, an analogous histogram for the $N=3$ case looks qualitatively similar to the $N=4$ non-pA case, except that the maximum Floquet exponents are on average smaller by over two orders of magnitude.
	
\begin{figure}[htb]
 \centering
\large
\resizebox{\linewidth}{!}{\includegraphics{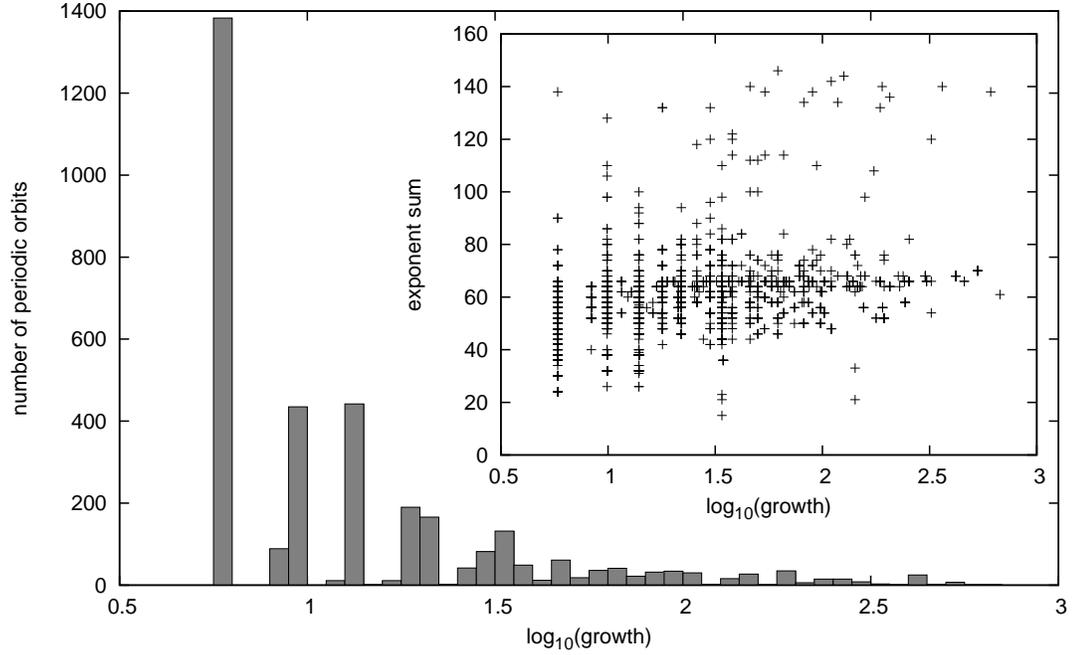}}
\normalsize
    \caption{A histogram showing the number of periodic orbits that have different pA growth/expansion factors, $\lambda_{pA}$.  The $log_{10}$ of the growth factor is shown for convenience, and the boxes are of width 0.05.  Inset is a graph of the exponent sum against the $log\lrp{\lambda_{pA}}$, which shows that there are families of different braid types with the same $\lambda_{pA}$.}
    \label{graph:POhistogrampAGrowth}
\end{figure}
	What about the distribution of pA expansion factors, $\lambda_{pA}$?  In Fig.~(\ref{graph:POhistogrampAGrowth}) we can see that most of the periodic orbits have small values of $\lambda_{pA}$.  There are gaps in this histogram, which indicate that $\lambda_{pA}$ take on discrete values.  This is perhaps clearer in the inset graph, which also shows the exponent sum of each braid.  Since this is a conjugacy invariant, the vertical sets of points indicate that many different braid types can share the same value of $\lambda_{pA}$.  Indeed it is worse than this as many different braid types have identical pA expansion factors and exponent sums ($e.s.$).  Thus, the pair $\lrp{\lambda_{pA},e.s.}$ is not a total topological invariant of the braid type.

\subsection{Chiral Braids}

\begin{figure}[htb]
 \centering
\large
\resizebox{\linewidth}{!}{\includegraphics{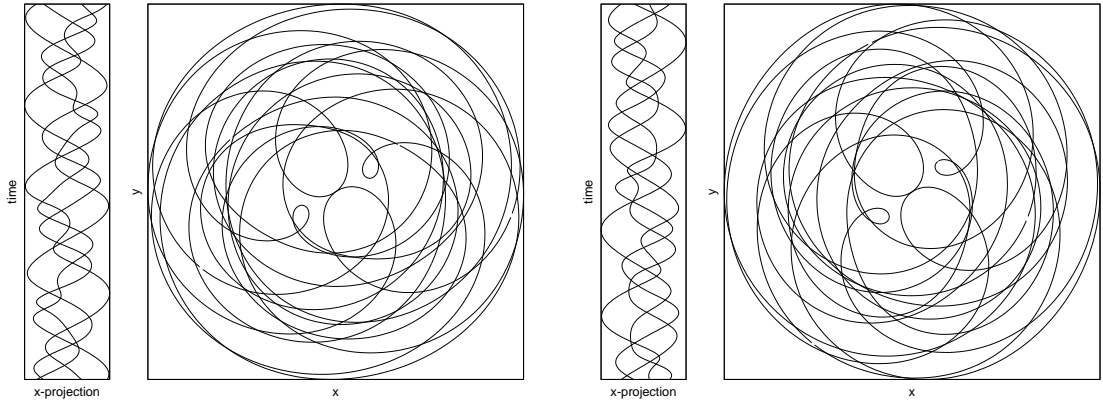}}
\normalsize
    \caption{These two periodic orbits comprise a chiral pair, and are conjugate to the braid $\lrp{\Delta^2}^4$.  Interestingly, unlike most chiral orbits, this pair has a symmetry of the projection on the $x-y$ plane.  However, unlike non-chiral orbits which have bilateral reflection symmetry, the $x-y$ projection of these orbits is symmetric to rotations by 180 degrees.}
    \label{graph:ChiralPair}
\end{figure}
	One surprising attribute of periodic orbits is their potential chirality.  For any given periodic orbit, we can transform it to a new periodic orbit by reflecting each vortex configuration about a line through the center of vorticity and then reversing the time ordering.  Since the PT transformation leaves the equations of motion unchanged, this new orbit is guaranteed to be a solution.  We can then compare these two periodic orbits, modulo rigid rotations and time translations, with the distance given in Eq.~(\ref{eq:FullDistance}), which we will call the chiral distance.  Often the two periodic orbits will have a chiral distance of zero.  In this case the two orbits are essentially the same and we say that this single periodic orbit is non-chiral.  We also have the case where the chiral distance is not zero and we must treat the two orbits as distinct solutions, though related.  In this case we refer to these solutions as chiral pairs.  An example of a chiral pair of periodic orbits is shown in Fig.~(\ref{graph:ChiralPair}).
	
	The first remarkable result is that there is a clear distinction between chiral and non-chiral orbits.  As shown previously in Fig.~(\ref{graph:POhistogramChiral}), there is an appreciable gap in between near zero chiral distances and non-zero chiral distances.  This gap appears to be insensitive to both the period and angular momentum of the orbits, which bolsters the idea that there is not only a quantitative difference between chiral and non-chiral periodic orbits, but also a fundamental qualitative difference between them.

\subsubsection{Chirality of Three and Four Vortex Orbits}	
	Another interesting result with chiral orbits concerns the circumstances necessary for their existence.  Much like pA orbits, chiral orbits only exist for the case of four vortices.  All periodic orbits of three vortices are non-chiral.  This is somewhat surprising at first, however a moment's thought will make it clear why this must be the case.  Consider the PT transformation from a geometric perspective, where it maps the submanifold $\mymanL$, the level set of $P=Q=H=0$ and $L = constant$, back onto itself.  In the $N=3$ case, this manifold is a torus, parameterized by two angle coordinates.  Thus, the PT transformation maps the torus back onto itself and is a torus automorphism.\footnote{The mapping class group of torus automorphisms is isomorphic to $GL\lrp{2,\mathbb{Z}}$, the group of $2\times2$ integer valued invertible matrices.  Since PT applied twice is the identity, the representative matrix must have $\det = \pm1$.  This leaves us with three essentially different viable matrices, the identity, that which switches the two directions characterizing the torus, and that which takes one direction to its negative.  The last two transformations can not represent the PT transformation, because they switch the orientation of the torus (which is not globally possible in the space foliated by $\mymanL$).  Thus, if the PT transformation is a torus automorphism, then it must be isomorphic to the identity.}  However, we can map any point on this torus to any other point through a combination of time translations and rigid rotations.  Since any two periodic orbits on this torus are equivalent in this manner, we can say that two periodic orbits connected by the PT transformation are themselves equivalent.  Therefore, all periodic orbits for three vortices are necessarily non-chiral.\footnote{One assumption which was implicit in this argument was that $\mymanL$ has only one path-connected component.  We could imagine the situation where there are two or more path-connected components, and the PT transformation maps one to another.  What the non-existence of chiral orbits for three vortices really tells us is that if there are multiple path connected components, then they are not related by the PT transformation.}  For the case of four vortices, the manifold $\mymanL$ is now four dimensional and not necessarily a 4-torus.  The existence of chiral orbits indicates that there are either multiple path connected components to $\mymanL$, pairs of which are mapped to each other under the PT transformation, or the PT transformation maps a single path connected component of $\mymanL$ to itself in a manner that precludes the new orbits from connecting to the old orbits through rigid rotations and time translations.

\begin{figure}[htb]
 \centering
\large
\resizebox{\linewidth}{!}{\includegraphics{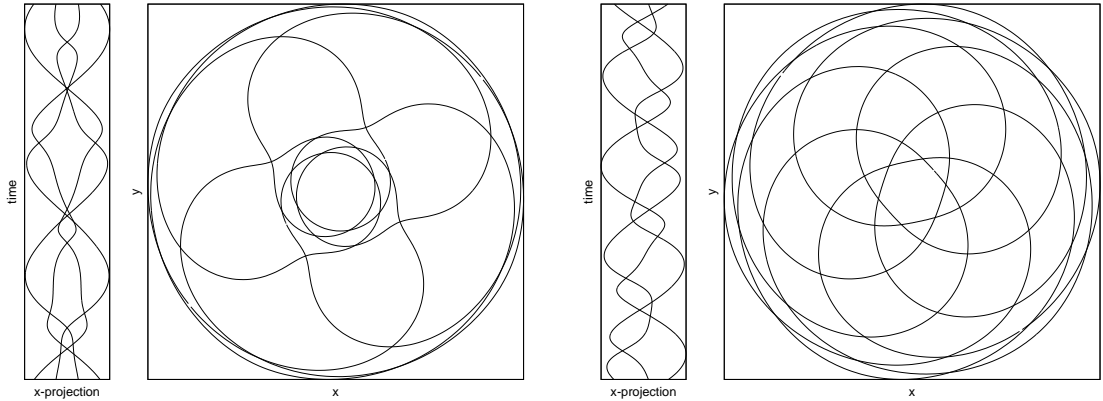}}
\normalsize
    \caption{These two orbits are both non-chiral, and therefore have at least one axis of reflection symmetry of the $x-y$ projection.  The orbit on the left is composed of four vortices and has four lines of reflection symmetry.  The orbit on the right is composed of three vortices and has six lines of reflection symmetry.}
    \label{graph:bisymm}
\end{figure}
\subsubsection{Bilateral Symmetry}	
	When we do have a non-chiral orbit, either in the $N=3$ or $N=4$ case, we can say something about the symmetry of the orbit.  In particular, the lack of chirality forces the $x-y$ projection of the orbit (the accumulated path on the plane in which the vortices move) to have at least one axis of bilateral reflection symmetry.  Consider the two transformations which comprise PT, the parity inversion P and time reversal T.  Since the parity inversion reflects each vortex configuration of an orbit about an axis through the center of vorticity, it results in an overall reflection of the $x-y$ projection.  The time reversal does not affect this projection.  For non-chiral orbits this reflected $x-y$ projection can be rotated through some angle, $\theta$, to coincide with the original.  This implies the existence of an axis of bilateral symmetry an angle, $\theta/2$, away from the original, and arbitrarily chosen, parity inversion axis.  The simplest example of this phenomena is seen for relative equilibria.  Every axis through the center of vorticity for relative equilibria is a refection symmetry axis, and they have $\infty$-fold symmetry.  Other non-chiral orbits will have a finite number of reflection axes.  Fig.~(\ref{graph:permutationOrbit1}) \& (\ref{graph:permutationOrbitComb}) show periodic orbits with one and two axes of reflection symmetry, while Fig.~(\ref{graph:pAfo}) \& (\ref{graph:bisymm}) show orbits with 16, 1, 4, and 6 axes of reflection symmetry.  On the other hand, chirality or its absence does not dictate rotational symmetry.  The chiral pair in Fig.~(\ref{graph:ChiralPair}) have rotational symmetry and no reflection symmetry, while the non-chiral orbit on the right side of Fig.~(\ref{graph:pAfo}) has reflection symmetry but no rotational symmetry.

\begin{figure}[htb]
 \centering
\large
\resizebox{\linewidth}{!}{\includegraphics{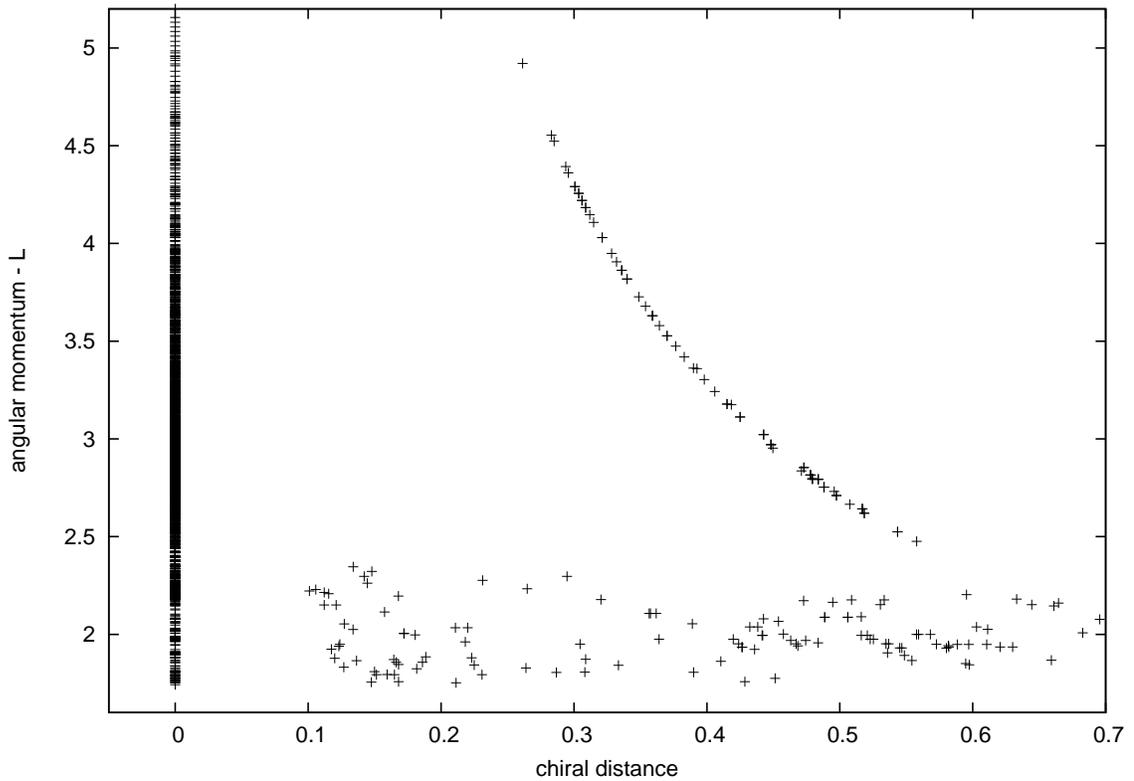}}
\normalsize
    \caption{The angular momentum is plotted vs. the chiral distance for the $T_2$ class of $\overline{4}_5$.  This class is formed of two pairs of close vortices which rotate about each other, and is expressed as $\lrp{2_1\uparrow 2_1}\uparrow 2_1$.  A one-parameter chiral family can be seen in the upper right.}
    \label{graph:ChiralFamily}
\end{figure}
\subsubsection{One-Parameter Families}
	In general, there is little to no connection between the chiral distance of a periodic orbit and this orbit's other geometric attributes.  One notable exception occurs for families of orbits from the $\overline{4}_5$ $T_2$ class, the class consisting of two pairs of close vortices which rotate about each other.  These families are distinguished by a definite functional relationship between the angular momentum and the chiral distance of an orbit.  In particular, orbits in these one-parameter families have decreasing chiral distances for increasing values of angular momentum, Fig.~(\ref{graph:ChiralFamily}).  We can describe each periodic orbit in this class by the triple of integers $\lrp{i,j,k}$, where $i$ describes the number of times the pairs of vortices obit about each other, and $\lrp{j,k}$ describe the the number of turns each vortex pair makes in one period.  For these one-parameter families (each family is indexed by a single $i$-value), we have observed that $k = j+1$.  That is to say that one of the vortex pairs rotates one extra time in a single period.  It is unclear why this relationship gives us a link between angular momentum and chiral distance, while other relationships between $j$ and $k$ do not.  However, we may gain some insight into the topology of the level-set submanifolds, $\mymanL$, merely from the existence of these families.  In the limit of very large values of angular momentum, each vortex in both vortex pairs will be very close to their partners.  The chiral distance of this configuration will necessarily trend toward zero as the angular momentum is increased.  This is, however, in opposition to the observation that there is a lower bound in chiral distance for chiral braids.  This indicates that the periodic orbits in these one-parameter families live on one or more path-connected components of the submanifold, $\mymanL$, which are separate from the components associated with the other chiral braids, and therefore not bound by the same topological restriction which cause the chiral distance lower bound.  From this we can infer the existence of multiple path-connected components of the submanifold $\mymanL$.
	
\subsubsection{Topological Chirality}
	As we have defined it, the difference between a chiral and a non-chiral periodic orbit is a geometric attribute, quantified by each orbit's chiral distance.  However, we can also view the notion of chirality from the topological perspective.  If we expand our concept of equivalence of periodic orbits beyond rigid rotations and time translations to include general deformations which isotopically preserve braid type, then we end up with a more restrictive condition that periodic orbits must satisfy to be considered chiral.  This topological chirality depends only on the braid type, and can be detected by determining whether the braid associated with a given periodic orbit is conjugate to the braid associated with the PT transformation of this periodic orbit.  Indeed, we could do away with the reference to the physical orbit and define the PT transformation directly on an algebraic braid in the following manner.  Consider a braid $\beta = \alpha_1\alpha_2\cdots\alpha_k$, where $\alpha_i$ indicate the specific Artin generator $\alpha_i \in \lrbk{\sigma^{\pm1}_1, \cdots, \sigma^{\pm1}_{N_v-1}}$ which is in the $i$-th position in the algebraic braid.  For a given generator $\alpha_i = \sigma^{\pm1}_{s}$, define the transformation $\overline{\alpha}_i = \sigma^{\pm1}_{N_v-s}$.  This just switches a generator symmetrically to the opposite side of the braid, without changing the over/under crossing.  The result of the PT transformation on $\beta$ can now be expressed as
\be
PT\lrp{\beta} = \overline{\alpha}_k\overline{\alpha}_{k-1}\cdots\overline{\alpha}_1.
\label{eq:PTbraidTrans}
\ee 
Thus, if we have a method for determining conjugacy, $\beta \simeq PT\lrp{\beta}$, we can identify topologically chiral braids and orbits.  Unfortunately, the TN braid tree structure is not a total conjugacy invariant, and can only indicate when two braids are not conjugate.  To find topologically chiral orbits, we would need to use the Garside algorithm, mentioned in ch.~(\ref{ch:braid}), to completely solve the conjugacy problem.  While we have not yet implemented this idea, the existence of topologically chiral orbits is an intriguing enough idea to warrant some further discussion.
	
	Note that topological chirality is a more restrictive concept than geometric chirality.  In particular, if a braid type is chiral, then any orbit which has this braid type is necessarily geometrically chiral.  However, the converse of this forcing is certainly not true.  For example, consider some orbits with braid type conjugate to $\lrp{\Delta^2}^4$.  The orbit on the right side of Fig.~(\ref{graph:permutationOrbit1}) is not geometrically chiral, which indicates that $\lrp{\Delta^2}^4$ can not be a topologically chiral braid type.  However, the orbits in Fig.~(\ref{graph:ChiralPair}) are geometrically chiral and of the same braid type.  Thus, geometric chirality does not force topological chirality.  One complicating idea to be aware of involves chiral knots.  Given a knot in three dimensions, we can ask whether a parity transformation of space will preserve this knot or produce a new knot that is not isotopic to the first.  This is tantalizingly close to our notion of topological chirality, especially since our braids are closed, and are therefore knots (links).  However the two notions of chirality are not the same!  For example, torus knots are a canonical example of chiral knots.  On the other hand, each braid which is conjugate to $\lrp{\Delta^2}^k$ has a closure which is a torus knot, yet is non-chiral with respect to our topological chirality.\footnote{For this example of $\lrp{\Delta^2}^k$, we can explicitly show that it is not topologically chiral.  We have that $\lrp{\Delta^2}^k = \lrp{\lrp{\sigma_1\sigma_2\cdots\sigma_N}\lrp{\sigma_1\sigma_2\cdots\sigma_{N-1}}\cdots\lrp{\sigma_1\sigma_2}\lrp{\sigma_1}}^k$.  Using the transformation on this gives $TP\lrp{\lrp{\Delta^2}^k} = \lrp{\lrp{\sigma_N}\lrp{\sigma_{N-1}\sigma_N}\cdots\lrp{\sigma_1\sigma_2\cdots\sigma_{N}}}^k$.  Both of these braids describe $k$ full twists of the strands, and we can even say that $TP\lrp{\lrp{\Delta^2}^k} = \Delta\lrp{\Delta^2}^k\Delta^{-1}$.  Thus $TP\lrp{\lrp{\Delta^2}^k} \simeq \lrp{\Delta^2}^k$, and this braid is non-chiral in the topological sense.}  The connections between geometric and topological chirality are certainly ideas worth pursuing in the future.

\FloatBarrier
\section{Classification Ideas}
\FloatBarrier
	In the previous section we defined the broad binary categories of pseudo-Anosov or finite order, chiral or non-chiral, and permutation orbit or full periodic orbit, which comprise some of the fundamental differences between periodic orbits.  We continue to classify periodic orbits in this section, though now into categories with potentially more than two classes.  Since both ideas that comprise this section have been mentioned before in the context of individual periodic orbits, we will focus primarily on illustrating how these two concepts help to order the whole set of accumulated periodic orbits.  First is the connection between the angular momentum of relative equilibria and qualitative partitions of the level set submanifolds $\mymanL$, which was mentioned in section~(\ref{sec:IntMotSymmChaos}).  Second is the $T_2$ fine braid type classification, mentioned in section~(\ref{sec:TNbraidtree}), which will be particularly fruitful in distinguishing solutions with different qualitative properties.
\FloatBarrier
\subsection{Angular Momentum and Relative Equilibria}
\FloatBarrier
	Consider the topology of two submanifolds $\mymanL$, which correspond to two different constant values of the angular momentum, $L_1$ and $L_2$.  These two manifolds are topologically identical, and therefore have qualitatively similar periodic orbits, if there is no intermediate value of angular momentum, $L_1 \leq L^{\ast} \leq L_2$ for which the vector fields $\HamVec{H}$ and $\HamVec{L}$ are linearly dependent.  At this value, $L^{\ast}$, of the angular momentum, the dimension of the level set submanifold $\mymanLa$ is one less than the manifolds on either side, $L^{\ast}\pm\epsilon$.\footnote{These ideas are formally expressed in terms of Morse theory.  It is somewhat easier to consider the dual case, where we keep the angular momentum value constant and look at the one-parameter family of level set submanifolds, $\mymanH$, parameterized by the value of the Hamiltonian.  In this view we can mod out the rigid rotations generated by $\HamVec{L}$, resulting in a reduced phase space foliated by the submanifolds $\mymanH$.  In this context the Hamiltonian, $H$, is a Morse function, whose critical points in the reduced phase space, $\HamVec{H} = 0$, indicate where the topology of $\mymanH$ changes as a function of $H$.  These points happen at the relative equilibria, where the vector fields $\HamVec{H}$ and $\HamVec{L}$ are linearly dependent.}  Thus, these special values of the angular momentum break up the phase space into disjoint regions where the submanifolds $\mymanL$ are topologically the same (diffeomorphic to each other).  Fortunately for us, these special values of $L$ are easy to find, as they are the angular momentum values for which the periodic orbits are relative equilibria (rigidly rotating vortex configurations).  We have a complete listing of these orbits, see section~(\ref{sec:PtVM}), for both the $N=3$ and $N=4$ case.  In this section, we will investigate how these regions between relative equilibria dictate various qualitative properties of periodic orbits.
	
\FloatBarrier
\subsubsection{Case: $N=3$}
\begin{figure}[htb]
 \centering
\large
\resizebox{\linewidth}{!}{\includegraphics{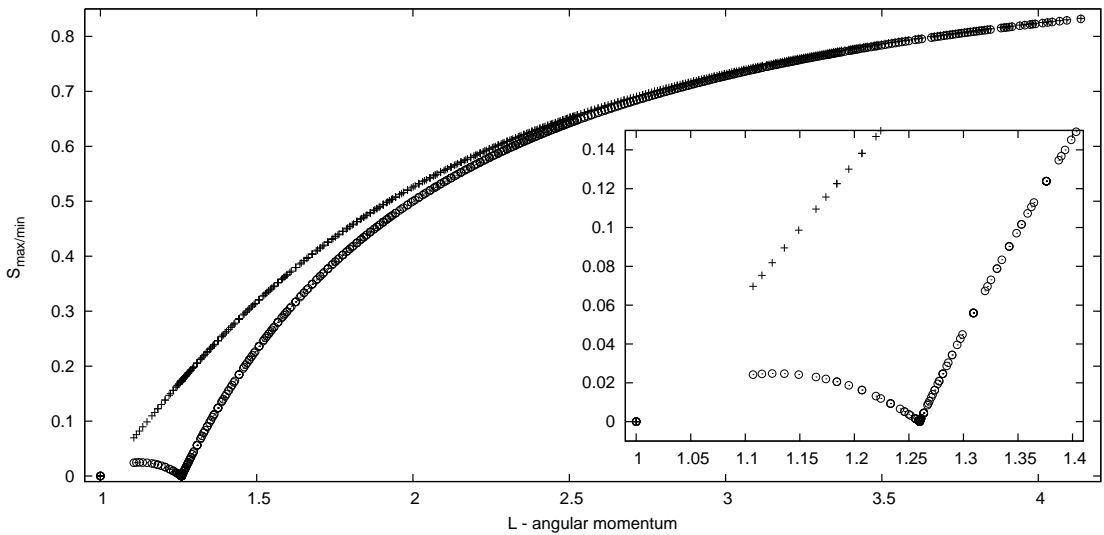}}
\normalsize
    \caption{Both $S_{min}$ and $S_{max}$ are plotted against the angular momentum for all $N=3$ periodic orbits.  Notice that the two minima occur exactly at the angular momentum values $L^{\ast}_1 = 1$ and $L^{\ast}_2 = 1.25992$, corresponding to the two relative equilibria.  This inset graph is simply a closer view of the area near the two relative equilibria.}
    \label{graph:N3LS}
\end{figure}

\FloatBarrier
	For three vortices, there are two relative equilibria: one where the vortices are fixed at the corners of an equilateral triangle, and one where the vortices are collinear.  The former has an angular momentum value of $L^{\ast}_1 = 1$ and the latter has $L^{\ast}_2 = 2^{\frac{1}{3}} = 1.25992$.  The best way to pick out these two periodic orbits from all the rest is to plot the maximum and minimum values of $S$, see Eq.~(\ref{eq:DiscreteLHprod}), along each orbit.  This quantity, $S$, measures the relative orientation of the vectors $\HamVec{H}$ and $\HamVec{L}$ for each point in time along an orbit.  It returns 0 if the two are completely antiparallel and 1 if they are perpendicular using the Euclidean metric.  Generic periodic orbits will have values of $S_{min}$ and $S_{max}$ which lie between these two extremes, while the relative equilibria have values of zero for both.  Figure~(\ref{graph:N3LS}) plots $S_{min}$ and $S_{max}$ against $L$ for all of the $N=3$ periodic orbits.
		
There are a few interesting things to note about this graph.  First of all, the relative equilibria, given by the zero values of $S_{min}$ and $S_{max}$, lie at the two predicted angular momentum values $L^{\ast}_1$ and $L^{\ast}_2$.  For all other periodic orbits, both $S_{min}$ and $S_{max}$ depend only on the value of angular momentum.  In particular it appears that the $S_{max}$ values are unaffected by the $L^{\ast}_2$ relative equilibrium, whereas the $S_{min}$ values near $L^{\ast}_2$ all trend to zero.  This indicates that near $L^{\ast}_2$ all periodic orbits have points at which the time flow and rotation flow directions come very close to being antiparallel.  Further from the relative equilibria, for larger angular momenta, both $S_{min}$ and $S_{max}$ increase monotonically.  However, they appear to be leveling off before reaching a value of 1, which indicates that the time flow and rotation flow directions are never exactly perpendicular, and certainly remain partially antiparallel.  

\begin{figure}[htb]
 \centering
\large
\resizebox{\linewidth}{!}{\includegraphics{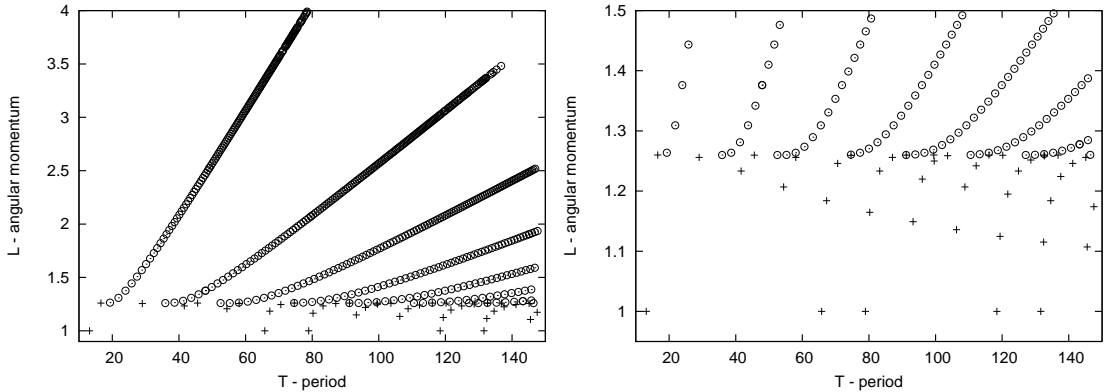}}
\normalsize
    \caption{Two views of the angular momentum vs. period for $N=3$ orbits.  The periodic orbits are marked differently based on whether they are in zone $Z_1$ or $Z_2$.  Notice the qualitative difference in the parameter space layout of the different families of orbits above and below $L^{\ast}_2$.}
    \label{graph:N3LTplot}
\end{figure}
Finally, these two relative equilibria angular momentum values, $L^{\ast}_1$ and $L^{\ast}_2$, divide the submanifolds $\mymanL$, and therefore periodic orbits, into three different classes.  Since there are no orbits in the region below $L^{\ast}_1$, we are left with two classes of periodic orbits in zones, $Z_1$ and $Z_2$, where $L^{\ast}_1 < Z_1 < L^{\ast}_2$ and $L^{\ast}_2 < Z_2$.  These two classes of orbits have large qualitative differences between them, which are related to the $T_2$ classification (more on this in subsequent subsections).  This difference can best be seen in a plot of the angular momentum against the period for all three-vortex orbits, Fig.~(\ref{graph:N3LTplot}).  Orbits in both zones cluster into discrete one-parameter families, however the families in $Z_2$ are much more densely populated.  Both classes of orbits lie on 2-toruses, though the angle variables in each case are quite different.  In zone $Z_1$, the two angles respectively represent the bulk rotation of the vortices about each other and the oscillations of the area of the triangle that they form.  In zone $Z_2$, the first angle represents the motion of the closer pair of vortices about each other, while the second angle coordinate represents the motion of this pair about the third vortex.  

\FloatBarrier
\subsubsection{Case: $N=4$}
\FloatBarrier

\begin{figure}[htbp]
 \centering
\large
\resizebox{\linewidth}{!}{\includegraphics{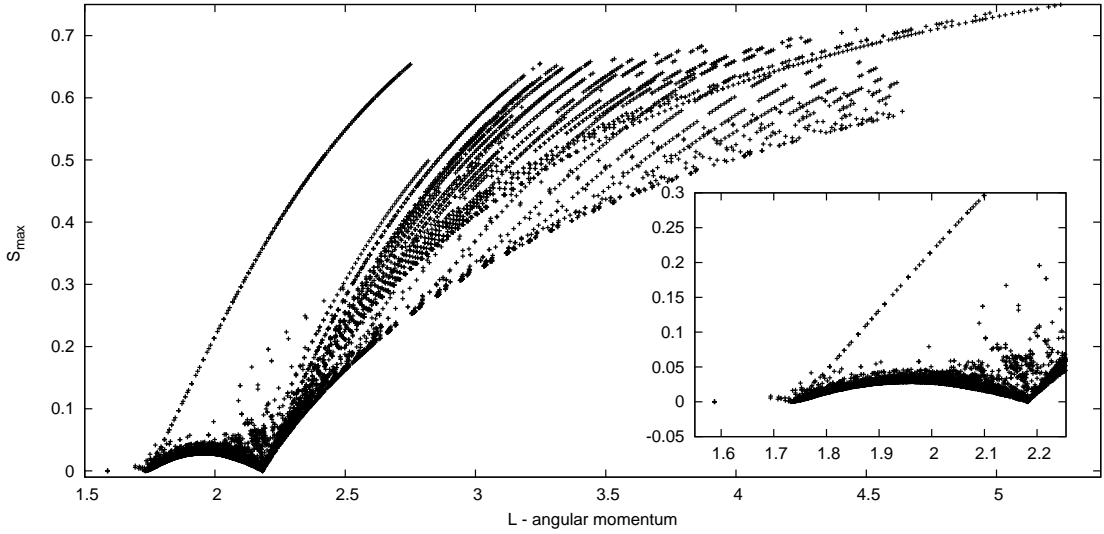}}
\normalsize
    \caption{$S_{min}$ is plotted against the angular momentum for all $N=4$ periodic orbits.  Notice that the three minima occur exactly at the angular momentum values $L^{\ast}_1 = 1.587$, $L^{\ast}_2 = 1.732$, and $L^{\ast}_3 = 2.182$ corresponding to the three relative equilibria.  The inset graph is a closer view of the area near the relative equilibria.}
    \label{graph:N4LS}
\end{figure}

When considering the case of four vortices, there are three relative equilibria to form the dividing lines between topologically different regions of phase space.  The first relative equilibrium is that of all four vortices on the corners of a square and occurs at an angular momentum value of $L^{\ast}_1 = 1.587$.  The second rigidly rotating solution is comprised of vortices at the corners and center of an equilateral triangle and has $L^{\ast}_2 = 1.732$.  The final relative equilibrium is the collinear solution, which shows up at $L^{\ast}_3 = 2.182$.  As before, we can easily see these three special solution in the plot of $S_{min}$ against the angular momentum, Fig.~{\ref{graph:N4LS}}, for all periodic orbits of four vortices.

This graph is certainly more complicated and has more structure than its analogous plot for $N=3$, however, we can still extract a few salient points from it.  First and most obvious is the dominating influence of the three relative equilibria, especially the two with larger angular momentum values.  Due to this influence, we can use these three relative equilibria to break up the phase space into three zones, $Z_1$, $Z_2$, and $Z_3$, with topologically different submanifolds $\mymanL$, in the angular momentum ranges $L^{\ast}_1 < Z_1 < L^{\ast}_2 < Z_2 < L^{\ast}_3 < Z_3$.   Most solutions which are reasonably close, in angular momentum, to these relative equilibria have at least one point on their orbit where the time flow and rigid rotation flow are aligned nearly antiparallel.  There is one obvious exception to this in the family of periodic orbits which originate near $L^{\ast}_2$ and have an $S_{min}$ value which keeps increasing with increasing angular momentum, and does not descend again in the area around $L^{\ast}_3$.  In particular, this suggests the interesting conclusion that the submanifolds $\mymanL$ have more than one path connected component.  Around $L^{\ast}_3$ the topology of $\mymanL$ does indeed change, though not all of the connected components participate in this change.  The previously mentioned family of periodic orbits appears to live on a component of $\mymanL$ which is separate from the component on which the rest of the solutions in the same range of $L$ exist.

Next, consider the $S_{min}$ value of the orbits in the $Z_2$ region.  While they are scattered around this plot somewhat randomly, there appears to be a smooth envelope constraining this $S_{min}$ value from below.  That this arching envelope is so smooth relative to the distribution of points above it suggests that it is a feature dictated by the topology of the submanifold $\mymanL$ in this zone.  A similar situation occurs in zone $Z_1$ of the three vortex case.  Finally, notice the qualitative nature of the distribution of points in Fig.~(\ref{graph:N4LS}) in the three different zones.  In $Z_3$, it appears that the orbits take discrete values of $L$ and $S_{min}$, and occur in separate families.  This will be born out later when we look at the braid types that are compatible with the topology of $\mymanL$ in this region.  In $Z_2$ the orbits take values of $L$ and $S_{min}$ seemingly from a continuum.  Here, the topology does not dictate the various geometric quantities which we can ascribe to periodic orbits nearly as stringently as is does in the $Z_3$ region.  There are many fewer orbits in the $Z_1$ region, and we will have less to say about their collective properties.

\begin{figure}[htb]
 \centering
\large
\resizebox{\linewidth}{!}{\includegraphics{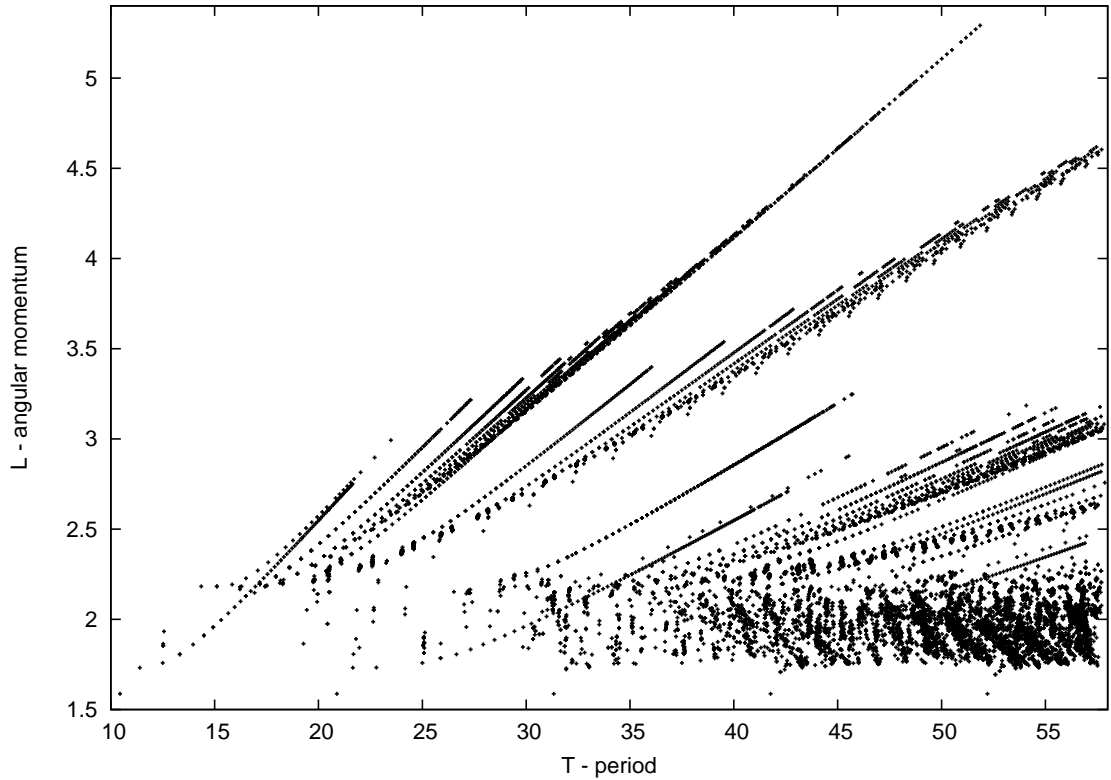}}
\normalsize
    \caption{A plot of the angular momentum vs. period for $N=4$ orbits.  This shows the $Z_3$ region of discrete families of orbits and the $Z_2$ region with continuous families of orbits.}
    \label{graph:N4LTplot}
\end{figure}
These three regions can also be seen in the angular momentum vs. period plot, Fig.~(\ref{graph:N4LTplot}), despite its relative complexity when compared to the analogous $N=3$ plot.  Above $L^{\ast}_3 = 2.182$, in zone $Z_3$, the periodic orbits form discrete families, in that the braid type determines the period, angular momentum, and other geometric quantities.  In the adjacent $Z_2$ region, below $L^{\ast}_3 = 2.182$ and above $L^{\ast}_2 = 1.732$, the periodic orbits form a continuum.  In both sections we see patterns which beg for explanation.  To disentangle these patterns it is very helpful to consider the different $T_2$ braid type classes of our periodic orbits.

\FloatBarrier
\subsection{Topological Classification}
\FloatBarrier

One of the defining features of each of our periodic orbits is its braid type.  Periodic orbits with the same braid type, and therefore conjugate braids, have very similar properties.  We can even ask whether there are multiple periodic orbits with the same braid type.  More loosely, braid types can be sorted into a number of different $T_2$ classes, mentioned in section~(\ref{sec:TNbraidtree}).  These classes encode information on both the most basic templates that pure braids can be formed with and whether they have pseudo-Anosov components or not.  In this section we show how the $T_2$ classification correlates with the distribution of periodic orbits in parameter space.
	
\FloatBarrier
\subsubsection{Case: $N=3$}
\FloatBarrier

For the case of three vortices, the $T_2$ classification provides a very definite division in the types of periodic orbit.  The first class, $\overline{3}_1$, corresponds to the finite order braids which are conjugate to some power of the full twist, $\lrp{\Delta^2}^k$.  Therefore, each braid type in this class is uniquely identified by this power, $k$, alone.  On the other hand, $\overline{3}_3$, (there are no $\overline{3}_2$ braid types) corresponds to finite order braids which are conjugate to a close pair of strands twisting about each other and a third strand twisting a different number of times about the close pair.  Braid types in this class are uniquely characterized by a pair of integers $\lrp{j,k}$, indexing the power of the full twist for the outer scale (close pair treated as one strand twisting about the remaining strand) and the inner scale (that of the close pair) respectively.  Interestingly, these two braid types completely partition the periodic orbits into the two zones $Z_1$ and $Z_2$ mentioned previously.  All $\overline{3}_1$ periodic orbits have angular momentum values less than or equal to $L^{\ast}_2$, while all $\overline{3}_3$ orbits have larger angular momentum values.  Indeed, in Fig.~(\ref{graph:N3LTplot}), the two types of orbits distinguished by different symbols are exactly these two types.

Consider the two regions of Fig.~(\ref{graph:N3LTplot}) more carefully.  The upper portion of the plot breaks down into a number of discrete families of orbits, each of which shares a common value of $j$.  Within each discrete family, the $k$ values increase as both the period and angular momentum increase.  Thus, every periodic orbit in this zone, $Z_2$, is uniquely labeled by the pair $\lrp{j,k}$.  This is the rare case where there is only one unique periodic orbit for every braid type in a given $T_2$ class.  On the other hand, the periodic orbits in the $Z_1$ zone, which have braid types in the class $\overline{3}_1$, do not have this property.  For example, there are multiple periodic orbits with braids conjugate to $\lrp{\Delta^2}^5$.  On the right side of Fig.~(\ref{graph:N3LTplot}), each family of periodic orbits (those that have braid types conjugate to a common power of the full twist) can be seen as a succession of points curving up and to the right.

These features can be connected back to the topology of the submanifolds $\mymanL$.  In the $\overline{3}_3$ case, $\mymanL$ is a 2-torus, and the angle coordinates correspond to the inner and outer rotations.  Thus, the two integers $\lrp{j,k}$ act as the winding numbers, counting the number of times the orbit winds around each direction of the torus before returning to the starting point.  For the $\overline{3}_1$ case, $\mymanL$ is also a 2-torus.  However, now there is only one integer that characterizes the braid type.  This single integer is not sufficient to characterize the winding numbers in both directions.  Indeed, only one of the angle variables corresponds to motions (rigid rotations) that we can distinguish from a topological perspective.  The other angle variable corresponds to the oscillations of the area enclosed by the vortex triangle, and is reflected in the geometric braids in a manner that is completely obscured by braid isotopy equivalence.
		
\FloatBarrier
\subsubsection{Case: $N=4$}	
\begin{figure}[htbp]
 \centering
\large
\resizebox{\linewidth}{!}{\includegraphics{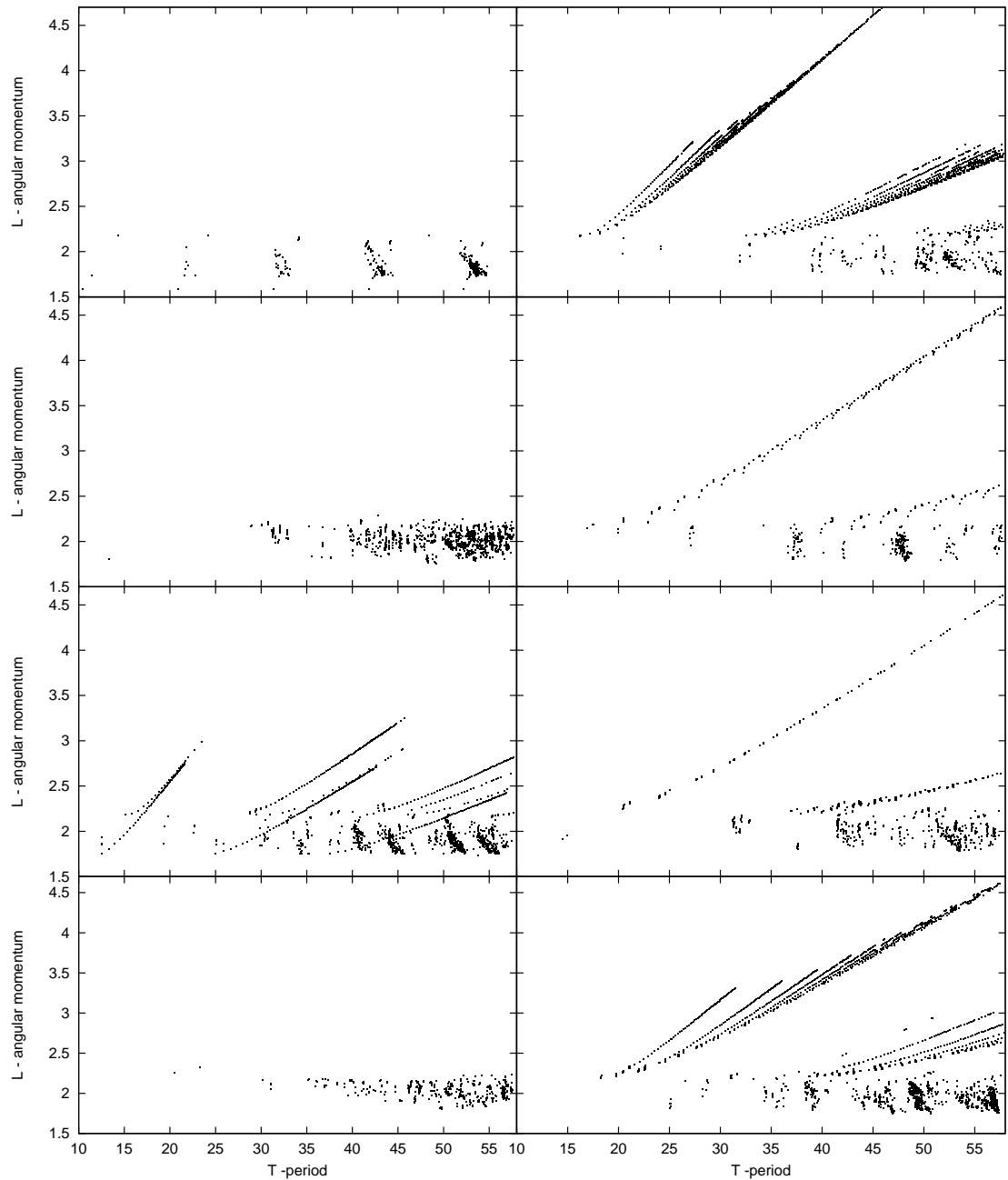}}
\normalsize
    \caption{Eight $L$ vs. $T$ plots corresponding to the eight $T_2$ classes for four-stranded pure braids.  On the left from top to bottom: $\overline{4}_1 = \lrp{4_1,fo}$, $\overline{4}_2 = \lrp{4_1,pA}$, $\overline{4}_3 = \lrp{4_2,fo}$, and $\overline{4}_4 = \lrp{4_2,pA}$.  On the right from top to bottom: $\overline{4}_5 = \lrp{4_3,fo}$, $\overline{4}_6 = \lrp{4_4,fo}$, $\overline{4}_7 = \lrp{4_4,pA}$, and $\overline{4}_8 = \lrp{4_5,fo}$.}
    \label{graph:N4LTX8}
\end{figure}

\FloatBarrier
For the case of four vortices, the $T_2$ classification is particularly helpful for making sense of the patterns in the $L$ vs. $T$ plot of Fig.~(\ref{graph:N4LTplot}).  In this case there are eight $T_2$ classes $\lrbk{\overline{4}_1,\cdots,\overline{4}_8}$.  For a picture of the representative reduction curves for these isotopy classes, refer to table~(\ref{table:RCclassification2}).  Similarly to the three vortex case, we can correlate the $T_2$ classes with the regions delineated by relative equilibria.  All $T_2$ classes have periodic orbits in the $Z_2$ zone.  However, only $\overline{4}_1$ orbits exist in zone $Z_1$, while $\lrp{\overline{4}_3,\overline{4}_5,\overline{4}_6,\overline{4}_7,\overline{4}_8}$ orbits lie in both zone $Z_2$ and $Z_3$.  The two remaining $T_2$ classes, $\overline{4}_2$ and $\overline{4}_4$, have orbits exclusively in $Z_2$.  All of these observations can be seen in Fig.~(\ref{graph:N4LTX8}), which plots $L$ vs. $T$ for each of the eight $T_2$ classes.

There are a number of ideas relating to Fig.~(\ref{graph:N4LTX8}), which we should emphasize.  First, notice again the qualitative differences in the families of periodic orbits in the two zones, $Z_2$ and $Z_3$.  In $Z_3$, the obits occur in discrete families, while in $Z_2$ they appear to form a continuum.  As we mentioned before, there is one exception to this general behavior in the discrete families that can be seen in zone $Z_2$ of the $\overline{4}_3$ plot.  This $T_2$ class is finite order, and consists of one pair of vortices which rotate about each other while the remaining two plus the pair considered as a single strand also rotate a different number of times about each other.  In this view there are two integers, $\lrp{j,k}$, each powers of the full twist, which describe the outer and inner rotations respectively.  Each of these discrete families which start at $L^{\ast}_2$ have a different value of $j$, the number of outer turns.  Within each discrete family, the different periodic orbits are distinguished by the number of inner turns or $k$ value.  Again, the coexistence of these discrete families with the continuum of solutions in zone $Z_2$ indicates that there are multiple path-connected components to the submanifold $\mymanL$ in this region.

Next, notice that the distribution of orbits within the $Z_2$ zone has a periodicity to it in the period direction for each of the $T_2$ classes with only finite order components.  The $T_2$ classes with pA components do not exhibit this same sort of periodicity in the $Z_2$ zone.  Another pattern, which is somewhat more explicable, is the distribution of orbits in the $Z_3$ zone for the last three $T_2$ classes, $\lrp{\overline{4}_6,\overline{4}_7,\overline{4}_8}$.  These three classes are related in that they are composed of the three stranded $T_2$ classes injected into the single two stranded $T_2$ class.  Indeed the $Z_3$ zone of their $L$ vs. $T$ plot can be constructed from multiple copies of the same plot for three vortices, Fig.~(\ref{graph:N3LTplot}).  It is interesting to note that there are pA $\overline{4}_7$ orbits despite the nonexistence of three vortex pA orbits.  In the $Z_3$ zone, these orbits occur right at the intersection of the $\overline{4}_6$ and $\overline{4}_8$ orbits.  This space is occupied in the $N=3$ case by multiples of the collinear relative equilibria, which is unsurprisingly the most unstable $N=3$ orbit (largest maximum Floquet exponent).  It would not be surprising if the $\overline{4}_7$ discrete pA orbits were closely related to the $N=3$ collinear relative equilibria.  A similar situation occurs with the $\overline{4}_4$ periodic orbits.  These are pA, despite the fact that, once again, it has a three stranded component which is pseudo-Anosov.  In this case there is a pair of close vortices which, when treated as a single vortex, form a pA braid in conjunction with the remaining two vortices.  In both of these cases a three stranded component of a four stranded braid is pseudo-Anosov, despite there not being any three stranded pA braids on their own.  The addition of the fourth vortex makes these two interesting cases possible.

Though relatively simple, the most important implication that should be taken away from Fig.~(\ref{graph:N4LTX8}) is that the topology, as seen in the eight $T_2$ classes, does a very good job of distinguishing qualitatively different periodic orbits.


\chapter{Future Work and Conclusions}
\label{ch:conc}

\section{Conclusions}

With over 13,000 orbits, we certainly achieved our ultimate goal of finding and classifying periodic orbits for the motion of three and four identical point vortices.  This was made possible by an understanding of the geometry and topology of phase space, and the algorithms it engendered.  We used the symplectic, Hamiltonian nature of the dynamics to construct an efficient Lie transform perturbation theory for better integrating point vortex motion~\cite{Smith:2011ey}.  This increased the efficiency and accuracy of generic numerical integrators when dealing with the point vortex equations of motion.  It also highlighted one of the main themes of this work: that groups of vortices can often be approximated, either analytically or qualitatively, by a single vortex.  Next we used the conserved integrals of motion to improve upon a reduction algorithm~\cite{Boghosian:2011ec} which takes arbitrary loops in phase space and reduces them to actual periodic orbits.  The reduction algorithm minimizes a functional on the space of all loops, while the integrals of motion indicate which descent directions should be restricted.  These ideas, coupled with a method for generating close return orbits, effectively extract periodic orbits from a representative region of phase space.  Finally, we use the braid type of each periodic orbit to sort, store, and check for uniqueness in orbits.  The TN braid tree conjugacy invariant, and the two related classification schemes, $T_1$ and $T_2$, are the main tools which we have developed to approximate the braid type.  These classification schemes decompose the periodic motion of vortices into various nested sets of vortices, the constituents of which always remain close to one another.  As mentioned before, this qualitative partitioning echoes the analytical equivalence between the motions of groups of vortices and that of a single vortex in our Lie transform methods.

These topological classification ideas are perhaps most fruitful when used to consider the qualitative differences between various periodic orbits.  In the case of three vortices, the $T_2$ classification cleanly divides the set of periodic orbits into two groups, which differ in their level-set submanifolds, $\mymanL$, of phase space.  Both sets have $\mymanL$ submanifolds which are topologically equivalent to a torus, however the angle variables, which quantify the winding in each torus direction, correspond to different bulk motions in each case.  Indeed, this qualitative division is also seen in the phase space regions defined by the relative equilibria.  There are two of these rigidly rotating periodic orbits, and the second, collinear solution has an angular momentum value which partitions phase space into exactly these two qualitative groups.

In the case of four vortices, the $T_2$ classification is invaluable.  The families of orbits seen in a plot of angular momentum vs. period significantly overlap, which complicates an analysis of the patterns that they form.  Fortunately, these individual patterns can be seen clearly when we restrict our attention to each of the eight $T_2$ classes.  The most immediate common pattern to emerge is the existence of a region with discrete families of orbits and a region with continuous families of orbits.  Orbits of the discrete families have well defined periods and angular momentum values for each braid type, whereas continuous-family orbits draw from a continuum of periods and angular momenta for each braid type.  These two regions are bounded by the angular momenta associated with the three rigidly rotating solutions.  Interestingly, there is one discrete family of orbits which exists in the otherwise continuous region.  This provides evidence that the submanifold, $\mymanL$, is composed of multiple path-connected components, some of which support qualitatively different types of orbits, and do not participate in every topological transition associated with the presence of relative equilibria.

In addition to the topological classification due to braid type and that due to relative-equilibria-delineated phase space regions, we also encountered a number of other ways to group periodic orbits into different bins.  Orbits can be composed of permutation orbits, they can have pseudo-Anosov components or just finite order, and they can be chiral or non-chiral.  Permutation orbits occur when a vortex configuration repeats after a non-trivial permutation of vortex indices.  Relative equilibria are prime examples of permutation orbits, which in general have a high degree of symmetry.  A most basic distinction between braids is whether they are composed of positive generators, negative generators, or some combination of the two in the most general case.  We have found that all braids generated by the dynamics of identical point vortices can be composed exclusively of positive braids.  This is quite surprising, and currently lacks a rigorous explanation.  

From a different point of view, these braid types can be composed of finite order component which are simple to describe, or they have at least one pseudo-Anosov component, in which case they can be very complicated.  The first noteworthy result regarding this division is that even though pA braids are allowed in the case of three vortices, we find no evidence for them.  We do, however, find many different pseudo-Anosov braids for four vortices.  Interestingly, we can correlate the existence of these pA component braids with the absence of stability in the corresponding periodic orbits.  It is certainly expected that the existence of pA braids would force some complexity in the Lagrangian chaos of advected fluid particles in the plane, however this observation implies that the pA braids also force a certain modicum of Eulerian chaos through the greater-than-zero Floquet exponents which determine the instability of periodic orbits.  This intriguing result certainly needs further investigation.  

Finally, periodic orbits can be either chiral or non-chiral.  This is a geometric notion of chirality which determines whether an orbit can be mapped back onto itself by way of rigid body rotations and time translations after undergoing a parity and time inversion.  As with pA braid components, three vortices alone are not sufficient to allow for geometrically chiral braids.  However, four vortices do exhibit both chiral and non-chiral orbits.  Most astonishing is perhaps the existence of a definite gap in the chiral distance, which indicates that topology might play some role in the existence of chiral orbits.  Indeed, there is one particular one-parameter family of chiral orbits which have a definite relationship between their angular momentum and their chiral distances.  An analysis of this indicates that these orbits strongly imply the presence of multiple path-connected components to the level-set sub manifold, $\mymanL$.  Non-chiral orbits are also interesting due to the degree of symmetry that they possess: each non-chiral orbits has at least one axis of bilateral symmetry.

We have investigated many different classification ideas, from geometric chirality, existence of pA components, and permutation orbits, to the TN braid type, $T_1$ \& $T_2$ schemes, and the phase space regions delineated by the relative equilibria.  Each concept has much to tell us about the structure of phase space and the existence of chaotic motion on its own, however, the common gestalt which they form is the idea that the topological attributes of periodic orbits inform, and in many cases force, their geometric attributes.

\section{Future Work}

The periodic orbits we have found are solutions to the conservative point vortex model.  However, as we showed in chapter~\ref{ch:phys}, there is a natural extension to the finite temperature case of driven dissipative vortex motion.  The equations of motion, Eq.~(\ref{eq:FiniteTemperatureVortexMotion}), are controlled by two parameters, both of which depend on temperature.  The movement of vortices depends exclusively on these parameters and the vortex positions when the underlying normal fluid is rotating rigidly.  Thus, the phase space has the same dimension as that of conservative point vortices.  It is natural to ask what happens to the periodic orbits when we adiabatically increase the temperature above absolute zero.  Which periodic orbits and topological classes disappear and at what temperatures?  For larger temperatures, we know that at least one of the relative equilibria is stable.  This might suggest that other orbits also remain solutions.  On the other hand, this divide might mirror the differences between fluid motion at high Reynolds numbers and purely conservative Eulerian fluid motion.  Despite the Euler equation being the limit of the Navier-Stokes equation for large Reynolds numbers, there is a qualitative difference between the fluid motions that result.  For us this might indicate that any finite amount of dissipation would fundamentally change the types of periodic solutions.  Regardless, this would be a natural extension of the current ideas to investigate.

Some of the motivation for considering periodic orbits came from the so called thermodynamic formalism of dynamic systems theory~\cite{Pollicott:2002vb}.  Here one can, roughly, replace time averages over a long trajectory with weighted averages over the set of periodic orbits.  This is a very successful theory for hyperbolic systems, systems where each direction in phase space is either expanding or contracting.  For our Hamiltonian system, there are also marginal directions which complicate the picture.  It would be fruitful to consider the extensions of this theory which help account for marginal directions.  In particular, this could help one calculate quantities such as the two point velocity correlation function, which is experimentally applicable measures of fluid turbulence.

This previous idea would be useful primarily in the limit of large numbers of vortices.  In this regime, the statistical ideas of Onsager, which led to the concept of negative temperature states~\cite{Eyink:1993vw}\cite{Eyink:2006jn}, would be applicable.  It would, therefore, be natural to find periodic orbits for larger numbers of vortices.  However, the size of phase space and the number of periodic orbits would quickly proliferate beyond our ability to find and store them.  Alternatively, we could focus on the search for relative equilibria exclusively.  Since these solutions help to topologically partition phase space, we might be able to extract general information on classes of large $N$ periodic orbits from this smaller set.

We would also like to know the combinatorial growth rate of the number of $T_1$ topological classes as the number of vortices increases.  This classification deals with the number of distinct ways in which the periodic motion of vortices can be partitioned.  The division of vorticity into different, non-interacting, regions is reminiscent of the concept of Lagrangian coherent structures in more realistic fluid dynamics.  It might be possible to apply the $T_1$ and $T_2$ classifications to areas of fluid with coherent structures.  In particular, this would give one a good idea of the length scales over which fluid mixing is occurring.

Another topological concept which needs some further work is the TN braid tree.  As defined, it is certainly a topological invariant of the braid conjugacy classes.  However it is not yet a total invariant, and does not completely solve the conjugacy problem.  We could easily implement the Garside algorithm, or some modern variant of it~\cite{Birman:1998tc}, to solve the conjugacy problem on pseudo-Anosov braids which have similar expansion factors and exponent sums.  This, along with a few simple ordering ideas, would elevate the TN braid tree to the level of a full conjugacy invariant for pure braids.

With this ability to uniquely determine conjugacy classes, we could immediately tackle a concrete question concerning chiral periodic orbits.  Our definition of chirality was a geometric concept, dependent on a space parity transformation and a time reversal.  We have also defined a topological version of this transformation, Eq.~(\ref{eq:PTbraidTrans}).  We would certainly like to know which periodic orbits are topologically chiral in addition to being geometrically chiral.  We could also investigate the connections between this notion of topological chirality on closed pure braids and the related, though different, concept of chiral knots.

Another topological notion to expand upon is braid ordering.  We used the Dehornoy~\cite{bib:Dehornoy} braid ordering to effectively solve the braid word problem.  For braids formed of three vortices in the $3_2$ class, this braid order has an additional intriguing utility: the order of braids closely mirrors the period, $T$, order of their corresponding periodic orbits.  While this idea does not easily extend to the case of four vortices, there are other types of braid orderings, many of which are based on familiar ideas of Thurston~\cite{Short:2000wi}.  Perhaps there exists an ordering which respects conjugacy classes and reflects the ordering of periodic orbits by their period.

Finally, we can consider the state of our Lie transform methods of section~\ref{sec:LieTrans}.  Since the motion of three vortices is integrable, we could consider a related Lie transform which treats three close vortices as a single vortex of larger circulation.  Indeed, we could analogously consider integrable groups of four or more vortices as well.  In general, we could have a Lie transform template for each $T_2$ class.  This idea is admittedly difficult, as even the simplest case of three integrable vortices would require the use of Jacobi elliptic functions.




\ssp 

\bibliography{thesisarxiv}

\end{document}